%% file: main.tex
\documentclass[bpam]{ipart}

\RequirePackage{hyperref}

\startlocaldefs

\endlocaldefs

\pubyear{2022}
\volume{0}
\issue{0}
\firstpage{1}
\lastpage{1}
\usepackage{soul}
\usepackage{tikz}
\usepackage{pgfplots}
\usetikzlibrary{calc}
\usepackage[section]{placeins}
\makeatletter
\AtBeginDocument{%
  \expandafter\renewcommand\expandafter\subsection\expandafter{%
    \expandafter\@fb@secFB\subsection
  }%
}
\makeatother
\usepackage[T1]{fontenc} 
\definecolor{darkgreen}{rgb}{0.0, 0.5, 0.0}
\usepackage{slashed}

\usepackage{url}

\usepackage{amssymb}  
\usepackage{float} 
\usepackage{xcolor} 
\usepackage{graphicx} 
\usepackage[normalem]{ulem}
\usepackage{bbm}
\usepackage{dsfont}
\usepackage[mathscr]{euscript}

\usepackage{cancel}
\input{LambdaMacros}

\input{WanMacrosKeep.tex}

\renewcommand{\hBo}{\blue{h}}

\renewcommand{\red}[1]{#1}
\renewcommand{\redc}{}
\renewcommand{\purple}[1]{#1}
\renewcommand{\blue}[1]{{#1}}
\renewcommand{\ptcheck}[1]{}
\begin{document}

\begin{frontmatter}

\title[Characteristic Gluing with $ \Lambda $ 1. Linearised EE on 4D spacetimes]{\boldmath Characteristic Gluing with $ \Lambda $ \\
 1. Linearised Einstein equations on four-dimensional spacetimes}

\begin{aug}

\author{\fnms{Piotr T.} \snm{Chru\'sciel}\ead[label=e1]{piotr.chrusciel@univie.ac.at}}
\address{University of Vienna, Faculty of Physics
  \\Boltzmanngasse 5, A 1090 Vienna, Austria\\\printead{e1}}
\and
\author{\fnms{Wan} \snm{Cong}%
       \ead[label=e3]{wan.cong@univie.ac.at}}
\address{University of Vienna, Faculty of Physics
  \\Boltzmanngasse 5, A 1090 Vienna\\\printead{e3}}
\end{aug}

\begin{abstract}
We establish a gluing theorem for linearised vacuum gravitational fields in Bondi gauge on  a class of characteristic hypersurfaces in  static vacuum four-dimensional backgrounds with cosmological constant $ \Lambda  \in \R$ and arbitrary topology of the compact cross-sections of
the null hypersurface. This generalises and complements, in the linearised case, the pioneering analysis of Aretakis, Czimek and Rodnianski, carried-out on light-cones in Minkowski spacetime.
\end{abstract}



\tableofcontents

\end{frontmatter}

\input{Introduction}

\input{0MatchingIn4D}

\input{Strategy4XI22mv.tex}

\input{DetailsOfGluing4XI22.tex}

\input{EndInduction}

\input{elluhab}
\input{LinearIsomorphism}

\input{PureChargeField}
\appendix

\input{Kappas}

\input{recursion}
\input{recursionWithAlpha}
\section{Operators on $\secN$}
    \label{App12XI22.1}

The aim of this appendix is to analyse the mapping properties of several operators acting on tensor fields defined on a compact orientable two-dimensional manifold $(\dmanif\equiv \secN,\zgamma)$ with constant Gauss curvature $\myGauss\in\{0,\pm1\}$.

\input{TensorHarmonics}

\input{CKV2d.tex}
\input{LandHatL}
\input{operators}

\input{PandLikes}

\input{TraceEqn.tex}

\input{lineariseTrace}
\input{huAcharges.tex}

\input{WanMiracle}
\input{factorisation2}

\newpage
\bibliographystyle{JHEP}
\bibliography{4dNullGluing-minimal
}



%
%

\end{document}

%% file: LambdaMacros.tex
\newcommand{\mrL}{\blue{\mathring{L}}}
\newcommand{\kerL}{\blue{[\Ker \mrL ]}}
\newcommand{\kerLp}{\blue{[(\Ker \mrL)^\perp ]}}
\newcommand{\kerLrp}{\blue{ \Ker \kchi{p}  }}
\newcommand{\kerLrpp}{\blue{(\kerLrp)^\perp}}
\newcommand{\imLrp}{\blue{ \im \kchi{p}  }}
\newcommand{\imLrpp}{\blue{(\imLrp)^\perp}}

\newcommand{\kgamma}{{\color{blue}{k_\gamma}}}
\newcommand{\kg}{\kgamma}

\newcommand{\Ck}{\blue{C^k_u\, C^\infty_{(r,x^A)}}}
\newcommand{\Ctwo}{\blue{C^2_u\, C^\infty_{(r,x^A)}}}

\newcommand{\genus}{{\blue{\mathfrak{g}}}}

\newcommand{\zmu}{\blue{\mathring{\mu}}}
\newcommand{\zlambda}{\blue{\mathring{\lambda}}}

\newcommand{\TTm}{{\blue{\mathring m}}}

\newcommand{\pref}[1]{\ref{#1}, p.~\pageref{#1}}

{\catcode `\@=11 \global\let\AddToReset=\@addtoreset}
\AddToReset{figure}{section}
\renewcommand{\thefigure}{\thesection.\arabic{figure}}

{\catcode `\@=11 \global\let\AddToReset=\@addtoreset}
\AddToReset{table}{section}
\renewcommand{\thetable}{\thesection.\arabic{table}}

\newcommand{\peqref}[1]{\eqref{#1}, p.~\pageref{#1}}

\newcommand{\harm}{{\blue{\text{H}}}}
\newcommand{\Harm}{\harm}

\DeclareMathOperator{\coker}{\blue{coker}}

\newcommand{\mym}{{\blue{m}}}
\newcommand{\myGauss}{{\blue{\twoscsign}}}
\newcommand{\myalpha}{{\blue{\alpha}}}
\newcommand{\myiota}{{\blue{\iota}}}

\newcommand{\Done}{{\purple{L_1 }}}
\newcommand{\Dtwo}{{\purple{L_2}}}

 \newcommand{\ddim}{{\blue{2}}}
 \newcommand{\dmanif}{\blue{{}^\ddim M}}
 \newcommand{\dmetric}{{\blue{\mathring{\gamma}}}}
 \newcommand{\dnabla}{\blue{\mathring{D}}}
 \newcommand{\dDelta}{\blue{\Delta_{\dmetric}}}
 \newcommand{\dR}{\blue{\mathring{R}}}

 \newcommand{\CKV}{{\blue{\text{CKV}}}}
\newcommand{\CKVp}{{\blue{\text{CKV}^\perp}}}
 \newcommand{\KV}{{\blue{\text{KV}}}}

\newcommand{\TSzlap}{\blue{\Delta_{\ringh}}}
\newcommand{\zDelta}{\TSzlap}

\newcommand{\TTt}{{\blue{\text{TT}}}}
\newcommand{\TTtp}{{\blue{\text{TT}^\perp}}}

\newcommand{\zdivtwo}{\blue{\operatorname{\mathring{\text div}_{(2)}}\,}}
\newcommand{\zDivtwo}{\zdivtwo}
\newcommand{\zdivone}{\blue{\operatorname{\mathring{\text div}_{(1)}}\,}}
\newcommand{\zDivone}{\zdivone}

\newcommand{\Lop}{\operatorname{\blue{L}}}
\newcommand{\hLop}{\widehat{\operatorname{\blue{ L}}}}

\newcommand{\TS}{\mathop{\mathrm{TS}}}

\newcommand{\guu}{\blue{g_{uu}}}
\newcommand{\twoscsign}{{\blue{\varepsilon}}}

\newcommand{\interph}{\blue{v}}

\newcommand{\overadd}[2]
{\overset{ (#1) }{#2}}

\newcommand{\psiP}{\blue{\psi}}

\newcommand{\tkfbd}{\blue{\mathrm{b.d.}|_{r_1}}}
\newcommand{\tk}{\tkfbd}

\newcommand{\secN}{\blue{\mathbf{S}}}

\newcommand{\Hgamma}{\blue{H_{k_\gamma}(\secN)}}

\newcommand{\wh}{\blue{w}}

\newcommand{\red}[1]{{\color{red} #1}}

\newcommand{\barh}{\blue{ h }}
\newcommand{\hBo}{\blue{ h^{\mathrm{Bo}}}}

\newcommand{\R}{\mathbb{R}}

\newcommand{\purple}[1]{{\purplec#1}}

\newcommand{\Ichi}{{\mu}}
\newcommand{\Ipsi}{{\lambda}}

\newcommand{\zhTB}{\blue{\check h}}

\newcommand{\zhTBW}{\blue{\gamma}}
\newcommand{\zzhTBW}{\blue{\mathring{\zhTBW}}}

\newcommand{\mcM}{{\mycal M}}
\newcommand{\mcN}{{\mycal N}}

\newcommand{\sectionofScri}%
{{ \,\,\,\,\mathring{\!\!\!\!\mcN}}}

 \newcommand{\hyp}{{\mycal S}}
\newcommand{\mcL}{{\mycal L}}

\newcommand{\ringh}{{  \zzhTBW }}%

\newcommand{\redc}{\color{red}}

\newcommand{\redca}[1]{\color{teal}\ptcr{change or addition or rewording on #1}}

\newcommand{\zR}{\blue{\mathring{R}}}

\newcommand{\rout}[1]{\red{\xout{#1}}}

\newcommand{\eeal}[1]{\label{#1}\end{eqnarray}}

\newtheorem{theorem}{\sc  Theorem\rm}[section]

\newtheorem{Theorem}[theorem]{\sc  Theorem\rm}

\newtheorem{lemma}[theorem]{\sc Lemma\rm}
\newtheorem{Lemma}[theorem]{\sc Lemma\rm}

\newtheorem{Proposition}[theorem]{\sc Proposition\rm}

\DeclareFontFamily{OT1}{rsfs}{}
\DeclareFontShape{OT1}{rsfs}{m}{n}{ <-7> rsfs5 <7-10> rsfs7 <10-> rsfs10}{}
\DeclareMathAlphabet{\mycal}{OT1}{rsfs}{m}{n}

\newcommand{\zgamma}{\blue{\mathring \gamma}}

\newcommand{\blue}[1]{{\color{blue} #1}}

\definecolor{applegreen}{rgb}{0.55, 0.71, 0.0}
\definecolor{armygreen}{rgb}{0.29, 0.33, 0.13}

\definecolor{caribbeangreen}{rgb}{0.0, 0.8, 0.6}

\newcounter{mnotecount}[section]

\renewcommand{\themnotecount}{\thesection.\arabic{mnotecount}}

\newcommand{\mnotex}[1]
{\protect{\stepcounter{mnotecount}}$^{\mbox{\footnotesize
$
\bullet$\themnotecount}}$ \marginpar{
\raggedright\tiny\em
$\!\!\!\!\!\!\,\bullet$\themnotecount: #1} }

\newcommand{\ptcr}[1]{\mnotex{{\bf ptc:} {\color{red} #1}}}

\newcommand{\bel}[1]{\begin{equation}\label{#1}}
\newcommand{\bea}{\begin{eqnarray}}
\newcommand{\bean}{\begin{eqnarray}\nonumber}
\newcommand{\beal}[1]{\begin{eqnarray}\label{#1}}
\newcommand{\eea}{\end{eqnarray}}

\newcommand{\nn}{\nonumber}

\newcommand{\qedskip}{\hfill $\Box$\medskip}
\newcommand{\be}{\begin{equation}}
\newcommand{\eeq}{\end{equation}}
\newcommand{\ee}{\end{equation}}
\newcommand{\beqa}{\begin{eqnarray}}
\newcommand{\eeqa}{\end{eqnarray}}
\newcommand{\beqan}{\begin{eqnarray*}}
\newcommand{\eeqan}{\end{eqnarray*}}
\newcommand{\ba}{\begin{array}}
\newcommand{\ea}{\end{array}}
\newcommand{\dt}{\mathfrak{d}}

\newcommand{\ptcheck}[1]{{\color{darkgreen}\mnotex{ptc : checked on #1}}}

\input{JHmacros}

\newcommand{\spaceD}{{ D}}
\newcommand{\zspaceD}{ {\mathring D}}

\newcommand{\mcE}{\mycal E}

\newcommand{\nobarg}{\blue{g}}

\newcommand{\tmcN}{\blue{\,\,\,\,\,\widetilde{\!\!\!\!\!\mcN}}}

\newcommand{\Lie}{{\mathcal{L}}}

\newcommand{\TSxip}{\blue{\zeta}}
\newcommand{\TSxi}{\blue{\xi}}
\newcommand{\TSr}{\blue{r}}
\newcommand{\TSu}{\blue{u}}

\newcommand{\TSoLie}{\blue{\mathring \Lie}}



%% file: JHmacros.tex
\def\d{\partial}

\def\dx{\text{d}\hspace{-0.06em}x}

\def\beq{\begin{eqnarray}}
\def\eeq{\end{eqnarray}}

\def\a{\alpha}
\def\b{\beta}

\def\d{\delta}

\def\x{\xi}
\def\z{\zeta}
\def\be{\begin{equation}}
\def\ee{\end{equation}}
\def\bea{\begin{eqnarray}}
\def\eea{\end{eqnarray}}

%% file: WanMacrosKeep.tex
\newcommand{\T}{\mathbb{T}}
\newcommand{\N}{\mathbb{N}}
\newcommand{\sm}{d\mu_{\zzhTBW}}
\newcommand{\ip}[2]{\langle #1, #2\rangle }

\newcommand{\Ker}{\ker}

\newcommand{\tdu}{u}
\newcommand{\tdr}{r}
\newcommand{\tdA}{A}
\newcommand{\tdB}{B}

\newcommand{\hkappafour}{ \hat{\kappa}_4}
\newcommand{\hkappafive}{ \hat{\kappa}_5}

\newcommand{\hkappaone}{ \hat{\kappa}_1}
\newcommand{\hkappasix}{ \hat{\kappa}_6}
\newcommand{\hkappa}{\hat{\kappa}}
\newcommand{\opsi}{\purple{\psi}}
\newcommand{\ochi}{\purple{\chi}}

\newcommand{\kphi}[1]{\blue{\overset{[#1]}{\hat\varphi}}{}}

\newcommand{\vphi}[1]{\blue{\overset{[#1]}{\varphi}}}
\newcommand{\kxi}[1]{\blue{\overset{[#1]}{\xi}}}
\newcommand{\kq}[1]{\blue{\overset{[#1]}{q}}{}}
\newcommand{\vp}[1]{#1^{[\TTt^\perp]}}

\newcommand{\okxi}[1]{\blue{\overset{(#1)}{\blue{\xi}}}{}}
\renewcommand{\kxi}[1]{\okxi{#1}}

\newcommand{\kchi}[1]{\blue{\overset{(#1)}{\blue{\chi}}}{}}

\newcommand{\kzeta}[1]{\blue{\overset{(#1)}{\blue{\zeta}}}}

\newcommand{\xiB}[1]{\blue{\overset{(#1)}{\blue{\xi}}_B}}
\newcommand{\xiA}[1]{\blue{\overset{(#1)}{\blue{\xi}}_A}}
\newcommand{\kQ}[2]{\blue{\overset{[#1]}{Q}_{#2}}{}}
\newcommand{\kC}[1]{\blue{\overset{[#1]}{\blue{C}}}}
\newcommand{\im}{\text{im}}
\newcommand{\qt}{\mathring s}

%% file: Introduction.tex
\section{Introduction}

In their  pioneering work~\cite{ACR1,ACR2,ACR3}, Aretakis, Czimek and Rodnianski presented a gluing construction, along a null hypersurface, of characteristic Cauchy data for the vacuum Einstein equations, for a class of asymptotically Minkowskian data. We wish to generalise their construction to null hypersurfaces with non-spherical sections, and to allow for a cosmological constant, in spacetimes of dimension four or  higher.

As a first step towards this, in this paper we consider four-dimensional vacuum Einstein equations, with a cosmological constant $\Lambda \in \R$, linearised at  Birmingham-Kottler metrics.
Recall, now, that the analysis in~\cite{ACR1,ACR2,ACR3} is based on the Christodoulou-Klainerman version of the Newman-Penrose formalism.
Recall also that there exists a well-studied Bondi-type parameterisation of the metric (see~\cite{MaedlerWinicour} and references therein), which generalises readily to any dimensions. So it appears of interest to analyse the problem in Bondi coordinates, to be able to exploit results which have already been established in the Bondi setting, with an eye-out both for possible advantages of the alternative setting
and for higher dimensions. Indeed,
while we are concerned with four-dimensional spacetimes in this work, we carry-out the higher dimensional construction in a companion paper~\cite{ChCong2}.

Interestingly enough, some more work needs to be done in other topologies and dimensions because of different properties  of the differential operators involved. In fact, the analysis on null three-dimensional hypersurfaces  with spherical cross-sections turns out to be somewhat simpler than the general case.
One of  key new aspects of other topologies or dimensions, when compared to null hypersurfaces with two-dimensional spherical sections, is the existence of non-trivial transverse-traceless two-covariant tensors. Their existence leads to  new difficulties which need to be addressed.  While the collection of TT-tensors is finite-dimensional on two dimensional manifolds, these tensors carry the bulk of information about the geometry in higher dimensions.

To make things precise, we consider the linearisation of the vacuum Einstein equations at a metric
\begin{equation}
{\nobarg}  = - \big(\twoscsign
 -{\alpha^2} r^2  {-\frac{2m}{r}}
  \big) du^2-2du \, dr
 + r^2 \ringh_{AB}dx^A dx^B
   \,,
   \label{23VII22.3intro}
\end{equation}
with
$$
  \alpha \in \{0,\sqrt{ {\Lambda} /3}\}\subset \R\cup \sqrt{-1}\R
  \,,
 \quad
   m \in \R
  \,,
$$
where
$\ringh_{AB}dx^A dx^B$ is a $u$ and $r$-independent metric with scalar curvature $2\twoscsign$, with $\twoscsign\in \{0,\pm1\}$.
Roughly speaking, the question addressed here is the following: given two smooth linearised solutions of the vacuum Einstein equations defined near the null hypersurfaces $\{u=0\,,\ r<r_1\}$ and  $\{u=0\,,\ r > r_2\}$, where $r_2>r_1$, can we find characteristic initial data on the missing region $\{u=0\,,\ r_1\le r\le r_2\}$ which,
when evolved to a solution of the linearised Einstein equations,  provide a linearised metric perturbation which coincides on $\{u=0\}$, together with $u$-derivatives up to order $k$, with the original data.
We refer to this construction as the $\Ck$-gluing.
 The resolution of this problem is presented in Theorem~\pref{T4XII22.2} below, which is the main result of this paper.
 This  theorem is the key step towards a nonlinear gluing~\cite{ChCongGray2}, where a suitable implicit function theorem is used.

An equivalent way of formulating the gluing problem, advocated in~\cite{ACR1}, is that of connecting two sets of ``sphere data'' using null-hypersurface data. This perspective can also be taken in our setting, with ``sphere data'' replaced by suitable linearised data on codimension-two spacelike manifolds, viewed as cross-sections of a null hypersurface.

It was found by Aretakis et al., in the case $\Lambda=0$ and $\myGauss=1$, that there exists a ten-parameter family of obstructions to do such a gluing-up-to-gauge, when requiring continuity of  two $u$-derivatives of the metric components along the null-hypersurface.
Our analysis shows that the analysis is affected both by the dimension, by the cosmological constant, by the topology of sections of the level sets of $u$ (which we assume to be compact), by the mass, and by the number of transverse derivatives which are required to be continuous. In the spherical
four-dimensional case with $m=0$ we provide an alternative proof of the corresponding result in \cite{ACR2} for $\Ctwo$-gluing, with the same number of obstructions.
Table~\ref{T26XI22.1}
\input{ChargeTable}
lists the obstructions which arise in the linearised gluing depending upon the geometry of the cross-sections of the initial data hypersurface and the mass parameter $m$. A key role in our construction is played by the
 radially-constant function  $\chi$
(cf.\ \peqref{24IX22.2} below), the existence of which has already been pointed-out in~\cite{ChHMS}, and the radially constant fields $q_{AB}$ and $\kQ{3,1}{}_{A}$
(cf.\ \peqref{3IX22.1} and \peqref{30VII23.1}), which do not seem to have been noticed so far in the Bondi gauge,
and which seem to be related to the radially constant fields
${\mathcal Q}_2$, ${\mathcal Q}_3$ and ${\mathcal Q}_4$ discovered in~\cite{ACR1}.
 We point out   a slightly different interpretation of the result, namely that the gluing can be performed without obstructions after adding fields,  which carry the missing radial charges and which we describe explicitly, to  the data on $\{r>r_2\}$.
We describe the additional obstructions
that arise  for  $\Ck$-gluing, $k\ge 3$, when linearising on a background with $m=0$, see Tables~\ref{T17XI22.1}, p.~\pageref{T17XI22.1}, and \ref{T17XI22.2}, p.~\pageref{T17XI22.2}.  We show, with a considerable amount of work, that the higher-order obstructions disappear  on   backgrounds with $m\ne 0$.

In their introduction, the authors of~\cite{ACR1} discuss several applications of their construction. The results presented here lead immediately to corresponding results for the linearised fields in our setting.

This work is organised as follows: In Section \ref{s26XI22.1} we introduce some of our notations. In Section~\ref{s3X22.1} we analyse the linearised Einstein equations in the Bondi gauge, following~\cite{ChHMS}. As already observed in~\cite{ACR1,ACR2,ACR3}, a key part of the gluing is played by  the residual gauges, discussed in Section~\ref{ss26XI22.1}.  The
main new element, as compared to~\cite{ChHMS}, is Section~\ref{sec:28VII22.2}, where inductive formulae for higher-order transverse derivatives are presented. The gluing construction is carried-out in Section~\ref{s12I22.1}. We present our strategy in Section~\ref{ss10IX22.1}, with further details provided in the remaining sections there. In Section~\ref{s13VII22.1} we reformulate our  gluing result as an unobstructed gluing-with-perturbation problem for the data on $\{r>r_2\}$.
 Various technical results are presented in the appendices.

We note that the first arxiv version of this work established solvability of the linearised gluing problem through an argument which was inadequate for the full gluing problem, addressed in~\cite{ChCongGray2}, because of insufficient differentiability of solutions of some of the equations used. The current, alternative version  resolves these issues.

\bigskip

\noindent{\sl Acknowledgements.} We are grateful to Finnian Gray for pointing out inaccuracies in a previous version of our manuscript. Useful discussions with Stefan Czimek are acknowledged. We thank the Beijing Institute for Mathematical Sciences and Applications for   hospitality and support during part of work on this paper.

\section{Notation}
 \label{s26XI22.1}

 Let $\zgamma=\zgamma_{AB}dx^A dx^B$ be a  metric on a $2$-dimensional, compact, orientable manifold $\secN$, with covariant derivative $\zspaceD$.
 We let $\zdivone\!$, respectively $\zdivtwo\!$, denote the divergence operator on vector fields $\xi$, respectively on  two-index  tensor fields $h$:
 \begin{equation}\label{12XI22.1}
   \zdivone \xi := \zspaceD_A \xi^A
   \,,
   \qquad
   (\zdivtwo h)_A := \zspaceD_B h^B{}_A
   \,.
 \end{equation}
Given a function $f$ we denote by $f^{[1]}$ the $L^2$-orthogonal projection of $f$ on the constants:
\begin{equation}\label{15XI22.1}
  f^{[1]}:= \frac{1}{|\secN|_{\zgamma}} \int_{\secN} f d\mu_{\zgamma}
  \,,
  \quad
  \text{where}\quad {|\secN|_{\zgamma}} = \int_{\secN}   d\mu_{\zgamma}
  \,.
\end{equation}
Note that  $f^{[1]}$ should not be confused with $f^{[=1]}$, which we use when decomposing a function or a tensor field in spherical harmonics on $S^2$.
We set
\begin{equation}\label{15XI22.2}
  f^{[1^\perp]}:= f -  f^{[1]}
  \,.
\end{equation}

 Let $\CKV$,
  respectively $\KV$,
  denote the space of conformal Killing vector fields on $\secN$, respectively Killing vector fields. Thus (cf.\ Appendix~\ref{App30X22}), $\CKV$ is six-dimensional on $S^2$, consists of covariantly constant vectors on $\T^2$, and is trivial on manifolds of higher genus. Given a vector field $\xi$ on $\secN$ we denote by $(\xi^A)^{[\CKV]}$  the $L^2$-orthogonal projection on the space $\CKV$, with
 $$
  (\xi^A)^{[\CKV^\perp]}
  := \xi^A -  (\xi^A)^{[\CKV]}
 \,,
 $$
 with a similar notation for $(\xi^A)^{[\KV]}$  and $(\xi^A)^{[\KV^\perp]}$.

We will denote by $\harm$ the space of harmonic 1-forms:
 \begin{equation}\label{12XI.22h}
   \harm=\{\xi_{A } \, | \,    \zspaceD^A \xi_{A } =0 =  \epsilon^{AB}\zspaceD_A \xi_{B } \}
   \,.
 \end{equation}
 By standard results (cf., e.g., \cite[Theorems~19.11 and 19.14]{ForsterRiemann} or \cite[Theorem~18.7]{ArminRiemann}), the space $\harm$ has dimension $2\genus$ on cross-sections $\secN$ with genus $\genus$, in particular it is trivial on spherical sections. We will denote by $\xi_A^{[\harm]}$ the $L^2$-orthogonal projection of $\xi_A$ on $\harm$, and by $\xi_A^{[\harm^\perp]}$ the projection on the $L^2$-orthogonal to $\harm$.

 Let $\TTt$ denote the space of transverse-traceless symmetric two tensors:
 \begin{equation}\label{12XI.22}
   \TTt=\{h_{AB} \, | \, h_{[AB]}= 0=\zgamma^{CD}h_{CD}= \zspaceD^E h_{EF}\}
   \,.
 \end{equation}
 Then $\TTt$ is trivial on $S^2$, consists of covariantly constant tensors on $\T^2$, and is $6(\genus-1)$-dimensional on
  two-dimensional manifolds of genus $\genus\ge 2$ (cf., e.g., \cite{Tromba} Theorem 8.2
  and the paragraph that follows, or \cite[Theorem~6.1 and  Corollary~6.1]{TamWan}).

 Given a tensor field $h=h_{AB}dx^Adx^B$ we denote by $h_{AB}^{[\TTt]}$ the $L^2$-orthogonal projection of $h $ on $\TTt$, and set
 \begin{equation}\label{12XI.23}
   h_{AB}^{[\TTt^\perp]}:= h_{AB} -h_{AB}^{[\TTt]}
   \,.
 \end{equation}
 Clearly, for two-covariant traceless symmetric tensors on $S^2$ it holds that $h_{AB}^{[\TTt^\perp]} = h_{AB}$, but this is not true anymore for the remaining two-dimensional compact manifolds.

 We will often follow terminology and notation from~\cite{ACR1}. In particular, scalar functions, vector fields, and traceless two-covariant symmetric tensors on  $S^2$ will be decomposed into spherical harmonics, see Appendix~\ref{ss20X22.1} for a summary. The notation $t^{[= \ell]}$ will denote the $L^2$-orthogonal projection of a tensor $t$ on the space of $\ell$-spherical harmonics. Then
 \begin{equation}\label{29X22}
   t^{[\le \ell]} = \sum_{i=0}^\ell t^{[=i]}
    \,,
    \qquad
   t^{[> \ell]} = t - t^{[\le \ell]}
   \,,
 \end{equation}
 with obvious similar definition of $ t^{[< \ell]}$, etc.

 One of the operators appearing below is the operator
 \begin{equation}\label{16IV23.2}
   \mrL : = \TSzlap (\TSzlap +2 \myGauss)
 \end{equation}
acting on functions. Its kernel consists of linear combinations of $\ell=0$ and $\ell=1$ spherical harmonics on $S^2$, and of constants on the remaining compact orientable two-dimensional manifolds. Given a function $f$ we will write $f^{\kerL}$ for the $L^2$-orthogonal projection on this kernel, and $f^{\kerLp}$ for the projection on its orthogonal.

%% file: ChargeTable.tex
\begin{table}[t]
  \centering
  \begin{tabular}{||c|c|c|c||}
  \hline
  \hline
      & $S^2$ & $\T^2$ & higher genus \\
  \hline
    $\kQ{1}{}$: $m=0$
        & 6
                &  2
                     & 0
\\
    \phantom{$\kQ{1}{}$,} $m\ne 0$
        & 3
                &  2
                     & 0
\\
  \hline
    $\kQ{2}{}$:  $m=0$ 
        & 4
                &  1
                     & 1
\\
    \phantom{$\kQ{2}{}$:}  $m\neq0$ 
        & 1
                &  1
                    & 1
\\
  \hline
    $\kQ{3,1}{}^{[\harm]}$: $m=0$
        & 0
                & coincides with  $\kQ{2}{}$
                     & $2\genus $
\\
     \phantom{$\kQ{3,1}{}^{[\harm]}$: } $m\neq0$
        & 0
                &  0
                     & 0
\\
  \hline
    $q_{AB}^{[\TTt]}$ : $m=0$, $\alpha =0$
        & 0
                &  2
                     & $6(\genus -1)$
\\
  \phantom{$q_{AB}^{[\TTt]}$ :} $m=0$, $\alpha \ne 0$
        & 0
                &  0
                     & 0
\\
     \phantom{$q_{AB}^{[\TTt]}$ : } $m\neq0$
        & 0
                &  0
                     & 0
\\
  \hline
    $\kQ{3,2}{}^{[\harm]}$: $m=0$
        & 0
                & 0
                     & $2\genus $
\\
     \phantom{$\kQ{3,1}{}^{[\harm]}$: } $m\neq0$
        & 0
                &  0
                     & 0
\\
  \hline
 $\overset{[2]}{q}{}_{AB}^{[\TTt]}$: $m=0$
        & 0
                &  2
                     & $6(\genus -1)$
\\
     \phantom{$\overset{[2]}{q}_{AB}^{[\TTt]}$:} $m\neq0$
        & 0
                &  0
                     & 0
  \\
  \hline
     together: $m=0$, $\alpha =0$
        & 10
                &   7
                     & $16\genus  -11$
\\
   \phantom{together:}  $m=0$, $\alpha \ne 0$
        & 10
                &  5
                     & $10\genus - 5$
\\
    \phantom{together:} $m\ne 0$
        & 4
                &  3
                     & 1
\\
  \hline
  \hline
\end{tabular}
  \caption{The dimension of the space of obstructions   for $\Ctwo$-gluing.  The radial charges  $\kQ{a}{}$, a=1,2, are defined in \eqref{24VII22.4}, p.~\pageref{24VII22.4} and \eqref{20VII22.1}, p.~\pageref{20VII22.1}; the radially-conserved tensor fields $q_{AB}$,
 $\kq2_{AB}$,
   and $\kQ{3,i}{} $ are defined in \peqref{3IX22.1}, \peqref{1XII22.1}, and \peqref{9IX22.9}; $\genus $ is the genus of the cross-sections of the characteristic initial data hypersurface; the superscripts $[\harm]$, respectively $[\TTt]$, denote
   the  $L^2$-orthogonal projection on the set of harmonic $1$-forms, respectively on   transverse-traceless tensors. On $S^2$ the four obstructions associated with $\kQ{2}{}$ correspond to spacetime translations, the three obstructions associated with $\kQ1{}$ when $m\ne0$ correspond to rotations of $S^2$, with the further three obstructions arising when $m=0$ corresponding to boosts. }
  \label{T26XI22.1}
\end{table} 

%% file: 0MatchingIn4D.tex
\section{Linearised characteristic constraint equations in Bondi coordinates}
 \label{s3X22.1}

Let $(\mcM ,g)$ be a $(3+1)$-dimensional spacetime. Locally, near a null hypersurface
for which the optical  divergence scalar is non-vanishing,
 we can  use Bondi-type coordinates $(u,r,x^A)$ in which the metric takes the form
\begin{align}
\label{23VII22.1}
    g_{\alpha \beta}dx^{\alpha}dx^{\beta}
   &=  -\frac{V}{r}e^{2\beta} du^2-2 e^{2\beta}dudr
\\
 &\qquad
   +r^2\zhTBW_{AB}\Big(dx^A-U^Adu\Big)\Big(dx^B-U^Bdu\Big)
    \nn
    \,,
\end{align}
where
\begin{equation}
    \det [\gamma_{AB} ] = \det[ \zzhTBW_{AB}]
     \,,
\end{equation}
with $\zzhTBW_{AB}(x^C)$ being a metric of constant scalar curvature $2\twoscsign$. In particular, $\det [\gamma_{AB}]$ is $r$ and $u$-independent, which implies
\begin{equation}
    \gamma^{AB}\partial_r\gamma_{AB} = 0\,,\qquad \gamma^{AB}\partial_u\gamma_{AB} = 0\,.
    \label{23VII22.2}
\end{equation}
As such, the inverse metric reads
\begin{equation}
    g^{\sharp} =  e^{-2\beta} \frac{V}{r}\, \partial_r^2 - 2 e^{-2\beta} \, \partial_u\partial_r - 2e^{-2\beta} U^A \,  \partial_r \partial_A +\frac{1}{r^2} \gamma^{AB} \, \partial_A\partial_B\,.
\end{equation}
Note that each surface $\{u=$ constant$\}$ is a null hypersurface with null normal proportional to $\partial_r$, and $r$ is a parameter which varies along the null generators. Finally, the $x^C$'s are local coordinates on the codimension-two  surfaces of constant $(u,r)$  which, as $r$ varies, foliate each null hypersurface of constant $u$.

The restriction of the Einstein equations (E.E.) to a null hypersurface gives a set of null constraint equations for the metric functions $(V,\beta,U^A,\gamma_{AB})$,  which lead to obstructions to the gluing of characteristic data.
In this work we will study the linearised problem around a null hypersurface in a
Birmingham-Kottler background, which includes a Minkowski, anti-de Sitter or de Sitter background.
 In Bondi coordinates the  background metrics can be written as
\be
{\nobarg}\equiv {\nobarg}_{\a \b} dx^\a dx^\b = \guu  du^2-2du \, dr
 + r^2 \ringh_{AB}dx^A dx^B
   \,,
   \label{23VII22.3}
\ee
with
$$
\guu :=
-
\big(\twoscsign
 -{\alpha^2} r^2  {-\frac{2m}{r}}
  \big)
\,,
 \quad
 \twoscsign \in \{0,\pm 1\}
 \,,
 \quad
  \alpha \in \{0,\sqrt{ {\Lambda} /3}\}
  \,,
 \quad
  {m \in \R}
  \,,
$$
where
$\ringh_{AB}dx^A dx^B$ is a $u$- and radially constant metric of scalar curvature $2\twoscsign$,
and  note that $\alpha \in \R\cup i \R$: a purely imaginary value of $\alpha$ is allowed to accommodate for a cosmological constant $ \Lambda  <0$. It holds that
$$
g^{\alpha\beta}\partial_\alpha\partial_\beta = -2 \partial_u\partial_r -  \guu  (\partial_r)^2
 + r^{-2}\zzhTBW^{AB}\partial_A\partial_B
 \,.
$$

Consider now a perturbation of the metric of the form
\begin{equation}
 \label{3VIII22.5}
 g_{\mu\nu} \rightarrow g_{\mu\nu} + \epsilon \hBo_{\mu\nu}
 \,,
\end{equation}
where $\epsilon$ should be thought as being very small.
The conditions on the linearised fields such that the perturbed metric is still in the Bondi form to $O(\epsilon)$ are,
\begin{equation}
\label{23VII22.4}
  h_{rA}=h_{rr}=\zzhTBW^{AB} h_{AB}=0
  \,.
\end{equation}
In what follows for perturbations around a Birmingham-Kottler background, we shall sometimes find it convenient to use fields
$\{\delta V,\delta \beta, \delta U_A
 := \zzhTBW_{AB} \delta  U^B \}$ to denote metric perturbations. These correspond respectively to
\begin{equation}\label{3X22.1p}
 \{\delta V + 2 V\delta \beta,\delta \beta, \delta U_A \}
 \equiv
 \{- r h_{uu}, - h_{ur} / 2, - h_{uA}/ r^2\}
 \,.
\end{equation}
  We will also use the notation
\begin{equation}\label{18XII22.1}
    \zhTB_{\mu\nu}:=\hBo_{\mu\nu}/r^2
 \,.
\end{equation}

\subsection{The linearised $\Ck$-gluing problem}

One of the key objects that arise in the characteristic gluing construction of~\cite{ACR1} are the  ``sphere data''. Roughly speaking, these are data that are needed on a cross-section of a characteristic surface  for the integration of the transport equations (see below).

Using a Bondi parameterisation of the metric, these data can be defined as follows. Let $\mcN_{I}$ be a null hypersurface $\{ u=u_0, r\in I\}$, where $I$ is an interval in $\R$, and let $\secN$ be a cross-section of $\mcN$, i.e. a two-dimensional submanifold of $\mcN$ meeting each null generator of $\secN$ precisely once.
Let $2\le k \in \N$ be the number of derivatives of the metric that we want to control at $\secN$. Using the  Bondi parameterisation of the metric, we define
linearised Bondi \emph{cross-section data} of order $k$ as the collection of fields
\begin{equation}\label{23III22.992}
  \dt_{\secN} = (\partial_u^{\ell}\partial_r^j \hBo_{AB}|_{\secN}
  ,\,\partial_u^{\ell}\partial_r^j \delta \beta|_{\secN},
  \, \partial_u^{\ell}\partial_r^j \delta U^A|_{\secN}
  ,\,  \partial_u^{\ell}\partial_r^j \delta V|_{\secN}
  )
  \,,
\end{equation}
for integers ${\ell},j$ such that ${\ell}+j\leq k$ \textcolor{red}{}.%
\footnote{The data $\dt_\secN$ are closely related to the data $\Psi_{\mathrm{Bo}}[\secN,k]$ of  \cite[Section~5]{ChCong0}. As discussed in more detail there,  some of the fields in \eqref{23III22.992} are not independent, but this is irrelevant for our purposes.}

For simplicity we assume that all the  fields in \eqref{23III22.992}
are smooth, though a finite sufficiently large degree of differentiability would suffice for our purposes, as can be verified by chasing the number of derivatives in the relevant equations; compare Section~\ref{ss25IV22.1} below.

A natural threshold for the gluing is $k=2$,
as then one expects
 existence of an associated space-time solving the vacuum Einstein equations when the fields are sufficiently differentiable in directions tangent to $\secN$ (cf.~\cite{LukRodnianski} for a small data result in a different gauge;  see~\cite{RendallCIVP,Luk,CCW} for existence without smallness restrictions under more stringent differentiability conditions).
In the linearised $\Ck$-gluing problem we start with two sections $\secN_1$ and $\secN_2\subset J^+(\secN_1)$ of a null hypersurface $\{u=0\}$ equipped with Bondi coordinates as in \eqref{23VII22.3}, each with constant $r$,
and their linearised Bondi cross-section data of order $k$, $\dt_{\secN_1}$ and
$\dt_{\secN_2}$. The goal is to interpolate between  $\dt_{\secN_1}$ and $\dt_{\secN_2}$ along a null hypersurface $\mcN_{[r_1,r_2]}$ such that (i) $\dt_{\secN_1}$ agrees with the restriction to $r_1$ of the interpolating field along $\mcN_{[r_1,r_2]}$;  (ii) $\dt_{\secN_2}$ agrees with the restriction  to $r=r_2$ of the interpolating field; and (iii) the constructed field satisfies the linearised null constraint equations. We shall see in Section \ref{sec:28VII22.1} how the linearised null constraint equations lead to obstructions to the gluing.

Since linearised Bondi data are defined up to linearised gauge transformations, we shall use these transformations to help us with the gluing.

\subsection{Gauge freedom}
 \label{ss26XI22.1}

Recall that linearised gravitational fields are defined up to a gauge transformation
\begin{equation}\label{4XII19.12}
  h\mapsto h+ \Lie_\TSxip g
\end{equation}
determined by a vector field $\TSxip$. Once the metric perturbation has been put into Bondi gauge, there remains the freedom to make gauge transformations which preserve this gauge:
\begin{eqnarray}
\Lie_{\TSxip} g_{\TSr \TSr}  &=&0
 \,,
 \label{4II20.3}
  \\
   \label{4II20.4}
  \Lie_{\TSxip} g_{\TSr A}  &=&0   \,,
\\
 g^{AB} \Lie_{\TSxip} g_{AB}
  &  =  &
  0
   \label{4XII19.14}
 \, ,
\end{eqnarray}
For the metric  \eqref{23VII22.3} this is solved by (cf., e.g., \cite{ChHMS})
\begin{eqnarray}
\TSxip^{\TSu}(u,r,x^A)&=&\TSxi^{\TSu}(\TSu,x^A)
 \,,
  \label{3XII19.t1}
\\
 \TSxip^{B}(u,r,x^A) &=&
   \TSxi^{B} (\TSu,x^A)
  -
   \frac{1}{\TSr} \zspaceD^B \TSxi^{\TSu}(\TSu,x^A)
    \,,
   \phantom{xxxxxx}
   \label{1VIII22.1}
   \\
      \TSxip^{\TSr}(u,r,x^A)
	&= &
-\frac 1	2 \TSr  \zspaceD_{B} \TSxi^{B} (u,x^A)
 +
   \frac{1}{2}  \TSzlap  \TSxi^{\TSu}(\TSu,x^A)
     \,,
   \label{3XII19.t2}
\end{eqnarray}
for some fields $\TSxi^{u} (\TSu,x^A)$, $\TSxi^{B} (\TSu,x^A)$, and where $\zspaceD_A$ and $\TSzlap$ are respectively the covariant derivative and the Laplacian operator associated with the two-dimensional metric $\ringh_{AB}$ appearing in \eqref{23VII22.3}.

We define
$$
 \TSoLie_\TSxip
$$
to be the Lie-derivation in the $x^A$-variables
with respect to the vector field $\TSxip^A\partial_A$.

The transformation~\eqref{4XII19.12} can be viewed as a result of linearised coordinate transformation to new coordinates $\tilde{x}^{\mu}$ such that
\begin{equation}
    x^{\mu} = \tilde{x}^{\mu} + \epsilon \zeta^{\mu}(\tilde{x}^{\mu})\,,
    \label{3VIII22.3}
\end{equation}
where $\epsilon$ is as in \eqref{3VIII22.5}.
Writing $\guu$ as
$$\guu =
- \twoscsign
 +{\alpha^2} r^2 +\frac{2m}{r}
  =: \varepsilon N^2\,, \ \mbox{where $\varepsilon \in \{\pm 1\}$,}
$$
under \eqref{3VIII22.3}, the linearised metric perturbation transforms as
\begin{align}
  \label{24IX20.1}
  \hBo_{uA}
   \to &
\ \tilde{\hBo}_{uA} = \barh_{uA} + \mcL_\zeta g_{uA}
  \\
  &
  =
  \barh_{uA} +
   \partial_{\tdA}(\epsilon N^2 \zeta ^u - \zeta ^r)
  + r^2 \ringh_{AB} \partial_{\tdu}\zeta ^B
  \nonumber
\\
&
 =
  \barh_{uA}
   -\frac{1}{2}\partial_{\tdA} \, [\, (\TSzlap \xi^u
    + 2 \twoscsign \xi^u)
      -r( \zspaceD_{\tdB} \xi^B - 2\partial_{\tdu} \xi^u)]
 \nonumber
 \\
  &
   \phantom{=}
 + r^2
 \big(
  \ringh_{AB} \partial_{\tdu} \xi^B
  +\big(  {\alpha^2
   + \frac{2m}{r^3}\big)
  } \partial_{\tdA} \xi^u
   \big)
    \,,
\nn
  \\
  \label{24IX20.23a}
  \hBo_{ur}
   \to & \
   \tilde{\hBo}_{ur} =
  \barh_{ur} + \mcL_\zeta g_{ur} =
  \barh_{ur}
  - \partial_{\tdu} \zeta^u
  + \epsilon N^2 \partial_{\tdr} \zeta^u
  - \partial_{\tdr} \zeta^r
\\
&
=
  \barh_{ur}
  - \partial_{\tdu} \xi^u
  + \frac{1}{2} \zspaceD^{\tdA}
   \xi_A
    \,,
    \nn
\\
      \hBo_{uu}
     \to & \
 \tilde \hBo_{uu} =  \hBo_{uu}  + \mcL_\zeta g_{uu} =
  \barh_{uu}
  + \epsilon \TSxip^{\TSr} \partial_{\tdr} N^2
	+2 \partial_{\tdu}(\epsilon N^2 \zeta^u - \zeta^r)
\label{24IX20.23x}
\\
&
=
\barh_{uu}-(2\twoscsign+\TSzlap)\partial_{\tdu}\xi^u
 + \tdr
  \big(
   \zspaceD_B \partial_{\tdu}\xi^B +
   \big(
    {\alpha^2 - \frac{m}{r^3}\big)
    } \TSzlap \xi^u
   \big)
 \nonumber%
\\
 &
   \phantom{=}
 +  \big({\alpha^2} \tdr^2 +\frac{2m}{r}\big)
  (2\partial_{\tdu}\xi^u )- \big({\alpha^2} r^2 -\frac{m}{r}\big) \zspaceD_{\tdB} \xi^B\, ,
 \nn
\\
  \hBo_{AB}
     \to  & \
    \tilde \hBo_{AB} =
  \barh_{AB} + \mcL_\zeta g_{AB} =
  \barh_{AB}
  +2 r \TSxip^{\TSr} \ringh_{AB}
	+
	r^2 \TSoLie_{\TSxip} \ringh_{AB} \,,
  \label{24IX20.23}
\end{align}
with
$$
 \TS [X_{AB}] := \frac 12 ( X_{AB}+X_{BA} - \ringh ^{CD} X_{CD} \ringh_{AB}
  )
$$
denoting
the traceless symmetric part of a tensor on a section $\secN$.

Given $\secN_{u_0,r_0}$ corresponding to a $\{u=u_0,r=r_0\}$ section of some $\mcN$, equations~\eqref{24IX20.1}-\eqref{24IX20.23} together with all their
$\tdu$- and $\tdr$-derivatives up to order $k$  define a new set of order-$k$ cross-section data
on
\begin{align*}
  \tilde{\secN}_{u_0,r_0}:&= \{\tilde u=u_0,\tilde r=r_0\}
   \\
   \nonumber
   & =\{u=u_0+ \epsilon \zeta^u(u_0,r_0,x^A),
   r=r_0+\epsilon \zeta^r(u_0,r_0,x^A)\}
 \,,
\end{align*}
a section lying close to the original $\secN_{u_0,r_0}$, in terms of the gauge fields
$$\{\partial^i_{\tdu}\xi^B|_{\tdu = u_0},\partial^i_{\tdu}\xi^u|_{\tdu = u_0}\}_{0\leq i\leq 3}$$
as well as the original metric perturbations evaluated on  $\tilde{\secN}_{u_0,r_0}$.

Equation \eqref{24IX20.23a} shows that we can always choose $\zeta$ so that
\begin{equation}\label{8III22.1}
  \tilde \hBo_{ur} = 0
  \,.
\end{equation}
After having done this, we are left with a residual set of gauge transformations,   defined by a $\tdu$-parameterised family of vector fields $\TSxi^A(\tdu,\cdot)$, and $\xi^u(\tdu,\cdot)$, with the condition
\begin{equation}\label{5XII19.1a}
	 \partial_{\tdu}  \xi^{u}(\tdu,x^A) =
\frac{ \zspaceD_{\tdB} \TSxi^{B}(\tdu, x^{A})}{2}
 \,
\end{equation}
needed to preserve the gauge $\tilde h_{ur} = 0$.

Under the residual gauge transformations with~\eqref{5XII19.1a}, the transformed fields take the form
\begin{align}
  \tilde \hBo_{uA}
   & =
  \hBo_{uA} -\frac{1}{2}\zspaceD_{\tdA} \TSzlap \xi^u + \epsilon N^2 \zspaceD_{\tdA} \xi^u + r^2 \partial_{\tdu} \xi_A
  \label{17III22.1}
\\
&
 =
  \hBo_{uA}-\frac{1}{2}\zspaceD_{\tdA} \, [\, (\TSzlap \xi^u
+ 2 \twoscsign \xi^u)
]
  \nonumber
\\
 &
 + r^2 \big[\ringh_{AB} \partial_{\tdu} \xi^B
    +
    {\big({\alpha^2} +\frac{2m}{r^3}\big)}
   \zspaceD_{\tdA} \xi^u
   \big]
 \,,
  \nn
\\
      \tilde \hBo_{uu}
    & =
\barh_{uu}
 + r\big[
   {\big({\alpha^2}  -\frac{ m}{r^3}\big)}
     \TSzlap \xi^u
      +   \zspaceD_{\tdB} \partial_{\tdu}\TSxi^{B}
 \big]
 \label{13III22.2}
 \\
 &\quad
   - \big(\twoscsign + \frac 12  \TSzlap  -\frac{3m}{r} \big) \zspaceD_{\tdB} \TSxi^{B}
\, ,
 \nn
\\
  \tilde \hBo_{AB}
    & =
 \barh_{AB}
  + 2 r^2
 \TS [\zspaceD_{\tdA}\xi _B]
  - 2 r \TS [\zspaceD_{A}\zspaceD_B \zeta^u]
 \,.
  \label{17III22}
\end{align}

Let $\dt_{\secN_1}$ and $\dt_{\secN_2}$ be linearised Bondi cross-section data of
order $k$
 on $\secN_1$ and $\secN_2$ respectively. Given gauge fields
$$
 \{\partial_{\tdu}^i\xi^B|_{\tilde\secN_a},\partial_{\tdu}^i\xi^u|_{\tilde\secN_a}\}_
 {0\leq i\leq k+1\,,1\leq a\leq 2}
\,,
$$
the associated transformed Bondi cross-section data are given by~\eqref{24IX20.1}-\eqref{24IX20.23} and their $\partial_{\tdu}$ and $\partial_{\tdr}$ derivatives. In the linearised gluing problem, we shall allow for such gauge transformations to the data; that is, we consider gluing along a null hypersurface of the transformed data $\tilde{\dt}_{\tilde{\secN}_1}$ and $\tilde{\dt}_{\tilde{\secN}_2}$ with the freedom of choosing gauge fields to achieve the gluing. We shall call this gluing-up-to-gauge.

To simplify notation we will write
\begin{align}
    \Done (\xi^u)_A &:=-\frac{1}{2}\zspaceD_{\tdA} \, [\, \TSzlap \xi^u + 2\twoscsign \xi^u]
    = -\zspaceD^B\TS[\zspaceD_A\zspaceD_B \xi^u]
    \,,
\\
    C(\zeta)_{AB}&:=\TS[\zspaceD_{\tdA}\zeta_B]
    \,,
     \label{18X22.41}
\\
    \Dtwo (\xi)&:=-\left(\twoscsign+\frac{1}{2}\TSzlap\right)
   \zspaceD_{\tdB}\xi^B\,.
\end{align}
For further convenience  we note the transformation laws, in this notation and for $ i \ge 1$,
\begin{align}
  \tilde \hBo_{uA}
   & =
  \hBo_{uA} +\Done (\xi^u)_A
 + r^2
    \big( \partial_{\tdu} \xi_A +
   \big({\alpha^2}   +\frac{2m}{r^3}\big)
    \zspaceD_{\tdA} \xi^u
     \big)
 \,,
  \label{26IX22.1}
\\
 \partial_u^i \tilde \hBo_{uA}
   & =
  \partial_u^i  \hBo_{uA} + \frac 12 \Done (\zspaceD_{B}  \partial_u^{i-1}\xi^B)_A
   \label{26IX22.1i}
\\
    & \quad
 + r^2 \big[
    \partial_u^{i+1} \xi_A
    + \frac{1}2
   {\big({\alpha^2} +\frac{2m}{r^3}\big)}
    \zspaceD_{\tdA} \zspaceD_{B}  \partial_u^{i-1}\xi^B
    \big]
 \,,
  \nn
\\
      \tilde \hBo_{uu}
    & =
\barh_{uu} + r\big[
   {\big({\alpha^2}  -\frac{ m}{r^3}\big)}
    \TSzlap \xi^u +   \zspaceD_{\tdB} \partial_{\tdu}\TSxi^{B}
 \big]
  + \Dtwo (\xi)
  \label{26IX22.2}
  \\
  &\quad
    +\frac{3m}{r} \zspaceD_{\tdB} \TSxi^{B}
\, ,
 \nn
\\
  \tilde \hBo_{AB}
    & =
 \barh_{AB}
 + 2 r^2 C(\zeta)_{AB}
\\
    & =
 \barh_{AB}
 + 2 r^2 C(\xi )_{AB} - 2 r \TS[\zspaceD_A\zspaceD_B \xi^u]
 \nn
 \,,
\\
  \partial_u^i
  \tilde \hBo_{AB}
    & =
  \partial_u^i
 \barh_{AB}
 + 2 r^2 C(
  \partial_u^i \xi )_{AB}
  \\
  &\quad
  -  r \TS[\zspaceD_A\zspaceD_B
  \zspaceD_C \partial_u^{i-1}\xi^C]
  \,,\nn
\\
  \zspaceD^A \tilde \hBo_{uA}
   & =
 \zspaceD^A \hBo_{uA}-\frac{1}{2}\TSzlap \,   (\TSzlap
+ 2 \twoscsign)  \xi^u
  \label{17III22.1ap}
\\
 &\quad
 + r^2 \big[
  \zspaceD_A  \partial_{\tdu} \xi^A +
    \big({\alpha^2} +\frac{2m}{r^3}\big)
   \TSzlap \xi^u
   \big]
 \,,
  \nn
\\
 \zspaceD^B \tilde \hBo_{AB}
    & =
\zspaceD^B \barh_{AB} +   r^2
   (\TSzlap + \twoscsign ) \xi _B
  -   r   \zspaceD_A (\TSzlap + 2\twoscsign )  \xi^u
  \label{17III22.1apr}
\\
    & =
\zspaceD^B \barh_{AB} +   r^2
   (\TSzlap + \twoscsign ) \xi _A
  + 2  r\Done (\xi^u)_A
   \,,
  \nn
\\
 \zspaceD^A \zspaceD^B\tilde \hBo_{AB}
    & =
\zspaceD^A \zspaceD^B\barh_{AB} +
    r^2  (\TSzlap + 2\twoscsign )
    \zspaceD_{\tdA}\xi^A
    \label{17III22p}
    \\
    &\quad
  -    r  \TSzlap (\TSzlap + 2\twoscsign )  \xi^u
   \nonumber
\\
    & =
\zspaceD^A \zspaceD^B\barh_{AB}
    -2 r^2  \Dtwo (\xi )
  -    r  \TSzlap (\TSzlap + 2\twoscsign )  \xi^u
 \,.
  \nn
\end{align}

\subsection{Null constraint equations}
\label{sec:28VII22.1}

We now turn our attention to  Einstein equations,
\begin{equation}
    G_{\mu\nu} := R_{\mu\nu}-\frac{1}{2}g_{\mu\nu}R = 8\pi T_{\mu\nu} - \Lambda  g_{\mu\nu}
    \label{2X22.2}
\end{equation}
and their linearisation in Bondi coordinates.

\subsubsection{$h_{ur}$}
 \label{CHGss29VII20.1}

The $G_{rr}$ component of the Einstein tensor, which we reproduce from~\cite{MaedlerWinicour}, reads:
\begin{equation}
         \label{CBCHG:beta_eq}
          \frac r 4 G_{rr} = \partial_{r} \beta - \frac{r}{16}\zhTBW^{AC}\zhTBW^{BD} (\partial_{r} \zhTBW_{AB})(\partial_r \zhTBW_{CD})
          \,.
\end{equation}
Since the right-hand side of \eqref{CBCHG:beta_eq} is quadratic in $\partial_r\gamma_{AB}$,  after linearising in vacuum we find

\begin{equation}\label{CHG28XI19.4aa}
 \partial_r \delta \beta =0  \quad
   \Longleftrightarrow
   \quad
   \delta \beta = \delta \beta (u,x^A)
  \,.
\end{equation}
Using a terminology somewhat similar to that of \cite{ACR1}, we thus obtain a pointwise radial conservation law for $\delta \beta$, and an apparent obstruction to gluing: two linearised fields can be glued together if and only if their Bondi functions $\delta \beta$ coincide.

However, it follows from \eqref{8III22.1} that we can always choose a gauge so that $\delta \beta \equiv 0$.
Thus, \eqref{CHG28XI19.4aa} does not lead to an obstruction for gluing-up-to-gauge.  Hence, when gluing, we will always use the gauge where $\delta \beta =0$.
As such,
 in the current section we will  \emph{not} assume $\delta \beta =0$
unless explicitly indicated otherwise.

\subsubsection{$h_{uA}$}
 \label{CHGss29VII20.2}

From the $G_{rA}$-component of the Einstein equations one has
\begin{eqnarray}
          &&  \partial_r \left[r^4 e^{-2\beta}\zhTBW_{AB}(\partial_r U^B)\right]
             =   2r^4\partial_r \Big(\frac{1}{r^2}\spaceD_A\beta  \Big)
                 \label{CBCHG:UA_eq} \\ &&\qquad
                 -r^2\zhTBW^{EF} \spaceD_E (\partial_r \zhTBW_{AF})
                  +16\pi r^2    T_{rA}
                  \,.
                            \nn
           \end{eqnarray}
The linearisation of $G_{rA}$ at a Birmingham-Kottler metric reads
\begin{eqnarray}
           2r^2  \delta G_{rA} &=&
           \partial_r \left[r^4  \ringh_{AB}(\partial_r \delta U^B)\right]
         -  2r^4\partial_r \Big(\frac{1}{r^2}\zspaceD_A\delta \beta  \Big)
    \label{CHG28XI19.5}
    \\
      &&     +
                    r^2
                   \partial_r \left(r^{-2}
                    \zspaceD^Bh_{AB}\right)
                 \,.
                        \nonumber
           \end{eqnarray}
The linearised vacuum Einstein equation thus gives
\begin{eqnarray}
          &&
           \partial_r \left[r^4  \partial_r(r^{-2} h_{uA}) + 2 r^2\zspaceD_A \delta \beta \right] =
                   8r\zspaceD_A\delta\beta+
                   \zspaceD^B r^2
                   \partial_r \left(r^{-2}
                    h_{AB}\right)
                 \,.
                 \label{24VII22.1}
           \end{eqnarray}
Integration of this transport equation gives us a representation formula for $\partial_r\hBo_{uA}$:
\begin{eqnarray}
          &&
           \left[s^4  \partial_s(s^{-2} h_{uA}) + 2 s^2\zspaceD_A \delta \beta \right]_{s=r_1}^r =
                   \int_{r_1}^r 8s\zspaceD_A\delta\beta+ \zspaceD^B s^2
                   \partial_s \left(s^{-2}
                    h_{AB}\right)\, ds
                 \,.
                 \label{24VII22.2}
           \end{eqnarray}
In the gauge $\delta\beta = 0$, and after performing an integration by parts on the right-hand side, this can be written as,
\begin{align}
    r^4\partial_r\zhTB_{uA}|_r = r_1^4 \partial_r\zhTB_{uA}|_{r_1}+\big[\zspaceD^B\hBo_{AB}\big]_{r_1}^{r} - 2\int_{r_1}^r \hat{\kappa}_1(s) \zspaceD^B\hBo_{AB}\,ds
    \label{25VII22.3}
\end{align}
where we have defined,
\begin{align}
    \hat{\kappa}_1(s):=\frac{1}{s}\,.
    \label{25VII22.2}
\end{align}
Given $\dt_{\secN_1}$ and $\dt_{\secN_2}$, equation \eqref{25VII22.3} evaluated at $r=r_2$ gives a condition for the field $\hBo_{AB}(r)$, where $ r\in(r_1,r_2)$ which has to hold when constructing the solution to the gluing problem on $\mcN_{[r_1,r_2]}$.

Now, the cokernel of the operator $\zdivtwo\!$
$$
 \zdivtwo:\varphi_{AB}\mapsto \zspaceD^B\varphi_{AB}
$$
acting on traceless symmetric tensors $\varphi_{AB}$, and which appears in \eqref{24VII22.1} in front of $h_{AB}$, is spanned by solutions of the system
\begin{equation}\label{6III22.3}
 \TS [\zspaceD_A \pi_B] =0
  \,,
\end{equation}
with $\pi_A= \pi_A(u, x^B)$. The space of solutions of \eqref{6III22.3} is the space of conformal Killing vector fields, which we denote by $\CKV$. This space is six-dimensional on $S^2$, and is isomorphic to the Lie algebra of the Lorentz group. On a two-dimensional torus $\T^2$, solutions of \eqref{6III22.3} belong to the two-dimensional space of  covariantly constant vectors. Finally, the space of solutions of \eqref{6III22.3} on a two-dimensional negatively curved compact manifold is trivial; cf~.~Appendix~\ref{App30X22}.

The projection of~\eqref{24VII22.1} onto $\pi_A$ in the gauge $\delta \beta = 0$ gives
\begin{align}
           \partial_r \int_{\secN}\pi^A \left[r^4  \partial_r(r^{-2} h_{uA})
           \right] \,\sm &=
                   \int_{\secN}\pi^A \zspaceD^B\big[ r^2
                   \partial_r \left(r^{-2}
                    \hBo_{AB}\right) \big]\,\sm
                    \label{24VII22.3}
\\
                    &=
                   \int_{\secN}\TS[\zspaceD^B\pi^A]\big( r^2
                   \partial_r \left(r^{-2}
                    \hBo_{AB}\right) \big)\,\sm
                    \nonumber
                    \\
                    &= 0
                 \,,
                 \nn
           \end{align}
and thus the integrals
\begin{eqnarray}
          &&
           \kQ{1}{}(\pi^A) [\secN] := \int_{\secN}\pi^A \left[r^4  \partial_r(r^{-2} h_{uA})\right] \,\sm
           \label{24VII22.4}
           \end{eqnarray}
form  a family of \textit{radially conserved   charges}, with
$$\partial_r \kQ{1}{} = 0$$
along any $u = $ constant null hypersurfaces with the gauge choice $\delta\beta = 0$.

This leads to a six-dimensional family of obstructions to gluing on $S^2$, two-dimensional on $\T^2$, and  no obstructions   on null surfaces with sections of higher-genus.

We shall denote the dependence of
$\kQ{1}{}$ on $\dt_{\secN}$ as $\kQ{1}{} = \kQ{1}{}[\dt_{\secN}]$. Thus in the gauge $\delta\beta =0$, \textit{to achieve gluing of $\dt_{\secN_1}$ and $\dt_{\secN_2}$, it must hold that}
\begin{equation}
    \kQ{1}{}[\dt_{\secN_1}] = \kQ{1}{}[\dt_{\secN_2}]\,.
    \label{31VII22.1}
\end{equation}

Indeed,
it follows from  Appendix~\pref{ss12XI22.2}
that \eqref{31VII22.1} is a necessary and sufficient condition for $r_2^4\partial_r\zhTB_{uA}|_{\secN_2}-r_2^4\partial_r\zhTB_{uA}|_{\secN_1}$ to lie in the image of the operator $\zdivtwo\!$ acting on traceless symmetric tensors, or equivalently, for the existence of a solution $\tilde{\varphi}_{AB}(x^C)$ to the equation
\begin{align}
    r_2^4\partial_r\zhTB_{uA}|_{\secN_2} &=
    r_1^4 \partial_r\zhTB_{uA}|_{\secN_1}-\zspaceD^B\hBo_{AB}|_{\secN_1}
    - \zspaceD^B\tilde{\varphi}_{AB} \,.
    \label{25VII22.6}
\end{align}
The gluing condition~\eqref{25VII22.3} evaluated at $r=r_2$ can thus be achieved by interpolating $\hBo_{AB}$ on $\mcN_{(r_1,r_2)}$ so that
\begin{equation}
    \ip{\hBo_{AB}}{\hat{\kappa}_1} = \tilde{\varphi}_{AB}\,,
\end{equation}
where $\tilde{\varphi}_{AB}$ is the solution to \eqref{25VII22.6},  and where  we write, for $f,h:(r_1,r_2)\rightarrow \R$,
$$
 \ip{f}{h}:=\int_{r_1}^{r_2}f(s)h(s)ds\,.
$$

Under the gauge transformation~\eqref{17III22.1}, $\kQ{1}{}$ transforms as
 \ptcheck{18XI22}
\begin{align}
        \int_{\secN} & \pi^A\left(r^{4}\partial_r \zhTB_{uA} \right)\,\sm
        \label{14VIII22.5}
        \\
        & \to
          \int_{\secN} \pi^Ar^{4}\partial_r \bigg( \zhTB_{uA}
          +\frac{1}{r^2} \Done (\xi^u)_A
        \nonumber
        \\
        &\qquad\qquad\qquad\qquad
        + (\ringh_{AB} \partial_u \xi^B +
        {\big(\alpha^2 + \frac{2m}{r^3}\big)} \partial_A \xi^u)
        \bigg)\,\sm
        \nonumber
\\
        &
          =
        \int_{\secN} \pi^A\bigg(r^{4}\partial_r  \zhTB_{uA}
        + {2}{r }\zspaceD^B \, \TS [\zspaceD_A\zspaceD_B\xi^u]
        -  {6 m \partial_A \xi^u}
        \bigg)\,\sm
        \nonumber
\\
        & =
        \int_{\secN}
         \big(
         \pi^A
         r^{4}\partial_r  \zhTB_{uA}
           + {6 m \xi^u} \zspaceD _A \pi^A
        \big)\,\sm\,.
         \nn
    \end{align}
 So on $S^2$, if $m=0$
  we see that  $\kQ{1}{}$ is gauge invariant, hence
$$
 \kQ{1}{}[\dt_{\secN_1}] = \kQ{1}{}[\dt_{\secN_2}]
 \quad
  \iff
  \quad
  \kQ{1}{}[\dt_{\tilde\secN_1}] = \kQ{1}{}[\dt_{\tilde\secN_2}]
 \,.
$$
If $m\neq 0$, $\kQ{1}{}$ is invariant under gauge transformations for which $\zspaceD _A \pi^A$ vanishes;  these generate rotations of $S^2$.

On the remaining topologies we have $\zspaceD_A \pi^A=0$, so that the charges  $\kQ{1}{}$ are gauge-invariant independently of whether or not the mass parameter $m$ vanishes.

Now, let $\psi_A$ denote (compare~\eqref{CHG28XI19.5}),
\begin{eqnarray}
         &
                 \psi_A:=
                 - 2r^4\partial_r \Big(\frac{1}{r^2}\spaceD_A\delta \beta  \Big)
           +
                    r^2
                   \partial_r \left(r^{-2}
                    \zspaceD^Bh_{AB}\right)
                 \,.
                  &
                            \label{CHG28XI19.7}
           \end{eqnarray}
Integrating~\eqref{CHG28XI19.5} in $r$ twice one obtains a representation formula for $\hBo_{uA}$:
\begin{eqnarray}
            h_{uA} (u,r,x^B)
            &
             =  & r^2 \Ichi_A(u,x^B)
 +  \frac{\Ipsi_A(u,x^B)}{r}
                 \label{CHG28XI19.6a}
\\ &  &
 -
                 r^2 \int_{r_1}^r \psi_A(u,s,x^B)\left(\frac{1}{3r ^3} - \frac{1}{3s^3}\right) ds
                 \,,
                            \nn
           \end{eqnarray}
%
with $\Ichi_A$ and $\Ipsi_A$ determined by $h_{uA} (u,r_1,x^B)$ and $\partial_r h_{uA} (u,r_1,x^B)$.


The part of \eqref{CHG28XI19.6a} involving $\hBo_{AB}$ can be viewed as the following map:
\begin{eqnarray}
         h_{AB}
         &\mapsto
         & -r^2 \int_{r_1}^r  s^2
                   \partial_s \left(s^{-2}
                    \zspaceD^Bh_{AB}\right)\left(\frac{1}{3r ^3} - \frac{1}{3s^3}\right) ds
                    \label{6III22.1a}
                    \\
         &=
         & -\frac{r^2}{3}\zspaceD^B
         \left[\int_{r_1}^r
                   \partial_s \left(s^{-2}
                    h_{AB}\right)
                    \left(\frac{s^2}{r^3} - \frac{1}{ s }\right) ds
                    \right]
                    \nonumber
                    \\
         &=
         & -\frac{r^2}{3}\zspaceD^B
         \Big[
                    h_{AB} (u,s,x^A)
                    \left(\frac{1}{r^3} - \frac{1}{ s^3 }\right)
                    \Big|_{r_1}^r
                   \phantom{\int_{r_1}^r}
                   \nonumber
                   \\
                   &&
         -
           \int_{r_1}^r
                   h_{AB}\left(\frac{2}{s r^3} + \frac{1}{ s^4 }\right) ds
                    \Big]
                 \,.
                            \nn
\end{eqnarray}
When $\delta \beta \equiv 0$ we thus obtain
\begin{align}
            h_{uA} (u,r,x^B)
            &
             =
              r^2 \Ichi_A(u,x^B)
 +  \frac{\Ipsi_A(u,x^B)}{r}
 \label{CHG28XI19.6b}
\\
 &  \quad
   +
          \zspaceD^B
                    h_{AB} (u,r_1,x^A)
                    \left(\frac{r }3  - \frac{r^2}{3r_1^3 }\right)
                  \nonumber
\\
 &  \quad
        +\frac{r^2}{3}
           \int_{r_1}^r
                   \zspaceD^Bh_{AB}\left(\frac{2}{s r^3} + \frac{1}{ s^4 }\right) ds
                 \,.
                            \nn
           \end{align}

For future use we will track the differentiability orders of the fields involved.
Denoting the Sobolev spaces over $\secN$ as $H_{k_{U}}$ for $\delta U^A\equiv - r^{-2} \ringh^{AB}h_{uB} $, and $ H_{k_{\gamma}}$ for $ \delta \gamma_{AB} \equiv \check h_{AB} =r^{-2}h_{AB} $,
Equation~\eqref{CHG28XI19.6b}  implies
\begin{equation}\label{25IV22.p1}
 k_{\gamma}\geq k_U+1
  \,.
\end{equation}
We emphasise that these spaces keep only track of the differentiability in directions tangent to $\secN$ at given $r$, with no information concerning the behaviour in the $r$-direction. But note that in our setup all fields on the interpolating initial data hypersurface are smooth in the $r$-variable.

\subsubsection{$\hBo_{uu}$}
 \label{ss29VII20.4}


To obtain the transport equation for the function $V$ occurring in the Bondi form of the metric, it turns out to be convenient to consider the expression for $ 2 G_{ur} + 2 U^A G_{rA} - V/r\, G_{rr} $:
 \begin{eqnarray}
            &&
               r^2 e^{-2\beta} (2 G_{ur} + 2 U^A G_{rA} - V/r\, G_{rr} )
                =
                R[\zhTBW]
                -2\zhTBW^{AB}  \Big[\spaceD_A \spaceD_B \beta
                \label{3X22.1}
\\
            &&\qquad
                + (\spaceD_A\beta) (\spaceD_B \beta)\Big]
               +\frac{e^{-2\beta}}{r^2 }\spaceD_A \Big[ \partial_r (r^4U^A)\Big]
               \nonumber
\\
            &&\qquad
                -\frac{1}{2}r^4 e^{-4\beta}\zhTBW_{AB}(\partial_r U^A)(\partial_r U^B)
                  - 2 e^{-2\beta} \partial_r V\,,
                  \nn
           \end{eqnarray}
(It follows directly from the definition of $G_{\mu\nu}$ and the Bondi parametrisation of the metric that $ r^2 e^{-2\beta}(2 G_{ur} + 2 U^A G_{rA} - V/r\, G_{rr} )$ can equivalently be written as $r^2 g^{AB}R_{AB}$; compare Appendix \ref{ss3X22.1}).
\ptcr{remove this, never needed nor used except for clarity, but has a term missing; alternatively add a commutator term}
 \rout{In vacuum
this can be rewritten as}
 \begin{eqnarray}
            &&
                 \partial_r(  V  -  \frac{ r^{2  }}{2}
                \spaceD_A U^A)
                =
                 \frac{e^{2\beta}}{2}
                 \Big\{
                R[\zhTBW]
                -2\zhTBW^{AB}  \Big[\spaceD_A \spaceD_B \beta
                + (\spaceD_A\beta) (\spaceD_B \beta)\Big]
                \label{27III2022.3}
 \\
                &&\qquad
                -\frac{1}{2}r^4 e^{-4\beta}\zhTBW_{AB}(\partial_r U^A)(\partial_r U^B)
                  -   2  \Lambda  r^{2}
                  \Big\}
                +
                 r\spaceD_AU^A
                  \,.
                   \nn
           \end{eqnarray}

Let $\zR _{AB} = \twoscsign \ringh _{AB}$ denote the Ricci tensor of the metric $\ringh_{AB}$. As $h_{AB}$ is $\ringh$-traceless we have
\begin{align}
 r^2\delta
               ( R[\zhTBW])|_{\zhTBW=\ringh}
  & =
   -\zspaceD^A\zspaceD_A( \ringh^{BC }h_{BC})+\zspaceD^A\zspaceD^B h_{AB}-\zR ^{AB}h_{AB}
\\
  & =  
  \zspaceD^A\zspaceD^B h_{AB}
  \,.\nn
\end{align}
Linearising~\eqref{27III2022.3}
 \ptcr{change crossref to  \eqref{3X22.1} if equation removed}
around a Birmingham-Kottler background thus gives
 \begin{align}
                 \partial_r( \delta V  -  \frac{ r^{2 }}{2}
                \zspaceD_A \delta U^A)
                = \
                &
                 \frac{1}{2}
                 \Big\{
                  \zspaceD^A\zspaceD^B\zhTB_{AB}
                -2\zhTBW^{AB}  \zspaceD_A \zspaceD_B \delta\beta\Big\}
                \label{27III2022.3b}
                \\
                &\quad
                +
                 r\zspaceD_A\delta U^A
                + 2(\myGauss - r^2  \Lambda) \delta \beta
                  \,.
                   \nn
           \end{align}
We note that since $\delta(G_{ur} + U^A G_{rA}) = \delta G_{ur}$, \eqref{27III2022.3b} is equivalent to the equation $r^2 \delta G_{ur}= r^2 \Lambda  \hBo_{ur} $.

In the $\delta\beta=0$ gauge, Equation~\eqref{27III2022.3b} provides another family of radially conserved   charges:
    \begin{align}
       \kQ{2}{}(\lambda):=  \int_{\secN}\lambda \bigg[\delta V - \frac{r}{2}\partial_r\bigg(r^2 \zspaceD^A \delta U_A\bigg)\bigg]\sm \,,
      \label{20VII22.1}
    \end{align}
where the functions $\lambda(x^A)$ are solutions of the equation
\begin{equation}
 \TS [\zspaceD_A \zspaceD_B \lambda] = 0\,.
    \label{24VII22.6}
\end{equation}
The only solutions of this equation on a torus or on a higher genus manifold are constants.  On $S^2$ such $\lambda$'s are linear combinations of $\ell=0$ or $\ell=1$ spherical harmonics~\cite{JezierskiPeeling}.
 We thus obtain another four-dimensional family of obstructions on $S^2$, and  a one-dimensional family of obstructions in the remaining topologies.

The conservation equation $\partial_r \kQ{2}{}=0$ is the consequence of
an identity, already observed in \cite{ChHMS}, of the form
\ptcheck{25IX22}
\begin{equation}
\delta G_{ur} - \frac 1{r}
  \zspaceD ^A \delta G_{rA}
  = \partial_r(....)
  \,,
   \label{5X22.1}
\end{equation}
 which can be derived as follows:
    \begin{align}
        \partial_r  \bigg[ \delta V
        &
        - \frac{ r}{2}\partial_r\bigg(r^2 \zspaceD^A \delta U_A\bigg)\bigg]
        =
        \partial_r \delta V
        - \partial_r \bigg(\frac{ r}{2}\partial_r\bigg(r^2 \zspaceD^A \delta U_A\bigg)\bigg)
        \label{24VII22.5}
\\
        &=
         \underbrace{\partial_r \delta V  -2  r\zspaceD_A \delta U^A -\frac{r^{2}}{2}\zspaceD_A\partial_r \delta U^A}_{= 1/2 \zspaceD^A\zspaceD^B \zhTB_{AB}}
        \nonumber
\\
        &\qquad
         \underbrace{-\frac{1}{2}( r^3 \partial_r^2 \zspaceD_A \delta U^A
         + 4 r^{2}\partial_r \zspaceD_A \delta U^A)}_{=r/2   \partial_r\big(\zspaceD^A\zspaceD^B\zhTB_{AB}\big)}
         \nonumber
\\
         & = \zspaceD^A\zspaceD^B\bigg[\frac{1 }{2} \zhTB_{AB}
         + \frac{ r}{2}  \partial_r\zhTB_{AB}\bigg]
         \nonumber
\\
        &  = \frac 1 2\partial_r \bigg( r \zspaceD^A\zspaceD^B \zhTB_{AB}\bigg) \,.
         \nn
    \end{align}
Hence
\begin{equation}
    \partial_r \kQ{2}{} = \frac{ 1}{2} \int_{\secN}\lambda
     \partial_r \big( r \zspaceD^A\zspaceD^B \zhTB_{AB}\big)\,\sm = 0.
\end{equation}
Under a gauge transformation the radial charge  $ \kQ{2}{}$ transforms as
 \ptcheck{25IX22 and the mass term added and checked 29X; corrected 13IV23}
    \begin{align}
       & \int_{\secN}\lambda\Big[ \delta V + \frac{ r}{2}\zspaceD^A \partial_r\hBo_{uA}\Big]\,\sm
    \label{20VII22.2}
   \\
   &\to
    \int_{\secN}\lambda\Big[
    \delta V
   +
   r\left(\twoscsign
    +\frac{1}{2}\TSzlap
    - \frac{3m}{r}\right)
    \big( \zspaceD_B\xi^B
      \big)
      \nonumber
      \\
      &\qquad
   - r^{2}
   \big[
   \zspaceD_B \partial_u\xi^B +
     {\big(\alpha^2 - \frac{m}{r^3}\big) }
    \TSzlap \xi^u
    \big]
      \nonumber
   \\
   &\qquad
    +\frac{ r}{2}\Big(\zspaceD^A \partial_r\hBo_{uA}
    + 2 r \big[
     \zspaceD_B \partial_u \xi^B +
    {\big(\alpha^2 - \frac{ m}{r^3}\big) }
    \TSzlap \xi^u
    \big]
    \Big)
    \Big]\,\sm
    \nonumber
\\
    &=\int_{\secN}\lambda\Big[ \delta V + \frac{ r}{2}\partial_r\zspaceD^A \hBo_{uA}\Big]\,\sm
      \nonumber
   \\
   &\qquad
    +\int_{\secN}\lambda\Big[   r
     \left(\twoscsign
      +\frac{1}{2}\TSzlap  -\frac{3m}{r} \right)
     \big(\zspaceD_B\xi^B
      \big)
    \Big]\,\sm\,.
    \nn
    \end{align}
    Taking $\zspaceD^A$ of~\eqref{24VII22.6} gives,
    \begin{equation}
        \zspaceD_B\TSzlap\lambda = -2\zR _{AB}\zspaceD^A\lambda\,,
    \end{equation}
    where $\zR _{AB}$ is the Ricci tensor of the metric $\ringh$.
    Inserting this into~\eqref{20VII22.2} gives
    \begin{align}
        \kQ{2}{}\mapsto \int_{\secN}\lambda\Big[ \delta V + \frac{ r}{2}\zspaceD^A \partial_r \hBo_{uA}\Big]\sm
        -3m \int_{\secN}\lambda \zspaceD_B\xi^B
      \,.
    \end{align}
Therefore, a) $\kQ{2}{}{}(\lambda)$ is gauge invariant  \underline{when $m=0$} for all $\lambda$'s satisfying \eqref{24VII22.6}, while b) \underline{when $m\neq 0$},  only $\kQ{2}{}{}(\lambda^{[1]})$ is gauge invariant.

\medskip

 As already pointed out, \eqref{24VII22.5} takes the form of a pointwise radial conservation law:
 $$\partial_r \chi = 0
 \,,
 $$ where
\begin{eqnarray}
    \chi
     & := &
       -\delta V + \frac{ r}{2}\partial_r\Big(r^2 \zspaceD^A \delta U_A\Big)+\frac{1}{2} r \zspaceD^A\zspaceD^B \zhTB_{AB}
       \label{24IX22.2}
\\
     &  = &
       -\delta V -  \frac{ r}{2}\partial_r  \zspaceD^A  h_{uA}
       +\frac{1}{2r}  \zspaceD^A\zspaceD^B h_{AB}       \,.
    \nn
\end{eqnarray}
We  note that \eqref{27III2022.3b} can be used to rewrite $\chi$ as (cf.\ \cite[Equation~(D.4)]{ChHMS})
\begin{equation}
    \chi = -r^2 \partial_r \hBo_{uu} + \zspaceD^A \hBo_{uA}\,.
    \label{27IX22.1}
\end{equation}

Under gauge transformations $\chi$ transforms as
 \ptcheck{13IV23; Corrected and rechecked}
\begin{equation}\label{24IX22.1}
  \chi \mapsto \chi
  -
   \frac{1}{2}  \underbrace{(\TSzlap +2\twoscsign)\TSzlap\TSxi^{\TSu}}_{=\mrL(\TSxi^u)}
  + 3m \zspaceD^B\xi_B
  \,.
\end{equation}
This shows that, on $S^2$, $\red{\chi^{\kerLp}=}\chi^{[\ge 2]}$
 can be made to achieve any desired value by
 a suitable choice of $(\TSxi^\TSu)^{[\geq 2]}$.
 For the remaining topologies this is the case for $\red{\chi^{\kerLp}=}\chi^{{[1^\perp]}}$ using $(\TSxi^\TSu)^{{[1^\perp]}}$.

\input{PointwiseCharge2}


Writing $H_{k_{V}}$ for the Sobolev space of the $\delta V$'s, \eqref{27III2022.3b} above implies
\begin{equation}\label{25IV22.p3}
 \mbox{$k_{\gamma}\geq k_V+2$ and $k_U\geq k_V + 1$.}
\end{equation}

\subsubsection{$\partial_u \hBo_{AB}$}
 \label{ss29VII20.3}

We continue with  $\partial_u \hBo_{AB}$, as determined from \cite[Equation~(32)]{MaedlerWinicour}:
%
\begin{align}\label{eq:30III22.1b}
    & \partial_r
    \Big[
     r  \partial_u \zhTBW_{AB}
     	 - \frac{1}{2}  V   \partial_r \zhTBW_{AB}
     	 -  \frac{1}{2r}  V    \zhTBW_{AB}
     \Big]
\\
         &\quad =
     	 -   \frac{1}{2} \partial_r(V /r)   \zhTBW_{AB}
          -
       \frac{1}{r}\TS\Big[
      e^{2\beta}  R[\gamma]_{AB} -2e^{\beta} \spaceD_A \spaceD_B e^\beta
      \nonumber
\\
 &\quad\quad
      +   \zhTBW_{CA} \spaceD_B[ \partial_r (r^{2}U^C) ]
          - \frac{1}{2} r^4 e^{-2\beta}\zhTBW_{AC}\zhTBW_{BD} (\partial_r U^C) (\partial_r U^D)
          \nonumber \\
       &\quad\quad
       +
               \frac{r^2}{2}  (\partial_r \zhTBW_{AB}) (\spaceD_C U^C )
              +r^2 U^C \spaceD_C (\partial_r \zhTBW_{AB})
                \nonumber
                \\
       &\quad\quad
        -
	r^2 (\partial_r \zhTBW_{AC}) \zhTBW_{BE} (\spaceD^C U^E -\spaceD^E U^C)
       +  \Lambda  e^{2\beta} g_{AB}
       -8\pi e^{2\beta}T_{AB}
       \Big]
       \,.
       \nonumber
\end{align}

The linearisation of \eqref{eq:30III22.1b} around a Birmingham-Kottler background
in vacuum
 reads, keeping in mind that $\TS [R[\gamma]_{AB}]=0$
 in dimension two,
\begin{align}
    0
    &= \frac{1}{r} \TS[\delta G_{AB}]
    \label{eq:31III22.3p0}
\\
    &
    =
    \partial_r \Big[
    r \partial_u \zhTB_{AB}
     - \frac{ 1}{2}  V  \partial_r \zhTB_{AB}
     -  \frac{1}{2 r}  V   \zhTB_{AB}
     -
            r \TS \big[\zspaceD_A   \zhTB_{uB}\big]
     \Big]
      \nonumber
\\&\quad
    \underbrace{
        +\frac{1}{2}  \partial_r  (  V/r)
        }_{ m r^{-2} - \alpha^2  r}   \zhTB_{AB}
         -
          r^{-1}
          \Big(
          2\zspaceD_A \zspaceD_B \delta\beta
           +
           r \TS \big[\zspaceD_A   \zhTB_{uB}
       \big]
       \Big)
        \,.
\nonumber
\end{align}
Integrating this equation gives
\begin{align}
     s  \partial_u \zhTB_{AB}
      \Big|_{r_1} ^r
      &=
       \Big[
      \frac{1}{2}  V  \partial_s \zhTB_{AB}
     +   \frac{V }{2s}       \zhTB_{AB}
     +
            s \TS \big[\zspaceD_A   \zhTB_{uB}\big]
     \Big]_{r_1} ^r
\label{31III22.3p1}
\\ &\quad
     	 + \int_{r_1}^{r}
     \Big[
       -\frac{1}{2}  \partial_s  (  V/s)  \zhTB_{AB}
                  \nonumber
           \\
           &\qquad\qquad\quad
         +
          1/s
          \Big(
          2\zspaceD_A \zspaceD_B \delta\beta
           + s \TS \big[\zspaceD_A   \zhTB_{uB})
       \big]
       \Big)
       \Big]
        ds
        \,.\nonumber
       \end{align}
Denoting $H_{k_{\partial_u \gamma}}\ni\partial_u \zhTB_{AB}$, \eqref{31III22.3p1} implies
\begin{equation}
\label{25IV22.p4}
 \mbox{$ k_{\beta}\geq k_{\partial_u \gamma}+2$ and $k_U\geq k_{\partial_u \gamma} + 1$.}
\end{equation}
When $\delta \beta =0$,
 using  \eqref{CHG28XI19.6b} in  the last term of \eqref{31III22.3p1}
 leads to
\begin{align}
      \partial_u \hBo_{AB}
      &=
      r \frac{\partial_u \hBo_{AB}|_{r_1}}{r_1}
     -\frac{r}{2}\Big[
        V  \partial_r \zhTB_{AB}
     +   \frac{1}{s}  V   \zhTB_{AB}
 \label{24VII22.61}
     \\
     &\qquad\qquad\qquad
     -
         \Big(\frac{ 1
            }{s^2}-\frac{1
            }{r^2}\Big)TS\Big[\zspaceD_A \zspaceD^C \hBo_{BC}
          \big]
     \Big]\Big|_{s=r_1}
     \nonumber
\\
 &
     +
        2 r \TS \Big\{  \frac{r_1^{ 2}(r_1^2-r^2)}{4r^2}
           \partial_s \zspaceD_A \zhTB_{uB}|_{r_1}
           -r_1\zspaceD_A \zhTB_{uB}|_{r_1}
            \nonumber
\\
&
       + r
          \zspaceD_A \Big[
          r^2 \Ichi_B(u,x^C)
        +  \frac{\Ipsi_B(u,x^C)}{r}
        \nonumber
        \\
        &\qquad\qquad\qquad
        -\frac{r^2}{3}
        \zspaceD^C
                    \hBo_{BC} (u,r_1,x^A)
                    \left(\frac{1}{ r_1^3 }-\frac{1}{r^{3}} \right)
                     \Big]
                      \Big\}
                     \nonumber
\\
&
  +
      r \Bigg[
     \frac{V }{2r}\Big[
         r \partial_r \zhTB_{AB}
     +    \zhTB_{AB}
     \Big]
     \nonumber
\\
&
            + \int_{r_1}^r
     \Big[
       \big(
       \frac{ {\alpha^2}}{s} - \frac{m}{s^4}
       \big)
        \hBo_{AB}
       +
          \Big(\frac{1}{ 3s r^{2}}
    +\frac{2 r}{3 s^4}\Big) \TS \big[
          \zspaceD_A   \zspaceD^C \hBo_{BC}
       \big]
       \Big]
        ds\Bigg]
        \,.
        \nonumber
\end{align}

Recall that
$$
 V = r \twoscsign - {\alpha^2} r^3  {-2m}
  \,.
$$
Let us write $\tkfbd$ for terms known from ``boundary data at $r_1$''.  We rewrite \eqref{24VII22.61} as
\begin{align}
\label{14VIII22.1}
      \partial_u &  h_{AB}
      =
      \tkfbd
      +
      r \Bigg[
     \frac{V }{2r}\Big[
         r \partial_r \zhTB_{AB}
     +    \zhTB_{AB}
     \Big]
      + \int_{r_1}^r
     \Big[
    \big( \frac{\alpha^2}{s} -\frac{m}{s^4} \big)
    \hBo_{AB}
\\
&\hspace{4cm}
       +
          \big(\frac{1}{ 3s r^{2}}
    +\frac{2 r}{3 s^4}\big)
 \underbrace{
 \TS \big[
          \zspaceD_A   \zspaceD^C \hBo_{BC}
       \big]
}_{=:P h_{AB}}
       \Big]
        ds
        \Bigg]
        \nonumber
\\
      &=
     \frac{r( \twoscsign r- {\alpha^2} r^3 -2m  ) }{2 }\Big[
         \partial_r(r^{-2} h_{AB})
     +   \frac{1}{r^3}    h_{AB}
     \Big]
     \nonumber
\\
&\quad
            +  \int_{r_1}^r
     \Big[
       \underbrace{(\frac{{\alpha^2}r }{s} - \frac{m r}{s^4})
        }_{\overadd{1,0}{\psi}} \hBo_{AB}
       +
          \underbrace{
           \big(\frac{1}{ 3s r }
    +\frac{2 r^2 }{3 s^4}
 \big)
    }_{\overadd{1,1}{\psi}(s,r)} P h  _{AB}
       \big]
       \Big]
        ds
       + \tk
        \,. \nonumber
\end{align}
For further use it is convenient to separate the terms involving $\alpha$ and $m$ from the remaining ones:
\begin{align}
\label{14VIII22.3}
      \partial_u &  h_{AB }
       =
     \frac{\twoscsign }{2 }\Big[
        \partial_r h_{AB}
     - \frac{1}{r }    h_{AB}
     \Big]   +  \int_{r_1}^r
           \bigg(\frac{1}{ 3s r }
    +\frac{2 r^2 }{3 s^4}\bigg)
     P h  _{AB}
     \,
        ds
\\
 &
     - (\frac{  {\alpha^2} r^2 }{2 }+\frac{ m}r )\Big[
        \partial_r h_{AB}
     - \frac{1}{r }    h_{AB}
     \Big]   +  \int_{r_1}^r
        (\frac{{\alpha^2}r }{s}   -\frac{mr}{s^4})  \hBo_{AB}
         \,
        ds
        + \tk
        \,.
        \nn
\end{align}

\input{duhABcharge}
\subsection{The remaining Einstein equations}
\label{sec:25VII22.1}

Let us start by recalling that the Einstein equations
$$
 \mcE_{\mu\nu}:=G_{\mu\nu} +  \Lambda  g_{\mu\nu} - 8 \pi T_{\mu\nu}
$$
can be split as
\begin{eqnarray}
&
   \mcE^u{}_{ \mu} =0\,, \qquad \mcE_{AB} - \frac{1}{2} g^{CD} \mcE_{CD} g_{AB}=0
   \,,
      &
       \label{7V22.1}
\\
      &
   g^{CD} \mcE_{CD} =0
    \,, &
       \label{7V22.2}
\\
&
  \partial_r(r^{2} e^{2\beta} \mcE^r{}_u )=0
  \,,
  \qquad
  \partial_r(r^{2} e^{2\beta} \mcE^r{}_A) =0
  \,,
       \label{7V22.3}
\end{eqnarray}
and the following holds (cf., e.g., \cite[Section~3]{MaedlerWinicour}): Suppose that \eqref{7V22.1} holds on a null hypersurface $\mcN$ and that
\begin{equation}\label{27IX22.1a}
   \partial_u \mcE^u{}_\mu|_{\mcN} = 0
    \,.
\end{equation}
Then a) \eqref{7V22.2} is satisfied automatically on $\mcN$, and b) the equations $\mcE^r{}_u|_{\mcN} =\mcE^r{}_A|_{\mcN} =0$ will hold if they are satisfied  at one single value of $r$. This follows from the observation that, in Bondi coordinates, we have the identity
\begin{equation}\label{27IX22.2}
  \nabla_\mu \mcE^\mu{}_\nu =
   \frac{1}{\sqrt{|\det g|}} \partial_\mu (\sqrt{|\det g|} \mcE^\mu{}_\nu )
   +
   \frac{1}{2}\mcE_{\mu\sigma} \partial_\nu g^{\mu \sigma}
   \,.
\end{equation}
In the current context this implies, using $\partial_\nu g^{u\mu} =0= \partial_u g^{\mu\sigma}$ and the divergence identity,
\begin{equation}\label{27IX22.2asdf}
 0=
   \frac{1}{\sqrt{|\det g|}} \partial_\mu (\sqrt{|\det g|} \delta  \mcE^\mu{}_\nu )
   +
   \frac{1}{2}\sum_{
   \mu, \sigma \ne u}
    \delta \mcE_{\mu\sigma}
     \underbrace{
       \partial_\nu g^{\mu \sigma}
       }_{0 \ \text{if} \ \nu = u}
   \,.
\end{equation}
Since   $\delta \mcE^u{}_\mu = -\delta \mcE_{r\mu}$, when the main equations \eqref{7V22.1} are satisfied
 \eqref{27IX22.2asdf}  becomes
\begin{eqnarray}
0
   &  = &
   \frac{1}{\sqrt{|\det g|}} \partial_\mu (\sqrt{|\det g|} \delta  \mcE^\mu{}_\nu )
   +
   \frac{1}{2}\sum_{
   \mu, \sigma \not \in \{ u,r\}}
    \delta \mcE_{\mu\sigma}
       \partial_\nu g^{\mu \sigma}
       \label{28IX22.1}
\\
   &  = &
   \frac{1}{\sqrt{|\det g|}} \partial_\mu (\sqrt{|\det g|} \delta  \mcE^\mu{}_\nu )
   +
   \frac{1}{2}
    \delta \mcE_{AB}
     \underbrace{
       \partial_\nu g^{AB}
       }_{0 \ \text{if} \ \nu = u}
   \,.
   \nn
\end{eqnarray}
In what follows we assume
\begin{equation}\label{28IX22.4}
\delta\mcE^u{}_\mu|_\mcN =0 = \partial_u\delta\mcE^u{}_\mu|_\mcN
\,.
\end{equation}
We review the standard argument, which is a somewhat simplified version of what needs to be done in our gluing.
Setting $\nu =r  $ in \eqref{28IX22.1} one obtains immediately
\begin{eqnarray}
  0 = - \frac{1}{ {r}}  g^{AB} \delta \mcE_{AB} |_\mcN
   \,,
   \label{28IX22.2}
\end{eqnarray}
hence the linearisation of \eqref{7V22.2} holds on $\hyp$. So the linearised version of the second equation in \eqref{7V22.1} is equivalent to
 $\delta \mcE_{AB} |_\mcN=0$. Then  $\delta \mcE^A{}_{B} |_\mcN=g^{AC}\delta \mcE_{CB} |_\mcN=0$, and  \eqref{28IX22.1} with $\nu = A$ becomes
\begin{eqnarray}
0
   &  = &
     \frac{1}{r^2} \partial_r (r^2 \delta  \mcE^r{}_A ) |_\mcN
   \,,
   \label{28IX22.1qr}
\end{eqnarray}
as desired. So, if $\mcE^r{}_A $ vanishes for some $r$ on $\mcN$, it will vanish throughout $\mcN$. Now,
   \eqref{28IX22.1} with  $\nu =u $  reduces to
\begin{eqnarray}
0
   &  = &
   \frac{1}{r^2} \partial_r (r^2 \delta  \mcE^r{}_u ) |_\mcN
    +\frac{1}{r^2} \partial_A(r^2 \delta\mcE^A{}_{u}) |_\mcN
   \,.
   \label{28IX22.1qa}
\end{eqnarray}
and what has been said about  $\delta\mcE^A{}_{u}|_\mcN$  gives the
result.

The above means that there is no need to integrate in $r$ these Einstein equations which have not been discussed so far, namely $g^{AB}\mcE_{AB}=0$,  $\mcE_{uA}=0$ and $\mcE_{uu}=0$, when \eqref{28IX22.4} holds. Indeed, once the already analysed equations \eqref{7V22.1} are solved, together with their first $u$-derivatives,
 the whole set of Einstein equations will be solved by ensuring that $\mcE^r{}_{A}=0 =\mcE^r{}_{u} $ holds at one value of $r$; this is equivalent to ensuring $\mcE_{uA}=0 =\mcE_{uu} $ at one value of $r$.

The same scheme applies to the set of equations obtained by further differentiating the Einstein equation in $u$ an arbitrary number of times.

\subsubsection{$\partial_u\partial_r\hBo_{uA}$}
\label{ss3VIII22.9}

The equations $ \mcE_{u A}=0$ are too long to be usefully displayed here.
Their linearisation $ \delta \mcE_{u A}\equiv - \delta \mcE^r{}_A +(\epsilon-{\alpha^2} r^2-\frac{2m}{r})\delta \mcE_{rA}$  in vacuum reads
\begin{align}
\label{9XI20.t1}
0= 2 \delta \mcE_{u A}
&=
\frac{1}{ r^2} \Bigg[
\zspaceD^{B}\zspaceD_{A}{h}_{uB}
-\zspaceD^{B}\zspaceD_{B}{h}_{uA}
+\partial_{u} \zspaceD^{B}{h}_{A B}
\\
&\qquad\quad
-r^{2}\bigg(
\left(\twoscsign-r^{2}\alpha^{2}-\frac{2m}{r}\right)\partial_{r}^{2}h_{uA}
 + (2\alpha^{2} +\frac{4m}{r^3})h_{uA}\nn
 \\
 &\qquad\qquad\qquad
  - r^{2}\partial_{r}\partial_{u}\left(\frac{h_{uA}}{r^{2}}\right)
 + \partial_{r}\zspaceD_{A}{h}_{uu}
\bigg)\Bigg]
\,.\nonumber
\end{align}
 This equation is satisfied both by $\dt_{\secN_1}$ and $\dt_{\secN_2}$ in vacuum.

 Assuming $\delta G_{rA}=0$, using the transport equation \eqref{24VII22.1}
 to eliminate $\partial_r^2\zhTB_{uA}$ and
the identity \eqref{27IX22.1} to eliminate $\partial_r\hBo_{uu}$, we can rewrite \eqref{9XI20.t1} as
\begin{align}
- r^{4}\partial_{r}\partial_{u}\left(\frac{h_{uA}}{r^{2}}\right)
& =
    \zspaceD^{B}\zspaceD_{A}{h}_{uB}
    -\zspaceD^{B}\zspaceD_{B}{h}_{uA}
    +\partial_{u} \zspaceD^{B}{h}_{A B}
\label{26IX22.w1}
\\
&\quad
    -r^{2}\bigg(
    \left(\twoscsign-r^{2}\alpha^{2} - \frac{2m}{r}\right)\partial_{r}^{2}h_{uA}
    \nn
    \\
    &\qquad\qquad
    + (2\alpha^{2}+\frac{4m}{r^3})h_{uA}
    + \partial_{r}\zspaceD_{A}{h}_{uu}
    \bigg)
\nonumber
\\
&
    =
    \zspaceD^B
    \Big[
    -2 \TS[\zspaceD_B \hBo_{uA}]
    +\partial_{u}  {h}_{A B}
    \nn
    \\
    &\qquad\quad
    + (\alpha^{2}r^2-\twoscsign + \frac{2m}{r})r^2
                   \partial_r \left(r^{-2}
                     h_{AB}\right)
                     \Big]
     + \zspaceD_A \chi
      \, .
\nn
\end{align}
Using the fact that $\partial_r\chi=0$ we obtain,   for any $\pi^A(x^B)$ satisfying $\TS[\zspaceD_A\pi_B]=0$,
\begin{equation}
 \partial_r \int_{\secN} \pi^A r^4\partial_u\partial_r\zhTB_{uA}\sm
 \equiv
  \partial_r \kQ{1,1}{}(\pi^A)=0\,,
\label{6X22.w1}
\end{equation}
where we recall from \eqref{26IX22.22hi} that for $0\leq i\in \N$,
\begin{equation}
 \kQ{1,i}{}(\pi^A):=\int_{\secN} \pi^A r^4\partial^i_u\partial_r\zhTB_{uA}\sm \,.
\label{6X22.w3}
\end{equation}
%
Clearly, by $u$-differentiating \eqref{26IX22.w1}, we conclude that $\partial^i_u \delta \mcE_{uA} = 0$ implies
\begin{equation}
 \partial_r\kQ{1,i+1}{}(\pi^A)=0 \,.
\label{6X22.w3b}
\end{equation}
for $i\geq 0$.

Denoting $H_{k_{\partial_u U}}\ni\partial_u \delta U^A$, \eqref{9XI20.t1} implies
\begin{equation}
\label{3VII22.w3}
 \mbox{$ k_{U} \geq k_{\partial_u U}+2$, $ k_{\partial_u\gamma} \geq k_{\partial_u U}+1$ and $k_V\geq k_{\partial_u U} + 1$.}
\end{equation}

\subsubsection{$\partial_u \hBo_{uu}$}
\label{ss3VIII22.8}

The equation $\mcE_{uu}=0$ is likewise too long to be usefully displayed here.
Its linearised version is shorter and, in vacuum, can be rewritten as an equation for the transverse derivative $\partial_u (r h_{uu}-  \zspaceD^{A} {h}_{u A})$:
\begin{align}
   0  & =
     2 \delta \mcE_{uu}
 \label{13VIII20.t3}
\\
 &     =
  \frac{1}{ r^2}\Big[2\Big(
      \partial_{u}
    +\big
    ({\alpha^2  r^2-\twoscsign + \frac{2m}{r}}
    \big)
     \partial_{r}
    + \frac{3m}{r^2} -\frac{\twoscsign}{r}
    \Big)
     \zspaceD^{A} {h}_{u A}
      \nonumber
 \\
  &\quad
    -  \zspaceD^{A} \zspaceD_{A} {h}_{u u}
-\big
    ({\alpha^2  r^2-\twoscsign + \frac{2m}{r}}
    \big) \Big(\frac{\zspaceD^{A} \zspaceD^{B} {h}_{A B}}{r^2}\Big)
     \nonumber
 \\
  &\quad
  -2 r \partial_{u} {h}_{u u}
-2 \big
    ({\alpha^2  r^2-\twoscsign + \frac{2m}{r}}
    \big)
     \partial_{r}(r {h}_{u u})
\Big]
\nn
\end{align}
%

 This must be satisfied by $\dt_{\secN_1}$ and $\dt_{\secN_2}$ when the linearised vacuum Einstein equations hold.

Denoting $H_{k_{\partial_u V}}\ni\partial_u \delta V$, \eqref{13VIII20.t3} implies
\begin{equation}
\label{3VII22.w4}
 \mbox{$ k_{U} \geq k_{\partial_u V}+1$, $k_V\geq k_{\partial_u V} + 2$ and $ k_{\gamma} \geq k_{\partial_u V}+2$.}
\end{equation}

\subsection{Regularity}
  \label{ss25IV22.1}

The regularity analysis carried-out so far is summarised by the following inequalities for the regularity of the  metric components:
\ptcheck{25IV22 together; this assumes that there is no loss of regularity when integrating in $r$ since we have integrated by parts; but boundary terms? $k_{\partial_r U}$? }
\begin{align}
    h_{uA} \text{ equation}: \quad & k_{\beta} \geq k_U+1\,,\quad k_{\gamma}\geq k_U +1 \,,
    \label{24IV22.2a}
\\
    h_{uu} \text{ equation}: \quad & k_{\gamma} \geq k_V + 2\,, \quad k_U \geq k_V +1 \,,
    \label{24IV22.2b}
\\
    \partial_u h_{AB} \text{ equation}: \quad & k_{\beta} \geq k_{\partial_u \gamma} + 2 \,, \quad k_U \geq k_{\partial_u \gamma} + 1\,,
    \label{24IV22.2c}
\\
    \partial_u\partial_r h_{uA} \text{ equation}: \quad & k_{U} \geq k_{\partial_u U} + 2\,, \quad k_V \geq k_{\partial_uU} +1
     \,,
    \label{24IV22.2d}
    \\
    & k_{\partial_u\gamma} \geq k_{\partial_u U} + 1 \,, \nn
\\
   \partial_u h_{uu} \text{ equation}: \quad & k_{U} \geq k_{\partial_uV} + 1\,, \quad k_V \geq k_{\partial_u V} + 2 \,,
    \label{24IV22.2e}
\\
 &  k_{\gamma} \geq k_{\partial_u V} + 2\,.\nn
\end{align}
A consistent scheme for the linearised equations will thus be obtained if we choose any field $h_{AB}$ such that $h_{AB}(r,\cdot) \in \Hgamma$,  for all  $r\in[r_1,r_2]$, with $k_\gamma \ge 4$ and
\begin{align}\label{25IV22.p6}
  k_\beta &= k_\gamma
  \,,\quad
  k_U = k_\gamma -1
  \,,\quad
  k_V = k_\gamma -2
  \,,\quad
  k_{\partial_u U} = k_{\gamma} - 3
  \,,\\
  k_{\partial_u V} &= k_\gamma - 4
  \,,\quad
  k_{\partial_u \gamma}  = k_\gamma -2
  \, .
\end{align}
Note that the question of regularity of $r$-derivatives of $\gamma$ has been swept under the rug using integration by parts. This question will need to be addressed when dealing with the nonlinear problem.

The regularity properties of the metric will be compatible with  gauge transformations~\eqref{24IX20.1}-\eqref{24IX20.23} if we assume, using obvious notation,
\begin{align}
        h_{uA} \text{ equation}: \quad &
        k_{\xi^u} \geq k_U+ 3\,,\quad
        k_{\partial_u\xi^u} \geq k_U+1\,,
    \label{24IV22.3a}
    \\
    &
        k_{\xi^A}\geq k_U + 2 \,,\nn
\\
    h_{ur} \text{ equation}: \quad &
        k_{\partial_u\xi^u} \geq k_\beta\,,\quad
        k_{\xi^A}\geq k_{\beta} + 1\,,
    \label{24IV22.3b}
\\
    h_{uu} \text{ equation}: \quad & k_{\partial_u\xi^u}\geq k_{V} + 2\,, \quad k_{\partial_u\xi^B}\geq k_V +1 \,,
    \label{24IV22.3c}
    \\
    & k_{\xi^u}\geq k_V +2\,, \quad k_{\xi^B}\geq k_V + 1\,,\nn
\\
    h_{AB} \text{ equation}: \quad & k_{\xi^A}\geq k_{\gamma} + 1\,,\quad k_{\xi^u} \geq k_{\gamma} + 2 \,,
    \label{24IV22.3d}
\\
    \partial_u h_{AB} \text{ equation}: \quad & k_{\partial_u\xi^A}\geq k_{\partial_u\gamma} + 1\,,\quad k_{\partial_u\xi^u} \geq k_{\partial_u\gamma} + 2 \,.
    \label{24IV22.3e}
\end{align}
A scheme consistent with \eqref{25IV22.p6}-\eqref{24IV22.3e} results by choosing
\begin{equation}\label{1VII23.2}
   k_{\xi^u}=k_\gamma+2\,,\quad
        k_{\partial_u\xi^u}=k_\gamma\,,\quad
        k_{\xi^A}= k_\gamma+1 \,,
\quad
        k_{\partial_u\xi^A}= k_\gamma-1 \,.
\end{equation}

\subsection{Further $u$-derivatives}
\label{sec:28VII22.2}

The representation formula for higher $u$-derivatives of the linearised metric components can be obtained by taking the $u$-derivatives of the existing equations. This gives, for $i\geq 0$, representation formulae of the form
\ptcheck{2VII22, in all dimensions in the file NullGluing}
\begin{align}
\label{18VI22.3}
 \partial_u^i \hBo_{AB}  &=
    \overadd{i}{\Psi}_{AB}(u,r, x^A)
     + \sum_{0\leq j+k\leq i,k\ne i} \overadd{i,j,k}{\psi}(r)\partial_r^j P^k \hBo_{AB}	
     \\
     &\quad
     +
      \int_{r_1}^{r}
       \sum_{j=0}^i
       \overadd{i,j}{\psiP} (s,r)P^j \hBo_{AB}
        \,
        ds
        \,,\nn
        \\
        \label{18VI22.1}
  \partial_u^i  \zhTB_{uA}  &=
    \overadd{i}{X}_{A}(u,r, x^A)
     	+
     \zspaceD^B
     \Big[
      \sum_{0\leq j+k\leq i,k\ne i}
           \overadd{i,j,k}{\chi}(r)\partial_r^j P^k \hBo_{AB}
           \\
           &\qquad\qquad\qquad\qquad\qquad
     	+
      \int_{r_1}^{r}
       \sum_{j=0}^{i }
       \overadd{i,j}{\chi} (s,r)P^j \hBo_{AB}
        \, ds
        \Big]
        \,,
         \nn
\end{align}
where $\overadd{i}{X}$ and $\overadd{i}{\Psi}$ depend only on data at $r_1$;
recall that $P$ denotes the operator
\begin{equation}\label{16V22.1}
  Ph_{AB}  = \TS [D_A D^C h_{BC}]
  \,.
\end{equation}
\input{zeroOrder}

We note that the terms involving $\overadd{i,j,k}{\psi}$ and $\overadd{i,j,k}{\chi}$ are innocuous at $r=r_2$, as they are determined by known boundary data at $r_2$. However, they are essential for the induction procedure for $r\ne r_2$, as they contribute to  the key terms $\overadd{i,j}{\psi}$ and $\overadd{i,j}{\chi}$ in the iteration.
This implies in particular that  the explicit form of
$\overadd{i,j,k}{\psi}$
etc.\ with the highest index $i=\ell$ is \emph{not} needed   when  gluing at order $\ell$.

Again by induction (cf.~Appendix~\ref{App14VIII22.2}), one shows the following:

\begin{enumerate}
  \item All the integral kernels  in \eqref{18VI22.3}-\eqref{18VI22.1}, depending upon  $r$ and $s$, are  polynomials in $s^{-1}$ with coefficients depending upon $r$;

\item  \underline{when $m=0$}, $ \overadd{i,0}{\psi}$ is  proportional to ${\alpha^2} s^{-1}$.

\item  The highest power of $1/s$ in $  \overadd{1,j}{\psi}$ is $s^{-4}$.

\item  The highest power of $1/s$ in $  \overadd{i,j}{\psi}$  with $1\le j\le i$
 is $s^{-(i+3)}$ \underline{when $m=0$}, and  this power is not larger than $s^{-2i +j - {3}}$  when $m\ne 0$; cf. Lemma~\ref{Ll6XI22.1}, Appendix~\ref{ss16XI22.2}.

 \item  It holds that
  \ptcheck{15VIII}
 \begin{equation*}
     \overadd{i+1,i+1}{\psiP} (s,r)
    =
    \int_s^r\overadd{i,i}{\psiP} (y,r)\overadd{1,1}{\psiP} (s,y)\,dy\,,
    \end{equation*}
with
\begin{equation*}
\overadd{1,1}{\psiP} (s,r)=\frac{2 r^2}{3 s^4}+\frac{1}{3 r s}
     \,,
 \end{equation*}
independently of $m$.

\item  The highest power of $1/s$ in  $  \overadd{i,j}{\chi}$ with $0\le j\le i$
 is $s^{-(i+4)}$\underline{when $m=0$}, and this power is not larger than $s^{-2i+ j-4}$ \underline{when $m\neq 0$}.

\end{enumerate}

In what follows we will often use the notation
\begin{align}
    \hat{\kappa}_i(s):=\frac{1}{s^i}\,.
\end{align}
We have collected the explicit formulae for all the integral kernels appearing in \eqref{18VI22.3}-\eqref{18VI22.1}, and needed for $\Ctwo$-gluing,  in Appendix~\ref{App14VIII22.2}.

\input{qabtt}

\section{Gluing up to gauge}
 \label{s12I22.1}

We now present a scheme for matching, up-to residual gauge,   the linearised fields
\begin{equation}\label{19XII22.1}
\{\hBo_{\mu\nu},
 \partial_u\hBo_{\mu\nu}\, \ldots\, \partial_u^k \hBo_{\mu\nu}\}
\end{equation}
in Bondi gauge,   with $2\le k < \infty$.
 We will assume, for simplicity,  that each of the fields $ \partial_u^i \hBo_{\mu\nu}\big|_{\{u=0\}}$, $0\le i \le k$, is smooth. The collection of fields of this differentiability class will be denoted by $\Ck$.

Let $0\le r_0<r_1<r_2<r_3\in \R $.
Consider two sets of vacuum linearised gravitational fields  in Bondi gauge,
 of  $\Ck$-differentiability class,
  defined in spacetime neighbourhoods of $\mcN_{(r_0,r_1]} $ and $\mcN_{[r_2,r_3)} $. Let us denote by $\secN_{1}$ the section of $\mcN_{(r_0,r_1]}$ at $r=r_1$. The linearised gravitational field near $\mcN_{(r_0,r_1]}$
 induces a set of Bondi cross-section data on $\secN_1$, which we denote
  as $\dt_{\secN_1}$. Similarly, we denote by $\secN_{2}$ the section of $\mcN_{[r_2,r_3)}$ at $r=r_2$ and the induced gluing data by $\dt_{\secN_2}$. Let us also denote by $\tilde{\secN}_1$ (resp. $\tilde{\secN}_2$) the codimension-two section obtained by
 gauge-transforming $\secN_{1}$ (resp. $\secN_2$) using arbitrary gauge fields $\okxi{1}^{\mu}$ (resp. $\okxi{2}^{\mu}$), the associated gluing data by $\tilde{\dt}_{\tilde{\secN}_1}$ (resp. $\tilde{\dt}_{\tilde{\secN}_2}$) and the outgoing null hypersurface on which it lies by $\tmcN _{(r_0,r_1]}$ (resp.
 $\tmcN_{[r_2,r_3)}$).

Of course, in the linearised gluing the initial hypersurface $\mcN_{(r_0,r_3)}$ does not change, thus $\tmcN_{(r_0,r_3)}=\mcN_{(r_0,r_3)}$ as a set, but the Bondi coordinates on either $\mcN_{(r_0,r_1]}$ or on $\mcN_{[r_2,r_3)}$ need to be ``infinitesimally deformed'' both in transverse and in tangential directions.    We use the symbol $\tmcN $ to emphasise the infinitesimal adjustment of  Bondi coordinates, as an adjustment of $\mcN_{(r_0,r_1]}$ or $\mcN_{[r_2,r_3)}$ is generically needed when passing to the nonlinear gluing both in our case and in~\cite{ACR3}.

 The  goal
is to glue $\tilde{\dt}_{\tilde{\secN}_1}$ and $\tilde{\dt}_{\tilde{\secN}_2}$
 along $\tmcN _{[r_1,r_2] }$ so that the resulting linearised field on $\tmcN _{(r_0,r_3)} $ provide smooth characteristic data for Einstein equations
 together with a matching of $k$ transverse derivatives.
Indeed, we claim:

\begin{Theorem}
 \label{T4XII22.2}
A  $\Ck$-linearised vacuum data set on $\mcN_{(r_0,r_1]}$ can be smoothly glued to another  such set on $\mcN_{[r_2,r_3)}$  if and only if the obstructions listed in Tables~\ref{T17XI22.1}-\ref{T17XI22.2} are satisfied.
\end{Theorem}

The rest of this section is devoted to the proof of this theorem.

\smallskip

Let $\interph _{AB}$ be any symmetric traceless tensor field  defined on a neighbourhood of $\mcN_{[r_1,r_2]}$ which interpolates between the original fields $ \hBo_{AB}|_{\mcN_{(r_0,r_1]}}$  and $  \hBo_{AB}|_{\mcN_{[r_2,r_3)}}$,
so that the resulting field on $\mcN_{(r_0,r_3)}$ is as differentiable as the original fields.
When attempting a $\Ck$-gluing, we can add to $\interph _{AB} $  a field $\wh_{AB}|_{[r_1,r_2]}$
which vanishes  smoothly
(i.e.\ together with  $r$-derivatives of all orders)
at the end cross-sections $\{r_1\}\times \secN$ and  $\{r_2\}\times \secN$ without affecting the gluing of $h_{AB}$.
To take into account the gauge freedom, let $\phi(\tdr)\ge 0$ be a smooth function which equals $1$ near $\tdr=r_1$ and equals $0$ near $\tdr=r_2$. Let $\kxi{1}^u$ and $\xiA{1}$ be gauge fields used to gauge the metric around $\mcN_{(r_0,r_1]}$, and let $\kxi{2}^u$ and $\xiA{2}$ be  gauge fields used to gauge the metric around $\mcN_{[r_2,r_3)}$. For $r_1\le \tdr \le r_2$ we set
\begin{equation}\label{16III22.2old}
  \tilde \hBo_{AB} = \interph _{AB}+ \wh_{AB}
   +  \phi   \tdr^2 \TS[
	\TSoLie_{\kzeta{1}} \ringh_{AB} ] + (1-\phi )  \tdr^2 \TS [
	\TSoLie_{\kzeta{2}} \ringh_{AB} ]
  \,.
\end{equation}
(Recall that $\TSxip^A=\xi^A -\zspaceD^A \xi^u/r$, cf.\ \eqref{1VIII22.1}.)

In the gluing problem, the gauge fields evaluated on $\tilde{\secN}_{1,2}$ and the field $\wh_{AB}$ on $\tmcN _{(r_1,r_2)}$ are \textit{free fields} which can be chosen arbitrarily. Our aim in what follows is to show how to choose these fields to solve the transport equations of Section \ref{sec:28VII22.1}-\ref{sec:28VII22.2} to achieve gluing-up-to-gauge.
When extending fields across $r_1$ by solving the transport equations, we will always choose initial data at $r_1$ which guarantee smoothness of the fields there.

For the $\Ck$-gluing we will need smooth functions
$$
 \kappa_i:(r_1,r_2)\rightarrow\R
 \,,
 \quad
    i\in\{0,\ldots,{k_{\red{[m]}}}+4\}
    \,,
$$
where $k_{\red{[m]}}=k$ \underline{when $m=0$} and $k_{\red{[m]}}=2k$ \underline{when $m\neq 0$}, satisfying
\begin{eqnarray}
  &
  \displaystyle
 \ip{\kappa_i}{\hat \kappa_j} \equiv \int_{r_1}^{r_2} \kappa_i(s) \hat \kappa_j(s) \, ds = 0
 \quad \mbox{for $i>j $}
  \,,
  &
  \label{13VIII22.1a}
   \\
   &
  \displaystyle
 \ip{\kappa_i}{\hat \kappa_i}  = 1
  \,,
  &
  \label{13VIII22.2a}
\end{eqnarray}
and vanishing   near the end points $r\in\{r_1,r_2\}$, which is possible since the $\hat{\kappa}_i$'s are linearly independent;
see Appendix~\ref{App13VIII22.1}.

The fields $ \wh_{AB}$ of \eqref{16III22.2old} will be taken of the following form: for  $s\in[r_1,r_2]$,
\begin{align}
   \wh_{AB}(s) = \sum_{  i=1}^{ {k_{\red{[m]}}}+4}\kappa_i(s)
     \vphi{i}_{AB}
   \,.
   \label{27VII22.1a}
\end{align}
We define
\begin{equation}
    \kphi{j}_{AB}:= \ip{\wh_{AB}}{\hkappa_j}
     \,.
   \label{27VII22.1a+}
\end{equation}
We will show how to construct the fields  $\{ \kphi{j}_{AB}\}_{j=1}^{k_{\red{[m]}}+4}$. In view of \eqref{13VIII22.1a}-\eqref{13VIII22.2a} we then have
\small
\begin{equation}\label{17III23.31}
   \left(
     \begin{array}{c}
       \kphi{1}_{AB} \\
       \kphi{2}_{AB} \\
        \vdots \\
       \kphi{k_{\red{[m]}}+4}_{AB} \\
     \end{array}
   \right)
   = \left(
       \begin{array}{cccc}
         1 & 0& \cdots  & 0 \\
         \ip{\kappa_1}{\hkappa_2}  & 1 & \cdots & 0 \\
         \vdots  &   & \ddots  & \vdots \\
         \ip{\kappa_1}{\hkappa_{k_{\red{[m]}}+4}} &  \ip{\kappa_2}{\hkappa_{k_{\red{[m]}}+4}} & \cdots & 1 \\
       \end{array}
     \right)
    \left(
     \begin{array}{c}
       \vphi{1}_{AB} \\
       \vphi{2}_{AB} \\
        \vdots \\
       \vphi{k_{\red{[m]}}+4}_{AB} \\
     \end{array}
   \right)
   \,,
\end{equation}
\normalsize
which allows one to determine    $\{ \vphi{j}_{AB}\}_{j=1}^{k_{\red{[m]}}+4}$ in terms of $\{ \kphi{j}_{AB}\}_{j=1}^{k_{\red{[m]}}+4}$ in the obvious way.

%% file: PointwiseCharge2.tex
It further follows from \eqref{24IX22.1} that the projection $\chi^{\red{[1]}}$ is gauge invariant for all topologies regardless of $m$. Moreover, when $m=0$, the projection $\chi^{[\le 1]}$ is also gauge-invariant on $S^2$.
These projections
are determined by the radial charge \eqref{20VII22.1}, i.e.\
    \begin{align}
       \kQ{2}{}(\lambda)&=
        r  \int_{\secN}\lambda \big[-h_{uu} + \frac{1}{2}\partial_r \zspaceD^Ah_{uA} \big]\sm
        \label{20VII22.1a}
        \\
        &
 =
       - r  \int_{\secN}\lambda h_{uu}  \sm
       +
        \frac r 2  \partial_r \int_{\secN}\lambda   \zspaceD^Ah_{uA} \sm
      \nn
      \,,
    \end{align}
where $\lambda $ is  a linear combination of $\ell=0$ and $\ell=1$ spherical harmonics
 on $S^2$,
and is a constant in the remaining cases,
 as follows:
Recall that
\begin{align}
           \kQ{1}{}(\zspaceD \lambda )
             &=
               \int_{\secN}\zspaceD^A \lambda  \left[r^4  \partial_r(r^{-2} h_{uA})\right] \,\sm
               \label{24VII22.4a}
               \\
               &
              = -\int_{\secN} \lambda  \left[r^4  \partial_r(r^{-2} \zspaceD^A h_{uA})\right] \,\sm
              \nonumber
\\
              &= 
                -r^4  \partial_r\big(
                 r^{-2} \int_{\secN} \lambda   \zspaceD^A h_{uA}
                  \,\sm
                  \big)
                  \,,
           \nn
           \end{align}
see \eqref{24VII22.4}.
Integrating \eqref{24VII22.4a} over $\secN$ shows that there exists a function $\kC{1}(u,x^A)$ such that
\begin{eqnarray}
            \int_{\secN} \lambda   \zspaceD^A h_{uA}
                  \,\sm
                  = \frac{\kQ{1}{}(\zspaceD \lambda )}{3r} + \kC{1} r^2
                  \,,
           \label{24VII22.4b}
           \end{eqnarray}
which is non-zero on $S^2$ only.
It then follows from \eqref{20VII22.1a} that
    \begin{align}
       \int_{\secN}\lambda h_{uu}  \sm
        &
          = - \frac{\kQ{2}{}(\lambda)}r
       +
        \frac 12  \partial_r \int_{\secN}\lambda   \zspaceD^Ah_{uA} \sm
        \label{20VII22.1b}
\\
 &
          = - \frac{\kQ{2}{}(\lambda)}r
       -
         \frac{\kQ{1}{}(\zspaceD \lambda )}{6r^2} + \kC{1} r       \,.
      \nn
    \end{align}
Hence,
whatever the topology,
\begin{eqnarray}
\int_{\secN}\lambda  \chi   \sm   &=&
\int_{\secN}\lambda  \zspaceD^{A} {h}_{u A}\sm
 -
   r^2 \partial_{r} \int_{\secN}\lambda  {h}_{u u} \sm
   =
    - \kQ{2}{}(\lambda)
  \,.
  \label{13VIII20.t12}
\end{eqnarray}
%

%% file: duhABcharge.tex
\subsection{A pointwise radial conservation law}
\label{ss26IX22.1}

In this section we show  that the equation
\begin{align}   
\TS\big(
   \frac{1}{r}\delta G_{AB}  +\zspaceD_A \delta G_{rB}
    \big)
                    =0
    \label{3XII22.1}
\end{align}
can be written as a radial conservation law, $\partial_r (....)=0$
 when $m=0=\alpha=\delta\beta$, where $P$ is as in
  \eqref{14VIII22.1}:
\begin{align}
 \label{11XI22.1}
 Ph_{AB} := \TS[\zspaceD_A \zspaceD^C h_{BC}]
    \,.
\end{align}
We further show that the equation obtained by taking $\zdivtwo\!$ of \eqref{3XII22.1},
\begin{equation}\label{28IX22.6}
\zspaceD^A \big[
  \TS\big(
   \frac{1}{r}\delta G_{AB}  +\zspaceD_A \delta G_{rB}
    \big)
  \big]
   =0
\end{equation}
can likewise be written as a radial conservation law
 when $m=0=\delta\beta$, for any $\alpha$.
This is likely to be related to the contracted Bianchi identity discussed in Section~\ref{sec:25VII22.1} below, but if and how is not clear.

Indeed, when $\delta\beta=0$, taking $\frac{1}{2 r^2}\times C$ of
 \eqref{24VII22.1} gives
\begin{eqnarray}
          &&
          \frac{1}{2 r^2} \partial_r \left[r^4  \partial_r(r^{-2} \TS[\zspaceD_Bh_{uA}]) \right]- \frac{1}{2 }
                   \partial_r \left(r^{-2}
                    P\hBo_{AB}\right) =0
                 \,.
                 \label{3IX22.w2}
           \end{eqnarray}
Subtracting \eqref{3IX22.w2} from \eqref{eq:31III22.3p0}  leads to
 \ptcheck{9IX22, mathematica file checking wan identity.nb}
\begin{align}
   &  \partial_r \Big[\underbrace{r \partial_u \zhTB_{AB}
     - \frac{  V}{2}   \partial_r \zhTB_{AB}
     -  \frac{ V }{2 r}   \zhTB_{AB}
     -\frac{\partial_r\big(r^2 \TS [\zspaceD_A\hBo_{uB}]\big)}{2r^{2}}
     +\frac{P\zhTB_{AB}}{2}}_{\blue{:=q_{AB}}}\Big]
      \label{3IX22.1}
\\
     	&\quad
      =
        \big(
        \frac{ {\alpha^2}}{r} - \frac{m}{r^4}
       \big)
        \hBo_{AB}
        \,.
        \nn
\end{align}
Hence $q_{AB}$ is radially conserved when $\alpha=m=0$.

Under a gauge transformation $q_{AB}$ transforms as
\begin{align}
    q_{AB}\mapsto
    q_{AB} -
   \Big[
   &
     \TS[\zspaceD_A\zspaceD_B \zspaceD_C\xi^C]
       -
    (P - \twoscsign + \alpha^2 r^2 -\frac{2m}{r}) C(\xi)_{AB}
    \big)
    \label{8XI22.w1}
\\
 &
   - (2 \alpha r + \frac{m}{r^2})\TS[\zspaceD_A\zspaceD_B \xi^u]
   \Big]
    \,.
    \nn
\end{align}

Since $C(X)^{[\TTt]} = 0$
for any vector field $X^A$ (cf.~Proposition~\ref{P30X22.2}, Appendix~\ref{App30X22} below),
 the field $q^{[\TTt]}_{AB}$  is gauge-independent and, when $\alpha = 0 = m$, gives a $2$-dimensional family of radially conserved  charges on $\T^2$, and a $6(\genus-1)$-dimensional family of such charges on sections with genus $\genus\geq 2$.

Next, taking the divergence of \eqref{3IX22.1} and using \eqref{24VII22.1} we find
\begin{align}
  & \partial_r (\zspaceD^B q_{AB})
     \label{31X22.1}
     =
    -
    \frac{{\alpha^2}}{2}\partial_r \left[r^4  \partial_r(r^{-2} h_{uA})- \zspaceD^B\hBo_{AB} \right]
      -  \frac{m}{  r^4}\zspaceD^B h_{AB}
     \,.
\end{align}
\input{duhABchargeAddMass}We define 
\begin{align} 
   \frac{\kQ{3,1}{A}}2 := &  r \zspaceD^B\partial_u \zhTB_{AB}
     - \frac{ 1}{2}  V  \partial_r \zspaceD^B\zhTB_{AB}
     -  \frac{1}{2 r}  V   \zspaceD^B\zhTB_{AB}
     \label{9IX22.11b}
     \\
     &
     -\frac{1}{2r^{2}}\partial_r\big(r^2 \zspaceD^B\TS [\zspaceD_A\hBo_{uB}]\big)
      \nonumber
     \\
     &
     +\frac{1}{2}\zspaceD^B P\zhTB_{AB}
    +
    \frac{{\alpha^2}}{2}
     \big(
      r^4  \partial_r(r^{-2} h_{uA})- \zspaceD^B\hBo_{AB} \big)
      \nonumber
      \\
      &
       + m (  3 \zhTB_{uA}  + r\partial_r \zhTB_{uA}
        - r^{-3}\zspaceD^B h_{AB})
    \,,
     \nn
\end{align}
with $\kQ{3,1}{A} $ being $r-$independent by 
\eqref{31X22.1b},
where the notation  $\kQ{3,1}{A}$ should be clear from \eqref{9IX22.9} below. Equivalently, the field
\ptcheck{26IX22}
\begin{align}
     \kQ{3,1}{A} = & \zspaceD^B
 \Big[
   2 r \partial_u \zhTB_{AB}
     -
      V  \partial_r \zhTB_{AB}
     -\frac{1}{ r^{2}}\partial_r\big(r^4
     \TS [\zspaceD_A\zhTB_{uB}]\big)
      \label{30VII23.1}
     \\
     &\qquad
    +
       \big(P
        -\twoscsign 
        \big)   \zhTB_{AB}
      \Big]
    +
     {{\alpha^2}r^4 }    \partial_r \zhTB_{uA}
     +  2 m (  3 \zhTB_{uA}  + r\partial_r \zhTB_{uA})
 \nn
\end{align}
is radially conserved.

To continue, for $\pi^A\in\CKV$  and  $i\ge 0$ we set
\begin{equation}
       \kQ{4,i}{}(\pi^A)
       :=\int_{\secN}
     \pi^A \bigg( 3 \partial_u^{i}\zhTB_{uA}+ r\partial_r \partial_u^{i}\zhTB_{uA}\bigg)
       \sm
       \,.
    \label{19I23.2b}
\end{equation}
It follows from the $u$-derivatives of \eqref{17I23.2b} that these are radially conserved charges.
%
Furthermore it holds that
\begin{align}
  \int _{\secN}  \pi^A \kQ{3,1}{A}
  \, \sm
   &
   = 
    \int _{\secN}  \pi^A
     (\alpha^2 r^4  \partial_r \zhTB_{uA} 
     + 2 m (  3 \zhTB_{uA}  + r\partial_r \zhTB_{uA})
     )\,
      \sm
      \label{26IX22.22}
      \\
    &
       = {\alpha^2} \kQ{1}{}(\pi^A)
       + 2m \kQ{4,0}{}(\pi^A)
    \,;
     \nn
\end{align}
recall that  $\kQ{1}{}(\pi^A)$ has been defined in \eqref{24VII22.4}.
Thus $\kQ{3,1}{}_{A}^{[\CKV]}$ is zero if $m=0=\alpha$, and is determined by $\kQ{1}{}$ and $\kQ{4,0}{}$ otherwise.

Under a gauge transformation $\kQ{3,1}{A}$ transforms as
\ptcheck{23X22 and rechecked 2XI22 for m and rechecked 20 VII 23} 
\begin{align}
    \kQ{3,1}{A} \mapsto
     &
     \kQ{3,1}{A}  +
     2
     \underbrace{
    \zspaceD^B
    \Big\{-
    \TS[\zspaceD_A\zspaceD_B\zspaceD_C\xi^C]
     +  \big(P
        -\twoscsign
        \big)
         \TS[\zspaceD_A\xi_B]
          \Big\}
        }_{=(\hLop \xi)_A}
        \label{26IX22.21b}
\\
 &
  + 6m \partial_u \xi_A
    \,,
     \nn
\end{align}
where   the operator $\hLop$ can be  written as
\begin{equation}\label{24X22.11}
 \hLop  = -  \zDivtwo  C \Lop
  \,,
  \quad
  \Lop :=
   \zspaceD \, \zDivone  -\zDivtwo  C + \twoscsign
 \,.
\end{equation}
%
It follows from Proposition~\ref{P30X22.1a}, Appendix~\ref{ss12XI22.11}, that the gauge transformations  
\eqref{26IX22.21b} associated with $\xi_A$ act transitively on $\kQ{3,1}{}_{A}^{{[(\CKV + \harm)^\perp]}}$.
\underline{In the case  $m\ne 0 $}, the gauge-transformations associated with $\partial_u \xi_A$ clearly act transitively on the collection  of all radial charges $\kQ{3,1}{}_{A}$.

On $S^2$ we have $\harm = \{0\}$, and \underline{when $m=0$} but $\alpha \ne 0$ we conclude that there the integrals
$$
 \kQ{3,1}{}_{A}^{[\CKV]}=\kQ{3,1}{}_{A}^{[<2]}=\kQ{3,1}{}_{A}^{[=1]}
  \,,
$$
%
provide
a 6-dimensional family of gauge-invariant radially-conserved charges.

{\ptcheck{30X; by
wan }
On $\T^2$ we have (compare \eqref{30X22.32}-\eqref{30X22.31} below)
\begin{align}\label{30X22.1}
  \hLop (\xi)_A &= -\frac 12 \zDelta(
   \zspaceD_A \zspaceD^C  \xi_C
   - {\frac{1}{2}}\zDelta \xi_A)
   \,,
   \\
  \Lop (\xi)_A &=
 \zspaceD_A \zspaceD^C \xi_C  -  \frac 12 \zDelta \xi_A
  \,, \label{30X22.1b}
\end{align}
with kernels and cokernels spanned by covariantly constant vectors.  So $\CKV=\KV=\harm$, and  \underline{when $m=0$} it follows that on a torus the gauge transformations  \eqref{26IX22.21b} act transitively on $\kQ{3,1}{}_{A}^{[\KV^\perp]}$,
and that $\kQ{3,1}{}_{A}^{[\KV]}$ gives a 2-dimensional family of gauge-invariant radially-conserved charges.

On negatively curved two dimensional manifolds with genus $\genus$  we have $\CKV=\{0\}$ so that $\CKV+ \harm=\harm$ and, again \underline{when $m=0$},  $\kQ{3,1}{A} $ leads to a $2\genus$-dimensional family of gauge-invariant radially-conserved charges $\kQ{3,1}{}^{[\harm]} $.

Summarising:
   we can  always choose $\kxi{2}_A$
so that
\begin{equation}\label{30XI22.21}
  \kQ{3,1}{ }[\dt_{\secN_1}] ^{[(\CKV+\harm)^\perp]}= \kQ{3,1}{ }[\dt_{\secN_2}]^{[(\CKV+\harm)^\perp]}
\end{equation}
holds.  \underline{When $m=0$} the equality
\begin{equation}\label{30XI22.22}
  \kQ{3,1}{ }[\dt_{\secN_1}] ^{[ \CKV+\harm ]}= \kQ{3,1}{ }[\dt_{\secN_2}]^{[ \CKV+\harm  ]}
\end{equation}
provides an obstruction to gluing;  \underline{when $m\ne 0$} it can be enforced by an appropriate choice of 
$ \partial_u\xi_A^{[\CKV+\harm]}$.
On $S^2$ and on $\T^2$ the condition \eqref{30XI22.22}  is trivially satisfied when $m=\alpha=0$, and reduces to the requirement of conservation of $\kQ1{}$ 
 and $\kQ{4,0}{}$ if  $m \alpha \ne 0$.

\medskip

It should be clear from the above that if we set, for $i \ge 1$,
\begin{align}
\label{9IX22.9}
        \kQ{3,i+1}{A} :=
      & \zspaceD^B
 \Big[
   2 r \partial_u ^{i+1} \zhTB_{AB}
     -    V  \partial_r (  \partial_u ^{i}\zhTB_{AB}  )
     -\frac{1}{ r^{2}}\partial_r\big(r^4
     \TS [\zspaceD_A\partial_u ^{i}\zhTB_{uB}]\big)
     \\
     &\quad
        +  \big(P
        -\twoscsign 
        \big)    \partial_u ^{i}\zhTB_{AB}
      \Big]
    +
      \alpha^2r^4    \partial_r  \partial_u ^{i}\zhTB_{uA}
       \nn
       \\
       &\quad
       + 2 m (  3 \partial_u^i\zhTB_{uA}  + r\partial_r \partial_u^i\zhTB_{uA})
    \,,
    \nn
\end{align}
then we have:

\begin{Lemma}
 \label{L10IX22.1}
Suppose that for  $i\ge 0$
the $i$'th $u$-derivative of \eqref{24VII22.1} and \eqref{eq:31III22.3p0} with $\delta \beta \equiv 0$ hold. Then
$$\partial_r \kQ{3,i+1}{A} = 0 
 \,.
$$
\hfill\qed
\end{Lemma}

Similarly to \eqref{26IX22.22}, for   conformal Killing vectors  $\pi^A$
we have
\begin{align}
  &\int _{\secN}  \pi^A \kQ{3,i}{A}
  \, \sm
   \label{26IX22.22hi}
   \\
   &
   = \int _{\secN}  \pi^A
     \big(
      {{\alpha^2}r^4  \partial_r \partial_u^{i-1}\zhTB_{uA} 
  + 2 m (  3 \partial_u^{i-1}\zhTB_{uA}  + r\partial_r \partial_u^{i-1}\zhTB_{uA})}
     \big)\,
      \sm
      \nonumber
      \\
    &
       =: {\alpha^2} \kQ{1, i-i}{}(\pi^A)
       + 2m \kQ{4, i-1}{}(\pi^A)
    \,,
     \nn
\end{align}
so that the left-hand side obviously vanishes if $\alpha = 0 = m$.   
In fact, for $i\ge 2$  there is no obstruction regardless  of $\alpha $ and $m$, 
as  our arguments below show that the right-hand side of the last equation is continuous at $r_2$ when the Einstein equations together with a sufficient number of their $u$-derivatives hold on $\mcN$.

Under gauge transformations, it follows from \eqref{26IX22.21b}  that
\begin{align}
    \kQ{3,i+1}{A}  \mapsto &
 \kQ{3,i+1}{A}
    +
    2 (\hLop \partial_u ^{i}\xi) _A
    + 6 m \partial_u^{i+1} \xi_A  
     \,.
     \label{9IX22.12}
\end{align}

%% file: duhABchargeAddMass.tex
Setting $\delta\beta = 0$ in \eqref{24VII22.1} we have  
 \ptcheck{2-VII23}
\begin{eqnarray}
        \partial_r \bigg ( 3 \zhTB_{uA}+ r\partial_r \zhTB_{uA}
        - r^{-3}\zspaceD^B
                  h_{AB}\bigg ) =
                  r^{-4} \zspaceD^B
                  h_{AB}
                 \,.
                 \label{17I23.2b}
           \end{eqnarray}

Equation \eqref{17I23.2b}
allows us to rewrite \eqref{31X22.1} as:
\begin{align}
  & \partial_r \Big[\zspaceD^B q_{AB} +
    \frac{{\alpha^2}}{2}\left(r^4  \partial_r(\zhTB_{uA})- \zspaceD^B\hBo_{AB} \right)
     \label{31X22.1b}
     \\
     &\qquad
    + m
    \big (  3 \zhTB_{uA}  + r\partial_r \zhTB_{uA}
        - r^{-3}\zspaceD^B h_{AB}
        \big)
     \Big]
    =
    0
     \,.
     \nn
\end{align}

%% file: zeroOrder.tex
The above is proved by induction (see Appendix~\ref{App14VIII22.2}), which is initialised with $i=0$ as follows:
\ptcheck{2VIII22 together}
\begin{enumerate}
  \item  Order zero   for   \eqref{18VI22.3} is trivial,   with
\begin{eqnarray}
   \overadd{0,0,0}{\psi}(r)  =1
   \,,
   \quad
     \overadd{0}{\Psi}_{AB}(u,r ,x^A)= 0 = \overadd{0,0}{\psiP} (s,r)
   \, .
   \quad
\end{eqnarray}
We note that order one  for   \eqref{18VI22.3} is obtained  from
   \eqref{14VIII22.1}, with
\begin{eqnarray}
  \displaystyle
    \overadd{1,0,0}{\psi}(r) & =&
      -\frac{\twoscsign}{2 r} + \frac{\alpha^2  r} 2  + \frac m {r^2}\,,
   \nonumber
   \\
   \overadd{1,1,0}{\psi}(r) & =& \frac{1}{2} \left(\twoscsign-{\alpha^2} r^2  - \frac{2m}{ r}\right)\,,
    \nonumber
\\
       \overadd{1,0,1}{\psi}(r)
       &=&
       0
    \,,
    \quad
   \overadd{1,1}{\psiP} (s,r)
     =
     \frac{2r^2}{3 s^4}
      +\frac{1}{3 s r }
     \,,
     \nonumber
     \\
   \overadd{1,0}{\psi}(s,r) &=&
     \frac{{\alpha^2}r }{s } - \frac{mr^2}{s^4}
   \, .
   \quad \nn
\end{eqnarray}
  \item
Order zero for   \eqref{18VI22.1} follows from  \eqref{CHG28XI19.6b}, where $\mu$ and $\lambda$ are determined from $h_{uA}|_{r_1}$ and $\partial_r h_{uA}|_{r_1}$, with
\begin{eqnarray}
   \overadd{0,0,0}{\chi}(r) =0
   \,,
   \quad
   \overadd{0,0}{\chi}(s,r) =\frac{1}{3}
    \left(
     \frac{2}{sr^3} + \frac 1 {s^4}
     \right)
   \,.
\end{eqnarray}
%
%

\end{enumerate}

%% file: qabtt.tex
\subsubsection{The transverse-traceless part}
 \label{sss2XII22.1}

For most of our further purposes, the  essential role is played by the integral kernels $ \overadd{i,j}{\chi}$ and $ \overadd{i,j}{\psi}$  appearing in \eqref{18VI22.3}-\eqref{18VI22.1}. However, it turns out that when $m=0$ the $\TTt$-part of $\partial_u^i h_{AB}$  leads to obstructions to gluing, in which case the boundary terms in \eqref{18VI22.3} become significant. This forces us to revisit the induction, as follows:

We first consider the $L^2$-projection of \eqref{24VII22.61} on $\TTt$, with $m=0$:
\ptcheck{1XII22}
\begin{align}
      \partial_u \hBo_{AB}^{[\TTt]}
      &=
      r \underbrace{\bigg[\frac{\partial_u \hBo_{AB}^{[\TTt]}\big|_{r_1}}{r_1}
      - \frac{1}{2}
      (\myGauss - \alpha^2 r_1^2)\bigg(\frac{1}{r_1}\partial_r\hBo_{AB}^{[\TTt]}\big|_{r_1}-\frac{1}{r_1^2}\hBo_{AB}^{[\TTt]}\big|_{r_1}\bigg) \bigg]}_{ q_{AB}^{[\TTt]}\big|_{r_1}}
     \label{24VII22.61t}
\\
&
    + \frac{r}{2}
      (\myGauss - \alpha^2 r^2)\bigg(\frac{1}{r}\partial_r\hBo_{AB}^{[\TTt]}-\frac{1}{r^2}\hBo_{AB}^{[\TTt]}\bigg)
            + {\alpha^2 r} \int_{r_1}^r
       \frac{ 1}{s}
        \hBo_{AB}^{[\TTt]} \, ds
        \,;
 \nn
\end{align}
equivalently,
\begin{eqnarray}
      &&\underbrace{
       \frac{1}{r}\partial_u \hBo_{AB}^{[\TTt]}- \frac{1}{2}
      (\myGauss - \alpha^2 r^2)\bigg(\frac{1}{r}\partial_r\hBo_{AB}^{[\TTt]}-\frac{1}{r^2}\hBo_{AB}^{[\TTt]}\bigg)
       }_{q_{AB}^{[\TTt]}\big|_{r}}
      =
      \label{24VII22.61tb}
\\
    &&
    \hspace{1cm}
       \underbrace{\bigg[\frac{\partial_u \hBo_{AB}^{[\TTt]}\big|_{r_1}}{r_1}
      - \frac{1}{2}
      (\myGauss - \alpha^2 r_1^2)\bigg(\frac{1}{r_1}\partial_r\hBo_{AB}^{[\TTt]}\big|_{r_1}-\frac{1}{r_1^2}\hBo_{AB}^{[\TTt]}\big|_{r_1}\bigg) \bigg]}_{ q_{AB}^{[\TTt]}\big|_{r_1}}
      \nonumber
      \\
      &&
    \hspace{1cm}
     + {\alpha^2 } \int_{r_1}^r
       \frac{ 1}{s}
        \hBo_{AB}^{[\TTt]} \, ds
        \,.
 \nn
\end{eqnarray}
This can of course also be derived directly from \eqref{3IX22.1}, but note that this calculation makes it clear how the tensor field $q_{AB}$ appears in the formalism.

It follows that when $\alpha = 0 = m$, the field $\overset{}{q}{}_{AB}^{[\TTt]}$ provides a 2-dimensional family of gauge-independent radially conserved charges on $\T^2$, and a $6(\genus-1)$-dimensional family of such charges on sections with genus $\genus\geq 2$.

When $\alpha \ne 0$ but $m$ remains zero,
taking  $u$-derivatives of \eqref{24VII22.61tb} leads to
 \begin{align}
    \overset{[p+1]}{q}{}_{AB}^{[\TTt]}\Big|_{r_1}^{r}
    &= \alpha^2 \int_{r_1}^{r} \hkappaone(s)\partial^p_u \hBo_{AB}^{[\TTt]} \, ds\,,
    \label{9XI22.w2u}
\end{align}
where, for $i\ge 1$,
\begin{equation}\label{1XII22.1}
    \overset{[i]}{q}_{AB}:=
  \frac{1}{r}\partial_u ^i \hBo_{AB}
   - \frac{1}{2}
      (\myGauss - \alpha^2 r^2)\big(
       \frac{1}{r}\partial_r \partial_u^{i-1} \hBo_{AB}
       -\frac{1}{r^2} \partial_u^{i-1}\hBo_{AB} \big)
 \,.
\end{equation}
Making use again of \eqref{24VII22.61t} we find
\ptcheck{2XII}
\begin{align}
    &\int_{r_1}^{r} \hkappaone(s)\partial_u\hBo_{AB}^{[\TTt]} \, ds
    \\
    &=
    \int_{r_1}^{r}
    \bigg[ q_{AB}^{[\TTt]}\big|_{r_1}
    + \frac{1}{2}
      (\myGauss - \alpha^2 s^2)\big (\frac{1}{s}\partial_s\hBo_{AB}^{[\TTt]}\big|_{s}
      -\frac{1}{s^2}\hBo_{AB}^{[\TTt]}\big|_{s}\big)
       \bigg]ds
      \nonumber
\\
     &\quad
     + {\alpha^2 } \int_{r_1}^{r} \int_{r_1}^s
       \bigg(\frac{ 1}{y}
        \hBo_{AB}^{[\TTt]}\big|_y\bigg)
        dy \, ds
        \nonumber
\\
    &=
    \bigg[s \ q_{AB}^{[\TTt]}\big|_{r_1}
    +  \frac{1}{2 s}
      (\myGauss - \alpha^2 s^2)\hBo_{AB}^{[\TTt]}\big|_{s}\bigg]_{r_1}^{r}
      + {\alpha^2 r}  \int_{r_1}^{r}
       \hkappaone(s)
        \hBo_{AB}^{[\TTt]} \ ds \,.\nn
\end{align}
It follows by induction that
{\ptcheck{2XII}
\begin{align}
    &\partial^p_u\int_{r_1}^{r} \hkappaone(s)\hBo_{AB}^{[\TTt]} \, ds
    \label{30XI22.w1}
    \\
    =&
    \sum_{k=0}^{p-1}(\alpha^2 r)^k
     \partial_u^{p-1-k}\bigg[s \ q_{AB}^{[\TTt]}\big|_{r_1}
    +  \frac{1}{2 s}
      (\myGauss - \alpha^2 s^2)\hBo_{AB}^{[\TTt]}\big|_{s}\bigg]_{r_1}^{r}
      \nn
\\
    &\quad
    + (\alpha^2 r)^p \int_{r_1}^{r} \hkappaone(s)\hBo_{AB}^{[\TTt]} \, ds\nn
    \\
    =&
    \sum_{k=0}^{p-1}(\alpha^2 r)^k
      \bigg[s \ \kq{p -k}_{AB}^{[\TTt]}\big|_{r_1}
    +  \frac{1}{2 s}
      (\myGauss - \alpha^2 s^2)
      \partial_u^{p-1-k}\hBo_{AB}^{[\TTt]}\big|_{s}\bigg]_{r_1}^{r}
      \nonumber
\\
   &\quad
    + (\alpha^2 r)^p \int_{r_1}^{r} \hkappaone(s)\hBo_{AB}^{[\TTt]} \, ds
    \,.
    \nn
\end{align}
This allows us to rewrite \eqref{9XI22.w2u} as
 \ptcheck{2XII; nice job}
\begin{align}
   & \overset{[p+1]}{q}{}_{AB}^{[\TTt]}\Big|_{r_1}^{r}
    \label{30XI22.w2}
    \\
    &=
    \alpha^2 \sum_{k=0}^{p-1}(\alpha^2 r)^k \bigg[
 s \ \kq{p-k}_{AB}^{[\TTt]}\big|_{r_1}
    +  \frac{1}{2 s}
      (\myGauss - \alpha^2 s^2)
       \partial_u^{p-1-k}
       \hBo_{AB}^{[\TTt]}\big|_{s}\bigg]_{r_1}^{r }
      \nn
\\
    &\quad
    +\alpha^{2(p+1)}   r ^p  \int_{r_1}^{r} \hkappaone(s)\hBo_{AB}^{[\TTt]} \, ds
    \,.
    \nn
\end{align}
}

%% file: strategy4XI22mv.tex
\subsection{Strategy}
 \label{ss10IX22.1}

A  collection of   fields $\{\partial_u^i h_{\mu\nu}\}_{0 \le i \le k}$ on a null hypersurface $\mcN$ will be called  \emph{characteristic $\Ck$-data for linearised vacuum Einstein equations on $\mcN$}, or simply \emph{$\Ck$-data},  if  the fields $\partial_u^i h_{\mu\nu}$ are smooth on $\mcN$ and satisfy on $\mcN$
the equations which are obtained by  differentiating the linearised vacuum Einstein equations in $u$ up to  $k$-times,  and in which no-more than $k$ derivatives of the $h_{\mu\nu}$'s with respect to $u$ occur. In Bondi gauge this means that the equations  $ \partial_u^i \mcE_{\mu\nu}= 0  $ should hold with $0\le i \le k-1$, and that in addition we also have     $ \partial_u^{k} \mcE_{rA}= 0 = \partial_u^{k}  \mcE_{rr} = \partial_u^{k}  \mcE_{ur}$.

We will say that $\Ck$-data are smooth if the  $\partial_u^i h_{\mu\nu}$'s are smooth   on $\mcN$.

We note that  the linearised Einstein equations are invariant under linearised gauge transformations. In our scheme we will perform gauge transformations which will be needed to ensure the continuity of the fields, but which will have no influence on the question whether or not the linearised Einstein equations hold.

A set of $\Ck$-data can be obtained by restricting a smooth solution of linearised vacuum equations, and its transverse derivatives, to a null hypersurface.
The converse is also true for null hypersurfaces with boundary, e.g.\ $\mcN_{[r_0,r_1)}$ or $\mcN_{[r_0,r_1]}$, in the following sense:  any such data set arises by restriction of (many) solutions of vacuum Einstein equations to $\mcN$. This can be realised by  solving a characteristic Cauchy problem with two null hypersurfaces intersecting transversally at $\{r=r_0\}$, and requires providing data on both hypersurfaces. We note that  losses of differentiability are unavoidable in the characteristic Cauchy problem when the data are not smooth:
solutions constructed from characteristic initial data which are of $C^k$-differentiability class will typically be of differentiability class  $C^{k-k_0}$, for some $k_0  \in \N$  which typically  depends upon $k$.
Compare~\cite{LukRodnianski,RendallCIVP}.

\input{LemmaStratgey}
As such, Lemma \eqref{L30IX22.1} will apply directly  at $r=r_2$, once we have shown that all desired equations hold for $r\in (r_0,r_2)$. However, the argument that we are about to present is more complicated because, within our construction, for $r\in [r_1,r_2]$ we can only solve some of the Einstein equations. Fortunately  the conditions of the Lemma are not independent, and the crux of the argument is to isolate and enforce the independent ones in a hierarchical way, proving as we progress both the continuity of the fields listed in the Lemma, and the satisfaction of the linearised Einstein equations, as well as their $u$-derivatives,  on $\mcN_{(r_0,r_2]}$.

Given $k\in \N$, $k\ge 2$, in order to carry out a $\Ck$-gluing  the smooth solution on $\mcN_{(r_0,r_1]}$ is extended to one on $\mcN_{(r_0,r_2]}$ using a smooth interpolating field $ \interph_{AB} $ as in \eqref{27VII22.1a} and smooth gauge fields $\kzeta{1}$ and $\kzeta{2}$ as in \eqref{16III22.2old}, with the $u$-derivatives extended using the equations in Section~\eqref{sec:28VII22.2}. This guarantees that some of the Einstein equations are satisfied. It now remains to show that we can choose $ \interph $, $\kzeta{1}$ and $\kzeta{2}$ to satisfy the remaining conditions of  Lemma~\ref{L30IX22.1} together with the Einstein equations on $\mcN_{(r_0,r_2]}$. This can be done in three steps:

\begin{enumerate}
  \item The requirement of continuity of the fields $\partial_u^p \tilde\hBo_{\mu\nu}$ for $0\leq p \leq k$ at $\tilde\secN_2$ imposes conditions on the given data $\dt_{\secN_1}$ and $\dt_{\secN_2}$, as well as on the gauge fields $\kxi{2}{}_A$ and $\kxi{2}^u$ and the gluing fields $\kphi{p}_{AB}$. We summarise these conditions here
      (cf.\ Tables~\ref{T17XI22.1}-\ref{T17XI22.2}), with further details presented in the next section:
      \input{SummaryTable}%
      %
      %
      \input{SummaryTableWithMass}
    \begin{enumerate}
        \item[i.] $\tilde\hBo_{uu}$: Continuity of $\tilde\hBo_{uu}$ at $\tilde\secN_2$ requires
        \begin{align}
            \chi[\dt_{\tilde\secN_1}] = \chi[\dt_{\tilde\secN_2}]\,.
        \end{align}
         The continuity of $\chi^{\kerL}$ at $r_2$ requires the radial-charge matching-condition
          \be
          \label{16IV23.1}
           \kQ{2}{}[\dt_{\secN_1}](\lambda)=\kQ{2}{}[\dt_{\secN_2}](\lambda)
           \,,
         \ee
    for all $\lambda$ with vanishing Hessian,
         where  $\mrL$ has been defined in \eqref{16IV23.2}.
       \underline{When $m=0$}, this condition is a gauge-invariant obstruction to gluing. The continuity of the remaining part  $\chi^{\kerLp}$ can be achieved using the gauge field
       $ (\kxi{2}^{u})^{\kerLp}$ (see \eqref{24IX22.1}).

         \underline{When $m\neq 0$}, the gauge-invariant obstruction is \eqref{16IV23.1} with constant $\lambda$. The remaining projection of $\chi$ can be made continuous at $r_2$
          using   the same gauge field as when $m=0$, together with \emph{proper} conformal Killing vectors $\kxi{2}_A^{[\CKV]}$ (which are zero unless we are on $S^2$).

        \item[ii.] $\partial_r \tilde\hBo_{uA}$: Continuity of $\partial_r\tilde\hBo_{uA}$ at $\tilde\secN_2$
        requires the radial-charge
        matching condition
            \begin{equation}\label{13XI22.2}
                \kQ{1}{}[\dt_{\tilde\secN_1}]=\kQ{1}{}[\dt_{\tilde\secN_2}]
                \,,
            \end{equation}
            as well as a suitable choice of the field $\kphi{1}_{AB}^{[\TTt^\perp]}$. This condition is equivalent to the obstruction
            \begin{equation}\label{13XI22.2b}
                \kQ{1}{}[\dt_{\secN_1}]=\kQ{1}{}[\dt_{\secN_2}]
                \,,
            \end{equation}
             %
             except if $\secN = S^2$ and $m\ne 0$.  In this last case we use the gauge field $(\kxi{2}^u)^{[=1]}$ to get rid of the obstructions associated with proper conformal Killing vectors in \eqref{13XI22.2b}.
        \item[iii.] $\tilde\hBo_{uA}$: Continuity of $\tilde\hBo_{uA}$ at $\tilde\secN_2$ is achieved by a suitable choice of $\kphi{4}{}^{[\TTt^\perp]}_{AB}$ and of
             $\partial_u\kxi{2}{}_A^{[\CKV]}$.
\input{item6}
        \item[v.] $\partial_u^p\tilde\hBo_{uA}$ for $1\leq p \leq k$: In the case  $m=0$, the continuity of $\partial_u^p\tilde\hBo_{uA}$ at $\tilde\secN_2$ determines $\kphi{p+4}_{AB}^{[\TTt^\perp]}$ and $\partial_u^{p+1}\kxi{2}{}_A^{[\CKV]}$,
          with additional obstructions coming from the kernel of the operator $ \sum_{j=0}^{p} \overset{(p,j)}{\ochi}_{p+4} P^j$. We provide an analysis of this kernel in Appendix \ref{ss20X22.1}.

        In the case  $m\neq 0$, the continuity of
        $\{\partial_u^p\tilde\hBo_{uA}\}_{p=1}^k$ at $\tilde\secN_2$ is obtained by solving a system of equations for the collections
          $\{\kphi{p}_{AB}^{[\TTt^\perp]}\}_{p=5}^{2k+4}$
          and $\{\partial_u^{p+1}\kxi{2}{}_A^{[\CKV]}\}_{p=1}^k$.

    \end{enumerate}
    \item Once the gauge fields and the fields $\kphi{p}_{AB}$ with $1\leq p \leq k+4$ in the case $m=0$, and with $1\leq p \leq 2k+4$ in the case $m\neq 0$, have been determined, we construct the fields $\partial_u^p\tilde\hBo_{\mu\nu}$ on $\tmcN_{[r_1,r_2)}$ by setting $\tilde h_{AB}$ according to \eqref{16III22.2old} and using this to solve the transport equations of Section \ref{sec:28VII22.1}-\ref{sec:28VII22.2}:

     \begin{enumerate}

     \item[i.] $\partial_u^p \tilde h_{ur}$ for $0\leq p\leq k$: We set $\partial_u^p \tilde h_{ur}|_\tmcN \equiv 0$, which guarantees both smoothness of $\tilde h_{ur}$ and the validity of  the equations, for all $i$,
    \begin{eqnarray}
    0
   &  = &
  \partial_u^i \delta \mcE _{rr}  |_\tmcN
  \equiv
  - \partial_u^i \delta \mcE ^u{}_{r}  |_\tmcN
  \equiv
  \partial_u^i \delta \mcE ^{uu}  |_\tmcN
   \,.
   \label{30IX22.11}
    \end{eqnarray}
         \item[ii.]
          $\partial_u^p\tilde\hBo_{uA}$ for $0\leq p \leq k$: Using the representation formulae \eqref{18VI22.1}, with all $h_{\mu\nu}$'s there replaced by $\tilde h_{\mu\nu}$'s. This guarantees that on $\tmcN_{[r_0,r_2)}$ we have
      \begin{equation}\label{15XI22.w1}
          \partial_u^p \delta  \mcE _{rA}|_{\tmcN_{[r_0,r_2)}}  \equiv -  \partial_u^p \delta \mcE ^u{} _{A}|_{\tmcN_{[r_0,r_2)}} = 0
         \,.
      \end{equation}
    It follows that
    \begin{equation}
        \partial^p_u \delta\mcE^A{}_B|_{\blue{\tmcN_{[r_0,r_2)}}} =  g^{AC}\partial^p_u\delta \mcE_{CB}|_{\blue{\tmcN_{[r_0,r_2)}}}\,.
        \label{15XI22.w2}
    \end{equation}
    The divergence identity
\begin{eqnarray}
  0 &\equiv & \nabla_\mu \delta \mcE ^{\mu}{}_A
    \label{2X22.6}
\\
 &= &
 r^{-2}\partial_r(r^2  \delta \mcE ^r{}_A)
   + \partial_u \delta \mcE ^u{}_A
    + \zspaceD_B \delta \mcE ^B{}_A
 \,,
 \nn
\end{eqnarray}
together with its $u$-derivatives, shows that we also have $\forall \
  0 \le i \le k-1  $,
\begin{equation}
   \Big(
  r^{-2}\partial_r(r^2  \partial_u^i\delta \mcE ^r{}_A)
    + \zspaceD_B \partial_u^i\delta \mcE ^B{}_A
    \Big)\Big|_{\tmcN_{[r_0,r_2)}}
     =0
 \,.
  \label{1X22.3}
\end{equation}
         \item[iii.]
         $\partial_u^p \tilde \hBo_{uu}$ for $0\leq p \leq k$: We impose  $\partial_r \overset{[p]}{\chi}|_{\tmcN_{[r_1,r_2)}} = 0$ with the initial conditions $\overset{[p]}{\chi}|_{r_1} = \overset{[p]}{\chi}[\dt_{\tilde\secN_1}]$, together with  the value  of $\partial_u^{p-1}\tilde\hBo_{uA}|_{\tmcN_{[r_1,r_2)}}$ determined in (ii) above. This ensures
      \begin{equation}\label{15XI22.w3}
           \partial^p_u\delta\mcE _{ru}|_{\blue{\tmcN_{[r_0,r_2)}}} - \frac 1{2r}
  \zspaceD ^A  \partial^p_u\delta \mcE _{rA}|_{\blue{\tmcN_{[r_0,r_2)}}}  = 0\,.
      \end{equation}
      Together with \eqref{15XI22.w1}, Equation~\eqref{15XI22.w3} ensures
      \begin{equation}\label{1X22.2}
    \partial_u^p \delta  {\mcE}_{ru}|_{\tmcN_{[r_0,r_2)}}
     \equiv
    -
    \partial_u^p \delta  \mcE^u{}_{ u}|_{\tmcN_{[r_0,r_2)}}
      = 0\,.
      \end{equation}
         \item[iv.] $\partial_u^p\tilde\hBo_{AB}$ for $1\leq p \leq k$:
         We use the representation formulae \eqref{18VI22.3}, with all $h_{\mu\nu}$'s replaced by $\tilde h_{\mu\nu}$'s. This ensures that
    \begin{equation}\label{15XI22.w4}
  \TS\big(
   \delta \partial^{p-1}_u \mcE _{AB}
    \big)
  \big|_{\tmcN_{[r_0,r_2)}}
   =0
    \,.
    \end{equation}
    The $u$ differentiated divergence identity \eqref{27IX22.2asdf} with $\nu = r$ reads
    \begin{align}
    \label{C21X22.5ii}
     0 &\equiv \partial^p_u  \delta  \mcE ^u{}_r
   +
   \frac{1}{r^2} \partial_r  (r^2  \delta  \partial^{p-1}_u\mcE ^r{}_r )
   \\
   &\quad
   +   \frac{1}{\sqrt{|\det \zgamma |}} \partial_A (\sqrt{|\det \zgamma |} \delta  \partial^{p-1}_u\mcE ^A{}_r )
   \nn\\
   &\quad
   -
   \frac{1}{r}
        g^{AB}
    \delta \partial^{p-1}_u\mcE _{AB}
   \,,
   \nn
    \end{align}
    so that, in view of \eqref{15XI22.w1} and \eqref{1X22.2}, we have now
    \begin{equation}\label{1X22.6}
    \forall \
     0 \le i \le k \qquad
      0 =
      \frac{1}{r}
        g^{AB}
      \partial_u^i  \delta \mcE _{AB}
        \big|_{\tmcN_{[r_0,r_2)}}
   \,.
    \end{equation}
    Together with \eqref{15XI22.w4}, it follows that
    \begin{equation}\label{2X22.4}
 \forall \
  0 \le i \le k-1 \qquad
   \partial_u^i \delta \mcE _{AB}
  \big|_{\tmcN_{[r_0,r_2)}}
   =0
    \,.
    \end{equation}
    Equation \eqref{1X22.3} then gives, $\forall \
      0 \le i \le k-1 $,
    \begin{align}
    \label{15XI22.w5}
        0&=r^{-2}\partial_r(r^2
       \partial_u^i \delta \mcE ^r{}_A) |_{\tmcN_{[r_0,r_2)}}
       \\
       &
        =
         - r^{-2}\partial_r(r^2
       \partial_u^i
       \delta \mcE _{uA}) |_{\tmcN_{[r_0,r_2)}}
       \,,\nn
    \end{align}
    where we have used
    \begin{align*} \partial_u^i\delta \mcE^{r}{}_A |_{\tmcN_{[r_0,r_2)}}
     &= - g_{uu}  \partial_u^i\delta \mcE_{r A}|_{\tmcN_{[r_0,r_2)}} -  \partial_u^i\delta \mcE _{u A}|_{\tmcN_{[r_0,r_2)}}
\\
      &=   -  \partial_u^i\delta \mcE _{u A}|_{\tmcN_{[r_0,r_2)}}
      \,;
    \end{align*}
    note that the last equality is justified by \eqref{15XI22.w1}. Continuity at $r_1$, where all the fields $
   \partial_u^i  \mcE _{\mu\nu} $, $i\in \N$, vanish when the data there arise from a smooth solution of linearised Einstein equations, together with \eqref{15XI22.w5} implies that
    \begin{equation}
     \forall \
      0 \le i \le k-1 \quad
       \partial_u^i  \delta \mcE ^r{}_A  |_{\tmcN_{[r_0,r_2)}}=0 = \partial_u^i  \delta \mcE _{uA} |_{\tmcN_{[r_0,r_2)}}
        \,.
         \label{2X22.8}
    \end{equation}

    Meanwhile, the divergence identity for the Einstein tensor with a free lower index $u$ now reduces to $\forall \
      0 \le i \le k-1 $,
    \begin{equation}
      0  \equiv  \partial_u^i \nabla_\mu  \delta \mcE ^{\mu}{}_u
        \big |_{\tmcN_{[r_0,r_2)}}
       = r^{-2} \partial_r (r^2 \partial_u^i \delta \mcE^r {} _u)
        \big |_{\tmcN_{[r_0,r_2)}}
       \,.
    \end{equation}
    Continuity and vanishing at $r_1$
     together with \eqref{30IX22.11}
     and  \eqref{1X22.2}
     implies that $\forall \
      0 \le i \le k-1$
    \begin{align}
    \label{2X22.7}
       0 &= \partial_u^i  \delta \mcE _{uu}
        \big |_{\tmcN_{[r_0,r_2)}}
         =  -  \partial_u^i  \delta \mcE ^r{}_{u}
        \big |_{\tmcN_{[r_0,r_2)}}
        \\
        &
         =   \partial_u^i  \delta \mcE ^{rr}
        \big |_{\tmcN_{[r_0,r_2)}}\,.\nn
    \end{align}
         \end{enumerate}
     \item
      \label{newpoint3}
     The construction above guarantees the continuity of $\tilde\hBo_{uu}$, $\partial_u\tilde\hBo_{AB}$, $\partial_u^p\tilde\hBo_{uA}$ with $0\leq p \leq k$, and $\partial_u^i\tilde\hBo_{AB}^{[\TTt]}$ with $2\leq i \leq k$ at $r_2$. Continuity of the fields $\partial_r\partial^p_u \tilde\hBo_{uA}$ and $\partial_u^p\tilde\hBo_{uu}$ for $1 \leq p \leq k$ and $\partial^i_u\tilde\hBo_{AB}^{[\TTt^\perp]}$ for $2\leq i \leq k$ at $r_2$ follows now
          by induction: The explicit form \eqref{9XI20.t1} of the equation $ \delta \mcE _{uA} =0$ together with the continuity of $\tilde\hBo_{uu}$,  $\tilde\hBo_{uA}$  and $\partial_u\tilde\hBo_{AB}$ at $r_2$ ensures the continuity of $\partial_r\partial_u\tilde\hBo_{uA}$ at $r_2$.
          This guarantees continuity of  $\kQ{1,1}{}$ and $\kQ{4,1}{}$, hence of  $\kQ{3,2}{}^{[\CKV]}$ in view of \eqref{26IX22.22hi} , namely
    \begin{align}
      \int _{\secN}  \pi^A \kQ{3,k}{A}
      \, \sm
       &
           = {\alpha^2} \kQ{1,k-1}{}(\pi^A)
           + 2m \kQ{4,k-1}{}(\pi^A)
        \,,
         \label{26IX22.22hip}
    \end{align}
    with $k=2$.
          \ptcheck{31VII}
          %
%
%
    This, together with point~\ref{olditemvi}{.iv}
ensures continuity of $\kQ{3,2}{A}$  at $r_2$, which in turn renders $\partial^2_u\tilde\hBo_{AB}^{[\TTt^\perp]}$ continuous there.

    Meanwhile the explicit form \eqref{13VIII20.t3} of $ \delta \mcE_{u u}=0$ together with smoothness  at $r_2$ of $\tilde\hBo_{uu}$, $\tilde\hBo_{uA}$  and $\partial_u\tilde\hBo_{AB}$, leads to continuity of $\partial_u\tilde\hBo_{uu}$ at $r_2$.

     Now, suppose that the continuity of the fields $\partial_u^p\tilde\hBo_{uu}$, $\partial_r\partial_u^p\tilde\hBo_{uA}$ and $\partial_u^p\tilde\hBo^{[\TTt^{\perp}]}_{AB}$ has been achieved up to $p=k-1$.
      It follows that we have $\partial^{k-2}_u\delta \mcE _{uA}|_{\tmcN} =0$ and thus $\partial_r \kQ{1,k-1}{}(\pi^A)|_{\tmcN}=0 = \partial_r \kQ{4,k-1}{}(\pi^A)|_{\tmcN}$ (compare \eqref{6X22.w3b}). Further, \eqref{26IX22.22hip} implies that the $(\kQ{3,k}{})^{[\CKV+\Harm]}$-part of the radial charges on $S^2$ and $\T^2$ are continuous. Meanwhile, recall that continuity of the radial charge $(\kQ{3,k}{})^{[(\CKV+\Harm)^\perp]}$ was ensured using the gauge field $\partial_u^{k-1}\kxi{2}_A$, while on higher-genus sections continuity of the charge $(\kQ{3,k}{})^{[\CKV+\Harm]}=(\kQ{3,k}{})^{[\Harm]}$ is achieved by $\partial_u^k\kxi{2}_A^{[\harm]}$ when $m\neq 0$ and is an obstruction whose continuity has to be assumed when $m=0$. We have now the continuity of $(\kQ{3,k}{}) = (\kQ{3,k}{})^{[(\CKV+\Harm)^\perp]} + (\kQ{3,k}{})^{[(\CKV+\Harm)]}$, which ensures the continuity of $\partial^k_u\tilde\hBo_{AB}^{[\TTt^\perp]}$ at $r_2$.

     Next, by differentiation of \eqref{9XI20.t1} we obtain the explicit form of \eqref{2X22.8} with $i=k-1$
    \begin{align}
    \nonumber
      r^2
       \partial_{r} \left( \frac{\partial_u^{k}h_{uA}}{r^{2}}\right) & =
     - \frac{1}{ r^2} \Big[
    \zspaceD^{B}\zspaceD_{A}\partial_u^{k-1}{h}_{uB}
    -\zspaceD^{B}\zspaceD_{B}\partial_u^{k-1}{h}_{uA}
    \\
    &\qquad\qquad\quad
     +\partial_{u} \zspaceD^{B}\partial_u^{k-1}{h}_{A B}\Big]
     \nn
     \\
    & \quad
    +
    \left(\twoscsign-r^{2}\alpha^{2} - \frac{2m}{r}\right)\partial_{r}^{2}\partial_u^{k-1}h_{uA}
     \nonumber
\\
 &\quad
     + (2\alpha^{2} +
    \frac{4m}{r^3})\partial_u^{k-1}h_{uA}
     + \partial_{r}\zspaceD_{A}{\partial_u^{k-1}h}_{uu}
    \,.
    \nn
    \end{align}
    This equation, together with the continuity of
    $\partial^{k-1}_u\tilde\hBo_{uu}$,
     $\partial_u^{k-1}\tilde \hBo_{uA}$ and $\partial^k_u\tilde\hBo_{AB}$,
       ensures the continuity of $\partial_r\partial_u^{k} \tilde\hBo_{uA}$ at $r_2$.

       Finally, the explicit form of \eqref{2X22.7} with $i=k-1$,
    i.e.
    \begin{align}
      0 & =   \partial_u^{k-1}  \delta \mcE _{uu}
        \big |_{\mcN_{[r_1,r_2)}}
 \nonumber 
    \\
     & =      \frac{1}{ r^2}
      \Big\{
       2\Big[
          \partial_{u}^{k}
        + \big({\alpha^2  r^2-\twoscsign + \frac{2m}{r}}
    \big) \partial_{r}
        + \frac{3m}{r^2} -\frac{\twoscsign}{r}
        \Big]
         \zspaceD^{A} {h}_{u A}
         \nonumber
         \\
    &
    \phantom{\frac{1}{ r^2}\Big[}
        -  \zspaceD^{A} \zspaceD_{A} \partial_u^{k-1}{h}_{u u}
    -({\alpha^2  r^2-\twoscsign + \frac{2m}{r}}
    \big)
     \big(\frac{\zspaceD^{A} \zspaceD^{B} \partial_u^{k-1}{h}_{A B}}{r^2}
      \big)
    \nonumber
\\
  &
    \phantom{\frac{1}{ r^2}\Big[}
    -2 r \partial_{u} ^{k}{h}_{u u}
    -2 ({\alpha^2  r^2-\twoscsign + \frac{2m}{r}}
    \big) \partial_{r}(r \partial_u^{k-1}{h}_{u u})
    \Big\}\,,
    \nn
    \end{align}
    together with smoothness  at $r_2$ of $\partial_u^{k-1}\tilde\hBo_{uu}$, $\partial_u^{k-1}\tilde\hBo_{uA}$  and $\partial^{k-1}_u\tilde\hBo_{AB}$, ensures the continuity of $\partial^k_u\tilde\hBo_{uu}$ at $r_2$.

 \end{enumerate} 

%% file: LemmaStratgey.tex
Our  gluing procedure for such fields rests on the following elementary result. Let
 $a<b<c$, and  let us for simplicity  assume that all fields
$\partial_u^i h_{\mu\nu}$, $i\in \N$, on $\mcN_{(a,b]}$  and $\mcN_{[b,c)}$  are smooth in all variables, up-to-and-including the common boundary  at  $b$; a similar result for finitely-differentiable fields, with distinct finite losses of differentiability for distinct fields, can be established using the results of Section~\ref{ss25IV22.1}, and is left as an exercise to a concerned reader.

\begin{Lemma}
 \label{L30IX22.1}
 Let $k\in \N$.
Two   $\Ck$ 
data  sets in Bondi gauge on $\mcN_{(a,b]}$ and  $\mcN_{[b,c)}$, with   $h_{AB}$ extending  smoothly across  $\{r=b\}$, extend to smooth $\Ck$ data  on $\mcN_{(a,c)}$  if and only if the fields
\begin{enumerate}
  \item
   \label{p30IX22.1n}
    $\partial_u^ih_{ur}$,   $\partial_u^ih_{uA}$,   $\partial_u^i h_{AB}$,  with   $0\le i\le k$,  as well as
  \item
  \label{p30IX22.2n}
  $ \partial_r h_{uA}$ and
 $ h_{uu} $
\end{enumerate}
extend by continuity  at $\{r=b\}$ to  continuous fields.
 \end{Lemma}

\proof
The necessity is obvious. The sufficiency
follows  from the equations  in Sections \ref{sec:28VII22.1}-\ref{sec:25VII22.1}, together with their $u$-derivatives, as follows:

Suppose that $\delta \beta$ extends by continuity at $b$, then \eqref{CHG28XI19.4aa} shows that $\delta \beta$ extends to a smooth function. Next, \eqref{24VII22.1} shows that continuity of
$ \partial_r(r^{-2} h_{uA}) $ at $b$ guarantees a smooth extension of $ \partial_r(r^{-2} h_{uA}) $.
But then, by another integration, continuity of $   h_{uA}  $ at $b$ guarantees smooth extendability. One can now use  \eqref{27III2022.3b} and \eqref{eq:31III22.3p0} to similarly show that
continuity, at $b$, of $ \delta V$ and $\partial_u h_{AB}$ leads to  smooth extensions of these fields.
In particular $\partial_u h_{AB}$ is now smooth on $\mcN_{(a,c)}$, and one can apply the argument just given to the equations obtained by $u$-differentiating the vacuum Einstein equations to obtain smoothness on $\mcN_{(a,c)}$ of $\partial_u h_{\mu\nu}$ and $\partial^2_u h_{AB}$.

Iterating this argument a finite number of times establishes the result.
\qedskip

%% file: SummaryTable.tex
\begin{table}[t]
\small
\begin{tabular}{||c|c|c|c||}
  \hline
  \hline
    & Gluing field & Gauge field & Obstruction\\
  \hline
    $h_{AB} $
        & $\interph_{AB} $
                &
                     &
  \\
  \hline
    $\partial^i_u\tilde\hBo_{ur}\,,\  i\ge 0$
        &
                &  $\partial_u^{i+1}\kxi{1}^{u }$
                and $\partial_u^{i+1}\kxi{2}^{u }$
                     &
  \\
  \hline
    $\tilde\hBo_{uu}$
        &
                &  $(\kxi{2}^{u})^{\kerLp}$
                     &     $\kQ{2}{}(\lambda)$
  \\
  \hline
    $\partial_r \tilde\hBo_{uA}$
        &   $\kphi{1}{}_{AB}^{[\TTt^\perp]}$  
            &
                &  $\kQ{1}{}(\pi)$
\\
  \hline
    $\tilde\hBo_{uA}$
        &  $\kphi{4}{}_{AB}^{[\TTt^\perp]}$
             & $\partial_u\kxi{2}{}_A^{[\CKV]}$
                 &
\\
  \hline
    $\partial_u\tilde\hBo^{[\TTt^\perp]}_{AB}$:
    $\genus\le1$
        &
            & $\kxi{2}{}_A^{[\CKV^\perp]}$
                &  $\kQ{1}{}(\pi)$
                if  $\alpha\neq 0$
\\
    \phantom{$\partial_u\tilde\hBo^{[\TTt^\perp]}_{AB}$,} $\genus\ge2$
        &
            & $\kxi{2}{}_A^{[\harm^\perp]}$
                &  $\kQ{3,1}{}^{[\harm]}$
\\
\hline
    $\partial_u\tilde\hBo^{[\TTt]}_{AB}$,   $\alpha \ne 0$
        &   $\kphi{1}_{AB}^{[\TTt]}$
            &
                &
\\
\hline
    $\partial_u\tilde\hBo^{[\TTt]}_{AB}$,   $\alpha =  0$
        & \phantom{$\kxi{2}$}
            &
                & $q^{[\TTt]}_{AB}$
\\
        &
            &
                &   (trivial on $S^2$)
\\
\hline
    $\partial_u^p\tilde\hBo_{AB}^{[\TTt^\perp]}$, $2\leq p \leq k$
        &
            &   $\partial_u^{p-1}\kxi{2}{}_{A}^{[(\CKV+\harm)^\perp]}$
                &   $\kQ{3,p}{}^{[\harm]}$ 
                if $\genus\ge2$
\\
\hline
    $\partial_u^p\tilde\hBo_{AB}^{[\TTt]}$,
      $\alpha =0$
        &
            &
                &
                   $ \kq{p}^{[\TTt]}_{AB}$
\\
\phantom{$\partial_u^p\tilde\hBo_{AB}^{[\TTt]}$,}
 $\alpha \ne 0$
        &
            &
                &   see \eqref{30XI22.w3}, involves
\\
        &
            &
                & $ \kq{p}^{[\TTt]}_{AB}[\dt_{ \secN_a}]$
\\
       $2\leq p \leq k$
        &
            &
                &
\\
  \hline
    $\partial_u^p\tilde\hBo_{uA}$,  $1\leq p \leq k-2$
        &  $\kphi{p+4}^{[\TTtp]}_{AB}$
            & $\partial_u^{p+1}\kxi{2}{}_A^{[\CKV]}$
                & $\ker \big(
                  \sum_{j=0}^{p} \overset{(p,j)}{\ochi}_{p+4} P^j
                \big)$
\\
           &
            &
                & (trivial if $\genus\ge 2$)
\\
\hline
 $\partial_u^p\tilde\hBo_{uA}$,  $k-1\leq p \leq k $
           & { $\kphi{p+4}^{[\TTtp]}_{AB}$}
            & {\redc
             only $S^2$ and $\T^2$
              }
                &  
\\
           & 
             & { $\T^2$: $\partial_u^{p+1}\kxi{2}{}_A^{[1]}$ 
            }
                &  
\\
        &
            &
                $S^2$: $\partial_u^{p+1}\kxi{2}{}_A^{[1 \leq \ell \leq p+1]}$ 
                &
\\
  \hline
    $\partial_u^p\tilde\hBo_{uu}$,  $1\leq p \leq k$
        &
            &
                &
\\
  \hline
    $\partial_u^p\partial_r\tilde\hBo_{uA}$,  $1\leq p \leq k$
        &
            &
                &
  \\
  \hline
  \hline
\end{tabular}
\caption{Fields used to ensure the continuity at $r_2$ when $m=0$.
  The continuity for the fields in the last two lines follows from Bianchi identities.
 The fields $\tilde\hBo_{\mu\nu}$ are the gauge-transformed fields
  $ \hBo_{\mu\nu}$
 using the gauge fields \protect$\kxi{1}$
 for $r\le r_1$ and $\kxi{2}{}$ for $r\ge r_2$, cf.\ Section~\ref{ss26XI22.1};
  the operator $\mrL$ has been defined in \eqref{16IV23.2};
  the fields $\interph_{AB}$ and $\kphi{k}_{AB}$
  are defined  in \eqref{16III22.2old} and \eqref{27VII22.1a}-\eqref{27VII22.1a+};  
  {\redc
  projections  such as
  $(\cdot)^{[\TTt]}$  and 
  $(\cdot)^{[\harm]}$ 
  (both trivial on $S^2$), 
  or $(\cdot)^{[\CKVp]}$ (identity on higher genus)
  are defined in Section~\ref{s26XI22.1};
  }
  the radial charges  $\kQ{a}{}$, a=1,2, are defined in \eqref{24VII22.4} and \eqref{20VII22.1}; the radially-conserved
  fields $\kq{i}{}_{AB}$ and $\kQ{3,i}{} $ are defined in \eqref{1XII22.1} and \eqref{9IX22.9};  the operator $P$ has been defined in \eqref{14VIII22.1}; the coefficients $\overset{(p,j)}{\ochi}_{p+4}$ are defined inductively in \eqref{18VI22.1} and \eqref{18VIII22.w1}.
}\label{T17XI22.1}
\end{table} 

%% file: SummaryTableWithMass.tex
%
\begin{table}
  \small
\begin{tabular}{||c|c|c|c||}
  \hline
  \hline
    & Gluing field & Gauge field & Obstruction\\
  \hline
    $h_{AB} $
        &  $\interph_{AB} $
                &
                     &
\\
  \hline
    $\partial^i_u\tilde\hBo_{ur}\,,\  i\ge 0$
        &
                &   $\partial_u^{i+1}\kxi{1}^{u }$
                and $\partial_u^{i+1}\kxi{2}^{u }$
                     &
    \\
         \hline
     $\tilde\hBo_{uu}$
        &
                & $(\kxi{2}^{u})^{\kerLp}$
                    & $\kQ{2}{}(\lambda^{\red{[1]}})$
    \\
        &
            &  and $\kxi{2}_A^{[\CKV]}$,
                    &
    \\
        &
                &    with $\zspaceD^A\kxi{2}_A\neq 0$
                     &
  \\
  \hline
    $\partial_r \tilde\hBo_{uA}$
        &   $\kphi{1}_{AB}^{[\TTt^\perp]}$
            &
        only on $S^2$:  $(\kxi{2}^u)^{[=1]}$
                &     $\kQ{1}{}(\pi^A)$
                \\
                 &
                    &
                        & $\pi^A$ -- KV of $\secN$
\\
  \hline
    $\tilde\hBo_{uA}$
        &  $\kphi{4}_{AB}^{[\TTt^\perp]}$
             & $\partial_u\kxi{2}{}_A^{[\CKV]}$
                 &
\\
        &
            &   only for $S^2$ and $\T^2$: 
                &
\\
    &
        & $\kxi{2}{}_A^{[\CKV^\perp]}$
            &
\\
\hline
    $\partial_u\tilde\hBo^{[\TTt]}_{AB}$
        &   $\kphi{1}_{AB}^{[\TTt]}$ and 
            &
                &
\\
    & $\kphi{4}_{AB}^{[\TTt]}$
        &
            &
\\
\hline
    $\{\partial^p_u\tilde\hBo_{AB}^{[\TTt^\perp]}\}_{1 \leq p \leq k }$
        &
            &  $ \{ \partial_u^{p}\kxi{2}{}_{A}^{[(\CKV+\harm)^\perp]}\}_{ 0 \leq p \leq k}$,
                &
\\
        &
            &  {\redc $ \{ \partial_u^{p}\kxi{2}{}_{A}^{[\harm]}\}_{ 0 \leq p \leq k}$}
                &
\\
\hline
    $\partial^p_u\tilde\hBo_{AB}^{[\TTt ]}$, $2\leq p \leq k$
        &  $ \kphi{2p+2}^{[\TTt]}_{AB}$ and
            &
                &
\\
    (cf.\ \eqref{8XI22.w2a})
        &  $ \kphi{2p+1}^{[\TTt]}_{AB}
                $  
            &
                &
\\
  \hline
    $ \{ \partial_u^p\tilde\hBo_{uA}$,  $1\leq p \leq k \}$
        &  $ \{\kphi{j}_{AB}^{[\TTt^\perp]} \}_{ 5 \leq j \leq 2k+4}$
            & $\partial_u^{p+1}\kxi{2}{}_A^{[\CKV]}$
                &
\\
  \hline
    $\partial_u^p\tilde\hBo_{uu}$,  $1\leq p \leq k$
        &
            &
                &
\\
  \hline
    $\partial_u^p\partial_r\tilde\hBo_{uA}$,  $1\leq p \leq k$
        &
            &
                &
  \\
  \hline
  \hline
\end{tabular}

\medskip

  \caption{Fields used to ensure the continuity at $r_2$ when $m\ne 0$. The notation, and the last two lines, are as in Table~\ref{T17XI22.1}. ``KV of $\secN$'' stands for ``Killing vector of  $(\secN, \ringh)$''. 
   }\label{T17XI22.2}
\end{table}

%% file: item6.tex
        \item[iv.]
        \label{olditemvi}
        $\partial_u^p\tilde\hBo_{AB}$ for $1\leq p \leq k$: This is addressed in Sections~\ref{s10IX22.1}, \ref{ss1VIII22.102} and \ref{ss31vi23.1}. The continuity of $\partial^p_u\tilde\hBo_{AB}^{[\TTt^\perp]}$ requires
        \begin{align}
        \label{15XI22.w6}
            \kQ{3,p}{A}[\dt_{\tilde\secN_1}] = \kQ{3,p}{A}[\dt_{\tilde\secN_2}]\,.
        \end{align}
        \underline{In the case  $m=0$}, the gauge fields $\partial_u^{p-1}\kxi{2}{}_{A}^{[(\CKV+\Harm)^\perp]}$ can be used to achieve the matching of $(\kQ{3,p}{})^{[(\CKV+\harm)^\perp]}$. $(\kQ{3,p}{})^{[\harm]}$ provide obstructions for gluing on negatively curved manifolds of $\genus \geq 2$.
        The requirement of continuity of $\partial_u^p\tilde\hBo_{AB}^{[\TTt]}$, which is non-trivial only on $\T^2$ and manifolds of higher genus, is subjected to the obstruction \peqref{30XI22.w3}, namely
\begin{align}
    \partial^p_u 
     &
      q_{AB}^{[\TTt]}|_{\tilde\secN_2} -\partial^p_u q_{AB}^{[\TTt]}|_{\tilde\secN_1}
   =
\alpha^2 \sum_{k=0}^{p-1}(\alpha^2 r_2)^k \bigg[
  s \ \kq{p-k}_{AB}^{[\TTt]}\big|_{r_1}
  \label{30XI22.w3m}
  \\
  &\quad
    +  \frac{1}{2 s}
      (\myGauss - \alpha^2 s^2)
       \partial_u^{p-1-k}
       \hBo_{AB}^{[\TTt]}\big|_{s}\bigg]_{r_1}^{r_2}
      \nonumber
\\
    &\quad
    + (\alpha^2 r_2)^p (q_{AB}^{[\TTt]}|_{\tilde\secN_2} -q_{AB}^{[\TTt]}|_{\tilde\secN_1})
    \,,
    \nn
\end{align}
which simplifies considerably when $\alpha=0$.

        \underline{In the case  $m\neq 0$}, the collection
        $\{\partial_u^{p}\kxi{2}_A^{[(\CKV+\Harm)^\perp]}\}_{p=0}^k$ of gauge fields can be used to match $\{\kQ{3,p}{}^{[(\CKV+\Harm)^\perp]}\}_{p=1}^k$. The gauge fields $\partial_u^{p}\kxi{2}{}_{A}^{[\harm]}$ can be used to achieve the matching of $\kQ{3,p}{}^{[\harm]}$.
        The requirement of continuity of $\partial_u^p\tilde\hBo_{AB}^{[\TTt]}$ is ensured by a suitable choice of
        \begin{equation}\label{17XI22.21}
     \overset{({p},0)}{\opsi}_{2p+2} (r_2)\kphi{2p+2}^{[\TTt]}_{AB}
    +  \overset{({p},0)}{\opsi}_{2p+1} (r_2) \kphi{2p+1}^{[\TTt]}_{AB}
     \,.
        \end{equation}

%% file: DetailsOfGluing4XI22.tex
We now pass to a more detailed presentation of some of the arguments above.

\subsection{Continuity at $r_2$}

\subsubsection{Gluing of $\delta\beta$}
 \label{ss1VIII22.1}

The sets of gauge functions $\partial_{\tdu}^i\okxi{1}^u|_{\tilde{\secN}_1}$ and $\partial_{\tdu}^i\kxi{2}^u|_{\tilde{\secN}_2}$
with  $i\leq k+1$ allow
 us to transform $\partial_u^j\delta\tilde{\beta}$ for $j\leq k $ to zero on $\tilde{\secN}_1$ and $\tilde{\secN}_2$,
  and hence, by invoking the $ur-$component of the linearised Einstein equation~\eqref{CHG28XI19.4aa}, on the whole $\tmcN _{[r_1,r_2]}$. In what follows, we assume that this gauge choice has been made, and set $\partial_u^j\delta\tilde\beta = 0$ for $j\leq k$ everywhere.

Furthermore, to simplify notation we omit the ``$|_{\tilde{\secN}_j}$'' on all gauge fields, with the understanding that all $\okxi{1}$ fields, and their $u$-derivatives, are evaluated on $\tilde{\secN}_1$, while all $\kxi{2}$ fields, and their $u$-derivatives, are evaluated on $\tilde{\secN}_2$, unless indicated otherwise.

\input{gaugefreeze}

\input{Gluinghuu}


\subsubsection{Continuity of $\partial_r\tilde h_{uA}$}
 \label{ss3VIII22.1}
Taking into account the allowed gauge perturbations to Bondi data, the gluing of $\partial_{\tdr}\tilde{\hBo}_{uA}$ requires $\tilde h_{AB}$ to satisfy on $\tmcN_{(r_1,r_2)}$,
\begin{align}
    r_2^4\partial_r\zhTB_{uA}|_{\tilde{\secN}_2}
    &=
     2r_2 \Done (\kxi{2}^u)_A + 2 r_2^2\zspaceD^{\tdB} C(\kzeta{2})_{AB}
    { - 6 m \zspaceD_A \kxi{2}^{u} }
     \label{4X122.w2}
\\
 &\quad
     +\Phi_A(x^C)
    +\zspaceD^{ B}\hBo_{AB}|_{\tilde{\secN}_2}
      -
      2\int_{r_1}^{r_2} \hat{\kappa}_1(s) \zspaceD^{\tdB}\tilde\hBo_{AB}\,ds
       \,.
       \nn
\end{align}
We have
$$
 \zspaceD_A(\kxi{2}^{u})^{\red{[1]}} = 0
 \,,
 \qquad
 \Done  \big(
  (\kxi{2}^{u})^{\kerL}
  \big ) =0
 \,,
 \qquad
  C(\kxi{2}{}^{[\CKV]})_{AB} =0
  \,,
$$
so that the gauge-part of the right-hand side of \eqref{4X122.w2}  involving $\kxi{2}^u$, except for the term explicitly involving $m$, depends only on
$(\kxi{2}^{u})^{\kerLp}$
and  has already been determined in terms of
the given data by \eqref{24IX22.3a}.   
 When $m \neq 0$ and $\secN\approx S^2$
 the remaining  part, which contributes to $-6m\zspaceD_A \kxi{2}^{u}$,  
is  already known from \eqref{18XI22.1}.  
For the remaining topologies  the kernel of $\mathring L$
  consists of constants, which do not contribute to the right-hand side. Thus in all cases, the terms in \eqref{4X122.w2} involving $\kxi{2}^{u}$ are
  either vanishing
  or already determined.

To clarify the freedom left, let us rewrite \eqref{4X122.w2} as an equation for $\kphi{1}_{AB}\equiv \ip{\hkappaone}{\wh_{AB}}$, where $\wh_{AB}$ is as in \eqref{16III22.2old}:
\begin{align}
 \zspaceD^B \kphi{1}_{AB}^{[\TTt^\perp]}
 = \
    &
  { \tilde{\Phi}}_A(x^C)
  +\zspaceD^{\tdB} \Big[ r_2^2C(\kxi{2}^{{[\CKVp]}})_{AB}
 \label{25VII22.7}
\\
 &
 \quad
      - 2
   \int_{r_1}^{r_2}
    \hat{\kappa}_1 (s)  (1-\phi  )s^2
     C(\kxi{2}^{{[\CKVp]}}(s))_{AB}  \,ds
     \Big]
   \,,
    \nn
\end{align}
where the already known fields such as $\zhTB_{uA}|_{\tilde\secN_2}$, $\interph_{AB}$  and $(\kxi{2}^{u})^{\kerLp}$,
 as well as the gauge fields $\okxi{1}^A$ and $\okxi{1}^u$ have been collected into the term $\tilde\Phi_A$.

Now,  the divergence operator on traceless symmetric two-tensors in two dimensions is elliptic; it has a cokernel spanned on conformal Killing vectors; on $S^2$ it has no kernel
 (see Appendix~\ref{ss12XI22.2}).
It follows that \eqref{25VII22.7}
determines a unique tensor field  $\kphi{1}_{AB}^{[\TTt^\perp]}$ on $S^2$  provided that the source term $\tilde{\Phi}$ is $L^2$-orthogonal to the cokernel. This orthogonality is guaranteed by the condition $\kQ{1}{}[\dt_{\secN_1}]=\kQ{1}{}[\dt_{\secN_2}]$ and either the gauge invariance of $\kQ{1}{}$
 in the case $m=0$, or by a suitable choice of
the gauge field $(\kxi{2}^u)^{[=1]}$ if $m\neq 0 $.
In other words, if {the radial charge} $\kQ{1}{}$ of the linearised field on  $\mcN|_{(r_0, r_1]} $ coincides with that  of the linearised field on  $\mcN|_{[r_2, r_3)} $,  the field $\kphi{1}_{AB}^{[\TTt^\perp]}$ satisfying \eqref{25VII22.7} exists, and is uniquely determined   in terms of the given data and the gauge field
$\kxi{2}{}_A^{[\CKVp]}$.

  By a similar analysis for the remaining topologies, \eqref{25VII22.7} determines $\vp{\kphi{1}}_{AB}$ uniquely in terms of the given data and the gauge field $\kxi{2}{}_A^{[{\CKVp} ]}$
  provided that the radial charges  $\kQ{1}{}$ at $r=r_1$ and $r=r_2$ coincide.

  \input{drhuA}

\subsubsection{Continuity of $\partial_u\tilde\hBo_{AB}$}
 \label{ss1VIII22.102}
\input{duhab}

\subsection{Higher derivatives}
 \label{ss26XI22.3}
Recall from Section \ref{sec:28VII22.2} that the terms $\overset{(i,j)}{\chi} (s,r)$ are linear combinations of $\hkappa_j(s)$'s with $1\leq j\leq i_{\red{[m]}}$, $j\neq 2,3$, with $i_{\red{[m]}}=i+4$  {when $m=0$} and $i_{\red{[m]}} = 2i+4$  {when $m\neq 0$},
where $i_{\red{[m]}}$ is not necessarily optimal unless  $i=0$; see Appendix~\ref{App14VIII22.2}. We shall henceforth write them as
\begin{equation}
\label{18VIII22.w1}
    \overset{(i,j)}{\chi} (s,r) = \sum_{\ell=1}^{i_{\red{[m]}}} \overset{(i,j)}{\ochi}_{\ell}(r)\, \hkappa_{\ell}(s)
    \,,
    \quad\text{with}
    \ \, \overset{(i,j)}{\ochi}_2=0= \overset{(i,j)}{\ochi}_3
    \,.
\end{equation}
Similarly we write, for $i\ge 1$,
\begin{equation}
\label{18VIII22.w1b}
    \overset{(i,j)}{\psi} (s,r) = \sum_{\ell=1}^{i_{\red{[m]}}-1} \overset{(i,j)}{\psi}_{\ell}(r)\, \hkappa_{\ell}(s)
    \,,
    \quad\text{with}
    \ \, \overset{(i,j)}{\psi}_2=0= \overset{(i,j)}{\psi}_3
    \,,
\end{equation}
where again the upper bound $i_{\red{[m]}}-1$ is not necessarily optimal unless $i=1$.

%% file: gaugefreeze.tex
\subsubsection{Freezing part of the gauge}
 \label{s10IX22.1}

 First, recall that the radial charge $\kQ{1}{}$ is gauge invariant except in the case $m \neq 0$ on $S^2$. In this case, we can use the gauge field $(\kxi{2}^{u})^{[=1]}$ for the matching of $\kQ{1}{}(\pi_A)$ when the conformal Killing field $\pi_A$ is such that $\zspaceD^A\pi_A \neq 0$, i.e.\ a proper conformal Killing vector field. According to \eqref{14VIII22.5}, this is achieved by choosing
 $(\kxi{2}^{u})^{[=1]}$ so that
\begin{align}
    \int_{\secN_2}\pi^A (r_2^4 \partial_r \zhTB_{uA}\big|_{r_2}
    - 6 m \zspaceD_A \kxi{2}^u)
    \sm
    =
    \int_{\secN_1}\pi^A (r_1^4 \partial_r \zhTB_{uA}|_{r_1})
    \sm \,.
    \label{18XI22.1}
\end{align}
However, for Killing vector fields the terms explicitly involving $m$ integrate-out to zero, and we obtain an obstruction to gluing.

Next, we determine the gauge fields needed to ensure continuity of $\chi$.
 For this, we evaluate the  function
 $\ochi $ of \eqref{24IX22.2} at $\secN_1$:
\begin{equation}
    \ochi [\dt_{\secN_1}] =
     \Big(-\delta V + \frac{ r}{2}\partial_r\big(r^2 \zspaceD^A \delta U_A\big) + \frac{1}{2} r \zspaceD^A\zspaceD^B \zhTB_{AB}
       \Big)
         \Big|_{\secN_1}\,,
          \label{7I23.3}
\end{equation}
and use the transformation law  \eqref{24IX22.1}, i.e.
\begin{equation}\label{24IX22.1x}
  \chi \mapsto \chi
  -
   \frac{1}{2}  (\TSzlap +2\twoscsign)\TSzlap\TSxi^{\TSu} 
  + 3m \zspaceD^B\xi_B
  \,,
\end{equation}
to do the gluing.   
As already pointed out, and clearly seen from this equation, the integral average of $\chi $ over $\secN$, i.e. $\kQ{2}{}{( \lambda=1)}$, is gauge-invariant regardless of topology or of the value $m$. Now, 
\underline{{when $m=0$}}, 
we can find a function
 $ {(\kxi{2}^{u})^{\kerLp}}$ so that 
\begin{align}
   -\frac{1}{2}  \mathring{L}
    {(\kxi{2}^{u})^{\kerLp}}
    &\equiv  -\frac{1}{2}  (\TSzlap +2\twoscsign)\TSzlap
    {(\kxi{2}^{u})^{\kerLp}} 
    \label{24IX22.3a}
    \\
    & = \ochi^{\kerLp}[\dt_{\secN_1}] - \ochi^{\kerLp}[\dt_{\secN_2}]
    \,.
   \nn
\end{align}
 The smooth fields $(\kxi{2}^{u})^{\kerL}$ are left arbitrary at this stage.
 
\underline{When $m\ne0$,}  to ensure the desired gluing we use  the field $(\kxi{2}^{u})^{\kerLp} $ determined in \eqref{24IX22.3a} together with the equation 
\begin{equation}
    (3 m \zspaceD _A\kxi{2}^A)^{\kerL} = \ochi^{ {\red{\kerL}}}[\dt_{\secN_1}] - \ochi^{ {\red{\kerL}}}[\dt_{\secN_2}]
    \,.
   \label{24IX22.3b}
\end{equation}
On $S^2$ this determines  the proper-conformal-Killing-vectors part of $ \kxi{2}_A^{[\CKV]}$, but does not provide a useful equation for the remaining topologies as there are no proper conformal Killing vectors there.

Next, we
determine the gauge fields $\partial^i_u\kxi{2}{}_A$ for $0\leq i\leq k$ as follows:
We evaluate the
 radially-constant covector field
$\kQ{3,i+1}{A}$ of \eqref{9IX22.9} at $\secN_1$:
\begin{align}
       \kQ{3,i+1}{A}[\dt_{\secN_1}]   =
  &
  \zspaceD^B \Big(
   2 r \partial_u ^{i+1} \zhTB_{AB}
     -    V  \partial_r (  \partial_u ^{i}\zhTB_{AB}  )
     \label{10IX22.9}
     \\
     &
     -\frac{1}{ r^{2}}\partial_r\big(r^4
     \TS [\zspaceD_A\partial_u ^{i}\zhTB_{uB}]\big)
    +
       (P
        -\twoscsign)   \partial_u ^{i}\zhTB_{AB}
      \Big)\Big|_{\secN_1}
      \nonumber
     \\
 &
  +
  \big( 
     {{\alpha^2}r^4 }    \partial_r  \partial_u ^{i}\zhTB_{uA}
      + {2 m (  3 \partial_u^i\zhTB_{uA}  + r\partial_r \partial_u^i\zhTB_{uA})}
      \big)\big|_{\secN_1}
    \,.
    \nn
\end{align}
The idea now is to use the transformation law \eqref{9IX22.12} to find gauge-vector fields such that
%
\begin{align}
    & 2 (\hLop \partial_u ^{i}\kxi{2}^{[(\CKV+\harm)^{\perp}]}) _A
    + 6 m \partial_u^{i+1} \kxi{2}_A
    \label{10IX22.8}
    \\
    &\qquad
      =
  \underbrace{ (\kQ{3,i+1}{A})[\dt_{\secN_1}] - (\kQ{3,i+1}{A})[\dt_{\secN_2}]
           } _{ =:\Delta \kQ{3,i+1}{A}}
     \,.
     \nn
\end{align}
%

We start by noting that the image of $\hLop$ is orthogonal to $\CKV+\harm$; (cf.~Proposition~\pref{P30X22.1a}). 
(The  spaces $\harm$ and $\CKV$ can be viewed as being orthogonal to each other, in that $\harm$ a) is trivial on $S^2$; b) can be set to $\{0\}$  on $\T^2$ being a subset of $\CKV$; c)  is non-trivial on the remaining topologies but $\CKV$ is trivial there.)

Note next that for $\pi^A\in \CKV$ we have  
\begin{align}
  &\int _{\secN}  \pi^A \kQ{3,i}{A}
  \, \sm\label{26IX22.22bck}
  \\
   &
   = 
    \int _{\secN}  \pi^A
     (\alpha^2 r^4  \partial_r \partial_u^{i-1}
      \zhTB_{uA} 
     + 2 m (  3 \partial_u^{i-1}\zhTB_{uA}  + r\partial_r \partial_u^{i-1} \zhTB_{uA})
     )\,
      \sm
      \nonumber
      \\
    & 
       = {\alpha^2} \kQ{1,i-1}{}(\pi^A)
       + 2m \kQ{4,i-1}{}(\pi^A)
    \,.
     \nn
\end{align}
Thus the continuity of $\kQ{3,i}{}_{A}^{[\CKV]}$ will be guaranteed by the continuity of $\kQ{1,i-1}{}$ and $\kQ{4,i-1}{}$, with the latter two being
addressed in point ~\pref{newpoint3}.
Hence we ignore for the moment  the condition  $\Delta\kQ{3,i}{}_{A}^{[\CKV]}=0$
and consider the system of equations
\begin{align} 
\displaystyle
      \small 2 (\hLop  \kxi{2}^{[(\CKV+\harm)^{\perp}]}) _A 
    + 6 m 
    \partial_u^{} \kxi{2}_A
   ^{[(\CKV+\harm)^{\perp}]}
       & =
      \big(
  \Delta \kQ{3,1}{A} 
   \big)^{[(\CKV+\harm)^{\perp}]}
 \,, 
   \label{21VIII23.90}
 \\  6 m 
    \partial_u \kxi{2}_A
   ^{[\harm]}
      & =
      \big(
  \Delta \kQ{3,1}{A} 
   \big)^{[\harm ]}
    \label{2VIII23.2}
 \,, 
 \\
 & \ \, \vdots
 \nonumber
 \\
 2 (\hLop  \partial_u^{k-2} \kxi{2}^{[(\CKV+\harm)^{\perp}]}) _A 
    + 6 m 
    \partial_u ^{k-1} \kxi{2}_A
   ^{[(\CKV+\harm)^{\perp}]}
       &=
      \big(
  \Delta \kQ{3,k-1}{A} 
   \big)^{[(\CKV+\harm)^{\perp}]}
 \,, 
   \label{21VIII23.97}
 \\  6 m 
    \partial_u^{k-1} \kxi{2}_A
   ^{[\harm]}
      & =
      \big(
  \Delta \kQ{3,k-1}{A} 
   \big)^{[\harm ]}
 \,, 
   \label{21VIII23.98}
 \\ 
      2 (\hLop  \partial_u^{k-1} \kxi{2}^{[(\CKV+\harm)^{\perp}]}) _A 
    + 6 m 
    \partial_u ^{k} \kxi{2}_A
   ^{[(\CKV+\harm)^{\perp}]}
       &=
      \big(
  \Delta \kQ{3,k}{A} 
   \big)^{[(\CKV+\harm)^{\perp}]}
 \,, 
   \label{21VIII23.99}
 \\  6 m 
    \partial_u^{k} \kxi{2}_A
   ^{[\harm]}
      & =
      \big(
  \Delta \kQ{3,k}{A} 
   \big)^{[\harm ]}
   \label{21VIII23.100}
 \,.
\end{align}
\underline{In the case $m\ne 0$} we set, e.g.,
\begin{equation}\label{2VIII23.1}
\partial_u ^{k} \kxi{2}_A
   ^{[(\CKV+\harm)^{\perp}]} =0
   \,,
\end{equation}
 and   we can solve  \eqref{21VIII23.99}-\eqref{21VIII23.100} to guarantee continuity 
of $ \kQ{3,k}{}_{A}^{[\CKV^{\perp}]} $ on $\T^2$ and $S^2$, and $ \kQ{3,k}{}_{A}^{[\harm]} $ on other topologies.
The solution of  \eqref{21VIII23.99} can then be inserted into  \eqref{21VIII23.97} to obtain a solution of  \eqref{21VIII23.97}-\eqref{21VIII23.98}. Continuing in this way, after a finite number of steps we obtain a solution  of the system 
   \eqref{21VIII23.90}-\eqref{21VIII23.100}.
   
Clearly the argument breaks down
\underline{when $m= 0$}, in which case  we can still solve \eqref{10IX22.8} 
for  $ \partial_u ^{i}\kxi{2}^{[(\CKV+\harm)^{\perp}]} _A$, but  the  equations 
$$ \big(
  \Delta \kQ{3,i}{A} 
   \big)^{[\harm ]} = 0
    \,,
     \quad
      i=0,\ldots,k
      \,,
  $$
  provide obstructions to gluing.

%% file: Gluinghuu.tex
\subsubsection{Continuity of $\tilde\hBo_{uu}$}
 \label{ss24IX22.1}
It follows from the pointwise radial conservation  of the function $\ochi$ defined in \eqref{24IX22.2} that the gluing of $\tilde\hBo_{uu}$ requires
\begin{equation}
    \ochi[\dt_{\tilde\secN_1}] = \ochi[\dt_{\tilde\secN_2}]\,.
\end{equation}
This is achieved by the condition $\kQ{2}{}[\dt_{\secN_1}]=\kQ{2}{}[\dt_{\secN_2}]$ together with  
formulae  \eqref{24IX22.3a}-\eqref{24IX22.3b} for the projected
gauge fields  $(\kxi{2}^{u})^{\kerLp} $ and 
$\kxi{2}_A^{[\CKV]}$.

%% file: drhuA.tex
\subsubsection{Continuity of $\tilde\hBo_{uA}$}
 \label{ss19XII22.1}
Taking into account the allowed gauge perturbations of Bondi data, it follows from~\eqref{17III22.1}
and \eqref{CHG28XI19.6b} that the continuity of $\tilde{\hBo}_{uA}$ at $r_2$ can be achieved by choosing  $\kphi{4}_{AB}^{{[\TTtp]}}$ so that
 \ptcheck{26VIII22, and XI22 with m}
\begin{align}
    \hBo_{uA}|_{\tilde{\secN}_2} &+ \Done (\kxi{2}^u)_A
     + r_2^2
     \big[
     \partial_{\tdu}\xiA{2}+(\alpha^2 + \frac{2m}{r^3_2})\zspaceD_{\tdA} \kxi{2}^u
     \big]
     \label{25VII22.8}
\\
     & =
       X_A(x^C)
         +\frac{1}{3}\zspaceD^{\tdB}
           \int_{r_1}^{r_2}
        \tilde{\hBo}_{AB}\left(\frac{2\hat{\kappa}_1(s)}{ r_2} + \hkappafour (s) r_2^2\right) ds
          \,,
                   \nn
\end{align}
where $X_A$ depends only on data at $r_1$.  More explicitly:
\ptcheck{15XI}
\begin{align}
 \frac{r_2^2}{3}
 &
 \zspaceD^{\tdB}\kphi{4}_{AB}^{ {[\TTtp]}}
   \equiv
   \frac{r_2^2}{3}\zspaceD^{\tdB}\ip{\hat{\kappa}_4}{\wh_{AB}}
 \label{29VII22.1}
\\
 & =
    r_2^2\partial_{\tdu}\xiA{2}
      -\tilde X_A(x^C)
  -
    \frac{2}{3r_2}\zspaceD^{\tdB}\kphi{1}_{AB}
    +
     (\alpha^2 + \frac{2m}{r^3})r_2^2 \zspaceD_A \kxi{2}^u
        \nonumber
\\
   & \quad
 -\frac{2}{3}
           \int_{r_1}^{r_2}
        \left(\frac{2\hat{\kappa}_1(s)}{ r_2} + \hkappafour (s) r_2^2\right)
 (1-\phi)s^2 \zspaceD^{\tdB}C(\kxi{2}^{[\CKVp]})_{AB} \,ds
         \nonumber
\\
   &
    =
    r_2^2\partial_{\tdu}\xiA{2}
      -\tilde X_A(x^C)
  -
    \frac{2r_2}{3}\zspaceD^{\tdB}C(\kxi{2}^{[\CKVp]})_{AB}
        \nonumber
\\
    & \quad
    -\frac{2}{3r_2}\tilde\Phi_A
     + (\alpha^2 + \frac{2m}{r^3})
      r_2^2 \zspaceD_A (\kxi{2}^{u})^{\kerL}
    \nonumber
    \\
    &\quad
 -\frac{2r_2^2}{3}
           \int_{r_1}^{r_2}
         \hkappafour (s)(1-\phi)s^2 \zspaceD^{\tdB}C(\kxi{2}^{[\CKVp]})_{AB} \,ds
         \,,
        \nn
\end{align}
where once again the already known fields such as $$
\mbox{$\hBo_{uA}|_{\tilde\secN_2}$, $\interph_{AB}$ and
$ (\kxi{2}^{u})^{\kerLp} $
}
$$
as well as the gauge fields $\okxi{1}^A$ and $\okxi{1}^u$ have been collected into the term $\tilde X_A$.
Since $\ker(C)= \im(\zdivtwo\!)^{\perp}$, we can use the freedom in choosing the smooth fields
\ptcheck{< 10XI22}
\begin{equation}\label{18X22.1}
 \big(
  \partial_u \kxi{2}{}_A
 +
     (\alpha^2+\frac{2m}{r^3})\zspaceD_A \kxi{2}^u
     )^{[\CKV]}
\end{equation}
 to arrange that the right-hand side of \eqref{29VII22.1} lies in the image of $\zDivtwo$.
 (Note that the first term in \eqref{18X22.1} has not been determined so far, while the remaining ones have only been determined if $m\ne0$.)
 It follows that \eqref{29VII22.1} can be solved uniquely for both $\vp{\kphi{4}}_{AB}$ and $\partial_{\tdu}\kxi{2}_A^{[\CKV]}$ in terms of $\partial_{\tdu}\kxi{2}_A^{[\CKVp]}$
 and $\kxi{2}_A^{[\CKVp]}$ when $\alpha =0=m$. For $\alpha,m\ne 0$ there remains some freedom
 in the gauge field $(\kxi{2}^{u})^{\kerL}$, made clear by \eqref{18X22.1}. On sections with higher genus, it follows from the surjectivity of $\zdivtwo\!$ (Lemma \ref{L30X22.1}, Appendix~\ref{ss12XI22.2}) that \eqref{29VII22.1} determines $\vp{\kphi{4}}_{AB}$ uniquely in terms of $\partial_u \kxi{2}{}_A$.

 A conserved, gauge dependent, radial charge involving  $h_{uA}$ and $\partial_r h_{uA}$ is derived in Appendix~\ref{sApp14IV23.1}.

%% file: duhab.tex
 \paragraph{\underline{The case $m=0$:}}  It follows from the pointwise radial conservation of $\kQ{3,1}{A}$ that the gluing of $\partial_u\vp{\tilde\hBo}_{AB}$ requires
\begin{equation}
    \kQ{3,1}{A}[\dt_{\tilde\secN_1}] = \kQ{3,1}{A}[\dt_{\tilde\secN_2}]\,.
\end{equation}
This is achieved on $S^2$ by the condition $\kQ{1}{}[\dt_{\secN_1}]=\kQ{1}{}[\dt_{\secN_2}]$ together with the expressions \eqref{10IX22.8} with $i=0$ for the gauge field $\kxi{2}{}^{[\CKVp]}_A$.

 For the remaining topologies
we use \eqref{3IX22.1} to obtain an equation for $q_{AB}^{[\TTt]}$:
 \begin{align}
      \partial_r q_{AB}^{[\TTt]} =  \partial_r \Big[r \partial_u \zhTB_{AB}^{[\TTt]}
     - \frac{ 1}{2}  V  \partial_r \zhTB_{AB}^{[\TTt]}
     -  \frac{1}{2 r}  V   \zhTB_{AB}^{[\TTt]}\Big]
      =
        \frac{ {\alpha^2}}{r}
        \hBo_{AB}^{[\TTt]}
        \,.
         \label{6VII23.2}
 \end{align}
Integrating, we obtain
\begin{align}
    q_{AB}^{[\TTt]}|_{\tilde\secN_2} -q_{AB}^{[\TTt]}|_{\tilde\secN_1}
    &= \alpha^2 \int_{r_1}^{r_2} \hkappaone(s)\tilde\hBo_{AB}^{[\TTt]} \, ds
    \nonumber
\\
    &=
    \alpha^2 \kphi{1}_{AB}^{[\TTt]}
    +
    \alpha^2 \int_{r_1}^{r_2} \hkappaone(s)\interph_{AB}^{[\TTt]} \, ds
    \,.
 \nonumber 
\end{align}
This provides an equation for $\kphi{1}_{AB}^{[\TTt]}$ when $\alpha \neq 0$:
\begin{align}
    \alpha^2 \kphi{1}_{AB}^{[\TTt]}
    &=  q_{AB}^{[\TTt]}|_{\tilde\secN_2} -q_{AB}^{[\TTt]}|_{\tilde\secN_1}
    -
    \alpha^2 \int_{r_1}^{r_2} \hkappaone(s)\interph_{AB}^{[\TTt]} \, ds
    \,.
    \label{9XI22.w2b2}
\end{align}

When $\alpha=0$, $\partial_u\hBo_{AB}^{[\TTt]}$ is  part of the radially conserved charge $q_{AB}^{[\TTt]}$ of \eqref{3IX22.1}. In this case, the continuity of $\partial_u\tilde\hBo_{AB}^{[\TTt]}$ at $r_2$ imposes the radial conservation law
\begin{align}
     q_{AB}^{[\TTt]}[\dt_{\secN_1}] =  q_{AB}^{[\TTt]}[\dt_{\secN_2}]\,.
\end{align}

\paragraph{\underline{The case $m\ne 0$:}}
Taking into account the allowed gauge perturbations of Bondi data, it follows from~\eqref{14VIII22.1} that we need to satisfy the equation
\ptcheck{19XI22; sign and coeff corrected; this propagates}
\begin{align}
   \partial_u \hBo_{AB}|_{\tilde{\secN}_2}  &=
   \overadd{1}{\tilde\Psi}_{AB}(r_2,x^A)
   -2 r_2^2C(\partial_u\kzeta{2})_{AB}
   \nn
   \\
   &\qquad +r_2 \left(\twoscsign-{\alpha^2} r_2^2-
   \frac{2m}{r_2}\right)C(\kxi{2})_{AB}
    \nonumber
\\
    &\qquad	+
     (\alpha^2 r_2+ \frac{1}{ 3 r_2 }P)\kphi{1}_{AB}
     -
    (mr_2 {-} \frac{2 r^2 }{3 }P)\kphi{4}_{AB}
    \nonumber
\\
    &\qquad	+
    2 \int_{r_1}^{r_2}(\alpha^2 r_2+ \frac{1}{ 3 r_2 }P)(1-\phi)\kappa_1(s)s^2C(\kxi{2})_{AB}
    \nonumber
\\
    &\qquad
     -
    2 \int_{r_1}^{r_2}(mr_2  {-} \frac{2 r^2 }{3 }P)\kappa_4(s)s^2C(\kxi{2})_{AB}
       \,.
 \nonumber 
\end{align}
Since $C(\xi)^{[\TTt]}=0$ for any vector field $\xi$, and similarly the image of $P$ is orthogonal to $\TTt$, 
continuity of $\partial_u\tilde\hBo_{AB}^{[\TTt]}$ at $\tilde\secN_2$ requires $\kphi{1}_{AB}^{[\TTt]}$ and $\kphi{4}_{AB}^{[\TTt]}$ to satisfy
 \ptcheck{19XI22}
\begin{align}
   \partial_u \hBo^{[\TTt]}_{AB}|_{\tilde{\secN}_2}  &=
   \overadd{1}{\tilde\Psi}{^{[\TTt]}}_{AB}(r_2,x^A)	+
     \alpha^2 r_2\kphi{1}^{[\TTt]}_{AB}
     -
    mr_2 \kphi{4}^{[\TTt]}_{AB}
       \,,
       \label{8XI22.w2}
\end{align}
which can be achieved by setting,  for example, $\kphi{1}^{[\TTt]}_{AB}=0$ and solving \eqref{8XI22.w2} for $\kphi{4}^{[\TTt]}_{AB}$.

%% file: EndInduction.tex
\subsubsection{Continuity of $\partial_u^{i}\tilde\hBo_{uA}$}
 \label{ss15X21}

While this is not needed for the current argument, we note that a  conserved, gauge dependent, radial charge involving  $\partial_u h_{uA}$   is derived in Appendix~\ref{sApp14IV23.2}.

To continue, let
 $k$ be the order at which we want to perform the gluing, i.e.\ the number of $u$-derivatives of $h_{\mu\nu}$ which we want to be continuous, and let $1\le {p}\le k$.

\paragraph{{\underline{The case $m=0$:}}} After performing a gauge transformation, Equation \eqref{18VI22.1} at order $i={p}$ together with \eqref{26IX22.1i} provides a gluing equation of the form,
\begin{align*}
    \partial_u^{{p}}
    \zhTB_{uA}|_{\tilde\secN_2}
    &=
     -
    \frac{1}{ 2r_2^2}
    \Done (\zspaceD^C\partial_u^{{p}-1}\kxi{2}{}_C)
    -\Big(\partial_u^{{p}+1}\kxi{2}{}_A
     +  \frac{\alpha^2}2
      \zspaceD_A\zspaceD^C\partial_u^{{p}-1}\kxi{2}{}_C\Big)
\\
    &
    + \overset{({p})}{X}_A
    \nonumber
    + \sum_{0\leq j+\ell \leq {p},\ell \ne {p}}
 \overset{({p},j,\ell)}{\chi}
 (r)\partial_r^j \zspaceD^B P^{\ell} \hBo_{AB}
\\
    &
    + \sum_{0\leq j+\ell \leq {p},\ell \ne {p}}
 \overset{ ({p},j,\ell) }{\chi}(r)
 \partial_r^j \zspaceD^B P^{\ell} \Big(2r^2C(\kzeta{2})_{AB}\Big)
               \nonumber
\\
    &\qquad
    + \sum_{j=0}^{{p}}\zspaceD^B\int_{r_1}^{r_2} \overset{({p},j)}{\chi}(s,r_2)P^j \tilde\hBo_{AB}(s)\, ds
    \nonumber
\end{align*}
which equals
\begin{align*}
    &
    \left\{
      \begin{array}{ll}
 \displaystyle
 -\partial_u^{{p}+1}\blue{\kxi{2}{}_A^{[\CKV]}} +
 \overset{({p})}{\tilde X}_A
    + \sum_{j=0}^{{p}} \sum_{\ell = 1}^{{p}+4}\zspaceD^B\int_{r_1}^{r_2} \overset{({p},j)}{\ochi}_{\ell}(r_2) \hkappa_{\ell}(s) P^j \wh_{AB}(s)\, ds
 \,;
\\
 \displaystyle
          -\partial_u^{{p}+1}\blue{\kxi{2}{}_A }
     +
         \overset{({p})}{\tilde X}_A
    + \sum_{j=0}^{{p}} \sum_{\ell = 1}^{{p}+4}\zspaceD^B\int_{r_1}^{r_2} \overset{({p},j)}{\ochi}_{\ell}(r_2) \hkappa_{\ell}(s) P^j \wh_{AB}(s)\, ds,
      \end{array}
    \right.
     \nonumber
\end{align*}
with the first case arising when $p\le k-2$ and the second for $p\in \{k-1,k\}$,
and    which we rewrite as
\begin{align}
    &\left\{
         \begin{array}{ll}
 \displaystyle
 -\partial_u^{{p}+1}\blue{\kxi{2}{}_A^{[\CKV]}}
 + \overset{({p})}{\tilde X}_A
    + \sum_{j=0}^{{p}} \sum_{\ell = 1}^{{p}+4} \overset{({p},j)}{\ochi}_{\ell} (r_2)\zspaceD^B P^j  \kphi{\ell}^{[\TTt^\perp]}_{AB} ,
   & \hbox{${p}\le k-2$;} \\
 \displaystyle
            -\partial_u^{{p}+1}\blue{\kxi{2}{}_A}
     + \overset{({p})}{\tilde X}_A
    + \sum_{j=0}^{{p}} \sum_{\ell = 1}^{{p}+4} \overset{({p},j)}{\ochi}_{\ell} (r_2)\zspaceD^B P^j  \kphi{\ell}^{[\TTt^\perp]}_{AB},\qquad
 & \hbox{${p}=k-1,k$.}
         \end{array}
       \right.
 \nonumber 
\end{align}
 \ptcheck{20X22}
Here we used the fact that  $P|_{\TTt} = 0$, that the fields   $\partial_u^{\ell }\kxi{2}{}_C^{{[\CKVp]}}$ with $\ell \le k-1$ are already known  from Section~\ref{s10IX22.1}, and  we  included them, together with all
other already known fields, in  $\overset{({p})}{\tilde X}_A =\overset{({p})}{\tilde X}_A(r,x^A) $.
Recall that $ \kphi{1}_{AB}$ has been determined in Section~\ref{ss3VIII22.1}, $ \kphi{4}_{AB}$  in  \eqref{29VII22.1}, and we further set
$$
\kphi{2}_{AB}=\kphi{3}_{AB}
 = 0
 \,.
$$
For the sake of induction, suppose  that the fields $ \kphi{\ell}_{AB}$ with $4\le \ell \le {p}+3$ are known.
 Together with $\overset{({p})}{\tilde X}_A  $ and $\partial_u^p\zhTB_{uA}|_{\tilde\secN_2}$ we collect them   into a term $\overadd{{p}}{\hat {\tilde X}}_{A}$, so that the requirement that $\partial_u^{{p}} \tilde h_{uA}$ be continuous at $r_2$  results in an equation of the form
\begin{align}
   \zspaceD^B\Big(
   \sum_{j=0}^{{p}}  \overset{({p},j)}{\ochi}_{{p}+4} (r_2)  P^j
   \kphi{{p}+4}^{[\TTt^\perp]}_{AB}
   \Big)
     	 &=
        \left\{
         \begin{array}{ll}
         \partial_u^{{p}+1}\blue{\kxi{2}{}_A^{[\CKV]}}
            - \overadd{{p}}{\hat {\tilde X}}_{A} ,
   & \hbox{${p}\le k-2$;} \\
 \displaystyle
             \partial_{\tdu}^{{p}+1} \xiA{2}
            - \overadd{{p}}{\hat {\tilde X}}_{A},
            \qquad
 & \hbox{${p}=k-1,k$.}
         \end{array}
       \right.
    \label{24X22.91}
\end{align}
Now, the operator at the left-hand side of this equation,
namely
\begin{equation}\label{11VII2232.3}
 \kchi{p} := \zdivtwo \circ \sum_{j=0}^{{p}}  \overset{({p},j)}{\ochi}_{{p}+4} (r_2)  P^j
   \kphi{{p}+4}^{[\TTt^\perp]}_{AB}
  \,,
\end{equation}
is elliptic,
of order $2p+1$, and
has a non-trivial cokernel on $S^2$ and $\T^2$.
(For example, the cokernel on $S^2$ is the space of spherical harmonic vectors with index $1\le \ell \le {p}+1$, see Appendix \ref{ss27XI22.1}; it is $30$-dimensional when ${p}=2$ and  $48$-dimensional when ${p}=3$.)
Recall that the fields $\partial_u^{p+1} \kxi{2}_A^{[\CKV]}$, $i\ge 2$, have not been determined so far.  We can use this freedom to arrange that $\partial_u^{{p}+1}\blue{\kxi{2}{}_A^{[\CKV]}}
            - \overadd{{p}}{\hat {\tilde X}}_{A} $ is in the image of $\zdivtwo\!$.
For all $p\le k$ we can therefore find a unique $ \kphi{{p}+4}^{[\kerLrpp]}_{AB}$ and $\partial_u^{{p}+1}\blue{\kxi{2}{}_A^{[\CKV]}}$
satisfying
\begin{equation}\label{10VII23.5}
   \kchi{p} (\kphi{{p}+4}^{[\kerLrpp]})
=
\partial_u^{{p}+1}\blue{\kxi{2}{}_A^{[\CKV]}}
  -
   \overadd{{p}}{\hat {\tilde X}}{}
    ^{[\imLrp]}
    \,.
\end{equation}
For $p\le k-1$ the $\imLrpp$-part of $\overadd{{p}}{\hat {\tilde X}}_{A} $ constitutes thus a further obstruction to the solvability of  \eqref{24X22.91}, as it is not clear whether or not the right-hand side is orthogonal to the cokernel.

However, if $p=k$ or $k+1$,  since the fields $ \partial_{\tdu}^{k } \xiA{2}$ and $ \partial_{\tdu}^{k+1} \xiA{2}$ are unconstrained so far, we can use \eqref{24X22.91} to choose these fields  so that continuity of $\partial_u^{k-1} \tilde h_{uA}$ and $\partial_u^{k} \tilde h_{uA}$ at $r_2$ holds:
\begin{equation}\label{10VII23.6}
 \imLrpp \ni
  \partial_{\tdu}^{{p}+1} \kxi{2}_A 
             :=
              \overadd{{p}}{\hat {\tilde X}}{}_{A}
              ^{[\imLrpp]}
               \,.
\end{equation}

\paragraph{The case $m\neq 0$:} After performing a gauge transformation, Equation \eqref{18VI22.1} at order $i={p}$ together with \eqref{26IX22.1i} provides a gluing equation of the form,
 \ptcheck{19XI22, and rechecked by wan 20XI}
\begin{align}
    \partial_u^{{p}}
    \zhTB_{uA}|_{\tilde\secN_2}
    &
    =
     -
    \frac{1}{ 2r_2^2}
    \Done (\zspaceD^C\partial_u^{{p}-1}\kxi{2}{}_C)_A
   -
     \partial_u^{{p}+1}\kxi{2}{}_A
 \label{6X22.w8b} 
     \\
     &\quad
      -  \big(\frac{\alpha^2}2 + \frac{m}{r_2^3}
     \big)
      \zspaceD_A\zspaceD^C\partial_u^{{p}-1}\kxi{2}{}_C
    + \overset{({p})}{X}_A
 \nn 
\\
    &\quad
    + \sum_{0\leq j+\ell \leq {p},\ell \ne {p}}
 \overset{({p},j,\ell)}{\chi}
 (r_2)\partial_r^j \zspaceD^B P^{\ell} \hBo_{AB}
 \nn
 \\
 &\quad
    + \sum_{0\leq j+\ell \leq {p},\ell \ne {p}}
 \overset{ ({p},j,\ell) }{\chi}(r_2)
 \partial_r^j \zspaceD^B P^{\ell}
 \Big(
  2r_2^2C(\kzeta{2})_{AB}\Big)
               \nonumber
\\
    &\quad
    + \sum_{j=0}^{{p}}\zspaceD^B\int_{r_1}^{r_2} \overset{({p},j)}{\chi}(s,r_2)P^j \tilde\hBo_{AB}(s)\, ds
    \nonumber
\\
    &=
    -\partial_u^{{p}+1}\kxi{2}{}_A
    +
 \overset{({p})}{\tilde X}_A
 \nn
\\
 &\quad
    + \sum_{j=0}^{{p}} \sum_{\ell = 1}^{{2p}+4}\zspaceD^B\int_{r_1}^{r_2} \overset{({p},j)}{\ochi}_{\ell}(r_2) \hkappa_{\ell}(s) P^j \wh_{AB}(s)\, ds
     \nonumber
\\
    &=
    -\partial_u^{{p}+1}\kxi{2}{}_A
    +
 \overset{({p})}{\tilde X}_A
    + \sum_{\ell = 1}^{{2p}+4} \underbrace{
      \sum_{j=0}^{{p}} \overset{({p},j)}{\ochi}_{\ell} (r_2)\zspaceD^B P^j
       }_{=:L_{p,\ell}}
      \kphi{\ell}_{AB}\,,
    \nn
\end{align}
where  we  have collected the fields $\partial_u^{j}\kxi{2}{}_A$ for $j\leq p$, together with all other already known fields, into $\overset{({p})}{\tilde X}_A$.

\input{newway}

%% file: newway.tex
For $k\ge 1$ we let
$$
 \Phi_k := \left(
             \begin{array}{c}
               \kphi{5}{} \\
               \vdots \\
                \kphi{4+k}{}  \\
             \end{array}
           \right)
           \,,
           \qquad
 \Psi_k := \left(
             \begin{array}{c}
               \kphi{5+k}{} \\
               \vdots \\
                \kphi{4+2k}{}  \\
             \end{array}
           \right)
           \,,
$$
and
$$
X_k := \left(
             \begin{array}{c}
                \partial_u
      \zhTB_{uA}|_{\tilde\secN_2}
   +\partial_u^{2}\kxi{2}{}_A
    -
 \overset{({1})}{\tilde X}_A \\
               \vdots \\
                \partial_u^{{k}}
      \zhTB_{uA}|_{\tilde\secN_2}
   +\partial_u^{{k}+1}\kxi{2}{}_A
    -
 \overset{({k})}{\tilde X}_A  \\
             \end{array}
           \right)
           \,.
 $$
 The system of equations \eqref{6X22.w8b} with $1 \leq p \leq k$ 
  takes the form
 \begin{equation}\label{26VII23.1}
   \zdivtwo \circ L_k \Phi_k + \zdivtwo \circ M_k \Psi_k = X_k
   \,,
 \end{equation}
 where
 \begin{equation}\label{26VII23.2}
   L_k =\left(
          \begin{array}{ccc}
            L_{1,5} & \ldots &  L_{1,k+4}\\
            \vdots &  \ddots & \vdots \\
             L_{k,5} & \ldots &  L_{k,k+4} \\
          \end{array}
        \right)
        \,,
        \quad
   M_k =\left(
          \begin{array}{ccc}
            L_{1,k+5} & \ldots &  L_{1,2k+4}\\
            \vdots &  \ddots & \vdots \\
             L_{k,k+5} & \ldots &  L_{k,2k+4} \\
          \end{array}
        \right)
        \,.
 \end{equation}
 Recall that the fields $\partial_u^i \kxi{2}_A^{[\CKV]}$, $i\ge 2$, have not been determined so far.  We can use this freedom to arrange that $    \partial_u
      \zhTB_{uA}|_{\tilde\secN_2}
   +\partial_u^{2}\kxi{2}{}_A
    -
 \overset{({1})}{\tilde X}_A$ is in the image of $\zdivtwo\!$. Inserting the result into $
 \overset{({2})}{\tilde X}_A$ we can arrange
 $    \partial^2_u
      \zhTB_{uA}|_{\tilde\secN_2}
   +\partial_u^{3}\kxi{2}{}_A
    -
 \overset{({2})}{\tilde X}_A$
 also lies in the image of $\zdivtwo\!$. Continuing in this way we obtain that all the entries in $X_k$ are in the image of $\zdivtwo\!$.

 We continue by noting that the operator $L_k$ is elliptic in the sense of Agmon, Douglis and Nirenberg (cf., e.g., \cite{MorreyNirenberg}, in particular see the estimates of Theorem~C there).
  Indeed, in each row of $L_k$   the operators $L_{i,j}$ are of order less than   $2i$, except for the  operator $L_{i,i}$ lying on the diagonal, which is elliptic and precisely of order $2i$. This shows that the Agmon-Douglis-Nirenberg condition on the orders $s_i+t_j$ of $L_{i,j}$ holds by setting $s_i=2i  $ and $t_j = 0$;
   ellipticity readily follows.

It is instructive to consider   explicitly the case $k=2$:
 \begin{align}
   L_2 &=\left(
          \begin{array}{cc}
            \frac{1}{4}(P+2\myGauss) &  -\frac{3m}{2}\\
             \frac{15m\alpha^2}{8 }  -  \frac{3 {m}}{8 r_2^2 }(P+2\myGauss) & \frac{9 {m}^2}{4 r_2^2}
 + \frac{1}{10}(P+2\myGauss)(P+5\myGauss)  \\
          \end{array}
        \right)
        \,,
       \label{26VII23.2b}
       \\
   M_2 &=\left(
          \begin{array}{cc}
            0 & 0\\
            -  \frac{7m}{4}(P + \frac{11  \twoscsign}{3} ) &   \frac{75 {m^2}}{8 } \\
          \end{array}
        \right)
        \,,\nn
 \end{align}
so that \eqref{26VII23.1} can be rewritten as
 \begin{align}\label{27VII23.1}
 \left(
       \begin{array}{cc }
          L_{\red{1,5}}(P)  &  -a\\
            L_{\red{2,5}}(P)  &  L_{\red{2,6}}(P) \\
          \end{array}
        \right)
          \left(
             \begin{array}{c}
               \kphi{5}{} \\
                \kphi{6}{}  \\
             \end{array}
           \right)
            +
 \left(
          \begin{array}{ccc}
          0 & 0\\
           \red{L_{\red{2,7}}}(P) & c \\
          \end{array}
        \right)
          \left(
             \begin{array}{c}
               \kphi{7}{} \\
                \kphi{8}{}  \\
             \end{array}
           \right)
           \\
         =    (\zdivtwo\!)^{-1}
          \left(
             \begin{array}{c}
              x_1 \\
               x_2 \\
             \end{array}
           \right)
        \,,\nn
 \end{align}
 with $a c \ne 0$,
  where the $L_{i,j}(P)$'s  are polynomials  in $P$,
and where $(\zdivtwo\!)^{-1}$ is the  inverse of $\zdivtwo\!$ viewed as a map from $(\ker \zdivtwo\!)^{\perp}$ to
$\im\,\zdivtwo=(\ker C)^\perp$.
It should be clear from this equation that for generic $m$ the operator $L_2$ is invertible from $(\ker \zdivtwo\!)^{\perp}$ to $(\ker C)^\perp$,
so that for such $m$'s we can solve the problem by setting
  $  \kphi{7}{} =
                \kphi{8}{}   \equiv 0$
                 and solving \eqref{27VII23.1} for
  $  \kphi{5}{} $ and $
                \kphi{6}{}  $.
                However, this genericity condition  is not needed; indeed, for any $m\ne 0$ we can proceed as follows: From the first equation of
\eqref{27VII23.1} we find
 \begin{equation}\label{27VII23.2}
                \kphi{6}{}  =    a^{-1}
                 \big(  L_{\red{1,5}}(P) \kphi{5}{}-
    (\zdivtwo\!)^{-1}  x_1
                \big)
        \,.
 \end{equation}
 Inserting into the second equation we obtain
 \begin{equation}\label{27VII23.3a}
       \underbrace{L_{\red{2,6}}(P)    L_{\red{1,5}}(P)}_{=:\mathfrak{L}_2}
        \kphi{5}{} + a   L_{1,2}(P)           \kphi{5}{}
         =   a
          \big(
    (\zdivtwo\!)^{-1}  x_2 - \red{L_{\red{2,7}}}(P)          \kphi{7}{}  - c          \kphi{8}{}
       \big)
        \,.
 \end{equation}
We can  choose smooth fields $   \kphi{7}{}$ and $ \kphi{8}{}$ so that
 $$
   \red{L_{\red{2,7}}}(P)          \kphi{7}{}  +c          \kphi{8}{} =
    \big(
       (\zdivtwo\!)^{-1} x_2
    \big)
    ^{[\im ( \mathfrak{L}_2)^\perp]}
    \in
    ( \im \mathfrak{L}_2)^\perp
 $$
(e.g., by setting $\kphi{7}{}=0$, but other choices might be more convenient); note that $ ( \im \mathfrak{L}_2)^\perp $ is finite dimensional and is spanned by smooth functions by ellipticity of $ \mathfrak{L}_2^\dagger$.
 We then solve the elliptic equation
 \begin{equation}\label{27VII23.3}
      \mathfrak{L}_2
        \kphi{5}{} + a   L_{1,2}(P)         \kphi{5}{}
         =   a \big (
    (\zdivtwo\!)^{-1} x_2
    \big)
    ^{[\im \mathfrak{L}_2]}
        \,,
 \end{equation}
 with a unique $  \kphi{5} \in (\ker \mathfrak{L}_2)^\perp$. Inserting into \eqref{27VII23.2} we have  thus obtained a desired solution of
 \begin{equation}\label{26VII23.1again}
   L_k \Phi_k + M_k \Psi_k = (\zdivtwo\!)^{-1}  X_k
   \,,
 \end{equation}
 with $k=2$.

%
%

Let us pass now to general $k$'s.
We note first that ellipticity and self-adjointness of $P$ implies existence of a complete set of smooth,
 pairwise $L^2$-orthogonal, eigenfunctions $\phi_\ell$ with a corresponding discrete set of eigenvalues $\lambda_\ell\to_{\ell \to \infty} \infty$.
We can therefore write
 \begin{equation}\label{29VII23}
   \Phi_k= \sum_{\ell} \Phi_{k,\ell} \phi_\ell
   \,, \quad
   \Psi_k= \sum_{\ell} \Psi_{k,\ell} \phi_\ell
   \,, \quad
   (\zdivtwo\!)^{-1}
  X_k = \sum_{\ell} X_{k,\ell} \phi_\ell
   \,.
 \end{equation}
 Equation \eqref{26VII23.1again} can be solved mode-by-mode:
 \begin{equation}\label{26VII23.1mode}
   L_k \Phi_{k,\ell}+ M_k \Psi_{k,\ell} = X_{k,\ell}
    \quad
    \Longleftrightarrow
    \quad
   L_k|_{P\mapsto \lambda_\ell} \Phi_{k,\ell}+ M_k|_{P\mapsto \lambda_\ell} \Psi_{k,\ell} = X_{k,\ell}
   \,,
 \end{equation}
 where $P\mapsto \lambda_\ell$ means that every occurrence of $P$ should be replaced by $\lambda_\ell$.

 Now, $\det L_k|_{P\mapsto \lambda_\ell}$ is a polynomial in $\lambda_\ell$.
Keeping in mind that in each line of the matrix $L_k$ the highest power of $P$ is on the diagonal, we see that $\det L_k|_{P\mapsto \lambda_\ell}$ is non-zero for $\ell$ large enough, and therefore there exists $N(k)$ such that we can find a unique solution of \eqref{26VII23.1mode} with  $\Psi_{k,\ell}=0$ for all $\ell > N(k)$. It remains to show that  \eqref{26VII23.1mode} can be solved in the finite dimensional space of $\Phi_k$'s and $\Psi_k$'s of the form
 \begin{equation}\label{29VII23.11}
   \Phi_k= \sum_{\ell\le N(k) } \Phi_{k,\ell} \phi_\ell
   \,, \quad
   \Psi_k= \sum_{\ell\le N(k) }  \Psi_{k,\ell} \phi_\ell
   \,, \quad
   (\zdivtwo\!)^{-1}
  X_k = \sum_{\ell\le N(k) }  X_{k,\ell} \phi_\ell
   \,.
 \end{equation}
This is equivalent to the requirement that all the linear maps  obtained by juxtaposing $L_k|_{P\mapsto \lambda_\ell}$ and $ M_k|_{P\mapsto \lambda_\ell}$ with $\ell < N(k)$ are surjective. (Note, by the way, that we have already established surjectivity for $\ell \ge N(k)$.)  This, in turn, is equivalent to the fact that the adjoints of these linear maps have no kernel.

 Let us denote by  $(L_k\, M_k)$ the relevant matrices. For simplicity write $L_{i,j}$ for $L_{i,j}(P)|_{P \mapsto \lambda_\ell}$.
 It follows from Appendix~\ref{ss16XI22.2}
  that  $(L_k\, M_k)$ is of the form
  \ptcheck{29VII23}
 \begin{align}
 \small
     \left(
          \begin{array}{ccccc}
                \begin{array}{|cc|}
                \hline
                  L_{1,4+1}   & L_{1,4+2} \\
                  L_{2,4+1}   & L_{2,4+2} \\
                  \vdots      & \vdots \\
                  \vdots      & \vdots \\
                  \vdots      & \vdots \\
                  L_{k,4+1}   & L_{k,4+2} \\
                \hline
                \end{array}
            & \hspace{-0.2cm}\ldots\hspace{-0.2cm}
            &
                \begin{array}{|cc|}
                \hline
                  0           & 0\\
                  \vdots      & \vdots \\
                  0           & 0   \\
                  L_{k-j,3+2(k-j)}   & L_{k-j,4+2(k-j)} \\
                  \vdots      & \vdots  \\
                  L_{k,3+2(k-j)}   & L_{k,4+2(k-j)} \\
                \hline
                \end{array}
            & \hspace{-0.2cm} \ldots \hspace{-0.2cm}
            &
                \begin{array}{|cc|}
                \hline
                  0           & 0 \\
                  \vdots      & \vdots \\
                  \cdot       & \cdot \\
                  \vdots      & \vdots \\
                  0           & 0 \\
                  L_{k,3+2k}   & L_{k,4+2k} \\
                \hline
                \end{array}
          \end{array}
        \right) \,,
 \nonumber %
 \end{align}
 where the $(i,j)$-th entry is $L_{i,4+j}$.  Note that each  of the two-columns pairs, as grouped above, has a specific vanishing-block structure. This gives the following adjoint matrix:
 \begin{align}
 \small
     \left(\begin{array}{l}
            \begin{array}{|cccccc|}
            \hline
               L_{1,4+1}            &  L_{2,4+1}  & \dots \quad \dots    & \ldots \quad \dots  & \dots\quad\dots\quad \dots\quad\dots       &  \, L_{k,4+1} \phantom{LL} \\
               \fbox{ $L_{1,4+2}$ } &  L_{2,4+2}  & \dots \quad \dots    & \ldots \quad \dots  & \dots\quad\dots\quad \dots\quad\dots       & \, L_{k,4+2} \phantom{LL}  \\
            \hline
            \end{array}
        \\
            \begin{array}{cccccc}
                 &  \phantom{L_{k,k}\qquad\qquad\qquad}   &   \phantom{L_{k}\qquad\qquad}   & \vdots &   &
            \end{array}
        \\
            \begin{array}{|cccccc|}
            \hline
              \phantom{L_{k,k}} \, 0  & \phantom{L_{k,k}} \quad  0  \,\dots  & 0    & L_{k-j,4+2(k-j)-1} & \dots\quad\dots\quad\dots  & L_{k,4+2(k-j)-1}\phantom{.} \\
              \phantom{L_{k,k}} \, 0  & \phantom{L_{k,k}} \quad  0  \,\dots  & 0    &  \fbox{$L_{k-j,4+2(k-j)}$} & \dots \quad\dots\quad\dots & L_{k,4+2(k-j)} \\
            \hline
            \end{array}
        \\
            \begin{array}{cccccc}
                 & \phantom{L_{k,k}\qquad\qquad}    &  \phantom{L_{k}\qquad\qquad\qquad}    & \vdots &   &
            \end{array}
        \\
            \begin{array}{|cccccc|}
            \hline
              \phantom{L_{k,k}} \, 0  & \phantom{L_{k,k}} \quad  0 \quad \dots  & \dots \quad \dots   & \ldots \quad \dots 0 &\, L_{k-1,4+2k-3} \quad  & L_{k,4+2k-3}\\
              \phantom{L_{k,k}} \, 0  & \phantom{L_{k,k}} \quad  0 \quad \dots  & \dots  \quad \dots  & \ldots \quad \dots 0 &\, \fbox{$L_{k-1,4+2k} $}\quad  & L_{k,4+2k-2} \\
            \hline
            \end{array}
        \\ \vspace{-.4cm}
        \\
            \begin{array}{|cccccc|}
            \hline
              \phantom{L_{k,k}} \, 0  & \phantom{L_{k,k}} \quad  0 \quad \dots  & \dots \quad \dots   & \ldots \quad \dots 0 \phantom{L_{kk}\dots} & 0 \,  & \phantom{L_{k,k}\quad} L_{k,4+2k-1}\\
              \phantom{L_{k,k}} \, 0  & \phantom{L_{k,k}} \quad  0 \quad \dots  & \dots  \quad \dots  & \ldots \quad \dots 0 \phantom{L_{kk}\dots} & 0 \,  & \phantom{L_{k,k}\quad} \fbox{$L_{k,4+2k}$} \\
            \hline
            \end{array}
            \end{array}
     \right)\,.
     \label{27VII23.w1}
 \end{align}
Equation~\eqref{9XI22.w5a1} shows
  that the  entries (boxed in the above) $L_{i,4+2i}$ are non-zero when $m\ne 0$.
This easily implies that  that the matrix  \eqref{27VII23.w1} has maximal rank, and thus has trivial kernel.

We have therefore proved:

\begin{theorem}
 \label{t29VII23.1}
The system  \eqref{6X22.w8b} with $p\in\{1,\ldots,k\}$ can be solved by a choice of gluing fields and gauge fields for any finite $k$. Its solutions are determined by an elliptic system, uniquely up to a finite number of eigenfunctions of $P$.
\end{theorem}

%% file: elluhab.tex
\subsubsection{Continuity of $\partial_u^p\tilde \hBo_{AB}$, $p\ge 2$}
 \label{ss31vi23.1}
 
 \paragraph{The case $m=0$:}  It follows from the pointwise radial conservation law of $\kQ{3,p}{}$ (cf.\ \eqref{9IX22.9}) that the continuity of  $\partial^p _u\tilde \hBo_{AB}^{[\TTt^\perp]}$ at $r_2$ requires
\begin{equation}
    \kQ{3,p}{}[\dt_{\tilde\secN_1}] = \kQ{3,p}{}[\dt_{\tilde\secN_2}]\,.
\end{equation}
The gauge field $\partial_u^{p-1}\kxi{2}{}_{A}^{[(\CKV+\harm)^{\perp}]}$ is used to achieve the matching of 
\newline
$(\kQ{3,p}{})^{[(\CKV+\harm)^{\perp}]}$ according to \eqref{10IX22.8}.

On $S^2$, this ensures the continuity of $\partial^p_u\tilde\hBo_{AB}$ at $r_2$ since then $\partial^p_u \vp{\tilde\hBo}_{AB} = \partial^p_u{\tilde\hBo}_{AB}$.

 For the remaining topologies we return to \eqref{30XI22.w2}.
Taking into account the gauge invariance of $\hBo_{AB}^{[\TTt]}$,  Equation~\eqref{30XI22.w2}
provides a necessary and sufficient condition for the continuity of $\partial_u^p\tilde\hBo_{AB}^{[\TTt]}$ at $r_2$ according to:
\ptcheck{2XII}
\begin{align}
    &\partial^p_u q_{AB}^{[\TTt]}|_{\tilde\secN_2} -\partial^p_u q_{AB}^{[\TTt]}|_{\tilde\secN_1}
    \label{30XI22.w3} 
    \\
    & =
\alpha^2 \sum_{k=0}^{p-1}(\alpha^2 r_2)^k \bigg[
 s \ \kq{p-k}_{AB}^{[\TTt]}\big|_{r_1}
    +  \frac{1}{2 s}
      (\myGauss - \alpha^2 s^2)
       \partial_u^{p-1-k}
       \hBo_{AB}^{[\TTt]}\big|_{s}\bigg]_{r_1}^{r_2}
      \nonumber
\\
    &\quad
    +\alpha^{2(p+1)}   r_2 ^p
    \bigg[\,
     \kphi{1}_{AB}^{[\TTt]}
    +  \int_{r_1}^{r_2} \hkappaone(s)\interph_{AB}^{[\TTt]} \, ds
    \,
    \bigg]
    \nonumber
\\
   &=
\alpha^2 \sum_{k=0}^{p-1}(\alpha^2 r_2)^k \bigg[
  s \ \kq{p-k}_{AB}^{[\TTt]}\big|_{r_1}
    +  \frac{1}{2 s}
      (\myGauss - \alpha^2 s^2)
       \partial_u^{p-1-k}
       \hBo_{AB}^{[\TTt]}\big|_{s}\bigg]_{r_1}^{r_2}
      \nonumber
\\
    &\quad
    + (\alpha^2 r_2)^p (q_{AB}^{[\TTt]}|_{\tilde\secN_2} -q_{AB}^{[\TTt]}|_{\tilde\secN_1})
    \,,
    \nn
\end{align}
where we   used the formula \eqref{9XI22.w2b2} for $\kphi{1}_{AB}^{[\TTt]}$ in the last step. Equation~\eqref{30XI22.w3} provides a further obstruction to be satisfied by the data. When $\alpha = 0$, the condition reduces to
\begin{equation}
 \label{20XI22.51}
 \forall \
  0 \le p \le k{-1}
 \qquad
 \overset{[p+1]}{q}{}_{AB}^{[\TTt]}[\dt_{ \secN_1}] = \overset{[p+1]}{q}{}_{AB}^{[\TTt]}[\dt_{ \secN_2}]
 \,.
\end{equation}

\paragraph{The case $m\neq0$:}

The continuity at $r_2$ of the $\TTtp$-
part of $\partial^p_u h_{AB}$ has already been addressed in Section~\ref{s10IX22.1}.
Taking into account the allowed gauge perturbations of the linearised gravitational field, it follows from~\eqref{18VI22.3} that 
it remains to satisfy the equation
\ptcheck{20XI} 
\begin{align}
   \partial^p_u \hBo_{AB}^{[\TTt]}|_{\tilde{\secN}_2}  &=
    \Big(
     \overadd{p}{\tilde\Psi}_{AB}(r_2, x^A)
    - 2 r_2^2 \TS [\zspaceD_{\tdA}\partial_{\tdu}^{p}\xiB{2}]
    \Big)^{[\TTt]}
        \label{9XI22.w3}
\\
    &\quad
    + \Big(\sum_{j=0}^{2} \sum_{\ell=2p+1}^{2p+2} \overset{({p},j)}{\opsi}_{\ell} (r_2)  P^j  \kphi{\ell}_{AB}
     \Big)^{[\TTt]}
       \,.
       \nn
\end{align}
Here, for the sake of induction, we treated the fields $\partial_u^{j}\kxi{2}{}_A$ for $0\leq j\leq p-1$ and $\kphi{\ell}_{AB}$ for $1\leq \ell\leq 2p$ as known, and  collected them together with the remaining known fields into the term $\overadd{p}{\tilde\Psi}_{AB}(r_2, x^A)$.
This equation is non-trivial only for $\T^2$ and for cross-sections $\secN$ of higher genus, 
    can be solved using a linear combination of  $\kphi{2p+2}_{AB}^{\blue{[\TTt]}}$ and $\kphi{2p+1}_{AB}^{\blue{[\TTt]}}$ :
\begin{align}
   \partial^p_u \hBo^{\blue{[\TTt]}}_{AB}|_{\tilde{\secN}_2}
     &=
   \overadd{p}{\tilde\Psi}{}^{\blue{[\TTt]}}_{AB}(r_2,x^A)	
   +
     \overset{({p},0)}{\opsi}_{2p+2} (r_2)\kphi{2p+2}^{\blue{[\TTt]}}_{AB}
    +  \overset{({p},0)}{\opsi}_{2p+1} (r_2) \kphi{2p+1}^{\blue{[\TTt]}}_{AB}
       \,.
       \label{8XI22.w2a}
\end{align}

%% file: LinearIsomorphism.tex
\section{An isomorphism theorem}
 \label{s7I23.1}

In Section~\ref{ss25IV22.1} we have verified  that  a consistent scheme for the linearised equations is  obtained
\begin{equation}
 \nonumber 
 \fbox{ if we assume that  $h_{AB}(r,\cdot) \in \Hgamma$,  for all  $r\in[r_1,r_2]$, with $k_\gamma \ge 4$}
\end{equation}
 and
\begin{align}\label{25IV22.p6b}
  k_\beta &= k_\gamma
  \,,\quad
  k_U = k_\gamma -1
  \,,\quad
  k_V = k_\gamma -2
  \,,\quad
  k_{\partial_u U} = k_{\gamma} - 3
  \,,
  \\
  k_{\partial_u V} &= k_\gamma - 4
  \,,\quad
  k_{\partial_u \gamma}  = k_\gamma -2
  \, . \nn
\end{align}
as well as
\begin{equation}\label{1VII23.2b}
   k_{\xi^u}=k_\gamma+2\,,\quad
        k_{\partial_u\xi^u}=k_\gamma\,,\quad
        k_{\xi^A}= k_\gamma+1 \,,
\quad
        k_{\partial_u\xi^A}= k_\gamma-1 \,.
\end{equation}

We wish  to check that this is consistent with the equations satisfied by the gluing functions, and that the gluing equations provide surjections in the relevant spaces, with splitting kernels, so that the implicit function theorem can be applied to the full nonlinear problem.

More precisely,
we assume that the gluing fields take the form as in \eqref{16III22.2old} and \eqref{27VII22.1a}. Assuming that the fields at $r_1$ and $r_2$ satisfy \eqref{25IV22.p6b}, and that all $r$-derivatives of the interpolating field $\interph_{AB}$ are in $H_\kg \equiv H_\kg(\secN)$, we want to show that  the gluing fields $\kphi{i}{}_{AB}$ will be in $H_\kg $, and that the gauge functions will satisfy \eqref{1VII23.2b}.

Note that within the scheme presented above,  the $r$-derivatives of all fields preserve their Sobolev  class.

Whether for the sake of induction when $m=0$, or for the analysis of the  elliptic system  \eqref{26VII23.1} when $m\ne 0$,
we assume that each $ u$-derivative of all fields obtained in a previous induction step has a loss of no more than two degrees of differentiability:
\begin{equation}\label{25IV22.p6bi}
  k_{\partial_u^i U} = k_{\gamma} - 1 - 2i
  \,,\quad
  k_{\partial^i_u V} = k_\gamma    - 2i
  \,,\quad
  k_{\partial^i_u \gamma}  = k_\gamma  - 2i
  \,,
\end{equation}
while for the gauge fields it holds that 
\begin{equation}\label{1VII23.2bi}
        k_{\partial^i_u\xi^u}=k_\gamma \red{+2} - 2i \,,\quad
        k_{\partial^i_u\xi^A}= k_\gamma \red{+1} - 2i  \,.
\end{equation}
{\redca{4IV24; implemented in skylatex all until this one}
(we have listed the condition on $\partial^i_u\xi^u$ with $i\ge 1$ for clarity of the arguments below, but in fact it follows from the one on $\partial^i_u\xi^A$ since we impose \eqref{5XII19.1a}, namely
\begin{equation}\label{5XII19.1awer}
	 \partial_{\tdu}  \xi^{u}(\tdu,x^A) =
\frac{ \zspaceD_{\tdB} \TSxi^{B}(\tdu, x^{A})}{2}
 \, .
\end{equation}
}

We start with the $ \beta$-equation, the linearised version of which, namely \eqref{CBCHG:beta_eq}, 
is clearly  consistent with the above.

Next, the field   $\ochi [\dt_{\secN_1}]$ given by \peqref{7I23.3} is in $H_{\kgamma -2}$. \underline{When $m=0$}, it follows by standard elliptic estimates that the solution  $
    {(\kxi{2}^{u})^{\kerLp}} $
of \eqref{24IX22.3a} is in $H_{\kgamma+2}$. Since the field $(\kxi{2}^{u})^{\kerL}$,
which is left undetermined at that stage of the argument,
is smooth, we conclude that
$$
    \kxi{2}^{u} \in H_{\kgamma+2}
   \,,
$$
as desired.  \underline{When $m \neq 0$}, we use again the  field $(\kxi{2}^{u})^{\kerLp} $ determined
by \eqref{24IX22.3a} and,
 on $S^2$,  the field $(\zspaceD _A\kxi{2}^A)^{[=1]}$, determined by  \eqref{24IX22.3b};
   this is implemented by a smooth field $ \kxi{2}\in \CKV$.

We pass now to the
   covector fields
$\kQ{3,i+1}{A}$ of \eqref{10IX22.9}, which are  in $H_{\kg - 2 i-3}$.
\underline{When $m=0$}
the vector fields $\partial_u^i \kxi{2}{}_A^{[(\CKV+\harm)^{\perp}]}$,
$i\ge 0$,
solve the fourth-order elliptic equation  \eqref{10IX22.8} (cf.\ Proposition~\pref{P30X22.1a}), hence we have
 $$
  \partial_u^i \kxi{2}{}_A\in H_{\kg - 2 i+1}
  \,, \quad i \ge 0 \,,
$$
 as desired.
\underline{When  $m\ne 0$,} the solution of the set of equations \eqref{21VIII23.90}-\eqref{21VIII23.100},  as obtained in Section~\ref{s10IX22.1}, is again of the desired differentiability class when the condition \eqref{2VIII23.1} is imposed.

As the next step, consider  the field  $\kphi{1}_{AB}\equiv \ip{\hkappaone}{\wh_{AB}}$ solving \eqref{25VII22.7}, cf.\ also \eqref{4X122.w2}.
Here a note on   the differentiability class of the boundary term appearing in \eqref{25VII22.7},
or in the equations that will be referred to in what follows, is in order. We have already checked that the differentiability classes in \eqref{25IV22.p6bi}-\eqref{1VII23.2bi} are consistent with all the equations. The boundary terms in all integrated equations are obtained by integration by parts in the $r$-variable,
which does not change the differentiability in the $\secN$-directions.
The boundary terms for the gauge equations only involve the known boundary data, which have the needed regularity by hypothesis.
The boundary terms for the gluing fields are either determined by the known boundary data,   or from the gauge fields.
\underline{When  $m=0$}, the hierarchical structure of our proof implies that the boundary terms arising at each further step of our analysis will have the right differentiability properties for the induction argument.
When $m\ne0$ the solutions are obtained by a global system of equations, with boundary terms determined by the known boundary data.

In any case,
one can directly chase through the derivatives of the already known fields which appear in \eqref{4X122.w2}-\eqref{25VII22.7},
 and use  that  on two-dimensional manifolds the operator $\zdivtwo\!$ is elliptic, and that the field $ \kphi{1}{}_{AB}^{[\TTt]} $
 solving \eqref{9XI22.w2b2}
  is smooth, to obtain that
 $$
  \kphi{1}{}_{AB} \in H_{\kg}
  \,.
$$
 as needed.


 \underline{When $m=0$,}
 we continue with \eqref{24X22.91}, where $p$ is an induction parameter. The operator appearing at the left-hand side is elliptic, of order $2p+1$.   By the arguments just given, or by a direct inductive calculation, the boundary terms $\overadd{{p}}{\hat {\tilde X}}_{A}$ are in $ H_{\kg-2p -3}$.
 For $p \le k-2$, where $k$ is the number of derivatives that we wish to glue, the operator recovers the number of derivatives lost by the right-hand side, which results in a field
\begin{equation}\label{10VI23.1b}
  \kphi{{p}+4}^{[\TTt^\perp]}_{AB} \in H_{\kg}
   \,.
\end{equation}
\underline{The case $m\ne0$} is taken care of by Theorem~\ref{t29VII23.1}.

Keeping in mind that all the operators considered in the analysis of the linearised equations have splitting kernels, we have proved:

\begin{theorem}
  \label{T1VII23.1}
The linearised gluing map, assigning to the collection
$$
\{\mbox{data at $r_1$, the gluing fields, and the gauge fields}\}
$$
 the data at $r_2$, as described above, is continuous and surjective in the Sobolev spaces defined in \eqref{25IV22.p6bi}-\eqref{1VII23.2bi}, with splitting kernel, modulo the obstructions listed in Tables~\pref{T17XI22.1} and \pref{T17XI22.2}.
\end{theorem}

%% file: PureChargeField.tex
\section{Unobstructed gluing to perturbed data}
 \label{s13VII22.1}

Given that there exist obstructions to glue  two arbitrary characteristic data sets of order $k$, the question arises whether something can be done about that. Since we are dealing with linear equations, the simplest solution is to add to the data another data set with charges chosen to compensate for the obstructions. This requires families of data sets with a sufficient number of radial charges to cover all obstructions.

Now, a static family of such data sets can be obtained by differentiating the Birmingham-Kottler metrics with respect to mass:
\be
\frac{d}{dm}
 \Big[\big(\twoscsign
 -{\alpha^2} r^2  {-\frac{2m}{r}}
  \big) du^2-2du \, dr
 + r^2 \ringh_{AB}dx^A dx^B
 \Big]
 =   -\frac{2}{r}
  du^2
   \,.
   \label{4XII22.1}
\ee
These metric perturbations can be used to compensate for the missing charge $\kQ{2}{}(\lambda)$ with $\lambda =1$.

Another such family is obtained by differentiating \eqref{23VII22.3intro} with respect to a parameter along  a curve  of metrics
$\lambda\mapsto \ringh_{AB}(\lambda)$ with constant scalar curvature:
\be
\frac{d}{d\lambda}
 \Big[\big(\twoscsign
 -{\alpha^2} r^2  {-\frac{2m}{r}}
  \big) du^2-2du \, dr
 + r^2 \ringh_{AB}dx^A dx^B
 \Big]
 =
 r^2 \frac{d\ringh_{AB}}{d\lambda}dx^A dx^B
   \,.
   \label{4XII22.2}
\ee
By~\cite[Theorem~8.15]{FischerTromba}
%
every $\TTt$-tensor, say $\TTm_{AB}$, is tangent to such a curve, and thus metric perturbations of the form
\be
 r^2 \TTm_{AB} dx^A dx^B
 \,,
  \ \mbox{with} \
  \zspaceD_A \TTm^{AB} = 0 = \ringh_{AB} \TTm^{AB}
   \label{4XII22.21}
\ee
provide the missing radial charges $\kq{i}_{AB}^{[\TTt]}$.

Yet another, time-independent, family is provided by differentiating the Kerr-(Anti) de Sitter metrics with respect to the angular-momentum. (Since there is no explicit formula for these metrics in Bondi coordinates, the associated linearised metrics can only be obtained by an indirect calculation.)
When calculated at the (A)dS solution (with $m\ne 0$), this leads to the metric perturbation \eqref{13VIII22.112} below, where $\zlambda^A\partial_A$ is a $u$-independent Killing vector on $S^2$, with the remaining fields  there vanishing.%
\footnote{We are grateful to Finn Gray for checking this.}

It turns out that we can  obtain a family of  metric perturbations compensating for all radial charges needed for $\Ctwo$-gluing by setting
\input{PureChargeField2.tex}

Keeping in mind that $\Ctwo$-gluing with $m\ne0$ needs only the matching of $\kQ1{}$ and $\kQ2{}$, we have proved:

\begin{Theorem}
 \label{T4XII22.1}
Any $\Ctwo$ linearised vacuum data on $\mcN_{(r_0,r_1]}$ can be  $\Ctwo$-glued  to any $\Ctwo$ linearised vacuum data on $\mcN_{[r_2,r_3)}$ after adding to one of them a suitable field of the form \eqref{13VIII22.112}.
\end{Theorem}

\proof
Indeed, when $m\ne 0$
we  only need (cf.\ Table~\pref{T26XI22.1})
\begin{eqnarray}
 \mathring h
  &
 =
  &
  \frac{\zmu}{r}
      du^2
      +
    \frac{\blue{\mathring \Ipsi}_A(x^C)}{r}
      dx^A  du
  \,,
 \label{13VIII22.112c}
\end{eqnarray}
with   $\zmu$  being  a constant   and $\zlambda_A$ being a combination of $\ell=1$ vector harmonics satisfying $\zspaceD^A \mathring \Ipsi_A = 0$ on   $S^2$; with constant $\zmu$ and covariantly constant $\zlambda_A$ on $\T^2$; with constant $\zmu$ and vanishing $\zlambda_A$ on higher genus manifolds. In all cases the fields are chosen so that the radial charges
 \begin{align}
     &\kQ{1}{}(\pi) = -3 \int_{\secN}\pi^A \mathring \Ipsi_A \,\sm \,,
     \qquad
      \kQ{2}{}(\lambda) = -\int_{\secN}\lambda \mathring \mu \,\sm
      \,,
       \label{9XII22.15p}
 \end{align}
 compensate for the difference of radial charges calculated from the fields at $r_1$ and at $r_2$.

 When $m=0$   we obtain the desired fields by choosing $\zmu$ and $\zlambda_A$ so that the radial charges in \eqref{9XII22.15p} compensate for the difference of the respective radial charges at $r_1$ and $r_2$  at $u=0$, and
by choosing
  \begin{equation}\label{9XII22.31}
  \TTm_{AB} \big|_{u=0}=0 =
    \qt_{AB} \big|_{u=0}
    \,.
  \end{equation}
The remaining fields vanish on $S^2$, in which case we are done.

Otherwise  recall the obstruction
\eqref{30XI22.w3} with $p=1$:
\begin{align}
    \kq{2}{}_{AB}^{[\TTt]}|_{\tilde\secN_2}\Big|_{r_1}^{r_2}
    &=
\alpha^2
  \bigg[
     s q _{AB}^{[\TTt]}\big|_{r_1}
    +  \frac{1}{2 s}
      (\myGauss - \alpha^2 s^2)
       \hBo_{AB}^{[\TTt]}\big|_{s}\bigg]_{r_1}^{r_2}
    + \alpha^2 r_2  q_{AB}^{[\TTt]} \Big|_{r_1}^{r_2}
     \nonumber
\\
    &=
\alpha^2
  \bigg[
     s q _{AB}^{[\TTt]}\big|_{s}
    +  \frac{1}{2 s}
      (\myGauss - \alpha^2 s^2)
       \hBo_{AB}^{[\TTt]}\big|_{s}\bigg]_{r_1}^{r_2}
    \,.
\nn 
\end{align}
So at, say,  $r=r_2$  we can compensate
all
 radial charge
  deficits by choosing the remaining fields as
 \begin{align}
 \partial_u \qt_{AB}^{[\TTt]}\big|_{u=0}
 &=
 \begin{cases}
     -  q_{AB}^{[\TTt]}\Big|^{r_2}_{r_1}
\\
    0 \ ,
 \end{cases}
 \nonumber %
\\
     \partial_u ^2\qt_{AB}^{[\TTt]}\big|_{u=0}
      &=
      \begin{cases}
     -  \kq{2}_{AB}^{[\TTt]}\Big|_{r_1}^{r_2}
\\
      -  \kq{2}_{AB}^{[\TTt]}\Big|^{r_2}_{r_1}
      +
    \alpha^2 \bigg[
     {s q _{AB}^{[\TTt]}\big|_{s}
    +
     \frac{1}{2 s}
      (\myGauss - \alpha^2 s^2)
       \hBo_{AB}^{[\TTt]}\big|_{s}\bigg]_{r_1}^{r_2}
     }
      \ ,
 \nonumber 
 \end{cases}
\end{align}
with the upper case being for $\alpha = 0$ and the lower case for $\alpha\neq 0$, and
\begin{align}
 (\zspaceD^B \partial_u \qt_{AB})^{[\harm]}\big|_{u=0}
 &=
 \begin{cases}
     0 \ , & \genus = 1\,,
\\
       - \frac 12
 \kQ{3,1}{}^{[\harm]}_A
       \Big|^{r_2}_{r_1} \ , & \genus \geq 2
       \,,
 \end{cases}
 \nonumber %
\\
 (\zspaceD^B \partial^2_u \qt_{AB})^{[\harm]}\big|_{u=0}
 &=
 \begin{cases}
     0 \ , & \genus = 1\,,
\\
      -\frac 12
  \kQ{3,2}{}^{[\harm]}_A
      \Big|^{r_2}_{r_1} \ , & \genus \geq 2 \,,
 \end{cases}
 \nonumber 
 \end{align}
 where $\big[f(r)\big]^{r_2}_{r_1} \equiv\big[f(s)\big]^{r_2}_{r_1} \equiv  f(s)\big|^{r_2}_{r_1} := f(r_2)-f(r_1)$, and where we have used that $\zlambda_A$ vanishes if $\genus \ge 2$.
 \qed

%% file: PureChargeField2.tex
%
\begin{eqnarray}
 \mathring h
  &
 =
  &
 \Big(
   \frac{\zmu(u,x^C)}{r}
    - \frac{\zspaceD^A \zlambda _A(u,x^C)}{2 r^2}
    +   \frac{1}{2r}\zspaceD^A\zspaceD^B \qt_{AB}(u,x^C)
    \Big)
      du^2
 \label{13VIII22.112}
\\
      &&
        +
  \Big(
    \frac{\blue{\mathring \Ipsi}_A(u,x^C)}{r}
     + \frac{1}{2}\zspaceD^B \qt_{AB}(u,x^C)
      \Big)dx^A  du
 \nonumber
\\
      &&
        +
   (r \qt_{AB}(u,x^C) + r^2 \TTm_{AB}(u,x^C))
    dx^A  dx^B
  \,,
  \nonumber
\end{eqnarray}
with symmetric $\ringh$-traceless tensors $\qt_{AB}$ and $\TTm_{AB}$. In addition, anticipating the fact that $\zspaceD^B \qt _{AB}$ plays a role in adjusting $(\kQ{3,i}{})^{[\harm]}$,
 we impose
\begin{equation}
\zspaceD^A \zspaceD^B \qt _{AB}(u,x^A) = 0 \,.
\label{9XII22.2}
\end{equation}
After using
\begin{equation}\label{9XII22.5}
   \zspaceD_A \Delta_{\ringh} \psi^A = ( \Delta_{\ringh}+ \myGauss)  \zspaceD_A  \psi^A
   \,,
\end{equation}
the linearised Einstein equations will hold if and only if
 \ptcheck{9XII}
\begin{align}
&
  \zspaceD^A \zspaceD^B \TTm_{AB}= 0 \,,
  \quad
  {\alpha^2 \qt_{AB} -\partial_u \TTm_{AB}= 0}
  \,,
  \quad
  3m \zspaceD^A\mathring \Ipsi_A = 0
\,,
  \label{13VIII22.12a12}
\\
 & \TS\big[\zspaceD_A    \mathring \Ipsi_{B} \big] + m \qt_{AB}
        = 0 \,,
       \label{13VIII22.12a22}
\\
   &
    3 \alpha^2  \zspaceD^A\zlambda _A   +2 \partial_u \zmu   =0
   \,,
      \quad
    3 \partial_u \zlambda _A -  \zspaceD_A  \zmu  +\frac{1}{2}(\Delta_{\ringh}-\myGauss) \zspaceD^B\qt_{AB}=0
   \,,
    \label{13VIII22.12b2}
\\
    &
    2 \myGauss \mathring \Ipsi_A + \zspaceD_A\zspaceD^B \mathring \Ipsi_B
    - \zspaceD^B\zspaceD_A \mathring  \Ipsi_B + \Delta_{\ringh} \mathring \Ipsi_A + 2 m \zspaceD^B\qt_{AB}= 0\,,
     \label{13VIII22.12c2}
    \\
    &
    \myGauss \zspaceD^A \mathring \Ipsi_A + \frac{1}{2}\Delta_{\ringh} \zspaceD^A \mathring \Ipsi_A = 0\,.
     \label{13VIII22.12c3}
\end{align}
For completeness we listed  above all conditions obtained from the linearised Einstein equations, cf.\ Sections~\ref{sec:28VII22.1} and \ref{sec:25VII22.1}, but we note that \eqref{13VIII22.12a12}-\eqref{13VIII22.12b2} suffice.
Indeed, taking $2\times\zdivtwo\!$ of \eqref{13VIII22.12a22} gives equation \eqref{13VIII22.12c2}, while equation \eqref{13VIII22.12c3} can be obtained by taking $\zdivone\!$ of \eqref{13VIII22.12c2} and by making use of \eqref{13VIII22.12a12}.

Equation~\eqref{13VIII22.12a12} implies that $\zdivone \mathring \Ipsi$ has to vanish when $m\ne 0$, and $\partial_u \TTm_{AB}$ has to vanish when $\alpha = 0$ or when we are on $S^2$. In addition, it follows from \eqref{13VIII22.12a22} that $\mathring \Ipsi_A$ has to vanish when $m=0$ and $\myGauss = -1$.

Equations~\eqref{13VIII22.12b2} together with their $u$-differentiated versions show that
\begin{align}
     2 \partial_u ^2 \mathring \mu &= - 3 \alpha^2  \Delta_{\ringh} \mathring \mu
      \,,
      \label{9XII22.12}
     \\
         \partial_u^2  \mathring \Ipsi _A
          & =
          - \frac {\alpha^2} 2
           \zspaceD_A \zspaceD^B
           \mathring \Ipsi_B
            -\frac{1}{6}(\Delta_{\ringh}-\myGauss) \zspaceD^B\partial_u \qt_{AB}
            \label{9XII22.13}
      \,.
\end{align}
When $m\ne 0$ we can use \eqref{13VIII22.12a22} to rewrite the last equation as
\begin{align}
         \partial_u^2  \mathring \Ipsi _A
          & =
          - \frac {\alpha^2} 2
           \zspaceD_A \zspaceD^B
           \mathring \Ipsi_B
            +\frac{1}{6m}(\Delta_{\ringh}-\myGauss) \zspaceD^B\partial_u
            \TS\big[\zspaceD_A    \mathring \Ipsi_{B} \big]
            \label{9XII22.13b}
      \,.
\end{align}
%
%
%
So, when $m\ne 0$,   Equations~\eqref{9XII22.12} and \eqref{9XII22.13b} provide evolution equations for $ \mathring \mu$ and $\mathring \Ipsi_A$, solutions of which determine the time-evolution of the remaining fields.

To continue, we note
 that equation \eqref{13VIII22.12c2} can be rewritten as,
 \ptcheck{9XII}
\begin{align}
    \frac{1}{2}   (\Delta_{\ringh} + \myGauss)\mathring  \Ipsi_A
    + m \zspaceD^B \qt_{AB} = 0 \,.
    \label{9XII22.1}
\end{align}
Next, the second equation in \eqref{13VIII22.12b2}, together  with \eqref{9XII22.1}, implies that
\begin{align}
   (\Delta_{\ringh} + \myGauss)
      \zspaceD_A  \zmu
    &=  3 (\Delta_{\ringh} +\myGauss)
     \partial_u \zlambda _A
     +
     \frac{1}{2}(\Delta_{\ringh} +\myGauss)
     (\Delta_{\ringh}-\myGauss) \zspaceD^B\qt_{AB}
     \nonumber
     \\
     &=
      - 6 m (\Delta_{\ringh} +\myGauss)
       \zspaceD^B\partial_u\qt_{AB}
     +
     \frac{1}{2}(\Delta_{\ringh} +\myGauss)
     (\Delta_{\ringh}-\myGauss) \zspaceD^B\qt_{AB}
     \,.
 \nonumber %
      \label{9XII22.2b}
\end{align}
Taking $\zdivone\!$ of this and making use of $\zspaceD^A\zspaceD^B \qt_{AB} = 0=\partial_u\zspaceD^A\zspaceD^B \qt_{AB}$  gives
 \ptcheck{9XII22}
\begin{align}
    \zspaceD^A(\Delta_{\ringh} + \myGauss)
      \zspaceD_A  \mathring \mu
    &=
   \Delta_{\ringh}  (\Delta_{\ringh} +2 \myGauss) \mathring \mu
   = 0\,.
 \nonumber 
\end{align}
In particular, when we are not on $S^2$,
the ``mass aspect function''  $\mathring \mu$ must be $x^A$-independent, while
on $S^2$ it is a linear combination of $\ell=0$ and $\ell=1$ spherical harmonic. It follows that
\begin{equation}
 \nonumber 
       \partial_u ^2 \zmu  =
     \left\{
       \begin{array}{ll}
         3   \alpha^2  \mathring \mu , & \hbox{$\secN=S^2$ and $\mathring \mu$ has no $\ell=0$ harmonics;} \\
         0, &  \hbox{$\secN\ne S^2$, or $\secN=S^2$ and $\mathring \mu$ has no $\ell=1$ harmonics.}
       \end{array}
     \right.
\end{equation}

Next,   when $m=0$ the space of $\mathring \Ipsi$'s satisfying
\eqref{13VIII22.12a22} is six-dimensional on $S^2$ and two-dimensional on $\T^2$; for negatively curved $\secN$ one finds $\mathring \Ipsi _A\equiv 0$.

 The tensor field ${\mathring h}_{AB}$ carries the full set of conserved  radial charges needed for $\Ctwo$-gluing
 when $\zmu$, $\zlambda _A$, $\qt_{AB} $ and $\TTm_{AB}$
run over the set of solutions of \eqref{13VIII22.12a12}-\eqref{13VIII22.12b2}:
 \ptcheck{9XII }
 \begin{align}
     &\kQ{1}{}(\pi) = -3 \int_{\secN}\pi^A \mathring \Ipsi_A \,\sm \,,
     \qquad
      \kQ{2}{}(\lambda) = -\int_{\secN}\lambda \mathring \mu \,\sm \,,
       \label{9XII22.15}
\\
     &\ q_{AB}^{[\TTt]} = -\frac{V}{2r}  \TTm_{AB}^{[\TTt]}
     +  \alpha^2 r  \qt_{AB}^{[\TTt]}
     +   \partial_u \qt_{AB}^{[\TTt]}
     \,,
\\
     &\ \kq{2}_{AB}^{[\TTt]} =
       -\frac{\alpha^2 V}{2r}
     \qt_{AB}^{[\TTt]}
     +  \alpha^2 r  \partial_u\qt_{AB}^{[\TTt]}
     +   \partial_u ^2\qt_{AB}^{[\TTt]}
     \,,
\\
     &
     (\kQ{3,1}{})^{[\harm]}_A = -{3}{\alpha^2}\mathring   \Ipsi^{[\harm]}_A
     + \frac{2m}{r }(\zspaceD^B \TTm_{AB})^{[\harm]}
    -\frac{m}{r^2}(\zspaceD^B \qt_{AB})^{[\harm]}
  \label{11XII22.98} 
    \\
    &\qquad\qquad\quad
    +2 (\zspaceD^B \partial_u \qt_{AB})^{[\harm]}\,,
    \nonumber
\\
     &
     (\kQ{3,2}{})^{[\harm]}_A
      = -{3}{\alpha^2} \partial_u \mathring \Ipsi^{[\harm]}_A
      + \frac{2m \alpha^2 }{r }(\zspaceD^B \qt_{AB})^{[\harm]}
    -\frac{m}{r^2}(\zspaceD^B \partial_u\qt_{AB})^{[\harm]}
     \label{7XII22.12}
    \\
    &\qquad\qquad\quad
    +2 (\zspaceD^B \partial^2_u \qt_{AB})^{[\harm]}\,,
    \nonumber
 \end{align}
 where $\pi$ and $\lambda$ satisfy, respectively,
 \eqref{6III22.3} and \eqref{24VII22.6}.
Note that when $\alpha=0=m$, we have $V/r = \myGauss$, in which case all expressions in \eqref{9XII22.15}-\eqref{7XII22.12} are $r$-independent, as they should be in this case.

%% file: Kappas.tex
\section{Constructing the $\kappa_i$'s}
 \label{App13VIII22.1}

Recall that $\hat \kappa_i(s)=s^{-i}$.
We wish to construct a  sequence  of smooth functions ${\kappa}_i$  compactly supported in $(r_1,r_2) $ satisfying
\begin{eqnarray}
  &
  \displaystyle
\ip{\kappa_i}{\hat \kappa_j} \equiv
\int_{r_1}^{r_2} \kappa_i(r) \hat \kappa_j(r) \, dr = 0
 \quad \mbox{for $j<i$}
  \,,
  &
  \label{13VIII22.1}
   \\
   &
  \displaystyle
 \ip{\kappa_i}{\hat \kappa_i}  = 1
  \,.
  &
  \label{13VIII22.2}
\end{eqnarray}
This can be done as follows:
  Let ${\chi}$ be any smooth  non-negative function supported away from neighborhoods of ${r}_1$ and ${r}_2$, with integral 1. Let
  $$
   {\kappa}_i  = c_i {\chi} f_i
   \,,
   $$
   where the $f_i$'s are constructed by a Gram-Schmidt orthonormalisation procedure from the family of monomials in $1/r$, namely $\{1,{r}^{-1},{r}^{-2},\ldots\}$, in the space $\mathbb{H}:= L^2([{r}_1,{r}_2],{\chi} d{r})$,
    so that
    the scalar product is
$$
 \langle \phi,\psi\rangle_{\mathbb{H}}  =
  \int_{{r}_1}^{{r}_2} (\phi \psi {\chi})(r) \, dr
  \,,
$$
and the $c_i$'s are constants  chosen so that
\eqref{13VIII22.2} holds; the possibility of doing so will be justified shortly.
Then, by construction, $f_i$ is a polynomial of order $i$ in $1/r$ which is $\mathbb{H}$-orthogonal to any such polynomial of order $j<i$; this  is \eqref{13VIII22.1}.
As for \eqref{13VIII22.2}, we note that each of the functions
${r}^{-i}$ can be decomposed in the basis $\{f_j\}_{j\in\N}$ as ${r}^{-i}= \sum_{j=0}^i a_{ij} f_j({r})$, with $a_{ii}\ne 0$ since otherwise the right-hand side would be a polynomial in $1/r$ of order less than or equal to $i-1$.
This  shows that
$$
 \int_{{r}_1}^{{r}_2}r^{-i} f_i(r) {\chi}(r) \, dr= a_{ii}\ne 0
 \,,
$$
so that we can indeed choose $c_i = 1/a_{ii}$ to fulfill  \eqref{13VIII22.2}.

%% file: recursion.tex
\section{Recursion formulae}
 \label{App14VIII22.2}

\input{C2recursionCoeff.tex}

Next, recall \eqref{14VIII22.3}:
%
\begin{align}
      \partial_u h_{AB }
      &=
     \frac{\twoscsign }{2 }\Big[
        \partial_r h_{AB}
     - \frac{1}{r }    h_{AB}
     \Big]   +  \int_{r_1}^r
           \bigg(\frac{1}{ 3s r }
    +\frac{2 r^2 }{3 s^4}\bigg)
     P h  _{AB}
     \,
        ds
\label{App14VIII22.3} 
\\
 &
     - (\frac{  {\alpha^2} r^2 }{2 }+{\frac{2m}r} )\Big[
        \partial_r h_{AB}
     - \frac{1}{r }    h_{AB}
     \Big]   +  \int_{r_1}^r
        (\frac{{\alpha^2}r }{s}  {-\frac{mr}{s^4})} \hBo_{AB}\, ds
        \nn
        \\
        &
        + \tk
        \,,
\nonumber 
\end{align}
where
$\tkfbd$ stands for terms known from  data at $r_1$.
 \ptcheck{14VIII22 }

\subsection{$\alpha =  m =  0$}
 \label{ss16XI22.1}
When $\alpha =  m =  0$, inserting \eqref{App14VIII22.3}
 into the $u$-derivative of \eqref{18VI22.3} leads to
\begin{align}
 \partial_u^{i+1} \hBo_{AB}
  =&
 \sum_{0\leq j+k\leq i,k\ne i} \overadd{i,j,k}{\psi}(r)\partial_r^j P^k \partial_u\hBo_{AB}
 \label{14VIII22.6} 
     \\
     &	
+
       \sum_{j=0}^i \int_{r_1}^{r}
       \overadd{i,j}{\psiP} (s,r)P^j \partial_u\hBo_{AB}\, ds
+  \tkfbd    \nonumber
\\
  =&
 \sum_{0\leq j+k\leq i,k\ne i} \overadd{i,j,k}{\psi}(r)\partial_r^j P^k
\Big[
  \frac{\twoscsign }{2 }\Big[
        \partial_r h_{AB}
     - \frac{1}{r }    h_{AB}
     \Big]
           \nn
           \\
          & \phantom{xxxxxxxxxxxxxxxxxxxxxx}
    +  \int_{r_1}^r
           \bigg(\frac{1}{ 3s r }+\frac{2 r^2 }{3 s^4}\bigg)
     P h  _{AB}\, ds\Big]
        \nonumber
\\
    & \quad
 +
       \sum_{j=0}^i \int_{r_1}^{r}
      \underbrace{
        \overadd{i,j}{\psiP} (s,r)P^j
       \Bigg[
  \frac{\twoscsign }{2 }\Big [
        \partial_s h_{AB}
        }_{\mbox{integrate by parts}}
     - \frac{1}{s }    h_{AB}
     \Big]_s
     \nn
           \\
          & \phantom{xxxxxxxzzzzz}
          +  \int_{r_1}^s
           \bigg(\frac{1}{ 3y s }
    +\frac{2 s^2 }{3 y^4}\bigg)
     P h  _{AB}|_y
        dy
        \Bigg]
        ds
        +  \tkfbd
         \nonumber
\\
  =&
 \sum_{0\leq j+k\leq i,k\ne i} \overadd{i,j,k}{\psi}(r)\partial_r^j P^k
\Big[
  \frac{\twoscsign }{2 }\Big[
        \partial_r h_{AB}
     - \frac{1}{r }    h_{AB}
     \Big]
        \nn
           \\
          & \phantom{xxxxxxxxxxxxxxxxxxxxxx}
          +    \int_{r_1}^r
           \bigg(\frac{1}{ 3s r }+\frac{2 r^2 }{3 s^4}\bigg)
     P h  _{AB}\, ds\Big]
        \nonumber
\\
    &\quad
    +
      \frac \twoscsign2  \sum_{j=0}^i
       \overadd{i,j}{\psiP} (s,r) \big|_{s=r}P^j
        h_{AB}\nn
        \\
        &\quad
 +
       \sum_{j=0}^i \int_{r_1}^{r}
      \big(-\frac \twoscsign 2 \partial_s \overadd{i,j}{\psiP} (s,r)
      - \frac {\twoscsign}{2 s}
       \overadd{i,j}{\psiP} (s,r)
       \big)
        P^j   h_{AB}\big|_s
        ds
        \nonumber
\\
        &\quad
        +
       \sum_{j=0}^i  \underbrace{ \int_{r_1}^{r}
         \overadd{i,j}{\psiP} (s,r)  \int_{r_1}^s
           \bigg(\frac{1}{ 3y s }
    +\frac{2 s^2 }{3 y^4}\bigg)
     P^{j+1} h  _{AB}|_y
        dy
        \, ds
        }_{
    \displaystyle
     \int_{r_1}^{r} \bigg(\int_{s}^r
     \bigg(\frac{1}{ 3y s }
    +\frac{2 y^2 }{3 s^4}\bigg)
     \overadd{i,j}{\psiP} (y,r)
        dy
        \bigg)
     P^{j+1} h  _{AB}|_s
        ds
        }
        +  \tkfbd
        \,.
\nonumber 
\end{align}
%

One finds that
a term
\begin{equation}\label{16VIII22.w1}
 a_{ki\ell} s^{-\ell} \ \mbox{ in $\overadd{k,i}{\psiP}$ }
\end{equation}
 with $\ell\not \in \{0,3\}$  induces  terms $s^{-1}$, $s^{-4}$ and
\begin{equation}\label{16VIII22.12}
 a_{ki\ell} \twoscsign
   \frac{\ell-1}{2}
 s^{-(\ell+1)}
 \ \mbox{ in $\overadd{k+1,i }{\psiP}$ and } \
 a_{ki\ell}
 \frac{\ell-1}{\ell(\ell-3)}
 s^{-(\ell+1)}
 \ \mbox{ in $\overadd{k+1,i+1}{\psiP}$;}
\end{equation}
see Figure~\ref{FtreeNoMass},
where we have anticipated the fact that the highest powers of $s^{-1}$ are not affected by $\alpha$.
\input{treeFigureNoMass}
We thus find
\begin{equation}\label{19VIII22.a2}
  \overset{(k,k)}{\psiP}(s,r) =
  \underbrace{\frac{2   r^2 }{ (k-1)!(k+2)  }
   }_{=:\overset{(k,k)}{\psiP}_{k+3}(r)}
    \frac{1}{ s^{k+3}} + \ldots
   \,,
\end{equation}
where $\ldots$ denotes a sum of lower-order powers of $s^{-1}$.

An  identical calculation applies to the $\overadd{k,i}{\chi}$'s, since  \eqref{18VI22.1} has an identical structure as \eqref{18VI22.3} from the point of view of induction.
In particular the recurrence relation  \eqref{16VIII22.w1}-\eqref{16VIII22.12} remains unchanged. After taking into account the initialisation of the recurrence, which is different for the $\chi$'s and $\psi$'s, one obtains
 \ptcheck{ 22VIII by wan}
\begin{equation}\label{19VIII22.2b}
  \overset{(k,k)}{\chi} =
  \underbrace{
   \frac{ 1 }{ (k )! (k+3) }
   }_{=:\overset{(k,k)}{\chi}_{k+4}}
    \frac{1}{ s^{k+4}} + \ldots
   \,.
\end{equation}

Let us write
\begin{equation}
\label{18VIII22.w1rc}
    \overset{(i,j)}{\chi} (s,r) = \sum_{\ell=1}^{i+4} \overset{(i,j)}{\ochi}_{\ell}(r)\, s^{-\ell}
    \,,
    \qquad
    \overset{(i,j)}{\psi} (s,r) = \sum_{\ell=1}^{i+4} \overset{(i,j)}{\psi}_{\ell}(r)\, s^{-\ell}
    \,.
\end{equation}
Since  (cf.~\eqref{25XI22.5}-\eqref{25XI22.6} with $m=0$, and regardless of $\alpha$)
\begin{align*}
    \overadd{2,0}{\opsi}_{5}(r)& = 0 =: 2r^2 \overadd{0,-1}{\ochi}_{5}(r) \,,\quad
    \overadd{2,1}{\opsi}_{5}(r) =2r^2 \overadd{1,0}{\ochi}_{5}(r)
    \,,\nn
    \\
    \overadd{2,2}{\opsi}_{5}(r) &= 2r^2\overadd{1,1}{\ochi}_{5}(r)\,,
\end{align*}
it follows by induction from \eqref{16VIII22.12} that
\begin{equation}
    \overset{(i,j)}{\opsi}_{i+3}(s,r) = 2
    r^2 \, \overset{(i-1,j-1)}{\ochi}_{i+3}(s,r)
    \,.
    \label{3IX22.w1}
\end{equation}

Next, using
\begin{equation}
    \overadd{1,0}{\chi}(s,r) = \frac{{\alpha^2} \hkappaone(s)}{2 r^2 }+\frac{\twoscsign\hkappafive(s) }{2}\,,
    \quad
    \overadd{1,1}{\chi} (s,r) = \frac{\hkappafive(s)}{4 }-\frac{\hkappaone(s)}{4 r^4 }\,,
\end{equation}
(cf.\ ~\eqref{25XI22.3}-\eqref{25XI22.6}) it follows by induction
that
\begin{equation}
    \overadd{i,j}{\ochi}_{i+3}(s,r) = 0\,.
    \label{2IX22.w1}
\end{equation}

%% file: C2recursionCoeff.tex
For ease of further reference we collect here all the integral kernels appearing in \eqref{18VI22.3}-\eqref{18VI22.1}, as needed for $\Ctwo$-gluing and for various induction arguments in the rest of this Appendix:
\begin{align}
\overadd{1,0,0}{\psi}(r) &=
     - \frac{1}{2r} \left(\twoscsign-{\alpha^2} r^2 {-\frac{2m}{r}}\right)
     \,,
     \\
     \overadd{1,1,0}{\psi}(r) &=
     \frac{1}{2} \left(\twoscsign-{\alpha^2} r^2 {-\frac{2m}{r}}\right) \,, \quad
   \overadd{1,0,1}{\psi}(r)
       =
       0
    \,,
\\
    \overadd{1,0}{\psi}(s,r)&= {{\alpha^2} r}  \hat{\kappa}_1(s)
    {-mr \hkappafour(s)}
    \,,\quad
    \overadd{1,1}{\psiP} (s,r)=\frac{2 r^2 \hkappafour(s)}{3 }+\frac{\hkappaone(s)}{3 r}
    \,,
    \label{25XI22.5}
\\
    \overadd{2,0}{\psi}(s,r)&=\frac{9 m^2 r}{2}\hkappa_6(s) + \frac{\alpha^2 (2 m + 4 r^3 \alpha^2)}{4 r}\hkappa_1(s)
 \label{25XI22.10} 
\\
    &\quad
    -  \frac{8 m^2 + 16 m r^3 \alpha^2}{16 r }\hkappafour(s)
    -  \frac{3 m r \myGauss}{2} \hkappafive(s)
   \,,
 \nonumber 
\\
    \overadd{2,1}{\psiP} (s,r)
    & =
    (r^2\twoscsign-\frac{3{m}r}{4})\hkappafive(s)
 +\frac{(9{m}-4r\twoscsign)\hkappaone(s)}{12 r^3}+\frac{\twoscsign r \hkappafour(s)}{3} 
   \label{25XI22.15} 
 \\
 & \quad -3{m}r^2\hkappa_6(s)
    \,,
\nonumber 
\\
      \overadd{2,2}{\psiP} (s,r)&=\frac{r^2\hkappafive(s)}{2 }-\frac{\hkappaone(s)}{6 r^2 }-\frac{r\hkappafour(s)}{3 }\,,
  \label{25XI22.15b}
\\
  \overadd{0,0,0}{\chi}(r) &=0
   \,,
   \quad
      \overadd{1,1,0}{\chi}(r) = \overadd{1,0,1}{\chi}(r) =0
   \,,
   \quad   \overadd{1,0,0}{\chi}(r)  = \frac{\twoscsign}{2 r^4}-\frac{{\alpha^2}}{2 r^2}
    \,,
\\
   \overadd{0,0}{\chi}(s,r)& =\frac{1}{3}
    \left(
     \frac{2\hkappaone(s)}{r^3} + \hkappafour(s)
     \right)
      \,,
\\
       \overadd{1,0}{\chi}(s,r) &=
  {-\frac{3m\hkappasix(s)}{2 }
    -\frac{m\hkappafour(s)}{2r^2 }
}
 +\frac{{\alpha^2} \hkappaone(s)}{2 r^2 }
 +\frac{\twoscsign\hkappafive(s) }{2}\,,
      \label{25XI22.3}
\\
    \overadd{1,1}{\chi} (s,r)& =
  \frac{\hkappafive(s)}{4 }-\frac{\hkappaone(s)}{4 r^4 }\,,
      \label{25XI22.6}
\\
 \overadd{2,0}{\chi}(s,r) &=
 {m}\bigg(\frac{75 {m}}{8 }\hkappa_8(s)
 -  \frac{77  \twoscsign}{12}\hkappa_7(s)
 + \frac{3}{8 }  (5 \alpha^2
  -  \frac{2 \twoscsign}{r^2})\hkappafive(s) \bigg)
 \label{25XI22.4} 
  \\
  &\quad
 + \frac{\alpha^2 (15 {m} + 8 r \twoscsign)}{24 r^4 }\hkappaone(s)
 -  \frac{15 {m}^2 + 8 r ({m}
 + r^3 \alpha^2) \twoscsign}{24 r^4 }\hkappa_4(s)
       \nonumber
\\
 &\quad
 + (\frac{9 {m}^2}{4 r^2}
 + \twoscsign^2)\hkappa_6(s)
 \,,
\nonumber 
\\
 \overadd{2,1}{\chi} (s,r) &=
 - \frac{7 {m}}{4 }\hkappa_7(s)
 -  \frac{3 {m}}{8 r^2 }\hkappa_5(s)
 + \frac{7 \twoscsign}{10}\hkappa_6(s)
 \label{25XI22.7} 
\\
 &\quad
 + \frac{6 {m} - 2 r^3 \alpha^2 + r \twoscsign}{6 r^3 }\hkappa_4(s)
  + \frac{15 {m} - 80 r^3 \alpha^2 + 16 r \twoscsign}{120 r^6}
      \hkappa_1(s)
 \,,
\nonumber 
\\
 \overadd{2,2}{\chi} (s,r)
 &= \frac{\hkappaone(s)}{15 r^5 }
 -\frac{\hkappafour(s)}{6 r^2}+\frac{\hkappasix(s)}{10 }
    \,.
    \label{3VIII22.2}
\end{align}
These  are all linear combinations of the $\hat \kappa_i$'s
with  $0\le i\le 8$, {$i\not\in \{2, 3\}$,}
with coefficients which might depend upon $r$.

%% file: treeFigureNoMass.tex
\begin{figure}
  \centering
\begin{tikzpicture}

\small

\def\x{1.3} 
\def\xy{0.3} 
\def\y{1.1} 
\def\z{0}
\def\l{0}
\def\d{0}
\def\dx{.5}

{\small

\node[align=center, name=045] at (0.4*\x -\dx ,5*\y +\d) {$ (i,j)$  };
\node[align=center, name=044] at (0.4*\x -\dx,4*\y +\d) {$ i=1 $};
\node[align=center, name=043] at (0.4*\x -\dx,3*\y +\d) {$2 $};
\node[align=center, name=042] at (0.4*\x -\dx,2*\y +\d) {3 };
\node[align=center, name=041] at (0.4*\x -\dx, \y +\d) {4 };
\node[align=center, name=040] at (0.4*\x -\dx, \d) {5 };
}

\node[align=center, name=161] at (\x,5*\y +\d) {$j=1$};
\node[align=center, name=162] at (2*\x,5*\y +\d) {$2$};
\node[align=center, name=163] at (3*\x,5*\y +\d) {$3$};
\node[align=center, name=164] at (4*\x,5*\y +\d) {$4$};
\node[align=center, name=165] at (5*\x,5*\y +\d) {$5$};

\node[align=center, name=11] at (\x,0*\y - \z) {$s^{-8}$};
\node[align=center, name=12] at (\x,\y - \z) {$ s^{-7}$};
\node[align=center, name=13] at (\x,2*\y) {$ s^{-6}$};
\node[align=center, name=14] at (\x,3*\y +\d) {$ s^{-5}$};
\node[align=center, name=15] at (\x,4*\y ) {$ s^{-4}$};

\node[align=center, name=21] at (2*\x,0*\y - \z) {$ s^{-8}$};
\node[align=center, name=22] at (2*\x,\y - \z) {$ s^{-7}$};
\node[align=center, name=23] at (2*\x,2*\y) {$ s^{-6}$};
\node[align=center, name=24] at (2*\x,3*\y +\d) {$ s^{-5}$};

\node[align=center, name=30] at (3*\x,0*\y - \z) {$ s^{-8}$};
\node[align=center, name=31] at (3*\x,\y - \z) {$ s^{-7}$};
\node[align=center, name=32] at (3*\x,2*\y) {$ s^{-6}$};

\node[align=center, name=41] at (4*\x,0*\y - \z) {$ s^{-8}$};
\node[align=center, name=42] at (4*\x,\y - \z) {$ s^{-7}$};
\node[align=center, name=43] at (4*\x,2*\y) {$  $};

\node[align=center, name=51] at (5*\x,0*\y - \z) {$ s^{-8}$};


\draw[line width=0.2mm, ->, densely dotted,color=blue] ($(42.south) + (0.15,0)$) -- ($(41.north) + (0.15,0)$);

\draw[line width=0.2mm, ->, densely dotted,color=blue] ($(15.south) + (0.15,0)$) -- ($(14.north) + (0.15,0)$);
\draw[line width=0.2mm, ->, densely dotted,color=blue] ($(32.south) + (0.15,0)$) -- ($(31.north) + (0.15,0)$);
\draw[line width=0.2mm, ->, densely dotted,color=blue] ($(31.south) + (0.15,0)$) -- ($(30.north) + (0.15,0)$);

\draw[line width=0.2mm, ->,   dotted,color=blue] (14.south east) -- (23.north west);

\draw[line width=0.2mm, ->, densely dotted,color=blue] ($(14.south) + (0.15,0)$) -- ($(13.north) + (0.15,0)$);
\draw[line width=0.2mm, ->,   dotted,color=blue] (14.south east) -- (23.north west);

\draw[line width=0.2mm, ->, densely dotted,color=blue] ($(13.south) + (0.15,0)$) -- ($(12.north) + (0.15,0)$);
\draw[line width=0.2mm, ->,   dotted,color=blue] (13.south east) -- (22.north west);

\draw[line width=0.2mm, ->, densely dotted,color=blue] ($(12.south) + (0.15,0)$) -- ($(11.north) + (0.15,0)$);
\draw[line width=0.2mm, ->,   dotted,color=blue] (12.south east) -- (21.north west);

\draw[line width=0.2mm, ->, densely dotted,color=blue] ($(24.south) + (0.15,0)$) -- ($(23.north) + (0.15,0)$);
\draw[line width=0.2mm, ->,   dotted,color=blue] (24.south east) -- (32.north west);

\draw[line width=0.2mm, ->, densely dotted,color=blue] ($(23.south) + (0.15,0)$) -- ($(22.north) + (0.15,0)$);
\draw[line width=0.2mm, ->, dotted,color=blue] (23.south east) -- (31.north west);

\draw[line width=0.2mm, ->, densely dotted,color=blue] ($(22.south) + (0.15,0)$) -- ($(21.north) + (0.15,0)$);
\draw[line width=0.2mm, ->,   dotted,color=blue] (22.south east) -- (30.north west);

;
\draw[line width=0.2mm, ->,   dotted,color=blue] (31.south east) -- (41.north west);


\draw[line width=0.2mm, ->,   dotted,color=blue] (15.south east) -- (24.north west);
\draw[line width=0.2mm, ->,   dotted,color=blue] (24.south east) -- (32.north west);
\draw[line width=0.2mm, ->,   dotted,color=blue] (32.south east) -- (42.north west);
\draw[line width=0.2mm, ->,   dotted,color=blue] (42.south east) -- (51.north west);

\scriptsize

%
%
%
%


\small

%
%

\end{tikzpicture}
\caption{Highest powers of $s^{-1}$ in
$\protect \overadd{i,j}{\psi}$
when $  m= 0$,  $\alpha \in \R$.   The structure of the tree for
$\protect \overadd{i,j}{\chi}$ is identical after replacing $(i,j)$ in the table by $(i-1,j-1)$, thus $(1,1)$ becomes $(0,0)$, etc.
}
   \label{FtreeNoMass}
\end{figure}
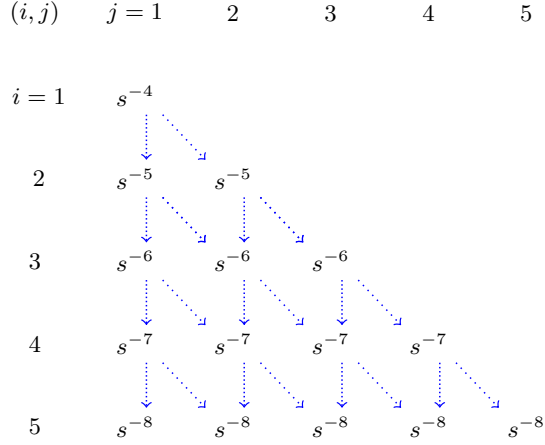

%% file: recursionWithAlpha.tex
\subsection{The general case}
 \label{ss16XI22.2}
When $\alpha\ne0$ and $m\ne0$, we will have instead
 \ptcheck{6XI22 with $m$}
\begin{align}
 &\partial_u^{i+1} \hBo_{AB}
 \nn
 \\
  &= \mbox{right-hand side of \eqref{14VIII22.6} }
  \nonumber
\\
 &\quad
 +
 \sum_{0\leq j+k\leq i,k\ne i}
 \overadd{i,j,k}{\psi}
  (r)\partial_r^j P^k \Bigg[
 - (\frac{  {\alpha^2} r^2  }{2 }+{\frac{m}{r}})\Big[
        \partial_r h_{AB}
     - \frac{1}{r }    h_{AB}
     \Big]
     \nn
     \\
     &\phantom{xxxxxxxxxxxxxxxxxx}
     +  \int_{r_1}^r
        (\frac{{\alpha^2}r }{s}   -\frac{mr}{s^4})    \hBo_{AB}
        {\, ds}
        \Bigg]
   \nonumber
     \\
     &\quad
+
       \sum_{j=0}^i
       \underbrace{
         \int_{r_1}^{r}
       \overadd{i,j}{\psiP} (s,r)P^j \Bigg[
   - (\frac{  {\alpha^2} s^2  }{2 }+{\frac{m}{s}})\Big[
        \partial_sh_{AB}\big|_{s}
         }_{\text{integrate by parts}}
     - \frac{1}{s }    h_{AB}\big|_{s}
     \Big]
     \nn
     \\
     &\phantom{xxxxxxxxxxxxxxxxxx}
     +  \int_{r_1}^s
        (\frac{{\alpha^2}s }{y}  {-\frac{ms}{y^4}})\hBo_{AB}\big|_{y}
        dy
        \Bigg]
        {\, ds}
        \nonumber
\\
  &= \mbox{right-hand side of \eqref{14VIII22.6} }
  \nonumber
\\
 &\quad
 +
 \sum_{0\leq j+k\leq i,k\ne i}
 \overadd{i,j,k}{\psi}
  (r)\partial_r^j P^k \Bigg[
 - (\frac{  {\alpha^2} r^2  }{2 }+{\frac{m}{r}})\Big[
        \partial_r h_{AB}
     - \frac{1}{r }    h_{AB}
     \Big]
     \nn
     \\
     &\phantom{xxxxxxxxxxxxxxxxxx}
     +  \int_{r_1}^r
        (\frac{{\alpha^2}r }{s}   -\frac{mr}{s^4})    \hBo_{AB}
        {\, ds}
        \Bigg]
  \nonumber
\\
 &\quad
         -
            \sum_{j=0}^i
      \big(
      (\frac{  {\alpha^2} r^2  }{2 }+{\frac{m}{r}})  \overadd{i,j}{\psiP} (s,r)
      \big) \big|_{s=r}   P^j h_{AB} \big|_{r}
   \nonumber
     \\
     &\quad
      +   \sum_{j=0}^i \int_{r_1}^{r}
      \bigg(
      \underbrace{
        \partial_s
      \big[
     (\frac{  {\alpha^2} s^2  }{2 }+{\frac{m}{s}}) \overadd{i,j}{\psiP} (s,r)
      \big]
     + (\frac{{\alpha^2}s }{2}  {+\frac{m}{s^2}}) \overadd{i,j}{\psiP} (s,r)
      }_{
     =
     (
     \frac{  { \alpha^2} s^2  }{2 }+{\frac{m}{s}}) \partial_s \overadd{i,j}{\psiP} (s,r)
     +  \frac{{3\alpha^2}s }{2}  \overadd{i,j}{\psiP} (s,r)
     }
     \bigg)    P^j h_{AB} \big|_{s}
        {\, ds}
   \nonumber
     \\
     &\quad	
 \displaystyle +
       \sum_{j=0}^i
       \underbrace{
       \int_{r_1}^{r}
       \overadd{i,j}{\psiP} (s,r)  \int_{r_1}^r
       { (\frac{\alpha^2s }{y} -\frac{ms}{y^4}}) P^j \hBo_{AB} \big|_{y}
        dy
         \,
        ds
        }_{
    = \int_{r_1}^{r}  \big(\int_{s}^r
     { (\frac{\alpha^2 y }{s} -\frac{m y }{s^4}})
     \overadd{i,j}{\psiP} (y,r)
        dy
        \big)
     P^{j } h  _{AB}|_s
        {\, ds}
        }
        \nonumber
\\
 &\quad
        +  \tkfbd
        \,,
 \nonumber 
\end{align}
It follows that,
in addition to \eqref{16VIII22.12},  a term
 \ptcheck{16VIII22, together, without $m$, and with m 6XI}
$$
 a_{ki\ell } s^{-\ell }   \ \mbox{ in $\overadd{k,i}{\psiP}$, }
$$
 with $k\ge 1$ and $0\le \ell\ne  2$, induces terms involving $1/s$, {$1/s^4$}, and
a term
\begin{equation}\label{16VIII22.13}
   a_{ki\ell}
   \frac{(1-\ell) }{2(2-\ell)}
 \left(
 {\alpha^2}
 (4-\ell)
 s^{-\ell+1}
  + 2 m  (1-\ell)  s^{-\ell-2}
\right)
 \
  \mbox{in $\overadd{k+1,i}{\psiP}$;}
\end{equation}
cf.\ Figures~\ref{FtreeAll} and \ref{Ftree}.
\input{treeFigureArrowsChi}%
\input{treeFigureOnlyHighest}%
This shows in particular that the recursion formulae \eqref{19VIII22.a2} and \eqref{19VIII22.2b}, established with $\alpha=0$, remain valid for $\alpha{,m} \in \R$;
 but  e.g.\ \eqref{3IX22.w1} does not hold anymore when $m\neq 0$.

To continue, it is convenient to set
\begin{equation}\label{24XI22.41}
   \overadd{k ,-1 }{\psi}_\ell
   =
   0
   \,.
\end{equation}
Using this notation, putting together \eqref{16VIII22.12} with \eqref{16VIII22.13}
we find  the recursion formula,  for {$k\ge i\ge 0$ and $k\ge 1$,
}
\ptcheck{26XI, corrected}
\begin{align}
  \overadd{k+1,i}{\psiP} (s,r)
  =
  &
   \overadd{k+1,i}{\psiP}_0(r)
   +
    \frac{\overadd{k+1 ,i}{\psiP}_{1}(r)}{s }
   +
    \frac{\overadd{k+1 ,i}{\psiP}_{4}(r)}{s^4 }
    \nonumber
\\
  &
   +
   \sum_{\ell=4}^{k+3}
   \Big[
    \frac{(1-\ell)}{2 (2-\ell)s^\ell}
     \Big(
      \alpha^2 (4-\ell) {s}
        - \frac{(\ell-2)\twoscsign}{s}
      + \frac{2m(1-\ell)}{s^2}
      \Big)
      \overadd{k ,i}{\psiP}_{\ell}(r)
 \nonumber
\\
 &
 \phantom{ +
   \sum_{\ell=4}^{k+3}
   \Big[  }
      +\frac{( \ell-1)}{ \ell(\ell-3){s^{\ell+1}}}
       \overadd{ { k} ,i  -1  }{\psiP}_{\ell}(r)
      \Big]
      \nonumber
\\
  =
  &
   \overadd{k+1,i}{\psiP}_0(r)
   +
    \frac{\overadd{k+1 ,i}{\psiP}_{1}(r)}{s }
   +
    \frac{\overadd{k+1 ,i}{\psiP}_{4}(r)}{s^4 }
    \nonumber
\\
  &
   +
   \sum_{\ell=4}^{k+3}
   (\ell-1)
    \bigg[
    \frac{\alpha^2 (4-\ell)}{2 (\ell-2)}
      \frac{\overadd{k ,i}{\psiP}_{\ell}(r)}{s^{\ell-1}}
     -
    \frac{ m(\ell-1)}{   \ell-2 }
    \frac{
      \overadd{k ,i}{\psiP}_{\ell}(r)}{s^{\ell+2}}
 \nonumber
\\
 &
 \phantom{
    +   \sum_{\ell=4}^{k+3}   \Big[
    }
 + \Big(
         \frac{\twoscsign}2
  \overadd{k ,i}{\psiP}_{\ell}(r)
      +\frac{1}{ \ell(\ell-3) }
     \,  \overadd{  {k} ,i-1}{\psiP}_{\ell}(r)
      \Big)
     \frac{1 }
        { s^{\ell+1}}
      \bigg]
      \,,
 \nonumber 
\end{align}
An identical formula holds for $\chi$
with
$k\ge i \ge 0$
after setting
\begin{equation}\label{24XI22.42}
   \overadd{k ,-1 }{\chi}_\ell
    =0
  \,.
\end{equation}

One is   led to:

\begin{lemma}
 \label{Ll6XI22.1} The integral kernels $
 \overadd{i,j}{\psi}$ and  $
 \overadd{i,j}{\chi}$   are polynomials in $1/s$  with coefficients  depending upon $r$,   with no terms   $1/s^2$ and $1/s^3$. Moreover \ptcheck{21XI22}
\begin{enumerate}
  \item[a)]
  When $m=0$,
  the integral kernels $
 \overadd{i,j}{\psi}$,
   $i\ge 1$,
 $1\le j \le i$,
   are polynomials in $1/s$ of order $i+3$,
 with $\overset{(i,0)}{\opsi}  $   of order 1.

  \item[b)]  When $m=0$,
    the integral kernels $
 \overadd{i,j}{\chi}$,  $0\le j \le i$,  are polynomials in $1/s$ of order   $i+4$.

 \item[c)]   When $m\ne 0$,
  the integral kernels $
 \overadd{i,j}{\psi}$,
   $i\ge 1$,
 $0\le j \le i$,
   are polynomials in $1/s$ of order not larger than $2i+ {3}-j$, with $
 \overadd{i,0}{\psi}$  and $
 \overadd{i,1}{\psi}$   of order  $2i+2 $, and   $
 \overadd{i,i}{\psi}$  of order    $i+3$.

 \item[d)] When $m\ne 0$,
  the integral kernels $
 \overadd{i,j}{\chi}$,  $0\le j \le i$,  are polynomials in $1/s$    of order not larger than $2i+4-j$, with   $
 \overadd{i,0}{\chi}$    of order  $2i+   4  $, and   $
 \overadd{i,i}{\chi}$  of order    $i+4$.

\end{enumerate}
\end{lemma}

\proof
We summarise the arguments so far, and add some details:

\begin{enumerate}
  \item The functions that initialise the induction for $\partial_u h_{AB}$ involve only $1/s$ and $1/s^4$ terms, and  the functions   that initialise the induction for $\partial_u h_{uA}$ involve only $1/s$ and $1/s^5$ terms.

  \item One then applies the recursion formulae \eqref{16VIII22.12} and \eqref{16VIII22.13}; cf.\  Figures~\ref{FtreeNoMass} and \ref{Ftree}.
      We note that $\ln r$- and/or $\ln s$-terms could \emph{a priori} arise in the induction from $1/s$ terms in some integrals, but the multiplicative coefficients $(\ell-1)$ which appear in the second and third lines in \eqref{16VIII22.12} and \eqref{16VIII22.13} guarantee that there will be no $s^{-2}$ terms in any of the integral kernels, which in turns guarantees that no logarithmic terms will occur.

      \item
  When $m=0$, the fact that $ \overadd{i,j}{\psi}$  is of order  $i+3$  in $s^{-1}$ follows from  \eqref{26X22.w1}.

  \item
  Point  a)  together with the equality $\overset{(i,j)}{\opsi}_{i+3}(s,r) = 2 r^2 \, \overset{(i-1,j-1)}{\ochi}_{i+3}(s,r)$ (cf.\ \eqref{3IX22.w1}) establishes b).

  \item It follows from \eqref{19VIII22.a2} that $
 \overadd{i,i}{\psi}$ is of order $s^{-i-3}$ when $m=0$, and Figure~\ref{Ftree} makes it clear that this is not affected by the non-vanishing of $m$.

   \item    By following the dashed-dotted arrows in Figure~\ref{Ftree}  starting from the $(1,1)$ entry makes it clear that
        $
 \overadd{i,1}{\psi}$ is of order $2i+2$ in $s^{-1}$ when $m\ne 0$. The same holds  for $
 \overadd{i,0}{\psi}$ since the recursion formulae do not depend upon the index $j$ of  $
 \overadd{i,j}{\psi}$, and both initialising polynomials  $
 \overadd{1,0}{\psi}$  and  $
 \overadd{1,1}{\psi}$ are of order 4.
  In fact one checks that
\begin{eqnarray}
&
 \displaystyle
    \overset{(k,0)}{\ochi}_{2k+4} =  \frac{(- m)^k }{3}\frac{\bigl(( 2 k+1)!\bigr)^2}{2^{3k} (k!)^3}
    \,,
    &
    \label{9XI22.w5a1}
\\
&
 \displaystyle
    \overset{(k,0)}{\opsi}_{2k+2} = \frac{ r (-m)^k}{2^{3 k-1} (k - 1)! }\bigg(\frac{(2 k)!}{k!}\bigg)^2
    =
    -\frac{3m}{2r}\overset{(k,1)}{\opsi}_{2k+2}\,,
    &
    \label{9XI22.w5a2}
\end{eqnarray}
which further implies
  \ptcheck{XI22; see ChiPsiHighestCoeff.nb}
\begin{align}
    \overset{(k,0)}{\ochi}_{2k+4} = -\frac{1}{3mr} \overset{(k+1,0)}{\opsi}_{2k+4} = \frac{1}{2r^2}\overset{(k+1,1)}{\opsi}_{2k+4}\,.
    \label{9XI22.w5}
\end{align}
\end{enumerate}
    \qedskip

We finish this section with the following relations, needed for  \eqref{9XI22.w4a}:

\begin{Lemma}
 \label{L25XI22.1}
For $k\ge 2$ we have
 \ptcheck{26XI including proof;\\--\\wan : checked 26XI}
\begin{align} \label{8XI22.w4a1}
    \overset{(k-1,0)}{\chi}_{2k+2}\overset{(k,0)}{\psi}_{2k+1} &= \overset{(k-1,0)}{\chi}_{2k+1}\overset{(k,0)}{\psi}_{2k+2}
    \,,
\\
    \overset{(k-1,0)}{\chi}_{2k+2}\overset{(k,1)}{\psi}_{2k+1} &= \overset{(k-1,1)}{\chi}_{2k+1}\overset{(k,0)}{\psi}_{2k+2}
    +
    \overset{(k-1,0)}{\chi}_{2k+1}
    \overset{(k,1)}{\psi}_{2k+2}
    \,,
    \label{8XI22.w4}
\\
    \overset{(k-1,0)}{\chi}_{2k+2}\overset{(k,2)}{\psi}_{2k+1}
     &= \overset{(k-1,1)}{\chi}_{2k+1}\overset{(k,1)}{\psi}_{2k+2}
    \,.
    \label{8XI22.w4b}
\end{align}
\end{Lemma}

\proof
\input{E19}

%% file: treeFigureArrowsChi.tex
\begin{figure}
  \centering
\begin{tikzpicture}

\small

\def\x{3.1} 
\def\xy{0.3} 
\def\y{1.3} 
\def\z{0}
\def\l{0}
\def\d{0}


\node[align=center, name=11] at (\x,0*\y - \z) {
$\overadd{ 5,1 }{\psiP}$ or $\overadd{4,0 }{\chi}$
 };
\node[align=center, name=12] at (\x,\y - \z) {
$\overadd{ 4,1}{\psiP}$ or $\overadd{3,0 }{\chi}$
 };
\node[align=center, name=13] at (\x,2*\y) {
$\overadd{ 3,1 }{\psiP}$ or $\overadd{2,0 }{\chi}$
 };
\node[align=center, name=14] at (\x,3*\y +\d) {
$\overadd{ 2,1 }{\psiP}$ or $\overadd{1,0 }{\chi}$
 };
\node[align=center, name=15] at (\x,4*\y ) {
$\overadd{ 1,1 }{\psiP}$ or $\overadd{0,0 }{\chi}$
};


\node[align=center, name=21] at (2*\x,0*\y - \z) {
$\overadd{ 5,2 }{\psiP}$ or $\overadd{4,1 }{\chi}$
 };
\node[align=center, name=22] at (2*\x,\y - \z) {
$\overadd{ 4,2 }{\psiP}$ or $\overadd{3,1 }{\chi}$
 };
\node[align=center, name=23] at (2*\x,2*\y) {
$\overadd{ 3,2 }{\psiP}$ or $\overadd{2,1 }{\chi}$
 };
\node[align=center, name=24] at (2*\x,3*\y +\d) {
$\overadd{ 2,2 }{\psiP}$ or $\overadd{1,1 }{\chi}$
  };

\node[align=center, name=30] at (3*\x,0*\y - \z) {
$\overadd{ 5,3 }{\psiP}$ or $\overadd{4,2 }{\chi}$
 };
\node[align=center, name=31] at (3*\x,\y - \z) {
$\overadd{ 4,3 }{\psiP}$ or $\overadd{3,2 }{\chi}$
 };
\node[align=center, name=32] at (3*\x,2*\y) {
$\overadd{ 3,3 }{\psiP}$ or $\overadd{2,2 }{\chi}$
 };

\node[align=center, name=41] at (4*\x,0*\y - \z) {
$\overadd{ 5,4 }{\psiP}$ or $\overadd{4,3 }{\chi}$
 };
\node[align=center, name=42] at (4*\x,\y - \z) {
$\overadd{ 4,4 }{\psiP}$ or $\overadd{3,3 }{\chi}$
 };
\node[align=center, name=43] at (4*\x,2*\y) {$  $};

\node[align=center, name=51] at (5*\x,0*\y - \z) {
$\overadd{ 5,5 }{\psiP}$ or $\overadd{4,4 }{\chi}$
 };

\def\ssh{0.3}
\draw[line width=0.2mm, dashed, ->,color=red] ($(14.south)-   (0.15,0)$)  -- ($ (13.north west)+(0.7,0)$);
\draw[line width=0.2mm, dashed, ->,color=red] ($(13.south)-   (0.15,0)$)  -- ($ (12.north west)+(0.7,0)$);
\draw[line width=0.2mm, dashed, ->,color=red]($(12.south)-   (0.15,0)$)  -- ($ (11.north west)+(0.7,0)$);

\draw[line width=0.2mm, dashed, ->,color=red] ($(24.south)-   (0.15,0)$)  -- ($ (23.north west)+(0.7,0)$);
\draw[line width=0.2mm, dashed, ->,color=red] ($(23.south)-   (0.15,0)$)  -- ($ (22.north west)+(0.7,0)$);
\draw[line width=0.2mm, dashed, ->,color=red]($(22.south)-   (0.15,0)$)  -- ($ (21.north west)+(0.7,0)$);

\draw[line width=0.2mm, dashed, ->,color=red] ($(32.south)-   (0.15,0)$)  -- ($ (31.north west)+(0.7,0)$);
\draw[line width=0.2mm, dashed, ->,color=red]($(31.south)-   (0.15,0)$)  -- ($ (30.north west)+(0.7,0)$);

\draw[line width=0.2mm, dashed, ->,color=red]($(42.south)-   (0.15,0)$)  -- ($ (41.north west)+(0.7,0)$);
\draw[line width=0.2mm, ->, dashdotted,color=violet] (42.south) -- (41.north);
\draw[line width=0.2mm, ->, densely dotted,color=blue] ($(42.south) + (0.15,0)$) -- ($(41.north) + (0.15,0)$);

\draw[line width=0.2mm, ->, dashdotted,color=violet] (15.south) -- (14.north);
\draw[line width=0.2mm, ->, densely dotted,color=blue] ($(15.south) + (0.15,0)$) -- ($(14.north) + (0.15,0)$);
\draw[line width=0.2mm, ->, dashdotted,color=violet] (32.south) -- (31.north);
\draw[line width=0.2mm, ->, densely dotted,color=blue] ($(32.south) + (0.15,0)$) -- ($(31.north) + (0.15,0)$);
\draw[line width=0.2mm, ->, dashdotted,color=violet] (31.south) -- (30.north);
\draw[line width=0.2mm, ->, densely dotted,color=blue] ($(31.south) + (0.15,0)$) -- ($(30.north) + (0.15,0)$);

\draw[line width=0.2mm, ->, dashdotted,color=violet] (14.south) -- (13.north);
\draw[line width=0.2mm, ->, densely dotted,color=blue] (14.south east) -- (23.north west);

\draw[line width=0.2mm, ->, densely dotted,color=blue] ($(14.south) + (0.15,0)$) -- ($(13.north) + (0.15,0)$);
\draw[line width=0.2mm, ->, densely dotted,color=blue] (14.south east) -- (23.north west);

\draw[line width=0.2mm, ->, dashdotted,color=violet] (13.south) -- (12.north);
\draw[line width=0.2mm, ->, densely dotted,color=blue] ($(13.south) + (0.15,0)$) -- ($(12.north) + (0.15,0)$);
\draw[line width=0.2mm, ->, densely dotted,color=blue] (13.south east) -- (22.north west);

\draw[line width=0.2mm, ->, dashdotted,color=violet] (12.south) -- (11.north);
\draw[line width=0.2mm, ->, densely dotted,color=blue] ($(12.south) + (0.15,0)$) -- ($(11.north) + (0.15,0)$);
\draw[line width=0.2mm, ->, densely dotted,color=blue] (12.south east) -- (21.north west);

\draw[line width=0.2mm, ->, dashdotted,color=violet] (24.south) -- (23.north);
\draw[line width=0.2mm, ->, densely dotted,color=blue] ($(24.south) + (0.15,0)$) -- ($(23.north) + (0.15,0)$);
\draw[line width=0.2mm, ->, densely dotted,color=blue] (24.south east) -- (32.north west);

\draw[line width=0.2mm, ->, dashdotted,color=violet] (23.south) -- (22.north);
\draw[line width=0.2mm, ->, densely dotted,color=blue] ($(23.south) + (0.15,0)$) -- ($(22.north) + (0.15,0)$);
\draw[line width=0.2mm, ->, densely dotted,color=blue] (23.south east) -- (31.north west);

\draw[line width=0.2mm, ->, dashdotted,color=violet] (22.south) -- (21.north);
\draw[line width=0.2mm, ->, densely dotted,color=blue] ($(22.south) + (0.15,0)$) -- ($(21.north) + (0.15,0)$);
\draw[line width=0.2mm, ->, densely dotted,color=blue] (22.south east) -- (30.north west);

;
\draw[line width=0.2mm, ->, densely dotted,color=blue] (31.south east) -- (41.north west);


\draw[line width=0.2mm, ->, densely dotted,color=blue] (15.south east) -- (24.north west);
\draw[line width=0.2mm, ->, densely dotted,color=blue] (24.south east) -- (32.north west);
\draw[line width=0.2mm, ->, densely dotted,color=blue] (32.south east) -- (42.north west);
\draw[line width=0.2mm, ->, densely dotted,color=blue] (42.south east) -- (51.north west);

\scriptsize

%
%
%
%


\small

%
%

\end{tikzpicture}
\caption{Recursion tree for the integral kernels.
The dash-dotted lines describe the contributions from the mass parameter $m$, increasing each power by 2. The dotted lines describe the contributions from the Gauss curvature $\myGauss$, increasing each power by 1. The dashed lines arise from the cosmological constant, and are slanted to the left to visualise the fact that they decrease  powers by 1.
}
   \label{FtreeAll}
\end{figure}
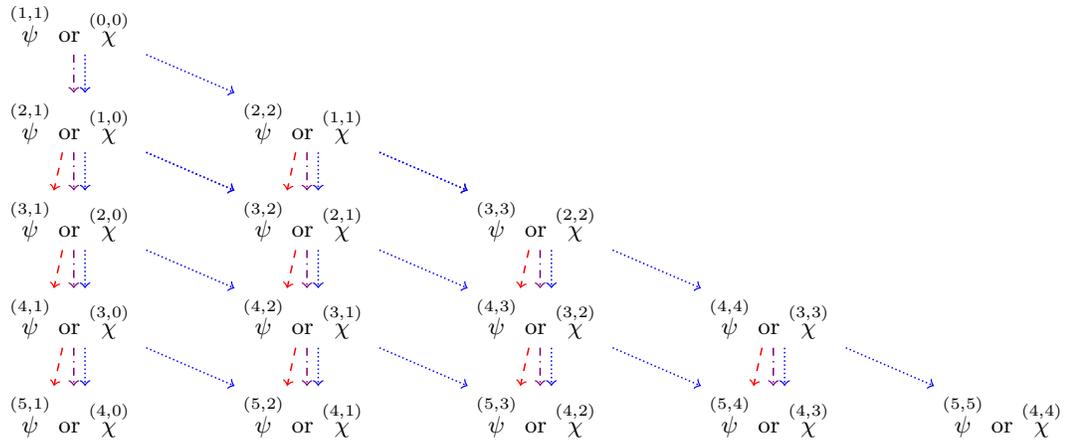

%% file: treeFigureOnlyHighest.tex
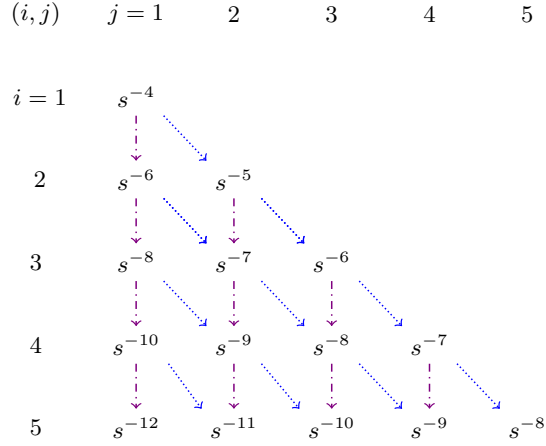
\begin{figure}
  \centering
\begin{tikzpicture}

\small

\def\x{1.3} 
\def\xy{0.3} 
\def\y{1.1} 
\def\z{0}
\def\l{0}
\def\d{0}
\def\dx{.5}

{\small

\node[align=center, name=045] at (0.4*\x -\dx ,5*\y +\d) {$ (i,j)$  };
\node[align=center, name=044] at (0.4*\x -\dx,4*\y +\d) {$ i=1 $};
\node[align=center, name=043] at (0.4*\x -\dx,3*\y +\d) {$2 $};
\node[align=center, name=042] at (0.4*\x -\dx,2*\y +\d) {3 };
\node[align=center, name=041] at (0.4*\x -\dx, \y +\d) {4 };
\node[align=center, name=040] at (0.4*\x -\dx, \d) {5 };
}

\node[align=center, name=11] at (\x,0*\y - \z) {$s^{-12}$};
\node[align=center, name=12] at (\x,\y - \z) {$ s^{-10}$};
\node[align=center, name=13] at (\x,2*\y) {$ s^{-8}$};
\node[align=center, name=14] at (\x,3*\y +\d) {$ s^{-6}$};
\node[align=center, name=24] at (2*\x,3*\y +\d) {$ s^{-5}$};
\node[align=center, name=15] at (\x,4*\y ) {$ s^{-4}$};

\node[align=center, name=161] at (\x,5*\y +\d) {$j=1$};
\node[align=center, name=162] at (2*\x,5*\y +\d) {$2$};
\node[align=center, name=163] at (3*\x,5*\y +\d) {$3$};
\node[align=center, name=164] at (4*\x,5*\y +\d) {$4$};
\node[align=center, name=165] at (5*\x,5*\y +\d) {$5$};

\node[align=center, name=21] at (2*\x,0*\y - \z) {$ s^{-11}$};
\node[align=center, name=30] at (3*\x,0*\y - \z) {$ s^{-10}$};
\node[align=center, name=22] at (2*\x,\y - \z) {$ s^{-9}$};
\node[align=center, name=31] at (3*\x,\y - \z) {$ s^{-8}$};
\node[align=center, name=23] at (2*\x,2*\y) {$ s^{-7}$};
\node[align=center, name=32] at (3*\x,2*\y) {$ s^{-6}$};

\node[align=center, name=41] at (4*\x,0*\y - \z) {$ s^{-9}$};
\node[align=center, name=42] at (4*\x,\y - \z) {$ s^{-7}$};
\node[align=center, name=43] at (4*\x,2*\y) {$  $};

\node[align=center, name=51] at (5*\x,0*\y - \z) {$ s^{-8}$};
 

\draw[line width=0.2mm, ->, dashdotted,color=violet] (42.south) -- (41.north);

\draw[line width=0.2mm, ->, dashdotted,color=violet] (15.south) -- (14.north);

\draw[line width=0.2mm, ->, dashdotted,color=violet] (32.south) -- (31.north);

\draw[line width=0.2mm, ->, dashdotted,color=violet] (31.south) -- (30.north);

\draw[line width=0.2mm, ->, dashdotted,color=violet] (14.south) -- (13.north);
\draw[line width=0.2mm, ->, densely dotted,color=blue] (14.south east) -- (23.north west);

\draw[line width=0.2mm, ->, densely dotted,color=blue] (14.south east) -- (23.north west);

\draw[line width=0.2mm, ->, dashdotted,color=violet] (13.south) -- (12.north);
\draw[line width=0.2mm, ->, densely dotted,color=blue] (13.south east) -- (22.north west);

\draw[line width=0.2mm, ->, dashdotted,color=violet] (12.south) -- (11.north);
\draw[line width=0.2mm, ->, densely dotted,color=blue] (12.south east) -- (21.north west);

\draw[line width=0.2mm, ->, dashdotted,color=violet] (24.south) -- (23.north);
\draw[line width=0.2mm, ->, densely dotted,color=blue] (24.south east) -- (32.north west);

\draw[line width=0.2mm, ->, dashdotted,color=violet] (23.south) -- (22.north);
\draw[line width=0.2mm, ->, densely dotted,color=blue] (23.south east) -- (31.north west);

\draw[line width=0.2mm, ->, dashdotted,color=violet] (22.south) -- (21.north);
\draw[line width=0.2mm, ->, densely dotted,color=blue] (22.south east) -- (30.north west);

;
\draw[line width=0.2mm, ->, densely dotted,color=blue] (31.south east) -- (41.north west);

\draw[line width=0.2mm, ->, densely dotted,color=blue] (15.south east) -- (24.north west);
\draw[line width=0.2mm, ->, densely dotted,color=blue] (24.south east) -- (32.north west);
\draw[line width=0.2mm, ->, densely dotted,color=blue] (32.south east) -- (42.north west);
\draw[line width=0.2mm, ->, densely dotted,color=blue] (42.south east) -- (51.north west);

\end{tikzpicture}
\caption{Highest powers of $s^{-1}$ in
$\protect \overadd{i,j}{\psi}$
when $m\ne 0$.
The dash-dotted lines describe the contributions from the mass parameter $m$, corresponding to an increase of the highest power by 2. The dotted lines describe the contributions from \eqref{16VIII22.12}, increasing the power by 1.
The tree for
$\protect \overadd{i,j}{\chi}$ is identical after replacing $(i,j)$ in the table by $(i-1,j-1)$.
}
   \label{Ftree}
\end{figure} 

%% file: E19.tex
We start by noting the
following recursion formulae, which can be read off \eqref{24XI22.41}-\eqref{24XI22.42}, for $k\ge 1$, $k\ge i\ge 0$, and $n\ge 5$:
\begin{eqnarray}
 \nonumber
&
   \overset{(k,i)}{\chi}_{n}
   =
    \mym_n\overset{(k-1,i)}{\chi}_{n-2}
    + \myGauss_n \overset{(k-1,i)}{\chi}_{n-1}
    + \myalpha_n\overset{(k-1,i)}{\chi}_{n+1}
    + \myiota_n\overset{(k-1,i-1)}{\chi}_{n-1}
    \,,
    &
\\
&
 \nonumber
   \overset{(k,i)}{\psi}_{n}
   =
    \mym_n\overset{(k-1,i)}{\psi}_{n-2}
    + \myGauss_n \overset{(k-1,i)}{\psi}_{n-1}
    + \myalpha_n\overset{(k-1,i)}{\psi}_{n+1}
    + \myiota_n\overset{(k-1,i-1)}{\psi}_{n-1}
    \,,
    &\label{24XI22.1a+a}
\end{eqnarray}
where $\mym_n$ arises from the mass $m$, $\myGauss_n$ from the Gauss curvature $\twoscsign$ of $\zgamma$,   $\myalpha_n$ from the cosmological constant encoded in $\alpha$, and $\myiota_n$ is associated with the term containing a shift in $i$:
\begin{align}
 \nonumber 
    \mym_n
    &=
    - m \frac{(n-3)^2}{n-4}
    \,,
    \quad
    \myGauss_n
    =
    \twoscsign \frac{n}{2}
    \,,
    \quad
    \myalpha_n
    =
    -
     \alpha^2 \frac{  n (n-3) }{2(n-1)}
    \,,
    \\
    \myiota_n
    &=
    \frac{n-2}{(n-1)(n-4)}
    \,.\nn
\end{align}
By  Lemma~\ref{Ll6XI22.1}, the coefficients $\overset{(i,j)}{\chi}_{\ell}$ vanish for $\ell+j>2i+4$, and   the coefficients $\overset{(i,j)}{\psi}_{\ell}$ vanish for $\ell+j>2i+3$. Thus, for $k\ge 2$ we can write
\begin{align*}
    &
     \displaystyle
      \overset{(k ,0)}{\psi}_{2k+2} =
    \mym_{2k+2} \, \overset{(k-1 ,0)}{\psi}_{2k}
    \,,
    \qquad
      \overset{(k ,1)}{\psi}_{2k+2} =
    \mym_{2k+2} \, \overset{(k-1 ,1)}{\psi}_{2k}
    \,,
    \\
    &
     \displaystyle
    \overset{(k,i)}{\psi}_{2k+1}
   =
     \mym_{2k+1}\overset{(k-1,i)}{\psi}_{2k-1}
      + \myGauss_{2k+1}\,  \overset{(k-1,i)}{\psi}_{2k}
    + \myalpha_{2k+1}\overset{(k-1,i)}{\psi}_{2k+2}
    \\
    &\qquad\qquad\quad
        + \myiota_{2k+1}\overset{(k-1,i-1)}{\psi}_{2k}
      \,,
\\
    &
     \displaystyle
    \overset{(k,2)}{\psi}_{2k+1}
   =
     \mym_{2k+1}\overset{(k-1,2)}{\psi}_{2k-1}
      + \underbrace{
       \myGauss_{2k+1}\,  \overset{(k-1,2)}{\psi}_{2k}
    + \myalpha_{2k+1}\overset{(k-1,2)}{\psi}_{2k+2}
        }_{=0}
       \\
    &\qquad\qquad\quad
    + \myiota_{2k+1}\overset{(k-1,1)}{\psi}_{2k}
      \,,
\\
    &
     \displaystyle
    \overset{(k,1)}{\psi}_{2k+1}
   =
     \mym_{2k+1}\overset{(k-1,1)}{\psi}_{2k-1}
      +
      \myGauss_{2k+1}\, \overset{(k-1,1)}{\psi}_{2k}
    +
    \myalpha_{2k+1}
     \underbrace{
     \overset{(k-1,1)}{\psi}_{2k+2}
     }_{=0}
     \\
    &\qquad\qquad\quad
        +
        \myiota_{2k+1}\overset{(k-1,0)}{\psi}_{2k}
        \,,
\\
    &
     \displaystyle
    \overset{(k,0)}{\psi}_{2k+1}
   =
     \mym_{2k+1}\overset{(k-1,0)}{\psi}_{2k-1}
      +
      \myGauss_{2k+1}\, \overset{(k-1,0)}{\psi}_{2k}
\\
    &\qquad\qquad\quad
    +
     \underbrace{
    \myalpha_{2k+1}
     \overset{(k-1,0)}{\psi}_{2k+2}
        +
        \myiota_{2k+1}\overset{(k-1,-1)}{\psi}_{2k}
     }_{=0}
      \,.
\end{align*}
Similarly for $k\ge 3$ we have
\begin{align}
\nonumber
    &
     \displaystyle
      \overset{(k-1,0)}{\chi}_{2k+2} =
    \mym_{2k+2} \, \overset{(k-2,0)}{\chi}_{2k}
    \,,
    \\
    &
     \displaystyle
    \overset{(k-1,1)}{\chi}_{2k+1}
   =
     \mym_{2k+1}\overset{(k-2,1)}{\chi}_{2k-1}
      +
     \underbrace{
      \myGauss_{2k+1}\, \overset{(k-2,1)}{\chi}_{2k}
    +
    \myalpha_{2k+1}
     \overset{(k-2,1)}{\chi}_{2k+2}
     }_{=0}\nn
\\
    &\qquad\qquad\quad
    +
        \myiota_{2k+1}\overset{(k-2,0)}{\chi}_{2k}
      \,,
\nonumber
\\
    &
     \displaystyle
    \overset{(k-1,0)}{\chi}_{2k+1}
   =
     \mym_{2k+1}\overset{(k-2,0)}{\chi}_{2k-1}
      +
      \myGauss_{2k+1}\, \overset{(k-2,0)}{\chi}_{2k}\nn
\\
    &\qquad\qquad\quad
    +
     \underbrace{
    \myalpha_{2k+1}
      \overset{(k-2,0)}{\chi}_{2k+2}
        +
        \myiota_{2k+1}\overset{(k-2,-1)}{\chi}_{2k}
     }_{=0}
      \,.
\nonumber 
\end{align}

We now check  that \eqref{8XI22.w4a1}-\eqref{8XI22.w4b} hold with $k=2$:
 \ptcheck{26XI22}
\begin{align}
      \underbrace{
    \overset{(1,0)}{\chi}_{6}
    }_{-\frac{3m}{2}
      \ \text{by \eqref{25XI22.3}};
     }
    \underbrace{
     \overset{(2,0)}{\psi}_{5}
      }_ {
     - \frac{3mr \myGauss}{2}
      , \ \text{by \eqref{25XI22.10}};
      }
       &=
         \underbrace{
          \overset{(1,0)}{\chi}_{5}
          }_{\frac{\twoscsign  }{2}
          \ \text{by \eqref{25XI22.3}};
           }
         \underbrace{
          \overset{(2,0)}{\psi}_{6}
      }_ {
       \frac{9 m^2r}{ 2}
      \ \text{by \eqref{25XI22.10}};
      }
    \,,
\nonumber 
\\
     \underbrace{
    \overset{(1,0)}{\chi}_{6}
    }_{-\frac{3m}{2}
      \ \text{by \eqref{25XI22.3}};
     }
     \underbrace{
      \overset{(2,1)}{\psi}_{5}
    }_{
    r^2\twoscsign-\frac{3mr}{4}
      \ \text{by \eqref{25XI22.15}};
     }
      &=
     \underbrace{
      \overset{(1,1)}{\chi}_{5}
      }_{
     \frac 14
       \ \text{by \eqref{25XI22.6}};
     }
         \underbrace{
          \overset{(2,0)}{\psi}_{6}
      }_ {
       \frac{9 m^2r}{ 2}
      \ \text{by \eqref{25XI22.10}};
      }\nn
\\
    &\qquad
    +
         \underbrace{
          \overset{(1,0)}{\chi}_{5}
          }_{\frac{\twoscsign  }{2}
          \ \text{by \eqref{25XI22.3}};
           }
         \underbrace{
           \overset{(2,1)}{\psi}_{6}
           }_
           {
     -3mr^2
      \ \text{by \eqref{25XI22.15}};
      }
    \,,
\nonumber  
\\
    \underbrace{
     \overset{(1,0)}{\chi}_{6}
      }_{ {-\frac{3m}{2}} \ \text{by \eqref{25XI22.3}};}
      \underbrace{\overset{(2,2)}{\psi}_{5}}_{ { \frac{r^2}{2}
      \ \text{by \eqref{25XI22.15b}};}}
     &=
     \underbrace{
     \overset{(1,1)}{\chi}_{5}
    }_{ { \frac{1}{4}} \ \text{by \eqref{25XI22.6}};
      }
   \underbrace{
          \overset{(2,1)}{\psi}_{6}
          }_
          { { -3 mr^2
     \ \text{by \eqref{25XI22.15}};}
     }
    \,.
\nonumber  
\end{align}
To continue, let $k\ge3$ and assume that \eqref{8XI22.w4a1} holds with $k$ replaced by $k-1$, then:
\begin{eqnarray}
 \lefteqn{
    \overset{(k-1,0)}{\chi}_{2k+2}\overset{(k,0)}{\psi}_{2k+1}
    -
    \overset{(k-1,0)}{\chi}_{2k+1}\overset{(k,0)}{\psi}_{2k+2}
    }
    &&
 \nonumber
\\
   &= &
    \mym_{2k+2} \, \overset{(k-2,0)}{\chi}_{2k}
    \big(
     \mym_{2k+1}\overset{(k-1,0)}{\psi}_{2k-1}
      +
     \cancel{ \myGauss_{2k+1}\, \overset{(k-1,0)}{\psi}_{2k}}
       \big)
  \nonumber
\\
 &&
-
 \big( \mym_{2k+1}\overset{(k-2,0)}{\chi}_{2k-1}
      +
     \cancel{ \myGauss_{2k+1}\, \overset{(k-2,0)}{\chi}_{2k}}
     \big)
    \mym_{2k+2} \, \overset{(k-1 ,0)}{\psi}_{2k}
     \nonumber
\\
   &= &
    \mym_{2k+2} \, \mym_{2k+1} \,
    \big( \overset{(k-2,0)}{\chi}_{2k}
    \overset{(k-1,0)}{\psi}_{2k-1}
-
  \overset{(k-2,0)}{\chi}_{2k-1}   \, \overset{(k-1 ,0)}{\psi}_{2k}
  \big)
     \nonumber
\\
&&
 = 0
    \,.
\nonumber 
\end{eqnarray}
Next, we assume that \eqref{8XI22.w4} holds with $k$ replaced by $k-1$. Then
\begin{align}
&
    \overset{(k-1,0)}{\chi}_{2k+2}\overset{(k,1)}{\psi}_{2k+1} - \overset{(k-1,1)}{\chi}_{2k+1}\overset{(k,0)}{\psi}_{2k+2}
   -
    \overset{(k-1,0)}{\chi}_{2k+1}\overset{(k,1)}{\psi}_{2k+2}
    \nonumber
\\
& =
    \mym_{2k+2} \, \overset{(k-2,0)}{\chi}_{2k}
    \overset{(k,1)}{\psi}_{2k+1} - \overset{(k-1,1)}{\chi}_{2k+1}
    \mym_{2k+2} \, \overset{(k-1 ,0)}{\psi}_{2k}
    \nonumber
    \\
    &\quad
   -
    \overset{(k-1,0)}{\chi}_{2k+1}
    \mym_{2k+2} \, \overset{(k-1 ,1)}{\psi}_{2k}
    \nonumber
\\
& =
    \mym_{2k+2} \Big[
     \overset{(k-2,0)}{\chi}_{2k}
    \Big(
    \mym_{2k+1}\overset{(k-1,1)}{\psi}_{2k-1}
      +
     \xcancel{
      \myGauss_{2k+1}\, \overset{(k-1,1)}{\psi}_{2k}
      }
      \nonumber
      \\
      & \qquad\qquad\qquad\qquad\quad
        +
        \cancel{
        \myiota_{2k+1}\overset{(k-1,0)}{\psi}_{2k}
        }
         \big)
         \nonumber
\\
 &
 \phantom{ \mym_{2k+2} \Big[}
    -
    \Big(
     \mym_{2k+1}\overset{(k-2,1)}{\chi}_{2k-1}
        +
        \cancel{
        \myiota_{2k+1}\overset{(k-2,0)}{\chi}_{2k}
        }
        \Big)
     \overset{(k-1 ,0)}{\psi}_{2k}
         \nonumber
\\
 &
 \phantom{ \mym_{2k+2} \Big[}
   -
    \Big(
     \mym_{2k+1}\overset{(k-2,0)}{\chi}_{2k-1}
      +
     \xcancel{
      \myGauss_{2k+1}\, \overset{(k-2,0)}{\chi}_{2k}
      }
      \Big)
     \overset{(k-1 ,1)}{\psi}_{2k}
     \Big]
      \nonumber
\\
& =
    \mym_{2k+2} \mym_{2k+1}
    \Big[
     \overset{(k-2,0)}{\chi}_{2k}
   \overset{(k-1,1)}{\psi}_{2k-1}
    -
    \overset{(k-2,1)}{\chi}_{2k-1}
     \overset{(k-1 ,0)}{\psi}_{2k}
     \nonumber
     \\
     &\qquad\qquad\qquad\qquad
     -
     \overset{(k-2,0)}{\chi}_{2k-1}
     \overset{(k-1 ,1)}{\psi}_{2k}
     \Big]
     \nonumber
\\
 & =
  0
\nonumber 
    \,.
\end{align}
Finally, suppose that \eqref{8XI22.w4b} holds with $k$ replaced by $k-1$. Then
 \ptcheck{23XI}
\begin{eqnarray}
   \lefteqn{\overset{(k-1,0)}{\chi}_{2k+2}\overset{(k,2)}{\psi}_{2k+1}
    - \overset{(k-1,1)}{\chi}_{2k+1}\overset{(k,1)}{\psi}_{2k+2}
    }
    &&
    \nonumber
\\
    &=&
    \mym_{2k+2} \overset{(k-2,0)}{\chi}_{2k}
     \,
    \big(
     \mym_{2k+1} \overset{(k-1,2)}{\psi}_{2k-1}
  +
  \cancel{
   \myiota_{2k+1}\overset{(k-1,1)}{\psi}_{2k}
   }
       \big)
   \nonumber
   \\
   &&\quad
   -
    \big(
     \mym_{2k+1} \overset{(k-2,1)}{\chi}_{2k-1}
     +
     \cancel{
       \mym_{2k+2} \overset{(k-2,0)}{\chi}_{2k}
       }
        \big)
       \myiota_{2k+1} \overset{(k-1,1)}{\psi}_{2k}
    \nonumber\\
    &=&
    \mym_{2k+2}  \mym_{2k+1}
    \big(
     \overset{(k-2,0)}{\chi}_{2k} \overset{(k-1,2)}{\psi}_{2k-1}
   -
      \overset{(k-2,1)}{\chi}_{2k-1}
        \overset{(k-1,1)}{\psi}_{2k}
       \big)
    \nonumber
   \\
   &=&
   0
    \nonumber
   \,.
\end{eqnarray}
The validity of \eqref{8XI22.w4a1}-\eqref{8XI22.w4b} follows thus by induction.
\qed

%% file: TensorHarmonics.tex
\subsection{Vector and tensor spherical harmonics}
 \label{ss20X22.1}

For integers $\ell\geq 1$, $-\ell\leq m \leq \ell$, let $Y^{(\ell m)}$ be the standard spherical harmonics on the unit sphere. Following the notations and conventions of \cite{ACR2,Czimek:2016ydb},
we define the  \emph{vector spherical harmonics}, as well as
trace-free symmetric \emph{$2$-tensor spherical harmonics}
on $S^2$ as:

\begin{enumerate}
    \item For $\ell \geq 1$, $-\ell\leq m\leq \ell$, define the vector fields
    \begin{align}
        E_A^{(\ell m)}:= -  \frac{1}{\sqrt{\ell(\ell+1)}} \zspaceD_A Y^{(\ell m)}
        \,,\
        H_A^{(\ell m)} := \frac{1}{\sqrt{\ell(\ell+1)}} \epsilon_{AB} \zspaceD^B Y^{(\ell m)}\,,
    \end{align}
    where $\epsilon_{AB}$ denote the volume two-form of $S^2$.
    \item For $\ell \geq 2$, $-\ell \leq m \leq \ell$, define the trace-free symmetric $2$-tensors
    \begin{align}
    \label{18X22.2w}
        \psi^{(\ell m)}_{AB}&:= -\frac{1}{\sqrt{\frac{1}{2}\ell(\ell+1)-1}} C(E^{(\ell m)})_{AB}
        \,,
        \\
        \phi_{AB}^{(\ell m)}& := -\frac{1}{\sqrt{\frac{1}{2}\ell(\ell+1)-1}}
        C(H^{(\ell m)})_{AB}
        \,, 
        \label{18X22.2wb}
    \end{align}
where the operator $C(\xi)_{AB} = TS(\zspaceD_A \xi_B)$ of \eqref{18X22.41}
corresponds to the operator  $-\slashed{\mathcal{D}}_2^*$
 of \cite{ACR2,Czimek:2016ydb}.
 \ptcheck{19X22}
\end{enumerate}

Let us summarise the properties of these
tensor harmonics, as needed in the main text. More details and proofs can be found   in \cite{Czimek:2016ydb}, see also~\cite{JezierskiPeeling}.

\begin{Lemma}
\label{l18X22.1}
The following holds.
\begin{enumerate}
    \item On $S^2$, $L^2$-integrable functions $f$, vector fields $\xi$ and trace-free symmetric $2$-tensors $\varphi$ can be decomposed as
    \begin{align}
        f &= \sum_{\ell\geq 0 }\sum_{-\ell\leq m\leq \ell} f^{\ell m} Y^{(\ell m)}\,,
        \nonumber
\\
        \xi_A &= \sum_{\ell\geq 1 }\sum_{-\ell\leq m\leq \ell}\xi^{(\ell m)}_E E^{(\ell m)}_A
        + \xi^{(\ell m)}_H H^{(\ell m)}_A\,,
        \nonumber
\\
        \varphi_{AB} &= \sum_{\ell\geq 2 }\sum_{-\ell\leq m\leq \ell}\varphi^{(\ell m)}_{\psi} \psi^{(\ell m)}_{AB}
        + \varphi^{(\ell m)}_{\phi} \phi^{(\ell m)}_{AB}\,,
\nonumber
    \end{align}
    where
    \begin{align}
        f^{(\ell m)} &:= \int_{S^2} f Y^{(\ell m)} \sm\,, & &
        \nonumber
\\
    \xi^{(\ell m)}_E &:= \int_{S^2} \xi^A E_A^{(\ell m)}
    \,, \qquad &
    \xi_H^{(\ell m)}&:= \int_{S^2} \xi^A H^{(\ell m)}_{A} \sm \,,
            \nonumber
\\
    \varphi_{\psi}^{(\ell m)} &:=\int_{S^2} \varphi^{AB} \psi_{AB}^{(\ell m)}
    \,, \qquad &
    \varphi_{\phi}^{(\ell m)}&:= \int_{S^2} \varphi^{AB} \phi^{(\ell m)}_{AB} \sm \,.
\nonumber
    \end{align}
    \item It holds that for $\ell \geq 2$,
    \begin{align}
    \label{18X22.1w}
        \zspaceD^A \psi_{AB}^{(\ell m)}=
          \sqrt{\frac{1}{2}\ell(\ell+1)-1}
           \,
             E_B^{(\ell m)}\,,
        \\
        \zspaceD^A \phi_{AB}^{(\ell m)} =
          \sqrt{\frac{1}{2}\ell(\ell+1)-1}
          \,
           H_B^{(\ell m)}\,.
           \label{18X22.1wb}
    \end{align}
    \item The space of conformal Killing vector fields on $S^2$ is spanned by $E_A^{(1m)}$ and $H_A^{(1m)}$.
\end{enumerate}
\end{Lemma}

%% file: CKV2d.tex
\subsection{The conformal Killing operator}
 \label{App30X22}

 Consider the   conformal Killing operator on a closed $\ddim$-dimensional Riemannian manifold $(\dmanif,\dmetric)$:
 \begin{equation}\label{30X22.CKV1a}
  \xi^A \mapsto  \dnabla _A \xi_B +  \dnabla _B \xi_A
    -    \dnabla ^C \xi_C
    \dmetric _{AB}
    \equiv  2  C(\xi)_{AB}
    \,.
 \end{equation}

We have

\begin{Proposition}
 \label{P12XI22.1}
 The conformal Killing operator on two dimensional manifolds is elliptic, with
 \begin{enumerate}
   \item six dimensional kernel and no cokernel on $S^2$;
   \item two dimensional kernel and cokernel on $\T^2$;
   \item no kernel and $6(\genus-1)$ dimensional cokernel on manifolds of genus $\genus\ge 2$.
 \end{enumerate}
 \end{Proposition}

 \proof
We first show that $C$ is elliptic. For this, let  $0\ne k\in T^*(\dmanif) $ and let $ \sigma (k)$ be the symbol of $C$, with kernel determined by the equation
\begin{align}
 \label{11XI22.3de}
 \big(\sigma (k)
 \big)_{AB} \equiv \frac 12 \big(
  k_A \xi_B  +
  k_B \xi_A -
   k^C \xi_{C} \zgamma_{AB}
   \big)
    =0
    \,.
\end{align}
Contracting with $k^A k^B$
one obtains
\begin{equation}\label{11XI22.4de}
  k^A k _{A} k^C \xi_C =0
  \qquad
  \Longrightarrow
   \qquad
 k^C \xi_C =0
  \,,
\end{equation}
Equation \eqref{11XI22.3de} becomes now
\begin{align}
 \label{11XI22.5}
  k_A \xi_B  +
  k_B \xi_A =0
    \,.
\end{align}
Contracting with $k^A $ one concludes that
\begin{align}
 \label{11XI22.5a}
 k^A k_A   \xi_B  =0
    \,.
\end{align}
Hence $\xi_{B} =0$, and  ellipticity of $C$ follows.

 Concerning the kernel in point 1.,
 we start by noting that the equation
 \begin{equation}\label{30X22.CKV1}
  \dnabla _A \xi_B +  \dnabla _B \xi_A
    -    \dnabla ^C \xi_C
    \dmetric _{AB} = 0
 \end{equation}
 is conformally invariant. Hence it suffices to analyse it on the unit round sphere.  Therefore, by Lemma~\ref{l18X22.1},  its solution are of the form
 $$
 \xi_A = \dnabla_A \varphi  + \epsilon_{AB} \dnabla^B \psi
 \,,
 $$
 where $\varphi$ and $\psi$ are linear combinations of $\ell=1$ spherical harmonics.  The $\varphi$-solutions are in one-to-one correspondence with the three generators of boosts of four-dimensional Minkowski space-time, while the $\psi$-solutions correspond to rotations.

 \medskip

 The statements about the kernel in points 2.\ and 3.\ follow from   Proposition \ref{P30X22.2} which we are about to prove.

 The statements about the cokernels follow from
 $$
  C^\dagger = - \zdivtwo
 $$
 where $\zdivtwo\!$ is the divergence operator on two-symmetric trace-free tensors,
 \begin{equation}\label{30X22.CKV5-b}
   (\zdivtwo h)_A:= \dnabla^B h_{AB}
   \,,
 \end{equation}
 together with the results in Section~\ref{ss12XI22.2} below.
 \qedskip

 Recall that we use the symbol $\CKV$ to denote the space of conformal Killing vectors, while $\TTt$ denotes the space of trace-free divergence-free symmetric two-tensors, and  orthogonality is defined in $L^2$. Then:

  \begin{Proposition}
   \label{P30X22.2}\ptcheck{21XI22, with proof crossched by Wan}
   \begin{enumerate}
     \item On $\T^2$ all  conformal Killing vectors  are covariantly constant, hence Killing.
     \item
  There are no nontrivial Killing vectors or conformal Killing vectors on higher genus two dimensional manifolds.
  \item {\rm$\im (\zdivtwo C) = \CKVp$.}
  \item For any vector field $\xi$ we have    $  C(\xi) ^{[ \TTt]} =0$.
   \end{enumerate}
  \end{Proposition}

  \proof
1.\ and 2.:
Taking the divergence of \eqref{30X22.CKV1} and commuting derivatives
leads to
 \begin{equation}\label{30X22.CKV3}
    \dnabla ^A \dnabla _A \xi_B   + \zR_{BC} \xi^C
    = 0
    \,.
 \end{equation}
  Multiplying by $\xi^B$ and integrating over $\dmanif$ one finds
 \begin{equation}\label{30X22.CKV4}
   \int ( | \dnabla   \xi|^2    -  \zR_{BC} \xi^B\xi^C)
    =  0
    \,.
 \end{equation}
 If $\zR_{BC} \le 0$ we find that $\xi$ is covariantly constant, vanishing if  $\zR_{BC} < 0$.

 \medskip

 3.
 Let  $\eta$ be $L^2$-orthogonal to the image of $\zdivtwo C$, thus for any   vector field  $\xi$ we have
 \begin{align}
  0
     &= \int_{\secN} \eta^A  \dnabla _B(\dnabla _A \xi_B +  \dnabla _B \xi_A
    -    \dnabla ^C \xi_C
    \dmetric _{AB} ) d\mu_{\dmetric}\nn
    \\
    &=2 \int_{\secN} \eta^A \dnabla ^B
    \big( \TS (\dnabla _A \xi_B  )
    \big)
    d\mu_{\dmetric}
     \nonumber
\\
     &=
      - 2 \int_{\secN} \dnabla^B \eta^A \TS (\dnabla _A \xi_B  ) d\mu_{\dmetric}
      =
      - 2 \int_{\secN} \TS (\dnabla^B \eta^A) \TS (\dnabla _A \xi_B  ) d\mu_{\dmetric}
      \,.
\nonumber
 \end{align}
 Letting  $\xi=\eta$ we conclude that $\eta$ is a conformal Killing vector.

 \medskip

 4. The field  $  C(\xi) ^{[ \TTt]} $ is obtained by $L^2$-projecting  $  C(\xi)    $ on $\TTt$. As such, for any $h\in \TTt$ we have
 \begin{eqnarray}
   \int_{\secN}  h^{AB}
   C(\xi)_{AB}
    \,
    d\mu_{\dmetric}
     &=&
   \int_{\secN}  h^{AB}
    \big( \TS (\dnabla _A \xi_B  )
    \big)
    \,
    d\mu_{\dmetric}
    \nonumber
\\
     &=&
     \int_{\secN} \TS (h^{AB})
     \dnabla _A \xi_B
    \,
    d\mu_{\dmetric}  =
     \int_{\secN}   h^{AB}
     \dnabla _A \xi_B
    \,
    d\mu_{\dmetric}
    \nonumber
\\
 &    =
  &
   -    \int_{\secN} \underbrace{ \dnabla _A  h^{AB}}_0
     \xi_B
    \,
    d\mu_{\dmetric}
    =0
      \,.
  \nonumber
 \end{eqnarray} 
 Hence $  C(\xi) ^{[ \TTt]} =0 $.
 \qed
 %

%
%
%

\subsection{\protect$\zdivtwo\!$}
 \label{ss12XI22.2}

We denote by $\zdivone\!$ the divergence operator  on vector fields:
 \begin{equation}\label{30X22.CKV5a}
    \zdivone \xi := \dnabla_A \xi^{A}
   \,.
 \end{equation}
and by  $\zdivtwo\!$ that on two-symmetric trace-free tensors,
 \begin{equation}\label{30X22.CKV5}
   (\zdivtwo h)_A:= \dnabla^B h_{AB}
   \,.
 \end{equation}

 As is well-known, $\zdivtwo\!$ is conformally covariant in all dimensions. In particular, in dimension two if $g_{AB} = e^\varphi \bar g_{AB}$ then
 \begin{equation}\label{18XI22.41}
   D_A h^{AB} = e^{-2\varphi} \bar{D}_A(e^{2\varphi} h^{AB})
   \,,
 \end{equation}
where $D$ is the Levi-Civita connection of $g$ and $\bar D$ that of $\bar g$.
It follows that it suffices to understand the kernel for metrics of constant Gauss curvature.

As already pointed out, on a two-dimensional closed negatively curved manifold of genus $\genus\ge 2$,
the operator $\zdivtwo\!$ has a $6(\genus-1)$-dimensional kernel; it has no kernel on $S^2$; on a flat torus  $\zdivtwo\!$  has a two-dimensional kernel  consisting of covariantly constant fields (cf., e.g., \cite{Tromba} Theorem 8.2
  and the paragraph that follows or \cite[Theorem~6.1 and  Corollary~6.1]{TamWan}).

 We claim that:

\begin{Lemma}
   \label{L30X22.1}
   Consider a two-dimensional Riemannian manifold $(\dmanif,\zgamma)$. Then the  operator {\rm $\zdivtwo\!$} acting on symmetric traceless tensors is elliptic, and it holds that
   {\rm
   $$
   \im \, \zdivtwo = \CKVp
   \,.
   $$
   }
   In particular if $\zR_{BC}<0$, the operator {\rm ${}\zdivtwo{}$} is surjective.
\end{Lemma}

\proof
We start with ellipticity.  For this, let  $0\ne k\in T^*(\dmanif) $ and let $ \sigma(k)$ be the symbol of $\zdivtwo\!$, with kernel determined by the equation
\begin{align}
 \label{11XI22.3d}
 \big(\sigma(k)h
 \big)_{A } \equiv   k^C h_{AC}
    =0
    \,.
\end{align}
In an orthonormal frame in which $k^2=0$ this is equivalent to
\begin{align}
 \label{11XI22.5bd}
 h_{11} =h_{12} =0
    \,.
\end{align}
For symmetric and traceless tensors $h_{AB}$ this is the same as $h_{AB} =0$. So $\sigma(k)$ has trivial kernel for $k\ne 0$, which is the definition of ellipticity.

Next, let $\xi$ be $L^2$-orthogonal to the  image of $\zdivtwo\!$, then for all smooth symmetric traceless tensors $h $ we have
 \begin{equation}\label{30X22.CKV5c}
  0 =  \int \xi^A \dnabla^B h_{AB}
  =  - \int \dnabla^B\xi^A  h_{AB} =
   - \int \TS (\dnabla^B\xi^A ) h_{AB}
   \,.
 \end{equation}
 This shows that $\TS (\dnabla^B\xi^A )=0$, hence $\xi^A$ is a conformal vector field.

 Since no such fields exist when the Ricci tensor is negative by Proposition~\ref{P30X22.2}, surjectivity for such metrics  follows.
 \qedskip

%% file: LandHatL.tex
\subsection{\protect$\hLop$ and \protect$\Lop$}
 \label{ss12XI22.11}

To continue, we wish to analyse the operators
\begin{equation}\label{24X22.11a}
 \hLop  = -  \zdivtwo  C \Lop
  \,,
  \quad
  \Lop  = (\zspaceD \zdivone  -\zdivtwo   C + \twoscsign)
 \,;
\end{equation}
recall that $\zdivone  \xi  = \zspaceD_A \xi^A$, $(\zdivtwo  h)_A = \zspaceD^B h_{AB}$,  and that $\twoscsign\in \{0,\pm 1\}$ is the Gauss curvature of $\zgamma$.

We consider  first  the operator $\xi \mapsto \zdivtwo C(\xi)$. One finds
\begin{equation}\label{30X22.32}
   \big(\zdivtwo C(\xi)\big)_A = \frac 12 (\dDelta  + \twoscsign) \xi_A
   \,,
\end{equation}
which is elliptic, self-adjoint,
with kernel and cokernel spanned by conformal Killing vectors.

Next, we turn our attention to   $\Lop$:
\begin{eqnarray}
 \phantom{xxx} \Lop (\xi)_A
  &  = &
  \blue{\zspaceD_A \zspaceD^C \xi_C}+\frac 12
   \big(
     \underbrace{
     \zspaceD_A \zspaceD^C \xi_C
     -
     \zspaceD^C ( \zspaceD_A \xi_C
     }_{-\mathring R^C{}_A \xi_C}
    + \zspaceD_C \xi_A
    )
    \big)
     +\twoscsign\xi_A
 \label{30X22.31} 
\\
  &  = &
 \zspaceD_A \zspaceD^C \xi_C  +
  \frac 12
   \big(
     -
     \zDelta
     +\twoscsign
    \big)\xi_A
    \,. 
\nonumber 
\end{eqnarray}
One readily checks that $\Lop$ is also elliptic and self-adjoint.

Applying $\dnabla^A$ to \eqref{30X22.31}, commuting derivatives, and using
\begin{equation}\label{30X22.CKV2c}
  \dR_{AB} = \twoscsign \dmetric _{AB}
\end{equation}
one finds that the kernel of $\Lop$ consists of vector fields satisfying
 \ptcheck{28XI:
 wan : checked for all $\myGauss$}
\begin{equation}\label{30X22}
 \frac 12  \dDelta \dnabla_A \xi^A = 0
  \,,
\end{equation}
hence $\dnabla_A \xi^A = c$ for some constant $c$. Integrating this last equality over $\dmanif$ shows that $c=0$.
It now follows that the kernel  of  $\hLop$ consists of vector fields satisfying
\ptcheck{1XI  by wan}
\begin{equation}\label{30X22.35}
  \big(
     -
     \zDelta
     +\twoscsign
    \big)\xi_A
    = 0
    \,,
    \qquad
    \dnabla_A \xi^A = 0
    \,.
\end{equation}
Recall the Hodge decomposition: on a compact two dimensional oriented manifold every one-form can be decomposed as
\begin{equation}\label{27XI22.31}
  \xi_A = \zspaceD_A \psi  +\epsilon_{AB} \zspaceD^B \phi + r_A
  \,,
\end{equation}
where $r_A$ is a harmonic one-form, i.e.\ a covector field satisfying
\begin{equation}\label{27XI22.32}
  \zspaceD^A r_A =0 =   \epsilon^{AB}\zspaceD_A r_B =  (
     -
     \zDelta
     +\twoscsign
    )r_A
  \,.
\end{equation}
On $S^2$ the forms $r_A$ vanish  identically, and   on manifolds with genus $\genus$ the space of  $r_A$'s is  $2\genus$-dimensional; cf., e.g., \cite[Theorems~19.11 and 19.14]{ForsterRiemann} or \cite[Theorem~18.7]{ArminRiemann}.

From the second equation in \eqref{30X22.35} together with \eqref{27XI22.31}-\eqref{27XI22.32}
 we find that the Laplacian of $\psi$ vanishes, hence $\psi$ is constant, and  the first equation in \eqref{30X22.35} gives
  \ptcheck{again with $r_A$ on 1XII}
\begin{equation}\label{30X22.356}
  \epsilon ^{AB} \dnabla_B
     \zDelta \phi
    = 0
    \,.
\end{equation}
It readily follows that $\phi$ is also constant, hence $\xi_A=r_A$, and we conclude that:

\begin{Lemma}
   \label{L30X22.1b}
   The operator $\Lop$ is elliptic, self-adjoint, with kernel and cokernel consisting of one-forms $r_A$ satisfying \eqref{27XI22.32}, hence of dimension equal to twice the genus of the compact, oriented, two-dimensional manifold.
\end{Lemma}

We are ready now to pass to the proof of:

\begin{Proposition}
   \label{P30X22.1a}
   The operator  $\hLop$ is elliptic, self-adjoint, with
   {\rm
   $$
    \ker\hLop = \coker \hLop=\CKV +\harm
    \,.
   $$
   }In particular:
   \begin{enumerate}
     \item on $S^2$ and on $T^2$ we have $\ker\hLop = \coker \hLop=\CKV$;
     \item on two-dimensional compact orientable manifolds of genus $\genus\ge 2$ both the kernel and cokernel of $\hLop$ are spanned by the $2\genus$-dimensional space of harmonic 1-forms.
   \end{enumerate}
\end{Proposition}

\proof
We first check that $\Lop$ and $-\zdivtwo C$ commute. In view of \eqref{30X22.32}-\eqref{30X22.31} it suffices to check the identity
\begin{equation}
 (\dDelta  + \twoscsign) \zspaceD_A \zspaceD^C \xi_C
=
  \zspaceD_A \zspaceD^C  (\dDelta  + \twoscsign)\xi_C
   \,,
\end{equation}
which follows from a straightforward commutation of derivatives. 
This shows that
$\hLop$ is the composition of two  commuting self-adjoint elliptic operators, hence elliptic and self-adjoint.

On $S^2$ the operator $\Lop$ is an isomorphism by Lemma~\ref{L30X22.1b}, hence the cokernel of $\hLop$ is determined by that of $-\zdivtwo C$. The claim on the kernel follows by duality.

It should be clear that in manifestly flat coordinates on  $\T^2$   the kernels of both $\Lop$ and $-\zdivtwo C$ consist  of covectors $\xi_A$ with constant entries, which span the space of conformal Killing vectors on $\T^2$. Self-adjointness implies the result for the cokernel.

In the higher genus case the operator $-\zdivtwo C$ is an isomorphism, so that the kernel of $\hLop$ coincides with the kernel of $\Lop$, as given by Lemma~\ref{L30X22.1b}. One concludes as before.
\qed

%% file: operators.tex
\subsection{$P$}
 \label{ss19XI22.1}

Consider the operator
\begin{align}
 \label{11XI22.2}
 Ph_{AB} := \TS[\zspaceD_A \zspaceD^C h_{BC}]
    \,.
\end{align}
of \eqref{11XI22.1}. where $h$ is symmetric and $\zgamma$-traceless.

We have:

\begin{Proposition}
 \label{P12XI22.2}
 The   operator $P$ is elliptic, self-adjoint and negative, with
 \begin{enumerate}
   \item six-dimensional cokernel and   kernel on $S^2$;
   \item two-dimensional kernel and cokernel on $\T^2$;
   \item   $6(\genus-1)$-dimensional cokernel and  kernel on manifolds of genus $\genus\ge 2$.
 \end{enumerate}
 \end{Proposition}

\proof
Note that
\begin{equation}\label{27XI22.11}
  P = C \circ \zdivtwo
\end{equation}
is a composition of elliptic operators, hence is elliptic.
Using
\begin{equation}\label{30XI22a}
  \zdivtwo\!\!^\dagger = - C
  \,,
  \quad
  C^\dagger = - \zdivtwo
  \,,
\end{equation}
we have
\begin{equation}\label{30XI22}
  P = - \zdivtwo \!\!^\dagger \circ \zdivtwo
  \,,
\end{equation}
from which self-adjointness  follows.

Finally, we have
\begin{equation}\label{30XI22-11}
   \int h^{AB}Ph _{AB}
   = -  \int h \,  \zdivtwo \!\!^\dagger \circ \zdivtwo h
   =  - \int |\zdivtwo h |^2 \le 0
   \,,
\end{equation}
hence all eigenvalues of $P$ are negative, and $Ph=0$ implies
$\zdivtwo h=0$.
%
\qed
%

%% file: PandLikes.tex
\subsubsection{$S^2$}
As already discussed in Section~\ref{ss20X22.1}, it follows from \cite{ACR2,Czimek:2016ydb} that  on $S^2$ we can write  symmetric trace-free $2$-tensors $\varphi_{AB}$ as
\begin{equation}
    \varphi_{AB} = \sum_{\ell\geq 2 }\sum_{-\ell\leq m\leq \ell}\varphi^{(\ell m)}_{\psi} \psi^{(\ell m)}_{AB}
        + \varphi^{(\ell m)}_{\phi} \phi^{(\ell m)}_{AB}\,.
\end{equation}
It follows from \eqref{18X22.2w}-\eqref{18X22.2wb} and \eqref{18X22.1w}-\eqref{18X22.1wb}
 that
the operator $P$ of \eqref{16V22.1}, namely
\begin{equation}\label{16V22.1app}
  P\varphi_{AB}  = TS[\zspaceD_A \zspaceD^C \varphi_{BC}]
  \equiv C (\zspaceD^C \varphi_{CD})_{AB}
  \,,
\end{equation}
 acts on $\varphi_{AB}$ as
\begin{align}
  \label{26X22.w3}
    P& \varphi_{AB} = \sum_{\ell\geq 2 }\sum_{-\ell\leq m\leq \ell}
    \varphi^{(\ell m)}_{\psi} C(\zspaceD^B\psi^{(\ell m)}_{AB})
        + \varphi^{(\ell m)}_{\phi} C(\zspaceD^B\phi^{(\ell m)}_{AB})
\\
 &
   =
    \sum_{\ell\geq 2 }\sum_{-\ell\leq m\leq \ell}
    \sqrt{\frac{1}{2}\ell(\ell+1)-1}\left(\varphi^{(\ell m)}_{\psi} C(E^{(\ell m)})_{AB}
        + \varphi^{(\ell m)}_{\phi} C(H^{(\ell m)}_{AB})\right)
        \nonumber
\\
    &=
    -\sum_{\ell\geq 2 }\sum_{-\ell\leq m\leq \ell}
    \underbrace{\left(\frac{1}{2}\ell(\ell+1)-1\right)}_{>0 \text{ for } \ell\geq 2}\left(\varphi^{(\ell m)}_{\psi}\psi^{(\ell m)}_{AB}
        + \varphi^{(\ell m)}_{\phi} \phi^{(\ell m)}_{AB}\right)\,.
     \nonumber 
\end{align}
In particular the operator $P$ is self-adjoint and has trivial kernel on $S^2$. On the other hand the operator $\zdivtwo(P+2)$, which appears in \eqref{24X22.91} with $p=1$ and $m=0$,
 acts according to
\begin{align}
   \zdivtwo
   &
   (P+2)(\varphi)_B  :=  \zspaceD^A(P+2)\varphi_{AB}
    \label{30XI22.1} 
\\
    &=
    -\sum_{\substack{\ell\geq 2 \\ |m|\leq \ell}}
    \left(\tfrac{\ell(\ell+1)}{2}-1-2\right)
    \zspaceD^A\left(\varphi^{(\ell m)}_{\psi}\psi^{(\ell m)}_{AB}
        + \varphi^{(\ell m)}_{\phi} \phi^{(\ell m)}_{AB}\right)
        \nonumber
\\
    &=
    -\sum_{\substack{\ell\geq 2 \\ |m|\leq \ell}}
    \underbrace{
    \left(\tfrac{\ell(\ell+1)}{2}-3\right)
    \sqrt{\tfrac{\ell(\ell+1)}{2}-1}}_{\substack{
    =0 \text{ for } \ell= 1,2 \\
    >0 \text{ for } \ell> 2
    }
    }
     \left(\varphi^{(\ell m)}_{\psi}E^{(\ell m)}_{B}
        + \varphi^{(\ell m)}_{\phi} H^{(\ell m)}_{B}\right)
         \,.
 \nn 
\end{align}
It follows that
the $L^2$-orthogonal
 $\big(\im(\zdivtwo(P+2))\big)^{\perp}$ of
 $\im(\zdivtwo(P+2)) $
 is spanned by conformal Killing vectors together with spherical harmonic vector fields with $\ell = 2$. Subsequently, for any covector field $X_A\in L^2$ the equation
\begin{equation}
    \zspaceD^B\left(P+2\right)\varphi_{AB}
     	 -
        \xi_A^{[\leq 2]}
            = X_{A}
\end{equation}
admits a unique solution
with a symmetric traceless $2$-tensor $\varphi_{AB}$ and a covector field $\xi_A^{[\leq 2]}$.
\ptcheck{19X22}

For a $\Ctwo$ gluing we need the operator
$$
 \zdivtwo( P ^2 + 7 \twoscsign P  +10 \twoscsign)
  \,,
$$
as determined from the coefficients of $\hkappa_6$ in the formulae \eqref{3VIII22.2} for $\overset{(2,i)}\chi$.
On $S^2$, a calculation  similar  to that in \eqref{30XI22.1} shows that its kernel consists of spherical harmonic tensors with $\ell=1, 2,3$, which results in a cokernel spanned on spherical harmonic vectors with $\ell=1, 2,3$.

\input{orderkkernel}

%% file: orderkkernel.tex
\subsubsection{Polynomials in $P$}
 \label{ss27XI22.1}

In this section we assume that $m=0$.

For $C^k$-gluing, the operator $\sum_{i=0}^{k}  \overset{(k,i)}{\ochi}_{k+4}  P^i $ appearing in \eqref{24X22.91} is of the form
\begin{align}
   \sum_{i=0}^{k}  \overset{(k,i)}{\ochi}_{k+4} (r_2)  P^i
   =
    \blue{\hat c_k}\prod_{i=1}^k (P+\twoscsign a_i )\,,\qquad
   a_i = \frac{1}{2}i(3+i)\,,
   \label{26X22.w1}
 \end{align}
where
 \ptcheck{28X}
$$\blue{\hat c_k}
= \frac{1}{ k!(k+3)}
\,.
$$
This can be verified by induction.

Indeed, when  $k=1$ this follows from \eqref{25XI22.3}-\eqref{25XI22.6} with $\hat c_1=1/4$. Using the recursion formula \eqref{16VIII22.12},
a straightforward calculation shows that the $\hkappa_{k+5}$ component of the $(k+1)$-order coefficient  are  given by
\ptcheck{27X22, and with $\myGauss$ and m=0 on 30XI}
\begin{equation}
    \overset{(k+1,i)}{\ochi}_{k+5} = \blue{c_k} \times \left\{
       \begin{array}{lll}
        \twoscsign a_{k+1} \overset{(k ,0)}{\ochi}_{k+4}\,,
          & \hbox{$i=0$}
           \,,
        \\[5pt]
          \twoscsign  a_{k+1} \overset{(k ,i)}{\ochi}_{k+4}  + \overset{(k,i-1)}{\ochi}_{k+4}\,,
           &  \hbox{$1\leq i \leq k$}
            \,,
        \\[5pt]
        \overset{(k ,k)}{\ochi}_{k+4}\,,
          &  \hbox{$i = k+1$}
           \,,
       \end{array}
     \right.
\end{equation}
with $\blue{c_k} = \frac{k+3}{(k+4)(k+1)}$. Therefore, assuming \eqref{26X22.w1}, the operator at order $k+1$ is actually $r_2$-independent and reads,
\ptcheck{27X22}
\begin{align}
    \sum_{i=0}^{k+1}  \overset{(k,i)}{\ochi}_{k+5} (r_2)  P^i
   &= \blue{c_k} \sum_{i=0}^{k}  \left(\overset{(k,i)}{\ochi}_{k+4}   P^{i+1}
   +
   \twoscsign a_{k+1}  \overset{(k,i)}{\ochi}_{k+4}   P^i \right)
   \nonumber
   \\
   &=
   \blue{c_k} \sum_{i=0}^{k}  \big(\overset{(k,i)}{\ochi}_{k+4}   P^{i} \big) (P+ \twoscsign a_{k+1}  )
      \nonumber
   \\
   &=
    \blue{c_k} \hat c_k  (P+ \twoscsign a_{k+1} ) \prod_{i=1}^k  (P+ \twoscsign a_i )
   \nonumber
   \\
    &=
    \blue{ \hat c_{k+1}}  \prod_{i=1}^{k+1} (P+\twoscsign a_i )\,,
    \qquad \text{with} \ \blue{\hat c_{k+1}}=  \blue{c_k} \hat c_k
    \,.
\nonumber %
\end{align}
It thus follows from \eqref{26X22.w3} that, on $S^2$, spherical harmonic vector fields with mode $\ell\ge 0 $ satisfying
\begin{align}
    0&=\prod_{i=1}^{k} \left(-\frac{1}{2}\ell(\ell+1)+ 1 + a_i\right)
    =
    \frac{1}{2} \prod_{i=1}^{k} (1+i-\ell)(2+i+\ell)
\end{align}
  belong to $\ker\left(\sum_{i=0}^{k}  \overset{(k,i)}{\ochi}_{k+4}  P^i \right)$. The corresponding values of $\ell$ are  $\ell = 2,...,k+1$.
 \ptcheck{27X}
 
For the remaining topologies, each of the operators
$$
 P + \myGauss a_i
$$
appearing in \eqref{26X22.w1} is negative.
On $\T^2$ its kernel, when acting on traceless tensors, is two-dimensional, consisting of covariantly constant tensors. Hence, in the toroidal case, the kernel of the left-hand side of \eqref{26X22.w1} is also two-dimensional, which can be seen e.g.\ by a Fourier-series decomposition.

On higher genus manifolds  $P + \myGauss a_i$ is strictly negative and therefore has no kernel. Hence so does the left-hand side of \eqref{26X22.w1}.

%% file: TraceEqn.tex
\section{A trace identity}
\label{ss3X22.1}

The aim of this appendix is to prove  the following curious consequence of Bianchi identities:
\ptcheck{12XI22, by Wan}
\begin{align}
    r^{-1} \gamma^{AB} \delta G_{AB} =  - \frac 12  \gamma^{AB}  \partial_r
          \delta R_{AB}
        +
        \zspaceD^A   \delta G_{rA}
     \,,
     \label{5X22.w4}
\end{align}
when $\partial_u^i\delta\beta=0$ (i.e., $\partial_u^i\delta G_{rr} = 0$) for $i=0,1$.

%% file: lineariseTrace.tex
\ptcheck{27X; old proof went to OldLineariseTrace}
For this, we start by noting that the operator $g^{AB}R_{AB}$ is related to that appearing in  \eqref{3X22.1}, which can be seen as follows:  From the definition \eqref{2X22.2} of the Einstein tensor $G_{\mu\nu}$ and the Bondi parametrisation of the metric \eqref{23VII22.1} we have
\begin{align}
\label{5X22.w1}
    G_{ur} = \frac{1}{2}e^{2\beta}g^{rr}G_{rr} - U^A G_{rA} + \frac{1}{2}e^{2\beta} g^{AB} R_{AB}\,.
\end{align}
%
Now, from the linearisation of \eqref{5X22.w1}, when $\delta \beta =0$, $G_{rr}=0$, and  $\partial_u^i \delta G_{rr}=0$, we have
\begin{align}
    \label{23X22.w1}
    \frac{1}{2}\ringh^{AB}\delta R_{AB} &= r^2 \delta G_{ur}
\quad
 \implies
 \quad
 \frac{1}{2}\ringh^{AB}\partial_r \delta R_{AB} = 2r \delta G_{ur} + r^2 \partial_r\delta G_{ur}\,,
\end{align}
and hence the identity \eqref{5X22.w4} is equivalent to
\begin{align}
    \zspaceD^A \delta G_{rA} - \frac{1}{r} \ringh^{AB}\delta G_{AB} = 2r \delta G_{ur} + r^2 \partial_r\delta G_{ur}\,.
    \label{23X22.w2}
\end{align}
Meanwhile, it follows from the divergence identity \eqref{27IX22.2} with $\nu=r$ that
\begin{align}
    0&= \frac{1}{\sqrt{|g|}}\partial_{\mu}(\sqrt{|g|}\mcE^{\mu}{}_r)
    + \frac12 \partial_r(g^{\mu\rho})\mcE_{\mu\rho}\,.
    \label{23X22.w3}
\end{align}
The linearisation of \eqref{23X22.w3} with $\partial_u  \delta G_{rr}=0$ gives,
\begin{align}
    0&=
    -\frac{1}{r^2}\partial_r(r^2 \delta G_{ur})
    +\frac{1}{r^2}\zspaceD^A\delta G_{rA}
    + \frac12 \partial_r\left(\frac{1}{r^2}\ringh^{AB}\right)\delta G_{AB}\,,
\end{align}
and hence,
\begin{align}
    2r \delta G_{ur} + r^2 \partial_r\delta G_{ur}&=
    \zspaceD^A\delta G_{rA}
    - \frac1r \ringh^{AB} \delta G_{AB}\,,
\end{align}
which agrees with \eqref{23X22.w2}.

%% file: huAcharges.tex
\section{Conserved charges for $\partial^i_u \zhTB_{uA}$}
 \label{App14IV23.1}

The aim of this appendix is to present an alternative way to obtain the obstructions for gluing of $\partial^i_u h_{uA}$, $i=0,1$, in terms of gauge-dependent radial charges, as suggested to us by S.~Czimek. We assume in what follows that
$$ m=0
 \,.
$$
Similar calculations can be done for $m\ne 0$ and for higher derivatives, but we have not pursued these ideas any further.

\subsection{$\zhTB_{uA}$}
 \label{sApp14IV23.1}

 Recall~\eqref{17I23.2b}:
 \ptcheck{14IV23}
\begin{eqnarray}
        \partial_r \bigg ( 3 \zhTB_{uA}+ r\partial_r \zhTB_{uA}
        - r^{-3}\zspaceD^B
                  h_{AB}\bigg ) =
                  r^{-4} \zspaceD^B
                  h_{AB}
                 \,.
                 \label{17I23.2}
           \end{eqnarray}
In particular we obtain a collection  of radially-conserved charges
\begin{equation}
    \partial_r \underbrace{ \int_{\secN}
     \pi^A \bigg( 3 \zhTB_{uA}+ r\partial_r \zhTB_{uA}\bigg)
       \sm
       }_{=:\kQ{4,0}{}(\pi^A)}
       = 0\,,
    \label{19I23.2forget}
\end{equation}
where the $\pi^A$'s are conformal Killing vectors of $(\secN,\ringh)$.
Keeping in mind our assumption that $m=0$,   under gauge transformations we have
\begin{equation}
    3 \zhTB_{uA}+ r\partial_r \zhTB_{uA} \mapsto 3 \zhTB_{uA}+ r\partial_r \zhTB_{uA}
    +  \frac{1}{r^2} \Done(\xi^u)_A
    + 3 (\partial_u \xi_A + \alpha^2 \zspaceD_A \xi^u)\,.
    \label{19I23.2x}
\end{equation}
Hence the gauge-dependent  radial charge $\kQ{4,0}{}$ transforms as,
\begin{equation}
    \kQ{4,0}{}(\pi^A) \mapsto\kQ{4,0}{}(\pi^A)
    + 3 \int \pi^A
    \big(
      \partial_u \xi_A
      + \alpha^2 \zspaceD_A \xi^u
      \big)
      \sm \,,
\end{equation}
since the $\Done$-gauge term  in \eqref{19I23.2x} has zero projection onto the $\pi^A$'s.
 \ptcheck{14IV23.1, the whole subsection above}

\subsection{$\partial_u\zhTB_{uA}$}
 \label{sApp14IV23.2}

Now, we hope to obtain a conserved radial charge involving $\partial_u\zhTB_{uA}$. For this, take $\partial_u$ of \eqref{17I23.2}:
\begin{equation}
    \label{17I23.3}
    \partial_r \bigg( 3 \partial_u\zhTB_{uA}+ r\partial_r \partial_u\zhTB_{uA}\bigg) =
                   \frac{1}{r}\zspaceD^B
                   \partial_r \left(\partial_u\zhTB_{AB}\right) \,.
\end{equation}
We can then achieve the goal by rewriting the RHS into terms of the form $\partial_r(\cdot)\, + $ (terms involving only $\partial_r^ih_{AB})$. For this, recall \eqref{eq:31III22.3p0} with $\delta \beta = 0$:
\ptcheck{14IV23}
\begin{align}
    0
   =&
    \partial_r \Big[
    r \partial_u \zhTB_{AB}
     - \frac{ 1}{2}  V  \partial_r \zhTB_{AB}
     -  \frac{1}{2 r}  V   \zhTB_{AB}
     -
            r \TS \big[\zspaceD_A   \zhTB_{uB}\big]
     \Big]
  \label{17I23.4} 
\\&\quad
        +\frac{1}{2}  \partial_r  (  V/r)
          \zhTB_{AB}
         -
          \TS \big[\zspaceD_A   \zhTB_{uB}
       \big]
        \,.
 \nn 
\end{align}
We wish to combine the above two equations by taking RHS of \eqref{17I23.3} $+$ $Ar^c \, \times \zspaceD^B$ \eqref{17I23.4} for some constants $A,c$ such that all
the $\partial_u\zspaceD^B\zhTB_{AB}$ terms collect into the form $B\partial_r(r^a\zspaceD^B\partial_u\zhTB_{AB})$, for some constants $B,a$. By a straightforward comparison of coefficients, we find that this is possible with $A = -\frac{1}{2} = -B$ and $c = -2$, $a = -1$. Indeed, we have,
\ptcheck{14IV23}
\begin{align}
    \partial_r \bigg( 3 \partial_u\zhTB_{uA}
    &
    + r\partial_r \partial_u\zhTB_{uA}\bigg) =
    \frac{1}{r}\zspaceD^B
                   \partial_r \left(\partial_u\zhTB_{AB}\right)
        -\frac{1}{2r^2} \partial_r \big(
    r \partial_u \zspaceD^B \zhTB_{AB}
     \big)
     \nonumber
\\
    &\quad
    +\frac{1}{2r^2} \zspaceD^B \partial_r \Big[
     \frac{ 1}{2}  V  \partial_r \zhTB_{AB}
     +  \frac{1}{2 r}  V   \zhTB_{AB}
     + r \TS \big[\zspaceD_A   \zhTB_{uB}\big]
     \Big]
     \nonumber
\\
     & \quad
    -\frac{1}{4 r^2}  \partial_r  (  V/r)
          \zspaceD^B \zhTB_{AB}
         + \frac{1}{2r^2}
          \zspaceD^B  \TS \big[\zspaceD_A   \zhTB_{uB}
       \big]
       \nonumber
\\
       &=
    \frac{1}{2} \partial_r\bigg(\frac{1}{r} \partial_u \zspaceD^B \zhTB_{AB}\bigg)
    \nonumber
\\
    &\quad
    +\frac{1}{2r^2} \zspaceD^B \partial_r \Big[
     \frac{ 1}{2}  V  \partial_r \zhTB_{AB}
     +  \frac{1}{2 r}  V   \zhTB_{AB}
     + r \TS \big[\zspaceD_A   \zhTB_{uB}\big]
     \Big]
     \nonumber
\\
     & \quad
    -\frac{1}{4 r^2}    \partial_r  (  V/r)
          \zspaceD^B\zhTB_{AB}
         + \frac{1}{2r^2}
          \zspaceD^B  \TS \big[\zspaceD_A   \zhTB_{uB}
       \big]\,.
\nonumber %
\end{align}
We rewrite this as
\ptcheck{14IV23}
\begin{align}
    \partial_r \bigg( 3 \partial_u\zhTB_{uA}
    &
    + r\partial_r \partial_u\zhTB_{uA}
    - \frac{1}{2r} \partial_u \zspaceD^B \zhTB_{AB}\bigg)
  \label{17I23.5} 
\\
    =  & \quad
    \frac{1}{2r^2} \zspaceD^B \partial_r \Big[
        \frac{ V}{2}    \partial_r \zhTB_{AB}
    +  \frac{V}{2 r}     \zhTB_{AB}  \Big]
      -\frac{1}{4 r^2}    \partial_r  \bigg( \frac{V}{r}\bigg)
          \zspaceD^B\zhTB_{AB}
     \nonumber
\\
     & \quad
   +\frac{1}{2r^2}  \partial_r \Big( r \zspaceD^B\TS \big[\zspaceD_A   \zhTB_{uB}\big]\Big)
         + \frac{1}{2r^2}
          \zspaceD^B  \TS \big[\zspaceD_A   \zhTB_{uB}
       \big]\,.
       \nn 
\end{align}
Next, we perform a similar trick, by taking RHS of \eqref{17I23.5} $+\, D \, r^d \times \zspaceD^B \TS[\zspaceD_A [$LHS of \eqref{17I23.2}$]$  to collect all the
$\zspaceD^B\TS[\zspaceD_A \zhTB_{uA}]$ terms into the form
$$\tilde{A} \partial_r \big( r^{\tilde a} \partial_r (r^{\tilde b}\zspaceD^B\TS[\zspaceD_A \zhTB_{uA}])\big)$$
for some constants $D,d,\tilde A,\tilde a,\tilde b$. By power matching
 and comparing  coefficients,
 we get $d=-1$ and $D = -3/8 = \tilde A = -1/\tilde b = 1 / \tilde a$. Explicitly, as a first step we write
\ptcheck{14IV23}
\begin{align}
    \frac{1}{2r^2}  \partial_r \Big( r
    &
    \zspaceD^B\TS \big[\zspaceD_A   \zhTB_{uB}\big]\Big)
         + \frac{1}{2r^2}
          \zspaceD^B  \TS \big[\zspaceD_A   \zhTB_{uB}
       \big]
       \nonumber
\\
      &\qquad
      - \frac{3}{8 r}
      \underbrace{\partial_r \bigg( 3 \zspaceD^B \TS[\zspaceD_A \zhTB_{uB}]+ r\partial_r \zspaceD^B \TS[\zspaceD_A \zhTB_{uB}]\bigg)}_{=
      \frac{1}{r} \partial_r \left(\zspaceD^B P \zhTB_{AB}\right) \text{ by \eqref{17I23.2}}}
       \nonumber
 \\
     &
    = -\frac38 \partial_r \big(r^{-8/3} \partial_r (r^{8/3}\zspaceD^B \TS[\zspaceD_A \zhTB_{uB}])\big)\,.
\nonumber %
\end{align}
Substituting this into \eqref{17I23.5} then gives,
\begin{align}
    \partial_r \bigg( 3 \partial_u\zhTB_{uA}
    &
    + r\partial_r \partial_u\zhTB_{uA}
    - \frac{1}{2r} \partial_u \zspaceD^B \zhTB_{AB}\bigg)  \nonumber
\\
   =  & \quad
    \frac{1}{2r^2} \zspaceD^B \partial_r \Big[
        \frac{ V}{2}    \partial_r \zhTB_{AB}
    +  \frac{V}{2 r}     \zhTB_{AB}  \Big]
      -\frac{1}{4 r^2}    \partial_r  \bigg( \frac{V}{r}\bigg)
          \zspaceD^B\zhTB_{AB}
     \nonumber
\\
     & \quad
   -\frac38 \partial_r \big(r^{-\frac83} \partial_r (r^{\frac83}\zspaceD^B \TS[\zspaceD_A \zhTB_{uB}])\big)
   + \frac{3}{8r^2} \partial_r \left(\zspaceD^B P \zhTB_{AB}\right)
   \,,
\nonumber 
\end{align}
which we rewrite as
\begin{align}
   \partial_r \bigg( 3 \partial_u\zhTB_{uA}
   &
   + r\partial_r \partial_u\zhTB_{uA}
    - \frac{1}{2r} \partial_u \zspaceD^B \zhTB_{AB}
    +  \frac38  r^{-\frac83} \partial_r \big(r^{\frac83}\zspaceD^B \TS[\zspaceD_A \zhTB_{uB}]\big)
    \bigg)
    \nonumber
    \\
       &=
    \frac{1}{2r^2} \zspaceD^B \partial_r \Big[
        \frac{ V}{2}    \partial_r \zhTB_{AB}
    +  \frac{V}{2 r}     \zhTB_{AB}  \Big]
      -\frac{1}{4 r^2}    \partial_r  \bigg( \frac{V}{r}\bigg)
          \zspaceD^B\zhTB_{AB}
          \nn
\\ &\quad
   + \frac{3}{8r^2} \partial_r \left(\zspaceD^B P \zhTB_{AB}\right)
     \nonumber
\\
     & =
     \partial_r
     \bigg( \frac{V}{4r^5}  \zspaceD^B h_{AB}
            + \frac{V}{4r^4}  \zspaceD^B \partial_r h_{AB}
           + \frac{3}{8r^4} P h_{AB}
     \bigg)
     \nonumber
     \\
     &\quad
     + \bigg( \frac{9V}{4r^6} - \frac{3 \partial_r V}{4r^5}
     \bigg) \zspaceD^B h_{AB}
     + \frac{3}{4r^5} \zspaceD^B P h_{AB}
   \,,
    \nonumber
\end{align}
or equivalently,
\begin{align}
  &  \partial_r \bigg( 3 \partial_u\zhTB_{uA}+ r\partial_r \partial_u\zhTB_{uA}
    - \frac{1}{2r} \partial_u \zspaceD^B \zhTB_{AB}
    +  \frac38  r^{-\frac83} \partial_r \big(r^{\frac83}\zspaceD^B \TS[\zspaceD_A \zhTB_{uB}]\big)
    \nonumber
    \\
    &\quad
    -\frac{V}{4r^5}  \zspaceD^B h_{AB}
            - \frac{V}{4r^4}  \zspaceD^B \partial_r h_{AB}
           - \frac{3}{8r^4} P h_{AB}
    \bigg)
    \nn
    \\
    &\quad
      =
     \frac{3}{4r^5} \zspaceD^B \bigg(
                        \underbrace{\frac{3V}{r} - \partial_r V }_{
                        = 2\twoscsign 
                        }
                        + P
                        \bigg)  h_{AB}
   \,.
   \label{17I23.11}
\end{align}
Let us denote by,
\begin{align}
  \kq{1}_A :=& 3 \partial_u\zhTB_{uA}+ r\partial_r \partial_u\zhTB_{uA}
    - \frac{1}{2r} \partial_u \zspaceD^B \zhTB_{AB}
    +  \frac38  r^{-\frac83} \partial_r \big(r^{\frac83}\zspaceD^B \TS[\zspaceD_A \zhTB_{uB}]\big)
    \nonumber
    \\
    &\quad
    -\frac{V}{4r^5}  \zspaceD^B h_{AB}
            - \frac{V}{4r^4}  \zspaceD^B \partial_r h_{AB}
           - \frac{3}{8r^4} P h_{AB}
   \,.
   \label{19I23.1}
\end{align}
%
Let $\mu_A \in \coker(\zdivtwo(P+2\twoscsign))$. Equation \eqref{17I23.11} gives
\begin{align}
    \partial_r \kQ{5}{}(\mu^A) = 0,
\end{align}
where
\begin{align}
    \kQ{5}{}(\mu^A) &:= \int_{\secN} \mu^A \kq{1}_A
    \sm \,.
\end{align}
\ptcheck{14IV23, everything up to here}
We have thus found a gauge-dependent  radial charge involving $\partial_u\zhTB_{uA}$. From the analysis in Appendix \ref{ss27XI22.1}, on $S^2$, $\coker(\zdivtwo(P+2\twoscsign))$ is the $16$-dimensional space of $\ell=1$ and $\ell=2$ spherical harmonic vectors; on $\T^2$,  $\coker(\zdivtwo(P+2\twoscsign))$ is the two dimensional space of covariantly constant vectors; on negatively curved manifolds of higher genus,  $\coker(\zdivtwo(P+2\twoscsign))$ is trivial. Note that $\coker(\zdivtwo(P+2\twoscsign))$ corresponds exactly to the space of obstructions for solving \eqref{24X22.91} with $p=1$ (compare also \eqref{30XI22.1}).

Under gauge transformations, $\kq{1}_A$ transforms as
\begin{align}
    \kq{1}_A &\mapsto \kq{1}_A - \frac{\twoscsign}{r^3}\Done(\xi^u)_A
    + \frac{3 \alpha^2}{2} ( \zspaceD_A\zspaceD_B\xi^B
                                + \zspaceD^B C(\xi)_{AB} )
        \nonumber
\\
    &\quad
    +3 \partial_u^2 \xi_A -\frac{3}{4r^2} \zspaceD^B(2\twoscsign + P)C(\xi)_{AB}\,.
  \nn 
\end{align}
Note that the $r$-dependent gauge terms in this equation vanish upon projection onto $\mu^A$. Hence, the charge $\kQ{5}{}$ transforms as,
\ptcheck{14IV23; if the previous one is correct this one also is}
\begin{align}
    \kQ{5}{}(\mu^A) \mapsto \kQ{5}{}(\mu^A)
                            + 3\int_{\secN}\mu^A
                            \bigg(
                            \frac{ \alpha^2}{2} ( \zspaceD_A\zspaceD_B\xi^B
                                + \zspaceD^B C(\xi)_{AB}
                                )
                                +\partial_u^2 \xi_A
                            \bigg) \sm \,.
\end{align}

%% file: WanMiracle.tex
\section{An identity}

One way of ensuring continuity of $\partial_u^{{p}}
    \zhTB_{uA}$ and
 $\partial_u^p\hBo_{AB}^{[\TTt^\perp]}$ at $r_2$ is through the fields $\{\kphi{2p+4}{}^{[\TTt^\perp]}_{AB}, \partial_{\tdu}^{p+1}\kxi{2}{}_C^{[\CKV]} \}$ and the gauge field $\partial_{\tdu}^{p}\kxi{2}{}_C^{[\CKVp]} $ respectively.
 This turns out not to be convenient from the perspective of nonlinear theory, as it involves losses of derivatives, which is avoided by the argument presented in the main body of the paper. We present this alternative calculation here as it involves some unexpected identities which might be useful for further applications.

We thus revisit \eqref{6X22.w8b}.
For the sake of the induction here we  assume that the fields
$\kphi{\ell}_{AB}$ for $\ell\leq 2p+2$ are known
  and collect them, together with $-\partial_u^p\zhTB_{uA}|_{\tilde\secN_2}$, into a new term $\overset{({p})}{\hat{\tilde X}}_A$, allowing us to rewrite
  the $L^2$-projections on $\CKV$ and $\CKVp$ of
  \eqref{6X22.w8b} respectively as
\begin{align}
 \partial_u^{p+1}\kxi{2}{}_A^{[\CKV]}
 & =
 \overset{({p})}{\hat{\tilde X}} {}^{[\CKV]}_A
    \,,
    \label{9XI22.w1a}
\\
 -  \overset{({p},0)}{\ochi}_{2p+4} (r_2)
 &\zspaceD^B  \kphi{2p+4}{}^{[\TTt^\perp]}_{AB}
  =
 -\partial_u^{{p}+1}\kxi{2}{}^{[\CKVp]}_A
    +
 \overset{({p})}{\hat{\tilde X}}{{}^{[\CKVp]}_A}
    \label{9XI22.w1}
\\
    &\quad
    + \zspaceD^B (\overset{({p},0)}{\ochi}_{2p+3} (r_2) + \overset{({p},1)}{\ochi}_{2p+3} (r_2) P)\kphi{2p+3}{}^{[\TTt^\perp]}_{AB}
    \,.
    \nonumber
\end{align}
An argument identical to that below \eqref{18X22.1} shows that,
   both on $S^2$ and $\T^2$,
  \eqref{9XI22.w1a} determines $\partial_u^{p+1}\kxi{2}{}_A^{[\CKV]}$ uniquely in terms of $\overset{({p})}{\hat{\tilde X}}{{}^{[\CKV]}_A}$ while \eqref{9XI22.w1} determines  $\kphi{2p+4}{}_{AB}^{[\TTt^\perp]}$ uniquely in terms of $\overset{({p})}{\hat{\tilde X}}{{}^{[\CKVp]}_A}$, $\kphi{2p+3}{}^{[\TTt^\perp]}_{AB}$ and $\partial_u^{p+1}\kxi{2}{}_A^{[\CKVp]}$; the fields $\kphi{2p+3}{}^{[\TTt^\perp]}_{AB}$ and $\partial_u^{p+1}\kxi{2}{}_A^{[\CKVp]}$ remain free to use for
  other gluing equations. On negatively curved sections with higher genus, \eqref{9XI22.w1} determines $\kphi{2p+4}{}^{[\TTt^\perp]}_{AB}$ in terms of $\overset{({p})}{\hat{\tilde X}}_A$, $\kphi{2p+3}{}^{[\TTt^\perp]}_{AB}$  and $\partial_u^{p+1}\kxi{2}{}_A^{[\CKVp]}=\partial_u^{p+1}\kxi{2}{}_A$, with the fields $\kphi{2p+3}{}^{[\TTt^\perp]}_{AB}$ and $\partial_u^{p+1}\kxi{2}{}_A$
   remaining free.

 Next, in order to take into account the dependence   of  $\kphi{2p+2} $ and $\kphi{2p+1} $ upon  $\partial_{\tdu}^{p}\kxi{2}{}_C^{[\CKVp]} $,
  we consider the equation obtained by acting with $\zdivtwo\!$ on \eqref{9XI22.w3}.
 There occur some miraculous cancellations, which are likely to have some simple origin:
\ptcheck{20XI, magical}
\begin{align}
       \label{9XI22.w4a}
   \zspaceD^A&
   \partial^p_u \hBo_{AB}^{[\TTt^\perp]}|_{\tilde{\secN}_2}
   =\zspaceD^A\overadd{p}{\tilde\Psi}{}^{[\TTt^\perp]}_{AB}(r_2, x^A)
    - 2 r_2^2 \zspaceD^A \TS [\zspaceD_{\tdA}\partial_{\tdu}^{p}\kxi{2}_B^{[\CKVp]}]
\\
    &\quad
    +
    \zspaceD^A (\overset{({p},0)}{\opsi}_{2p+2} (r_2) + \overset{({p},1)}{\opsi}_{2p+2} (r_2) P )\kphi{2p+2}{}^{[\TTt^\perp]}_{AB}
     \nonumber
\\
    &\quad
    +
     \zspaceD^A(\overset{({p},0)}{\opsi}_{2p+1} (r_2) + \overset{({p},1)}{\opsi}_{2p+1} (r_2) P + \overset{({p},2)}{\opsi}_{2p+1} (r_2) P^2 )\kphi{2p+1}{}^{[\TTt^\perp]}_{AB}
     \nonumber
\\
&=\zspaceD^A\overadd{p}{\tilde\Psi}{}^{[\TTt^\perp]}_{AB}(r_2, x^A)
    - 2 r_2^2 \zspaceD^A \TS [\zspaceD_{\tdA}\partial_{\tdu}^{p}\kxi{2}_B^{[\CKVp]}]
        \nonumber
\\
    &\quad
    + \overset{({p},0)}{\opsi}_{2p+2}/\overset{({p-1},0)}{\ochi}_{2p+2} \, \partial_u^{p}\kxi{2}{}_B^{[\CKVp]}
        \nonumber
\\
    &\quad
    + \overset{({p},1)}{\opsi}_{2p+2}/\overset{({p-1},0)}{\ochi}_{2p+2} \, \zspaceD^A C(\partial_u^{p} \kxi{2}{}^{[\CKVp]})_{AB}
    \nonumber
\\
    &=
    \zspaceD^A\overadd{p}{\tilde\Psi}{}^{[\TTt^\perp]}_{AB}(r_2, x^A)
    - 3m r_2 \partial_u^{p}\kxi{2}{}_B^{[\CKVp]}
\,
 \nonumber,
\end{align}
where in the second equality we made use of the expression for $\kphi{2p+2}^{[\TTt^\perp]}_{AB}$ from \eqref{9XI22.w1} at order $(p-1)$,
while the last equality uses \eqref{9XI22.w5}-\eqref{8XI22.w4b}, Appendix~\ref{App14VIII22.2}.
Thus, continuity of $\partial_u^p\hBo_{AB}^{[\TTt^\perp]}$ can be achieved by solving \eqref{9XI22.w4a} for $\partial_{\tdu}^{p}\kxi{2}{}_C^{[\CKVp]} $:
\begin{align}
     3m r_2 \partial_{\tdu}^{p}\kxi{2}{}_B^{[\CKVp]}
      &
    =
    -
     \zspaceD^A\partial^p_u \hBo_{AB}^{[\TTt^\perp]}|_{\tilde{\secN}_2}
     +
    \zspaceD^A\overadd{p}{\tilde\Psi}{}^{[\TTt^\perp]}_{AB}(r_2, x^A)
\,.
       \label{9XI22.w4c}
\end{align}

%% file: factorisation2.tex
In fact, Lemma~\pref{L25XI22.1} shows that we have the factorisation
\begin{align}
      \overset{({p-1},0)}{\ochi}_{2p+2} &
       (r_2)
       \big(
        \overset{({p},0)}{\opsi}_{2p+1}
     (r_2)
       + \overset{({p},1)}{\opsi}_{2p+1} (r_2) P + \overset{({p},2)}{\opsi}_{2p+1} (r_2) P^2
       \big)
        \nonumber
\\
\hspace{-11.5cm}
 = &
 \big(
  \overset{({p},0)}{\opsi}_{2p+2} (r_2)
  + \overset{({p},1)}{\opsi}_{2p+2} (r_2) P
    \big)
    \big(
     \overset{({p-1},0)}{\ochi}_{2p+1} (r_2) +
      \overset{({p-1},1)}{\ochi}_{2p+1} (r_2) P
    \big)
\,.
       \label{14VII23.p1}
       \nonumber
\end{align}